\documentclass[structabstract]{aa} 
\usepackage{booktabs}
\usepackage{graphicx}
\usepackage{txfonts}
\usepackage{natbib}
\usepackage{longtable}
\usepackage{lscape}
\usepackage{multirow}
\usepackage{pdfpages}
\usepackage{adjustbox}
\usepackage{ulem}
\usepackage[colorlinks=true, allcolors=blue]{hyperref}

\bibpunct{(}{)}{;}{a}{}{,}

\newcommand{\kms}{km\,s$^{-1}$}

\newcommand{\Msun}{M$_{\odot}$}
\newcommand{\Lsun}{L$_{\odot}$}

\newcommand{\nodata}{...}

\begin{document}
\title{ATLASGAL: 3-mm class I methanol masers in high-mass star formation regions}


\author{W.~Yang\inst{1},  Y.~Gong\inst{1},  K.~M.~Menten\inst{1}, J.~S.~Urquhart\inst{2},  C.~Henkel\inst{1,3,4},  F.~Wyrowski\inst{1}, T.~Csengeri\inst{5}, S.~P.~Ellingsen\inst{6},  A.~R.~Bemis\inst{7},  J.~Jang\inst{1} }
\offprints{W. Yang, \email{wjyang@mpifr-bonn.mpg.de}}

\institute{
Max-Planck-Institut f{\"u}r Radioastronomie, Auf dem H{\"u}gel 69, D-53121 Bonn, Germany
\and
Centre for Astrophysics and Planetary Science, University of Kent, Canterbury CT2 7NH, UK
\and
Astronomy Department, Faculty of Science, King Abdulaziz University, PO Box 80203, Jeddah, 21589, Saudi Arabia
\and
Xinjiang Astronomical Observatory, Chinese Academy of Sciences, Urumqi 830011, PR China
\and
Laboratoire d'astrophysique de Bordeaux, Univ. Bordeaux, CNRS, B18N, all{\'e}e Geoffroy Saint-Hilaire, 33615 Pessac, France
\and
School of Natural Sciences, University of Tasmania, Private Bag 37, Hobart, Tasmania 7001, Australia
\and
Leiden Observatory, Leiden University, P.O. Box 9513, 2300 RA Leiden, The Netherlands
}

\date{Received date ; accepted date}

\abstract
{Class I methanol masers are known to be associated with shocked outflow regions around massive protostars, indicating a possible link between the maser properties and those of their host clumps.} 
{The main goals of this study are (1) to search for new class I methanol masers, (2) to statistically study the relationship between class I masers and shock tracers, (3) to compare the properties between class I masers and their host clumps, also as a function of their evolutionary stage and, (4) to constrain the physical conditions that excite multiple class I masers simultaneously.}
{We analyzed the 3-mm wavelength spectral line survey of 408 ATLASGAL clumps observed with the IRAM 30m-telescope, focusing on the class I methanol masers with frequencies near 84, 95 and 104.3\,GHz.}
{We detect narrow, maser-like features towards 54, 100 and 4 sources in the maser lines near 84, 95 and 104.3\,GHz, respectively. Among them, fifty 84\,GHz masers, twenty nine 95\,GHz masers and four rare 104.3\,GHz masers are new discoveries. The new detections increase the number of known 104.3\,GHz masers from 5 to 9.
The 95\,GHz class I methanol maser is generally stronger than the 84\,GHz maser counterpart. We find 9 sources showing class I methanol masers but no SiO emission, indicating that class I methanol masers might be the only signpost of protostellar outflow activity in extremely embedded objects at the earliest evolutionary stage. Class I methanol masers that are associated with sources that show SiO line wings are more numerous and stronger than those without such wings. The total integrated intensity of class I methanol masers is well correlated with the integrated intensity and velocity coverage of the SiO (2--1) emission. The properties of class I methanol masers are positively correlated with the bolometric luminosity, clump mass, peak H$_2$ column density of their associated clumps but uncorrelated with the luminosity-to-mass ratio, dust temperature, and mean H$_2$ volume density.} 
{We suggest that the properties of class I masers are related to shocks traced by SiO. Based on our observations, we conclude that class I methanol masers at 84 and 95~GHz can trace a similar evolutionary stage as H$_2$O maser, and appear prior to 6.7 and 12.2 GHz methanol and OH masers. 
Despite their small number, the 104.3~GHz class I masers appear to trace a short and more evolved stage compared to the other class I masers.}


\titlerunning{class I methanol masers in ATLASGAL sources}

\authorrunning{Yang et al.}

\maketitle


\section{Introduction}

Astronomical methanol (CH$_3$OH) masers were first reported in the Orion Kleinmann-Low (KL) nebula \citep{1971ApJ...168L.101B}, shortly after CH$_3$OH emission had been discovered in the Galactic center \citep{1970ApJ...162L.203B}. Subsequent observations have resulted in the detection of numerous CH$_3$OH maser transitions which have been proven to be important tracers of star formation regions \citep[e.g.,][]{1991ASPC...16..119M}. These CH$_3$OH masers are divided into two categories, based on their different observational properties and pumping mechanisms \citep{1987Natur.326...49B,1991ASPC...16..119M}.
Class~I CH$_3$OH masers are thought to be collisionally pumped \citep[e.g.,][]{2016A&A...592A..31L}, and often found to be offset from embedded infrared sources and ultracompact H{\sc ii} (UCH{\sc ii}) regions.
In contrast, class~II CH$_3$OH masers are thought to be radiatively pumped \citep[e.g.][]{2005MNRAS.360..533C} by the infrared radiation emitted by massive young stellar objects in whose environments they are found \citep{Walsh1998}. Only a small portion of these produce detectable radio continuum emission that can ionize
UCH{\sc ii} regions \citep{Nguyen2022}.



Based on previous studies \citep[see Table~1 in ][for instance]{2016A&A...592A..31L}, the three methanol transitions at 84 GHz, 95 GHz, and 104.3 GHz are the brightest class I methanol masers in the 3 mm band. 
The methanol emission in the $J_k =$ $5_{-1}-4_0\,E$ line near 84 GHz belongs to the same line series of strong class I masers as the 36 GHz $4_{-1}-3_0\,E$ transition (i.e., they connect the same k-ladders).
Maser emission at 84 GHz was discovered by \cite{1988ApJ...329L.117B} and \cite{1991ASPC...16..119M} towards DR21(OH), NGC2264 and OMC-2.
Since 2001, several extensive searches for this transition have been conducted towards sources associated with other class I masers, and young bipolar outflows in low-intermediate-mass star formation regions \citep{2001ARep...45...26K,2006ARep...50..289K,2018IAUS..336..239R,2019MNRAS.484.5072B}. The detection rate of the 84\,GHz methanol line, which can show narrow maser-like features and broad quasi-thermal emission, in these targeted surveys is greater than 70\%.
Methanol emission in the 95 GHz  $J_K = $ $8_0-7_1\,A^{+}$ transition belongs to the same line series (common $K$ quantum numbers) as the strong and widespread $7_0-6_1\,A^{+}$ maser at 44\,GHz \citep{Haschick1990}. 
The number of known 95 GHz methanol masers has been significantly increased by systematic surveys, including towards (1) extended green objects (EGOs) identified in the $4.5 \mu$m band of the Spitzer Galactic Legacy Infrared Mid-Plane Survey Extraordinaire (GLIMPSE), whose emission is dominated by rotationally excited shocked H$_2$ \citep{2011ApJS..196....9C,2013ApJS..206....9C}; (2) molecular outflow sources \citep{2013ApJ...763....2G}; (3) many sources associated with both GLIMPSE point sources and 1.1 mm dust continuum emission detected in the Bolocam Galactic Plane Survey \citep[BGPS;][]{2012ApJS..200....5C}; (4) a large sample of BGPS sources \citep{2017ApJS..231...20Y,2020ApJS..248...18Y}; and (5) red Midcourse Space Experiment sources \citep{2018ApJS..236...31K}.
\cite{2019AJ....158..233L} compiled an online database of class I methanol masers and reported a total of 129 and 534 maser detections at 84 and 95 GHz, respectively. These facts suggest that both masers are widespread across the Milky Way.

On the other hand, the $11_{-1}-10_{-2}\,E$ methanol maser at 104.3 GHz is rarely seen.
Maser emission in this line was first predicted by theoretical calculations \citep{1999AstL...25..149V}, and was successfully detected towards W33-Met \citep{2005Ap&SS.295..217V}.
A follow-up Mopra 104.3 GHz survey led to only two detections out of 69 targets, suggesting that this maser line is rare \citep{2007IAUS..242..182V}. 
Theoretical calculations suggest that more energetic conditions (i.e., higher temperatures and densities) are required to produce these rare class I methanol masers \citep{2005IAUS..227..174S, 2012IAUS..287..433V} than the widespread class I methanol masers at 36, 44, 84, and 95~GHz. 
Prior to this work, only five sources (G019.61$-$0.23, G305.21+0.21, G357.97$-$0.16, IRAS16547$-$4247 and W33-Met) were known to harbor 104.3\,GHz methanol masers \citep[e.g.][]{2006MNRAS.373..411V}. However, the overall incidence of the 104.3\,GHz methanol masers in the Milky Way is still poorly constrained.

It has been established that class I methanol masers are closely related to shocked regions. 
Based on high-angular resolution observations towards DR21/DR21(OH), \citet{1990ApJ...364..555P} suggested that class I methanol masers trace the interface between outflows and ambient dense clouds. Subsequent observations supported this scenario and found a coincidence between class I masers and other molecular shock tracers (e.g. SiO, \citealt{2004ApJS..155..149K}; HNCO, \citealt{2018ApJ...856..134G}; shocked H$_2$ emission at 4.5~$\mu$m and 2.12~$\mu$m, \citealt{2009ApJ...702.1615C,2006MNRAS.373..411V}). However, the relationship between class I masers and shock tracers has not been well studied in a detailed statistical manner. 

The APEX telescope large area survey of the Galaxy (ATLASGAL) is an unbiased 870~$\mu$m sub-millimetre continuum survey of the inner Galaxy (300$^\circ$ $< l <$ 60$^\circ$, $\left| b \right| <$ 1.5$^\circ$) \citep{2009A&A...504..415S}. This survey provides a large inventory of dense molecular clumps \citep[over $\sim$10 000,][]{2013A&A...549A..45C,2014A&A...565A..75C,2014A&A...568A..41U}  that comprise a variety of early evolutionary stages related to high-mass star formation.
The evolutionary stages of a large number ($\sim$ 5000) of ATLASGAL clumps have been determined and their physical properties (e.g. distance, dust temperature, bolometric luminosity, clump mass and H$_2$ column density) have been studied in detail \citep{2017A&A...599A.139K,2017A&A...603A..33G,2018MNRAS.473.1059U,2022MNRAS.510.3389U}.
This well characterized sample allows us to carry out a targeted survey  searching for new class I methanol masers and perform a statistical analysis of the properties of detected masers and their associated star-forming clumps at different evolutionary stages.

Silicon monoxide (SiO) emission is an excellent tracer of shock interactions in star formation regions.
The abundance of SiO is enhanced by up to six orders of magnitude in shocked regions associated with molecular outflows with respect to that measured in quiescent gas \citep[e.g.][]{2008A&A...482..549J} through grain-grain collisions and sputtering, after which the released atomic silicon forms SiO \citep[e.g.][]{1992A&A...254..315M,1997A&A...322..296C}. The SiO emission towards our targeted ATLASGAL clumps has been well studied by \cite{2016A&A...586A.149C}. This provides us a unique opportunity to statistically study the relationship between the class I methanol masers and the shocked gas traced by SiO emission.

In this work, we focus on the three class I methanol maser transitions by analyzing 3~mm line survey data obtained using the Institut de Radioastronomie Millim\'etrique (IRAM) 30~m telescope on 408 ATLASGAL clumps.
The aims of this work are: (i) to search for new class I maser sources; (ii) to statistically study the relationships between class I masers and a shock tracer; (iii) to investigate the properties between class I masers and their host clumps, as well as a function of their evolutionary stages; and (iv) to provide strong constraints on environments in which multiple class I masers occur.
Also covered by our $\lambda$ $\sim$3~mm data  are six additional methanol transitions that have detected maser features at  that belong to class II. These include the 85.5, 86.6, 86.9, 104.1, 107 and 108 GHz lines. We will focus on these class II methanol maser transitions, especially the maser emission and absorption features at 107 GHz, towards the same 408 ATLASGAL sources in a subsequent paper.

The structure of this paper is as follows.
Our observations and data reduction are described in Sect.~\ref{Sec:obs}. The results are given in Sect.~\ref{Sec:result}. 
In Sect.~\ref{Sec:discuss}, we discuss the properties of the associated ATLASGAL clumps and SiO emission for our detections, and perform theoretical calculations to constrain the physical conditions of maser environments.
A summary of this work and highlighted conclusions are presented in Sect.~\ref{Sec:sum}.

\section{Observations and data reduction}\label{Sec:obs}

The methanol transition data were extracted from unbiased spectral surveys with a frequency coverage of 83.8$-$115.7 GHz using the Eight MIxer Receiver \citep[EMIR,][]{2012A&A...538A..89C} at 3 mm (E090) of the IRAM 30m telescope (see \citealt{2016A&A...586A.149C} for details, project ids: 181-10, 049-11 and 037-12), from 2010 May to 2012 October. 
Flux limited samples of ATLASGAL sources \citep{2009A&A...504..415S, 2014A&A...565A..75C} were selected, with additional infrared selection criteria that ensure to cover a large range of evolutionary stages and luminosity. The sample includes (i) $\sim$120 brightest ATLASGAL clumps that are infrared bright, (ii) $\sim$50 bona-fide massive young stellar objects that obey the \cite{2002MNRAS.336..621L} infrared color criteria for embedded, massive (proto)stars and are associated with ATLASGAL clumps, (iii) $\sim$120 brightest ATLASGAL clumps that are GLIMPSE 8~$\mu$m dark but contain 24~$\mu$m sources, and (iv) $\sim$120 brightest ATLASGAL clumps that are Spitzer 8 and 24~$\mu$m dark.

The properties and observational parameters of these transitions are summarized in Table \ref{Tab:freq}.
The FFTS backend with a frequency resolution of $\sim$200 kHz was used, resulting in a channel spacing of $\sim$0.7~km s$^{-1}$ (multiply by 1.16 to convert to velocity resolution, see \citealt{2012A&A...542L...3K}). 
Table~\ref{Tab:source} provides information on the observed sources. The half-power beam width (HPBW) is $\sim$30\arcsec\,at 84\,GHz.
The observations were performed in position switching mode with an offset of 10$\arcmin$ in right ascension and declination for the reference position. 
The data were processed using the GILDAS/CLASS package \citep{2005sf2a.conf..721P}.
To characterize the spectra, we employed a Gaussian fit method for each transition source by source, and a multi-Gaussian fit was used in case of multiple velocity components.

\begin{table*}[!hbt]
\caption{Class I methanol maser transitions at 3~mm. }\label{Tab:freq}
\normalsize
\centering
\begin{tabular}{lrrcccccccc}
\hline  \hline
CH$_3$OH Transition  & Rest frequency   & $E_{\rm up}$  & $n_{\rm crit}$ & beam size &$V_{\rm res}$  & rms  &  $\eta_{\rm f}$ &  $\eta_{\rm mb}$  & Scaling factor \\ 
   & (MHz)       &    (K) & (cm$^{-3}$)   & ($\arcsec$) &(km s$^{-1}$)  & (Jy)  & &  & (Jy/K)\\ 
\hline
$5_{-1}-4_0\,E$ & 84 521.169(10)  &40.4 & $1.04\times 10^{5}$ &29 &0.80 & 0.2 & 0.95 &0.81  &5.82\\
$8_0-7_1\,A^{+}$& 95 169.463(10)   &83.5 & $7.29\times 10^{5}$  &26 &0.71 & 0.2 & 0.95 &0.81  &5.82\\
$11_{-1}-10_{-2}\,E$ &104 300.414(7)   &158.6 & $1.49\times 10^{6}$ &24 &0.65 & 0.2  & 0.94 &0.78  &5.98\\
\hline \hline
\end{tabular}
\note{Columns 1--4 give the transition, rest frequency, upper energy level,  and critical density of each methanol line. The rest frequency is adopted from \citet{2004A&A...428.1019M}, and the critical density (calculated using a gas kinetic temperature of 100~K) is taken from \citet{2022A&A...658A.192Y}. 
Columns 5--10 give the beam size, velocity resolution, typical observed noise level, forward efficiency, main beam efficiency and scaling factor to convert antenna temperature to flux density, respectively.
Detailed information on the IRAM 30~m telescope can be found here: \url{https://publicwiki.iram.es/Iram30mEfficiencies}.
}
\normalsize
\end{table*}

\begin{table*}[!hbt]
\caption{Source information. }\label{Tab:source} 
\scriptsize
\centering
\setlength{\tabcolsep}{1.3pt}
\begin{tabular}{lccccccccccccccc}
\hline \hline
Source  & R.A.   & Dec. & ATLASGAL & $V_{\rm LSR}$ & Dist. & $R_{\rm GC}$ & Evolution         & $T_{\rm dust}$ & log[$L_{\rm bol}$] & log[$M_{\rm fwhm}$] & log[$N(\rm H_2)$] & log[$n_{\rm fwhm}(\rm H_2)$] & $L_{\rm bol}$/$M_{\rm fwhm}$ & \multicolumn{2}{c}{SiO} \\
\cmidrule(lr){15-16}
 name  & $\alpha$(J2000)  & $\delta$(J2000) & CSC name & (km s$^{-1}$) & (kpc) & (kpc) & type & (K) & (\Lsun) & (\Msun) & (cm$^{-2}$)  & (cm$^{-3}$) & (\Lsun/\Msun)   & (2-1) & (5-4) \\
(1)  & (2)  & (3) & (4) & (5) & (6) & (7) & (8) & (9) & (10) & (11) & (12)  & (13) & (14)   & (15) & (16) \\ 
\hline
G06.22$-$0.61       & 18:02:02.97 & $-$23:53:12.8 & AGAL006.216$-$00.609 & 18.5    & 3.0      & 5.4     & Protostellar     & 14.6    & 3.078      & 2.549        & 23.047    & 5.208     & 3.38      & Y        & \nodata \\
G08.05$-$0.24       & 18:04:35.26 & $-$22:06:40.4 & AGAL008.049$-$00.242 & 39.1    & 5.0      & 3.5     & PDR+Embedded     & 18.2    & 2.969      & 2.032        & 22.489    & 4.396     & 8.64      & \nodata       & \nodata \\
G08.68$-$0.37       & 18:06:23.03 & $-$21:37:10.9 & AGAL008.684$-$00.367 & 35.6    & 4.4      & 4.0     & Protostellar     & 24.5    & 4.355      & 2.738        & 23.032    & 4.845     & 41.41     & Y        & Y       \\
G08.71$-$0.41       & 18:06:36.65 & $-$21:37:16.4 & AGAL008.706$-$00.414 & 38.3    & 4.4      & 4.0     & Protostellar     & 11.9    & 3.162      & 3.184        & 22.927    & 4.357     & 0.95      & Y        & Y       \\
G10.08$-$0.19       & 18:08:38.57 & $-$20:18:55.9 & AGAL010.079$-$00.196 & 26.5    & 2.9      & 5.5     & Ambiguous        & 19.8    & 2.768      & 1.831        & 22.319    & 4.535     & 8.65      & \nodata       & \nodata \\
G10.11$-$0.41       & 18:09:31.77 & $-$20:23:57.8 & AGAL010.104$-$00.416 & 10.9    & 2.9      & 5.5     & Ambiguous        & 20.0    & 3.006      & 2.345        & 22.250    & 4.595     & 4.58      & \nodata       & \nodata \\
G10.15$-$0.34       & 18:09:21.21 & $-$20:19:27.9 & AGAL010.151$-$00.344 & 8.4     & 2.9      & 5.5     & Ambiguous        & 38.6    & 5.362      & 2.346        & 22.602    & 4.521     & 1037.23   & Y        & \nodata \\
G10.17$-$0.36A$\dagger$      & 18:09:26.75 & $-$20:19:02.9 & AGAL010.168$-$00.362 & 16.0    & 2.9      & 5.5     & HII              & 40.1    & 5.365      & 2.611        & 22.619    & 4.273     & 567.38    & Y        & \nodata \\
G10.17$-$0.36B$\dagger$      & 18:09:27.62 & $-$20:19:05.3 & AGAL010.168$-$00.362 & 16.0    & 2.9      & 5.5     & HII              & 40.1    & 5.365      & 2.611        & 22.619    & 4.273     & 567.38    & Y        & \nodata \\
G10.21$-$0.30       & 18:09:20.46 & $-$20:15:02.2 & AGAL010.214$-$00.306 & 11.3    & 2.9      & 5.5     & Quiescent        & 19.1    & 2.754      & 2.101        & 22.725    & 5.111     & 4.50      & Y        & \nodata \\
\hline \hline
\end{tabular}
\normalsize
\note{Columns 1--4 give the source name, coordinates and ATLASGAL CSC name. The ``$\dagger$" marks the sources with a nearby observed target within 30\arcsec.
Columns 5--14 give the velocity, distance, galactocentric distance, classification, dust temperature, bolometric luminosity, clump mass, H$_2$ column density, H$_2$ volume density and luminosity-to-mass ratio that are taken from \cite{2018MNRAS.473.1059U,2022MNRAS.510.3389U}. Columns 15 and 16 inform about the detection of SiO, information that is taken from \cite{2016A&A...586A.149C}. ``Y" indicates a detection, ``\nodata" indicates that it was either not observed or not detected.
Only the first ten sources in the table are shown here for guidance, the full table is available at the CDS.\\}
\end{table*}

\section{Results}\label{Sec:result}

We consider emission above the 3$\sigma$ level as a detection, which corresponds to a typical detection threshold of $\gtrsim$0.6~Jy for our three methanol transitions.
Table~\ref{Tab:overview} gives an overview of detections for individual sources in the three class I maser lines. 
Overall, we detect 282, 224 and 29 sources with methanol emission at 84\,GHz (282/404, 70\%), 95\,GHz (224/408, 55\%) and 104.3\,GHz (29/404, 7\%)\footnote{Four sources (G12.91$-$0.26B, G30.72$-$0.08, G30.73$-$0.06, G30.75$-$0.05) are only observed in the 95 GHz line, so the total number (408) of sources observed at 95 GHz is higher than for the other two transitions.}, respectively. 

\begin{figure}[!htbp]
\centering
\includegraphics[width=0.35\textwidth]{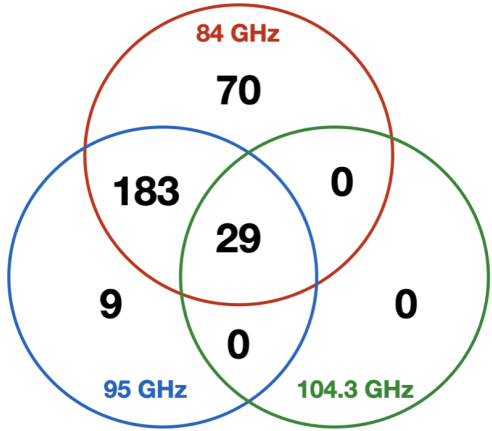}
\caption{Venn diagram showing the number of sources detected in the three class I maser lines. All detected 104.3 GHz emission
lines in our sample have associated methanol emission at both
84 and 95 GHz.
\label{fig:detnumber}}
\end{figure}

Figure~\ref{fig:detnumber} shows the number of sources detected in emission in the three class I methanol maser lines. A total of 212 sources (over 50\% of the observing targets) are detected with methanol emission at both 84 and 95 GHz (see Fig.~\ref{fig:detnumber}).
Seventy sources show methanol detections only at 84\,GHz and nine sources have detections only at 95\,GHz. All 29 detected 104.3 GHz emission lines in our sample have associated methanol emission at both 84 and 95 GHz, corresponding to 14\% of the 212 sources (see Fig.~\ref{fig:detnumber}). The detection rate decreasing from the 84 GHz transition to the 104.3 GHz transition, is readily explained by their relative energy levels and critical densities (see Table~\ref{Tab:freq}). 

Table~\ref{Tab:results} provides Gaussian fitting results of detected lines for each source.
Figure~\ref{fig:408spec} shows example spectra of the three class I (and the other six class II methanol maser transitions at 3~mm) towards the first six ATLASGAL sources in our sample.

The following subsections present the details of the detections of the widespread class I masers (at 84 and 95\,GHz) and rare class I masers (at 104.3\,GHz), respectively.

\begin{table}[!hbt]
\caption{Overview over the class I methanol maser detections.}\label{Tab:overview}
\small
\centering
\begin{tabular}{lccc}
\hline \hline
Source      & 84 & 95 & 104.3  \\
\hline 
G06.22$-$0.61  & Y      & Y      & N  \\
G08.05$-$0.24  & y      & Y      & N  \\
G08.68$-$0.37  & y      & y      & N  \\
G08.71$-$0.41  & y      & y      & N  \\
G10.08$-$0.19  & N      & N      & N  \\
G10.11$-$0.41  & N      & N      & N  \\
G10.15$-$0.34  & y      & N      & N  \\
G10.17$-$0.36A & y      & N      & N  \\
G10.17$-$0.36B & y      & N      & N  \\
G10.21$-$0.30  & Y      & Y      & N  \\
\hline \hline
\end{tabular}
\normalsize
\note{``Y" indicates a maser detection, lower-case ``y" indicates a maser candidate, ``N" indicates a non-detection.
Only a portion of the table is shown here for guidance, the full table is available at the CDS.}
\end{table}

\begin{table*}[!hbt]
\caption{Observational results of class I CH$_3$OH transitions.}\label{Tab:results}
\small
\centering
\setlength{\tabcolsep}{3.5pt}
\begin{tabular}{lrrrrrrr}
\hline \hline
Source      & Line & $V_{\rm pk}$ & $\Delta V$  & $S_{\rm pk}$ & $S$  & $\int S{\rm d}V$ & $L_{\rm CH_3OH}$ \\
&     (GHz)  & (km s$^{-1}$) & (km s$^{-1}$)  & (Jy)  & (Jy km s$^{-1}$)  & (Jy km s$^{-1}$) & (\Lsun)\\ 
\hline 
G06.22$-$0.61  & 84 & 19.61(0.02) & 1.57(0.08) & 15.20(0.42) & 25.47(0.82) & 25.47 & 2.0$\times$10$^{-5}$ \\
G06.22$-$0.61  & 95 & 19.65(0.00) & 1.21(0.01) & 38.00(0.25) & 49.01(0.33) & 49.01 & 4.4$\times$10$^{-5}$ \\
G08.05$-$0.24  & 84 & 38.89(0.12) & 2.32(0.35) & 2.09 (0.26) & 5.16 (0.56) & 5.16  & 1.1$\times$10$^{-5}$ \\
G08.05$-$0.24  & 95 & 38.48(0.07) & 1.34(0.17) & 1.59 (0.19) & 2.27 (0.26) & 2.27  & 5.6$\times$10$^{-6}$ \\
G08.68$-$0.37  & 84 & 37.98(0.03) & 4.54(0.09) & 9.06 (0.23) & 43.84(0.66) & 43.84 & 7.5$\times$10$^{-5}$ \\
G08.68$-$0.37  & 95 & 38.37(0.10) & 3.91(0.25) & 2.02 (0.18) & 8.41 (0.44) & 8.41  & 1.6$\times$10$^{-5}$ \\
G08.71$-$0.41  & 84 & 39.08(0.08) & 3.29(0.21) & 2.30 (0.18) & 8.05 (0.43) & 8.05  & 1.4$\times$10$^{-5}$ \\
G08.71$-$0.41  & 95 & 39.64(0.26) & 2.76(0.63) & 0.58 (0.16) & 1.69 (0.32) & 1.69  & 3.2$\times$10$^{-6}$ \\
G10.15$-$0.34  & 84 & 8.97 (0.14) & 4.66(0.34) & 1.32 (0.15) & 6.56 (0.42) & 6.56  & 4.9$\times$10$^{-6}$ \\
G10.17$-$0.36A & 84 & 14.35(0.55) & 6.54(1.87) & 0.54 (0.18) & 3.75 (0.72) & 3.75  & 2.8$\times$10$^{-6}$ \\
G10.17$-$0.36B & 84 & 15.63(0.48) & 5.85(0.94) & 0.55 (0.19) & 3.43 (0.54) & 3.43  & 2.5$\times$10$^{-6}$ \\
G10.21$-$0.30  & 84 & 11.45(0.21) & 6.79(0.79) & 1.71 (0.18) & 12.34(1.00) & 17.52 & 1.3$\times$10$^{-5}$ \\
G10.21$-$0.30  & 84 & 12.30(0.06) & 1.70(0.24) & 2.86 (0.18) & 5.18 (0.99) &       &                   \\
G10.21$-$0.30  & 95 & 11.62(0.13) & 4.27(0.36) & 1.39 (0.11) & 6.35 (0.45) & 11.12 & 9.3$\times$10$^{-6}$ \\
G10.21$-$0.30  & 95 & 12.53(0.03) & 1.34(0.07) & 3.35 (0.11) & 4.78 (0.40) &       &                   \\
\hline \hline
\end{tabular}
\normalsize
\note{Column 1 gives the source name. Column 2 gives the frequency of the each methanol transition. Columns 3--6 give the peak velocity $V_{\rm pk}$, the FWHM line width $\Delta V$, the peak flux density $S_{\rm pk}$, and the integrated flux density of each component estimated from Gaussian fits. Formal errors from the Gaussian fits are given in parentheses. Column 7 gives the total integrated flux density for each source and each methanol line, which is the sum of integrated flux densities over all velocity components for a given transition. Column 8 gives the isotropic luminosity of methanol emission estimated from the total integrated flux density.
Only a portion of the table is shown here for guidance, the full table is available at the CDS.}
\end{table*}

\subsection{84 and 95 GHz methanol emission}

\begin{figure*}[!htbp]
\centering
\mbox{
\begin{minipage}[b]{4.5cm}
\includegraphics[width=0.92\textwidth]{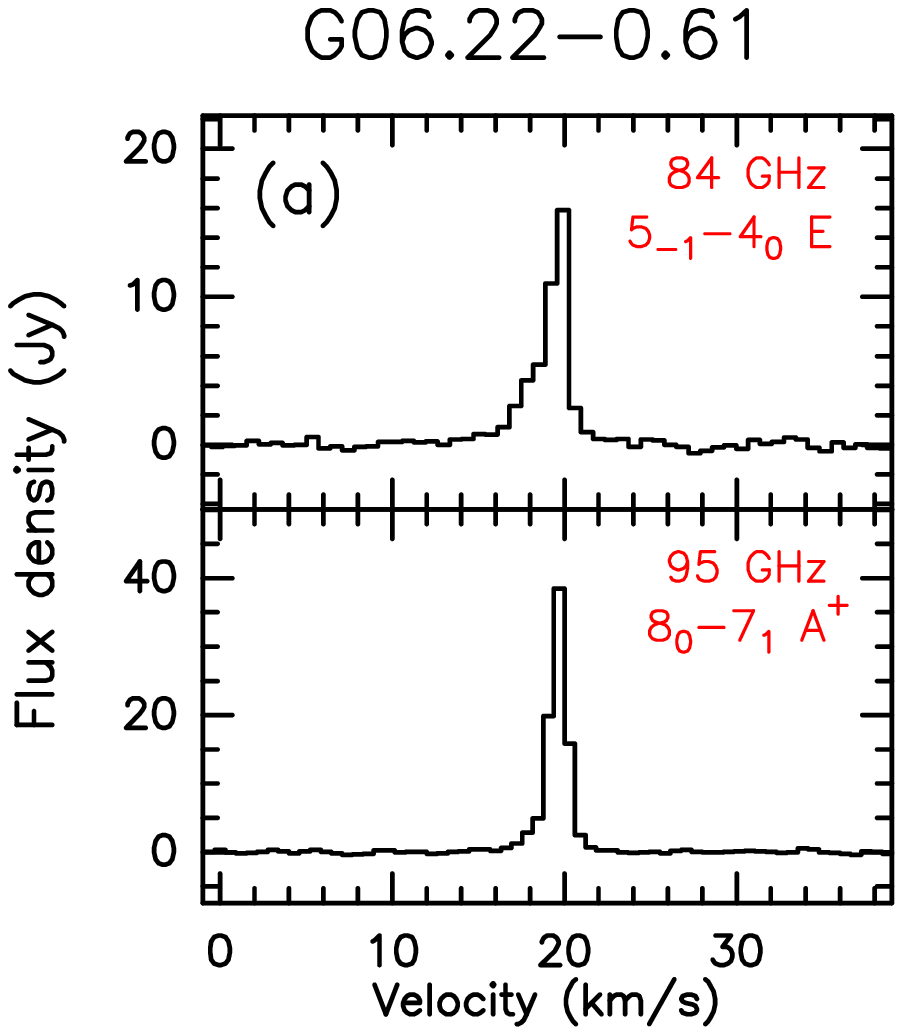}
\vspace{2mm}
\end{minipage}
\begin{minipage}[b]{4.5cm}
\includegraphics[width=0.92\textwidth]{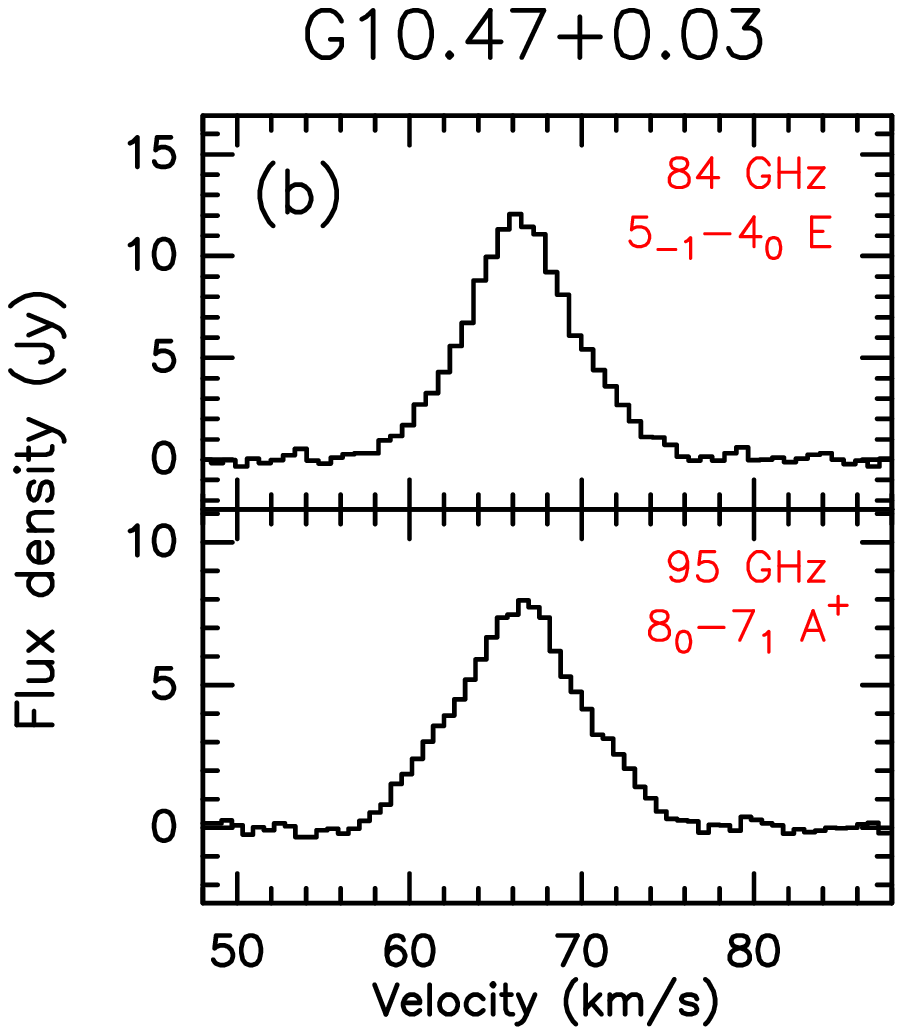}
\vspace{2mm}
\end{minipage}
\begin{minipage}[b]{4.5cm}
\includegraphics[width=0.92\textwidth]{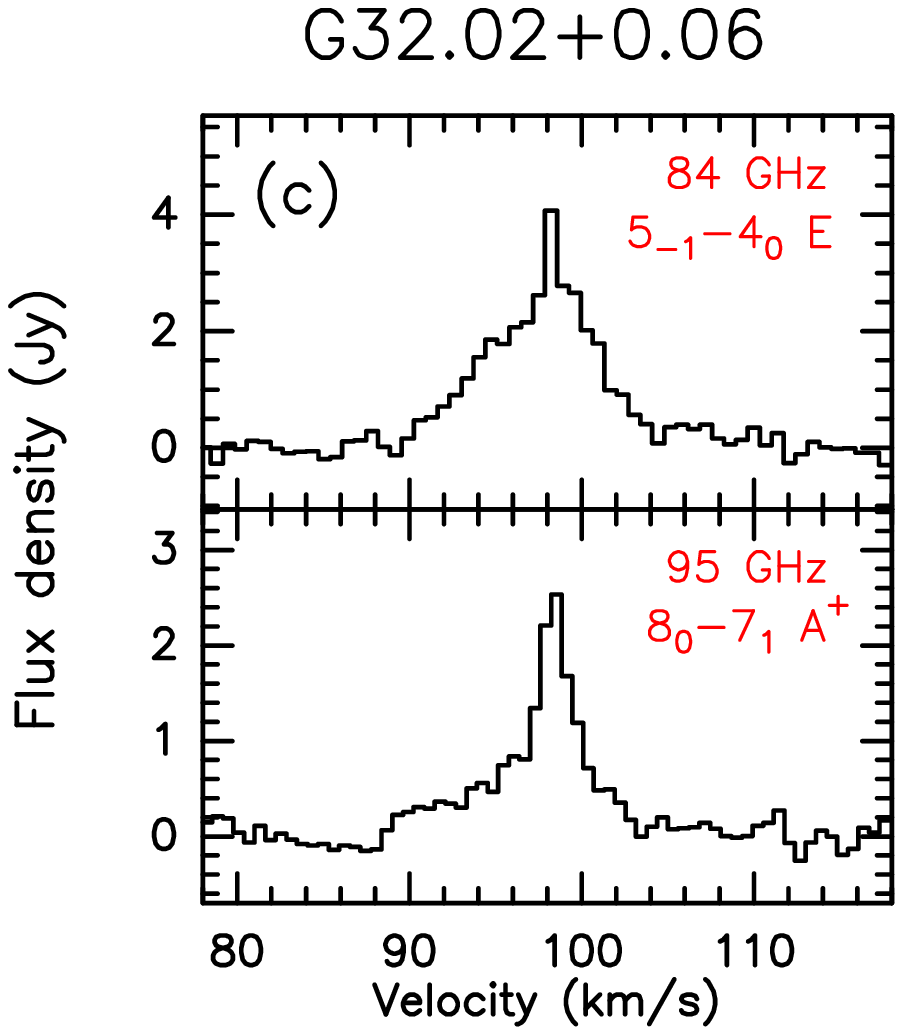}
\vspace{2mm}
\end{minipage}
}
\mbox{
\begin{minipage}[b]{4.5cm}
\includegraphics[width=0.92\textwidth]{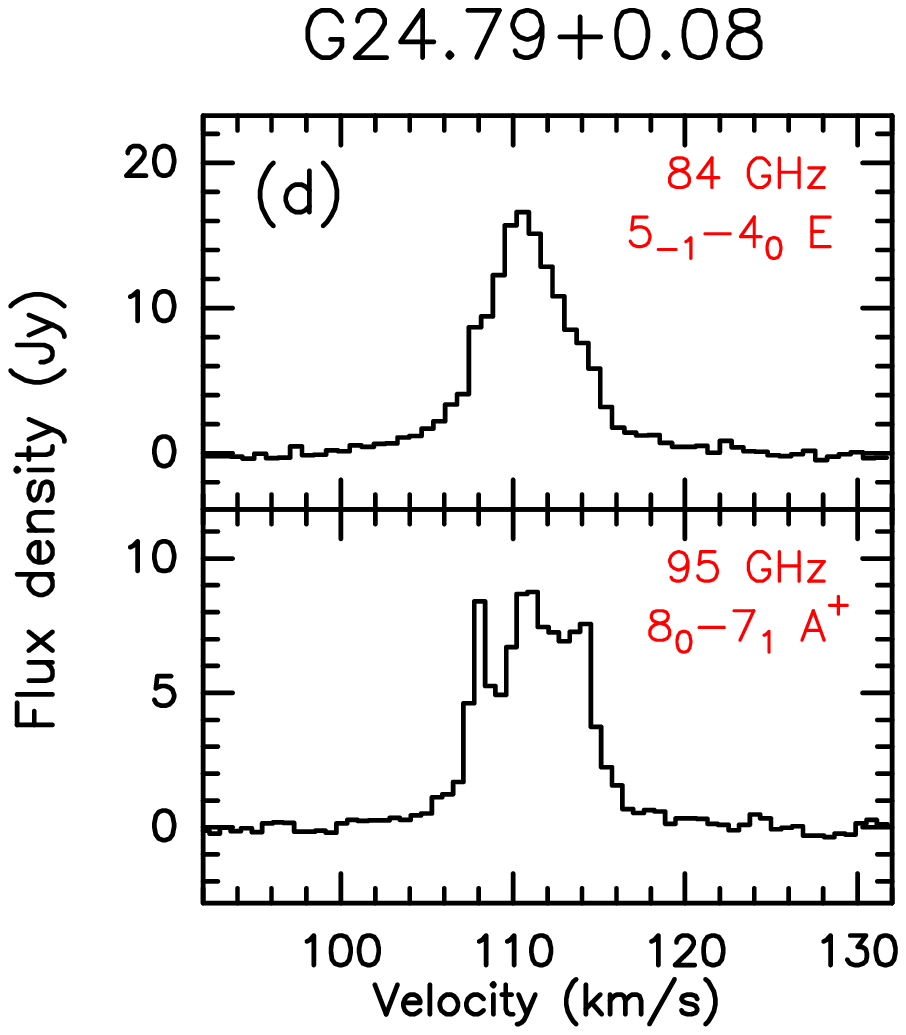}
\vspace{2mm}
\end{minipage}
\begin{minipage}[b]{4.5cm}
\includegraphics[width=0.92\textwidth]{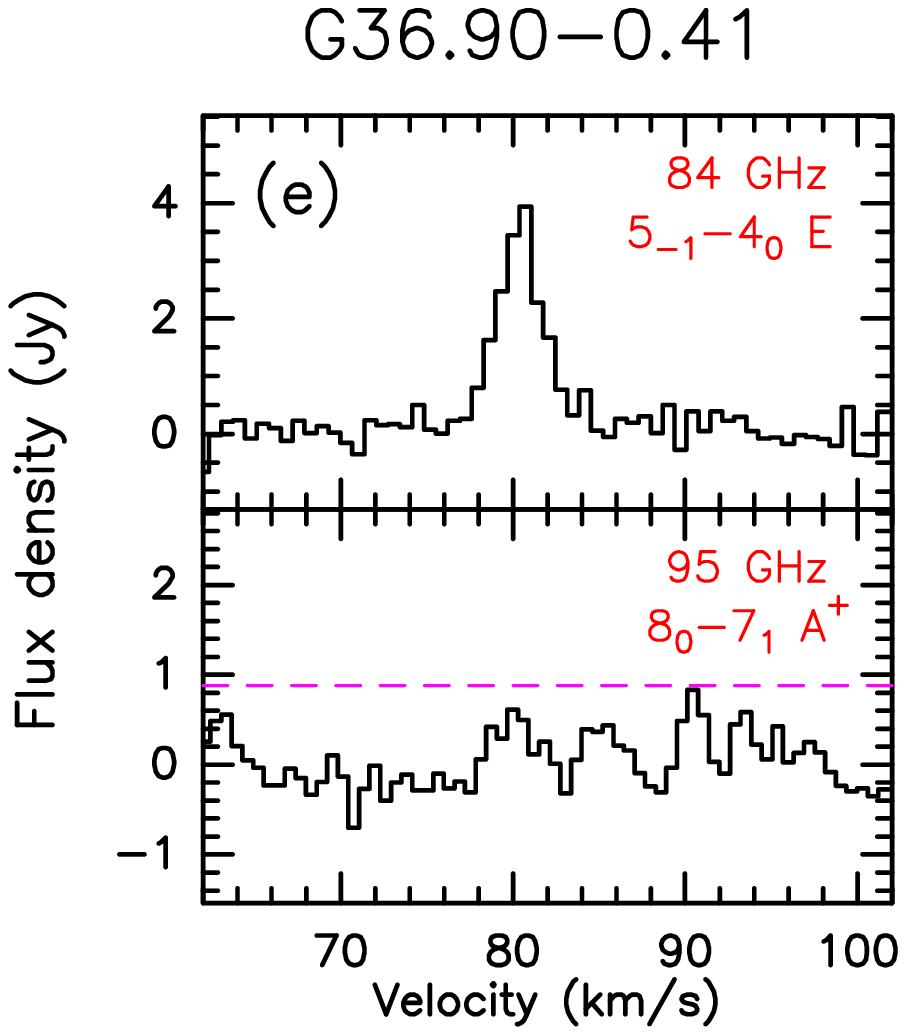}
\vspace{2mm}
\end{minipage}
\begin{minipage}[b]{4.5cm}
\includegraphics[width=0.92\textwidth]{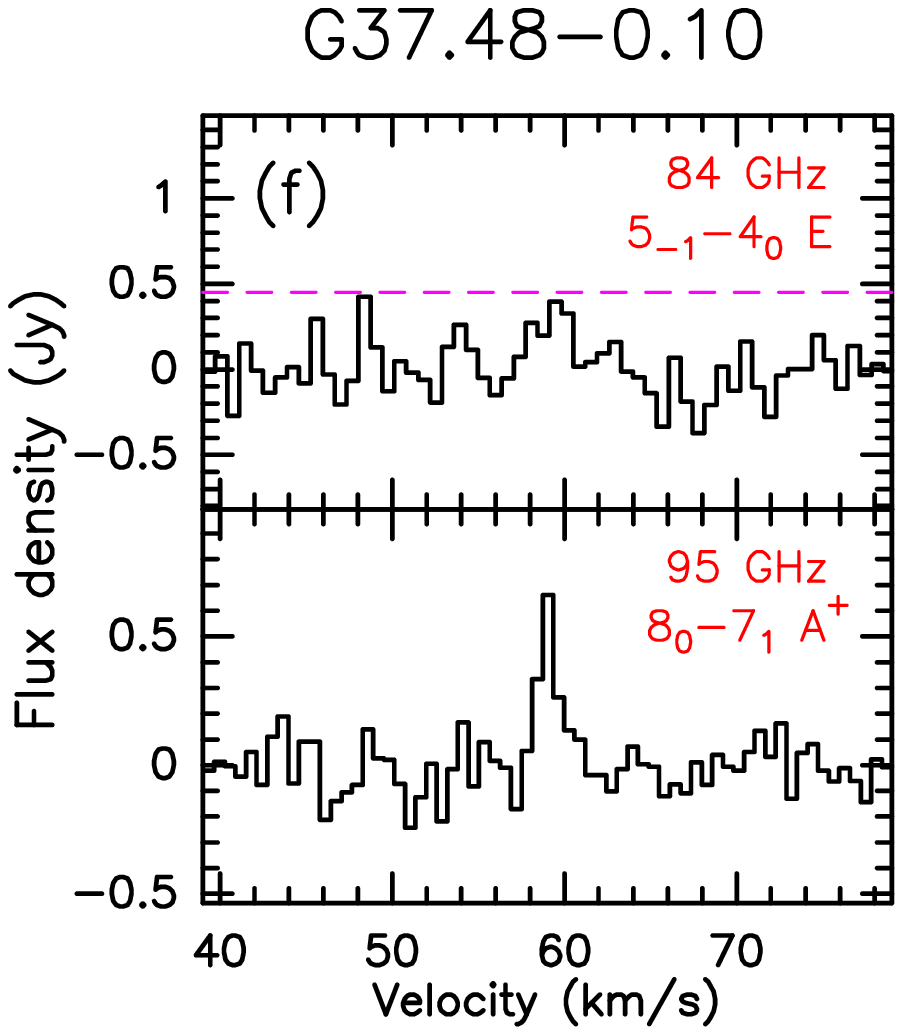}
\vspace{2mm}
\end{minipage}
}
\caption{Examples of diverse line profiles of observed class I methanol emission at 84 (upper panel) and 95 GHz (lower panel), depicting narrow maser-like features (in panel a), broad thermal-like features (in panel b) and a combination of the two (in panel c).
Line profiles of the two transitions are sometimes quite similar as shown in panels (a) to (c), and are sometimes very different as shown in panels (d) to (f).
The frequency and the corresponding quantum numbers of each transition are shown in the upper right corner of each panel for six sources.
The horizontal magenta dashed lines in G36.90$-$0.41 and G37.48$-$0.10 represent the 3$\sigma$ noise level for the undetected transitions.
\label{fig:ch3oh-8495}}
\end{figure*}

Figure~\ref{fig:ch3oh-8495} shows diverse 84 and 95\,GHz spectral profiles, from narrow maser-like features (e.g. G06.22$-$0.61, see Fig.~\ref{fig:ch3oh-8495}a) to broad (quasi-) thermal-like components (e.g. G10.47+0.03, see Fig.~\ref{fig:ch3oh-8495}b), and a combination of the two (e.g. G32.02+0.06, see Fig.~\ref{fig:ch3oh-8495}c).
The line profiles of the two transitions could be quite similar ( Fig.~\ref{fig:ch3oh-8495}a--c), or could be very different (e.g. see Fig.~\ref{fig:ch3oh-8495}d--f). 
G24.79+0.08 (see Fig.~\ref{fig:ch3oh-8495}d) is an example showing different numbers of narrow features in the two lines.
Emission from both transitions is not always detected towards a given source (e.g. G36.90$-$0.41 and G37.48$-$0.10 in Fig.~\ref{fig:ch3oh-8495}e--f).

Typical spectral maser profiles show strong and narrow emission features. However, with single-dish observations, we cannot definitively decide whether the observed sources show only maser emission or a combination of maser and thermal emission.
In this work, an object showing narrow ($\lesssim$2~\kms, i.e. 3 channels) and relatively strong ($\gtrsim$1~Jy, i.e. 5$\sigma$) features (e.g. Fig.~\ref{fig:ch3oh-8495}c for the 84\,GHz line toward G32.02+0.06) is likely to be a maser source (labelled as ``Y" in Table~\ref{Tab:overview}). Otherwise, objects showing other features are considered as maser candidates (labelled as ``y"). A lack of emission above the  3$\sigma$ level is rated as a non-detection and labelled as ``N".
For example, G06.22$-$0.61 shows narrow and strong maser features in both transitions (Fig.~\ref{fig:ch3oh-8495}a). On the other hand, the 84 GHz emission in G36.90$-$0.41 (Fig.~\ref{fig:ch3oh-8495}e) is broader than 2~\kms. Thus, this source is considered as a maser candidate.

We note that narrower emission ($<$1~\kms) could be better detected with higher spectral resolution. With our velocity resolution of $\sim$ 0.7~\kms, such narrow emission could be smoothed out over  channels, appearing broader and weaker.
For example, the narrow and strong component of 95\,GHz emission in G06.22$-$0.61 shows a line width of 0.8~\kms\ and a peak flux density of 59~Jy at a velocity resolution of 0.11~\kms\, significantly better than ours \citep{2013ApJS..206....9C}.
Assuming that this maser component is not variable, our observations nevertheless yield a  line width of 1.2~\kms\,and a peak flux density of 38~Jy, indicating a constant integrated flux density, but blurred peak and line width parameters.
Thus, it is not arbitrary but potentially stringent, to use a line width of 2~\kms\,as a criterion for determining narrow maser features in this work. Based on this criterion, a total of 54 and 100 sources show maser features at 84 and 95 GHz, respectively.

After cross-matching the maser database\footnote{\url{https://maserdb.net}} \citep{2019AJ....158..233L}, we find that 260 and 101 sources with methanol emission at 84 and 95\,GHz are detected for the first time, respectively. 
Among them, fifty 84\,GHz masers and twenty nine 95\,GHz masers are  new discoveries. For 84~GHz masers with known distances, 78\% of them are located within 5~kpc, 16\% of them are located between 5 and 10~kpc, and 6\% of them are located farther than 10~kpc. The situation is similar for the 95~GHz masers.
Since an angular resolution of 26$\arcsec$  corresponds to a linear scale of 0.6~pc at a distance of 5~kpc, our following analysis primarily focuses on the scales of $\sim$0.6 pc which we refer to as the clump scale.


Due to the close proximity of some sources to each other (within 30\arcsec, nearly the beam size at 3~mm), the observing beams may overlap and the detected emission could be from the same region.
Thirteen pairs of sources are in close proximity, marked by a ``$\dagger$" near their source names in Table~\ref{Tab:overview}.
Among them, four pairs (G14.19$-$0.19 and G14.20$-$0.19; G14.63$-$0.57 and G14.63$-$0.58; G34.40+0.23 and G34.40+0.23A; G59.50$-$0.23 and G59.50$-$0.24) show narrow maser features at 95 GHz.
However, their small number means that they make negligible contributions to the overall statistics.


Interferometric observations reveal that 95\,GHz class I maser emission is stronger than 84\,GHz maser emission detected in six spots towards IRAS16547$-$4247 \citep[see Figs. 2 and 3 in][]{2006MNRAS.373..411V}. Our observations allow us to figure out whether the 84\,GHz or 95\,GHz masers are brighter in a statistical manner using single-dish data.

\begin{figure}[htbp]
\center
\includegraphics[width=0.48\textwidth]{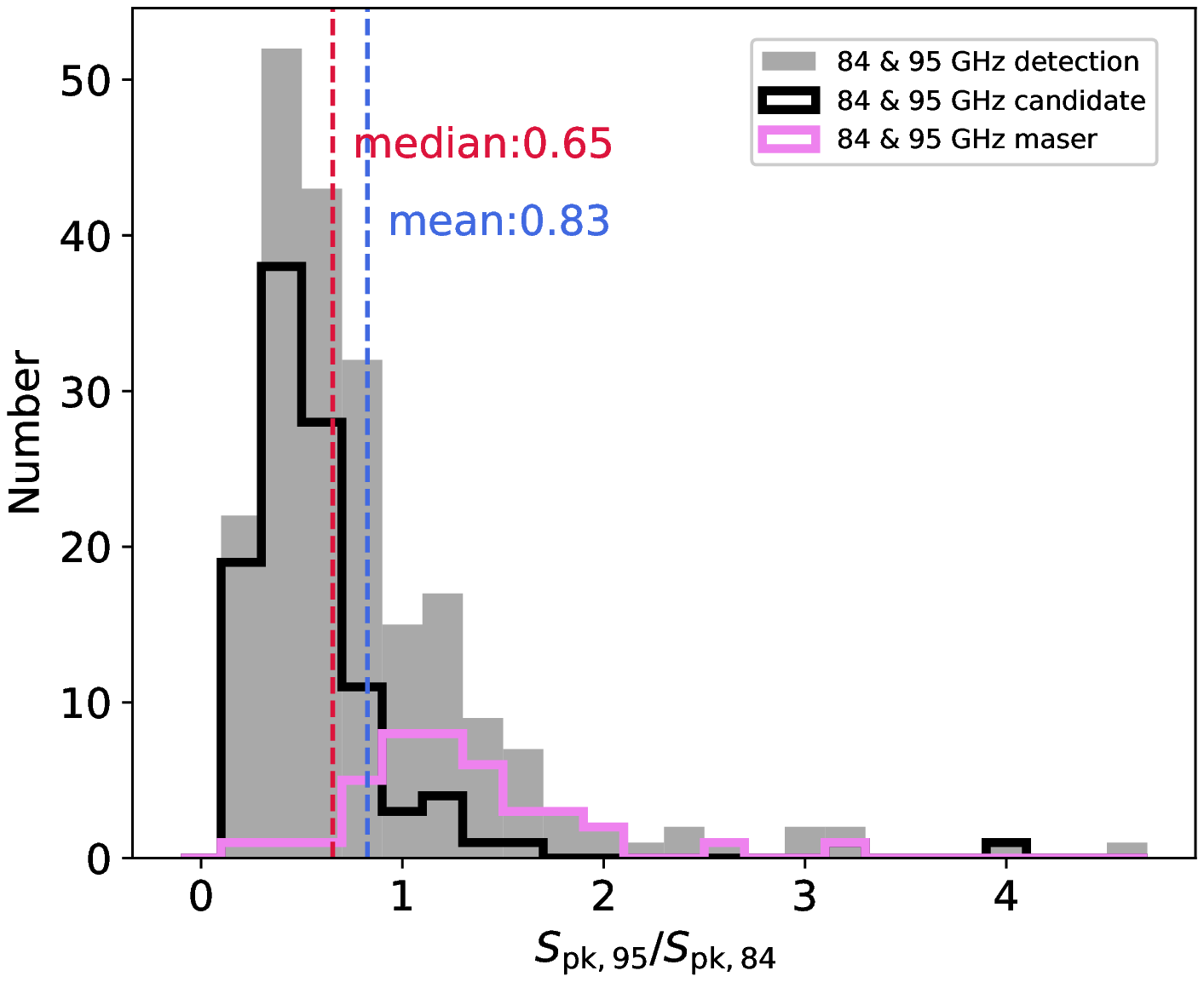}
\caption{Histogram of the peak flux density ratio of 95 and 84\,GHz emission. The grey filling represents the whole sample with both 95 and 84\,GHz detections, the pink line shows the sources hosting both 84 and 95\,GHz masers, and the black line shows the sources hosting maser candidates in both transitions. The red and blue dashed lines depict the median and mean values of the whole sample.
\label{fig:8495}}
\end{figure}

Figure~\ref{fig:8495} shows the histogram of the peak flux density ratios ($S_{\rm pk,95}/S_{\rm pk,84}$) of 95 and 84\,GHz methanol emission.
The median and mean values of peak flux density ratios for the whole 212 sources showing both emission are 0.65 and 0.83, respectively.
Half (107/212) of the whole sample have both maser candidates, and 92\% of the sources have stronger 84\,GHz emission than their 95\,GHz counterparts.
The $S_{\rm pk,95}/S_{\rm pk,84}$ ratios for these 107 sources range from 0.2 to 4.0 with a median value of 0.5 and a mean value of 0.6. If thermal emission dominates methanol detection in these sources, it is reasonable that 84\,GHz emission is stronger than 95\,GHz emission, since the upper level energy and critical density of the former are much lower than for the latter.

On the other side, forty sources host both masers, and show a different distribution of peak flux density ratios compared with the sources hosting both maser candidates in Fig.~\ref{fig:8495}.
Among them, we find that 32 (80\%) sources show stronger 95\,GHz maser emission.
The $S_{\rm pk,95}/S_{\rm pk,84}$ ratios range from 0.3 to 3.1 with a median value of 1.2 and a mean value of 1.3. This suggests that the 95 GHz methanol masers are usually stronger than their 84 GHz maser counterparts. The trend is different from the fact that we find for the maser candidates. It is probably because the emission at 84 GHz may contain more emission of thermal origin for the candidates in Fig.~\ref{fig:8495}. 


\subsection{104.3 GHz methanol emission}

\begin{figure*}[htbp]
\centering
\mbox{
\begin{minipage}[b]{3.8cm}
\includegraphics[width=0.96\textwidth]{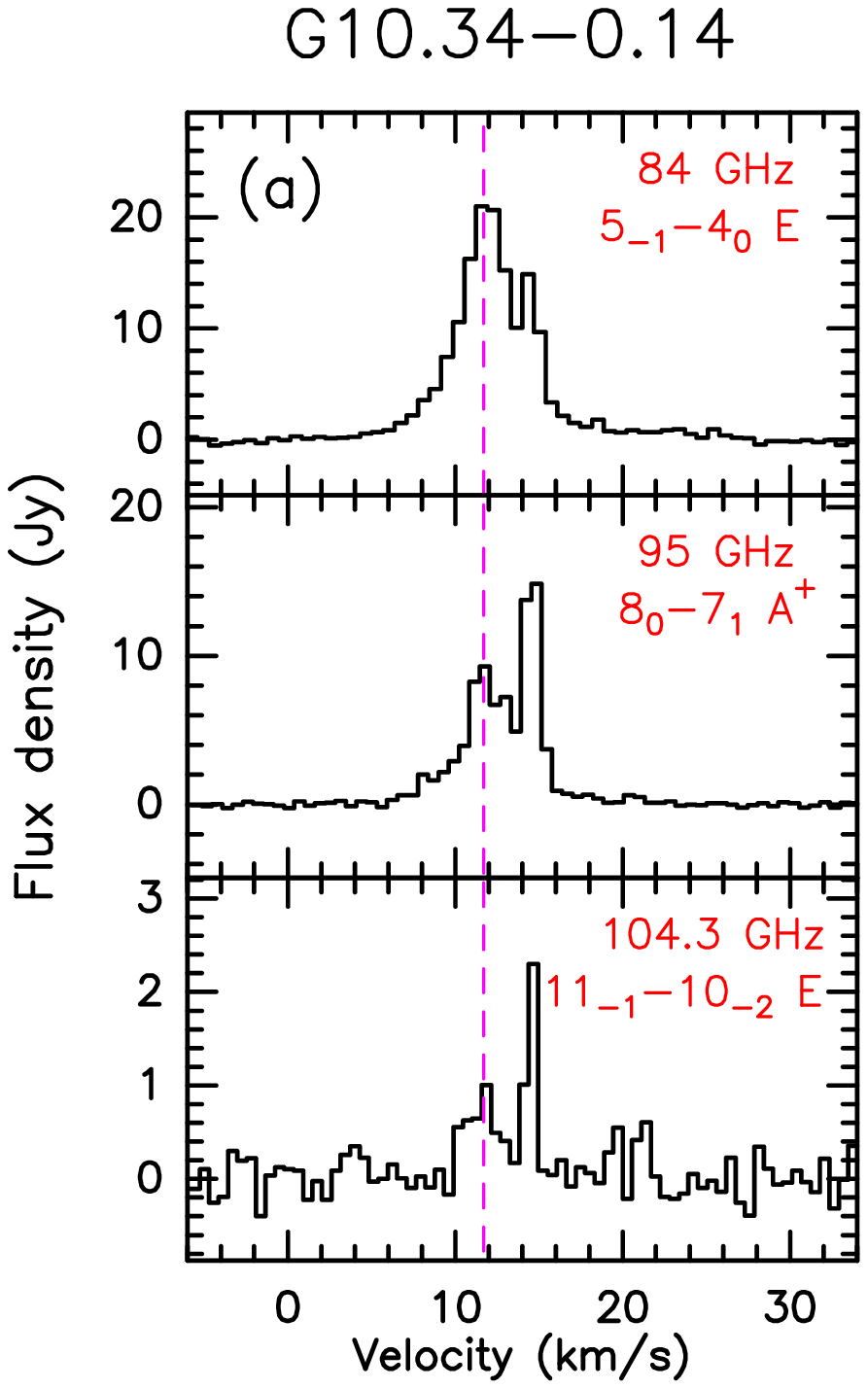}
\vspace{2mm}
\end{minipage}
\begin{minipage}[b]{3.8cm}
\includegraphics[width=0.96\textwidth]{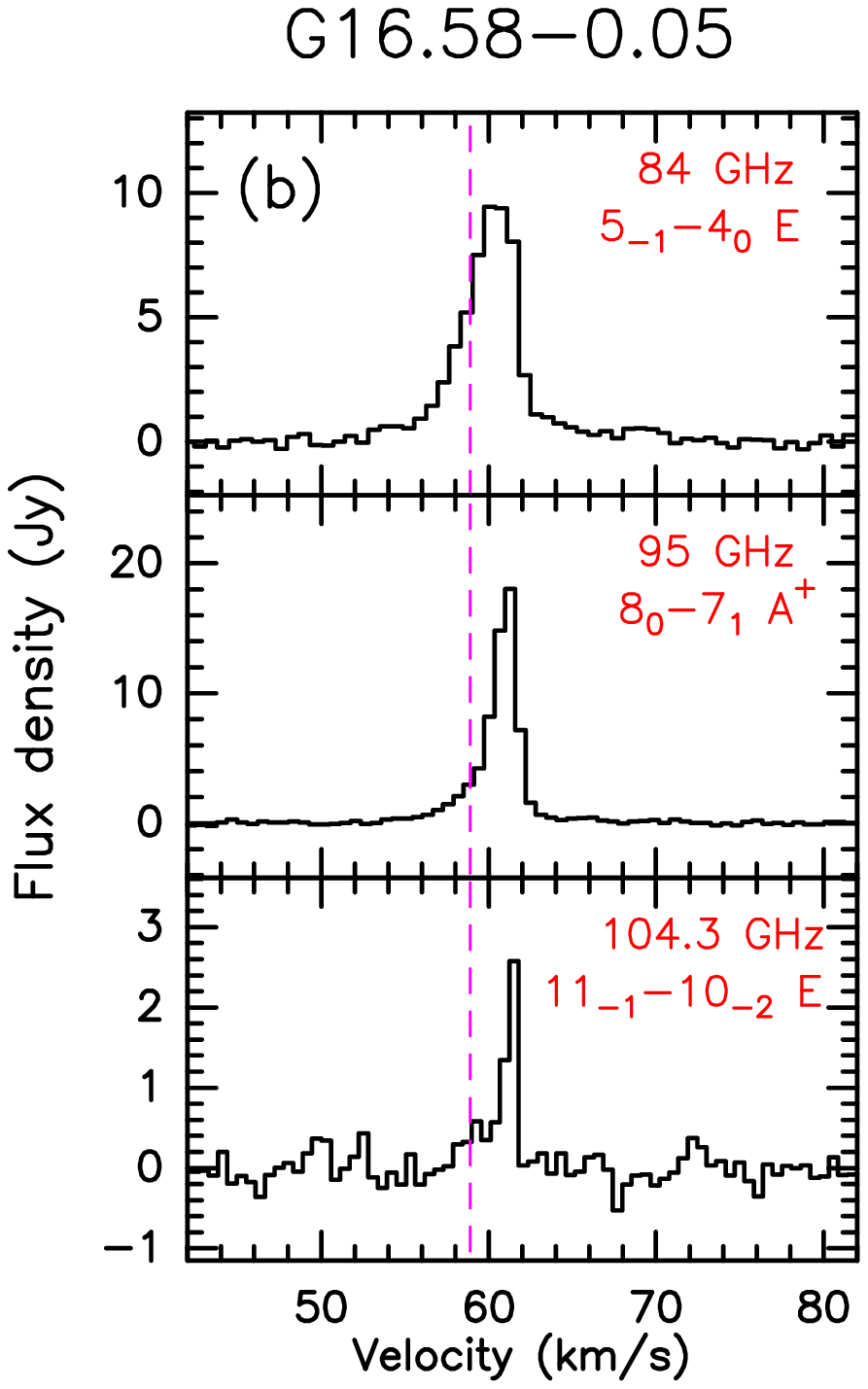}
\vspace{2mm}
\end{minipage}
\begin{minipage}[b]{3.8cm}
\includegraphics[width=0.96\textwidth]{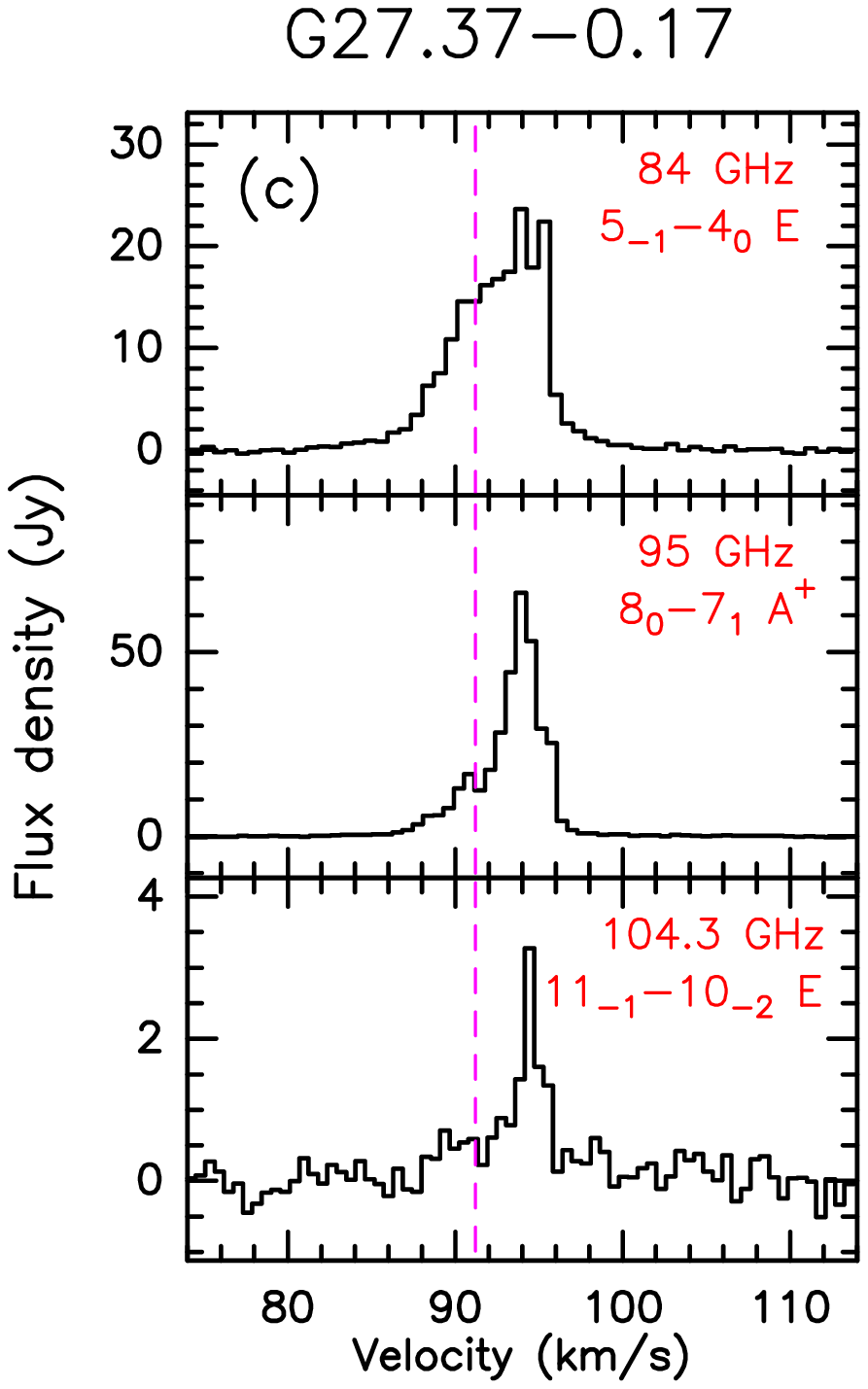}
\vspace{2mm}
\end{minipage}
\begin{minipage}[b]{3.8cm}
\includegraphics[width=0.96\textwidth]{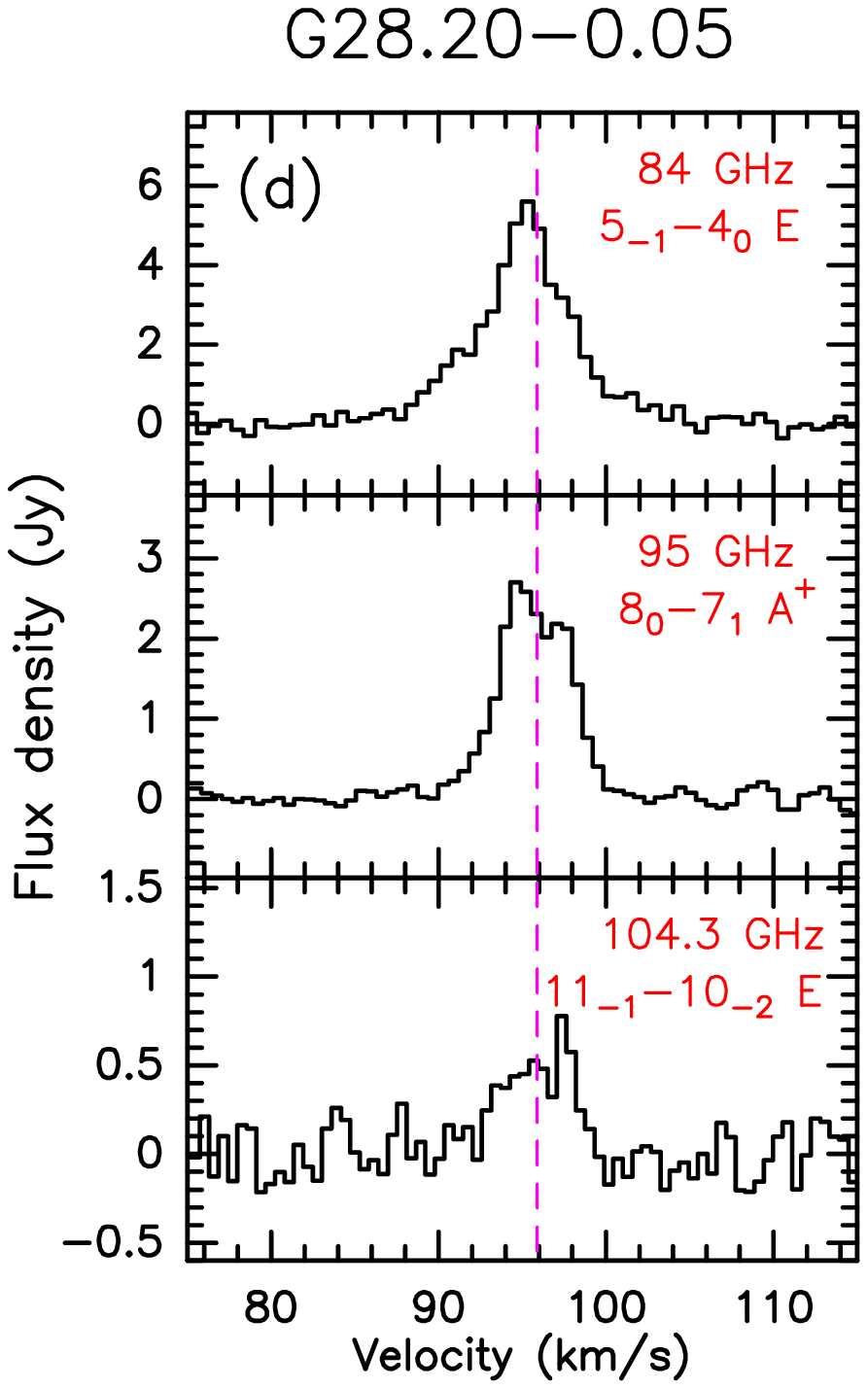}
\vspace{2mm}
\end{minipage}
}
\caption{Spectra of the 84, 95, 104.3 GHz lines (from top to bottom) for 4 sources showing 104.3 GHz methanol maser features.
The vertical magenta dashed lines represent the systemic velocity adopted from \cite{2022MNRAS.510.3389U}.
The frequency and the corresponding quantum numbers of each transition are shown in each panel.
\label{fig:ch3oh-104}}
\end{figure*}

We take the criteria of narrow ($\lesssim$2~\kms) and relatively strong ($\gtrsim$1~Jy) features to determine maser features in the 104.3 GHz spectra, which are the same as those for the 84 and 95 GHz transitions.
Among the 29 detected methanol sources at 104.3 GHz, four possess narrow and relatively strong maser characteristics (see Fig.~\ref{fig:ch3oh-104}), and the other 25 sources are regarded as maser candidates.
The four 104.3\,GHz masers are detected for the first time. Our work increases the number of known 104.3\,GHz masers from five to nine.

As can be seen from Fig.~\ref{fig:ch3oh-104}, the 104.3\,GHz maser emission generally align with the 84 and/or 95\,GHz maser emission in velocity, indicating that for all three lines, the  emission arise from the same source.
However, compared to the 84 and 95 GHz spectra, the 104.3 GHz maser spectra appear to show fewer components and narrower line widths, which is consistent with a previous study \citep{2007IAUS..242..182V}. An explanation may be that the spatial distribution of the 104.3 GHz masers is not as extensive as that of the 84 and 95 GHz masers.

This scenario is supported by interferometric observations towards IRAS16547$-$4247, where 
the rarer class I methanol maser emission in the 9.9, 25 and 104.3 GHz lines is confined to  only one of the six 84 and 95 GHz maser spots, which is associated with the brightest shocked H$_2$ 2.12~$\mu$m emission \citep[see Fig. 1 in][]{2006MNRAS.373..411V}.

Both G10.34$-$0.14 and G16.58$-$0.05 showing 104.3\,GHz maser emission have associated EGOs \citep{2008AJ....136.2391C} which are believed to be caused by shocked molecular gas in active outflows (see above).
\cite{2017ApJS..230...22T} detected class I methanol maser emission in the $5_2-5_1\,E$ line at 25\,GHz towards these two sources. 
We cross-match our sample with the EGO sample associated with 25 GHz methanol maser observations \citep{2017ApJS..230...22T}. We find three sources (G12.68$-$0.18, G14.33$-$0.64 and G22.04+0.22) showing 25\,GHz masers and 104.3\,GHz maser candidates, five sources (G10.29$-$0.12, G11.92$-$0.61, G12.90$-$0.03, G35.03+0.35 and G45.47+0.05) showing 25\,GHz masers but without 104.3\,GHz detections at the current sensitivity. This indicates that the 104.3\,GHz masers require more energetic conditions than the 25~GHz masers.  


The four detected 104.3\,GHz maser sources are associated with nearby 6.7\,GHz class II masers \citep[e.g.][]{2010MNRAS.409..913G,2015MNRAS.450.4109B,2016ApJ...833...18H}, and three of them also have a 12.2\,GHz class II maser association, with the exception of G28.20$-$0.05 \citep[e.g.][]{2014MNRAS.438.3368B,2016MNRAS.459.4066B,2022ApJS..258...19S}. 
Although G28.20$-$0.05 does not have an associated 12.2\,GHz maser \citep{2010MNRAS.401.2219B},  it has been classified as a source  in a later evolutionary stage due to the presence of an H{\sc ii} region \citep{2022MNRAS.510.3389U}. 
G16.58$-$0.05 and G27.37$-$0.17 are also classified as being in the  H{\sc ii} region stage, while G10.34$-$0.14 is classified as a PDR+Embedded source because of the complex infrared background \citep{2022MNRAS.510.3389U}.
\citet{2006MNRAS.373..411V} detected a 104.3\,GHz maser toward IRAS16547$-$4247 but no 6.7\,GHz class II maser, suggesting that this source is either too young or too old to have a 6.7\,GHz maser. Recently, \citet{2022MNRAS.510.3389U} classified its host dust continuum source (AGAL343.128$-$0.062) to be in the  H{\sc ii} region stage. Overall, these facts support that the 104.3~GHz masers are likely associated with dust clumps that may already host H{\sc ii} regions in \citet{2022MNRAS.510.3389U}. 

\subsection{Systemic velocity} \label{sec:velo}


\begin{figure}[htbp]
\center
\includegraphics[width=0.4\textwidth]{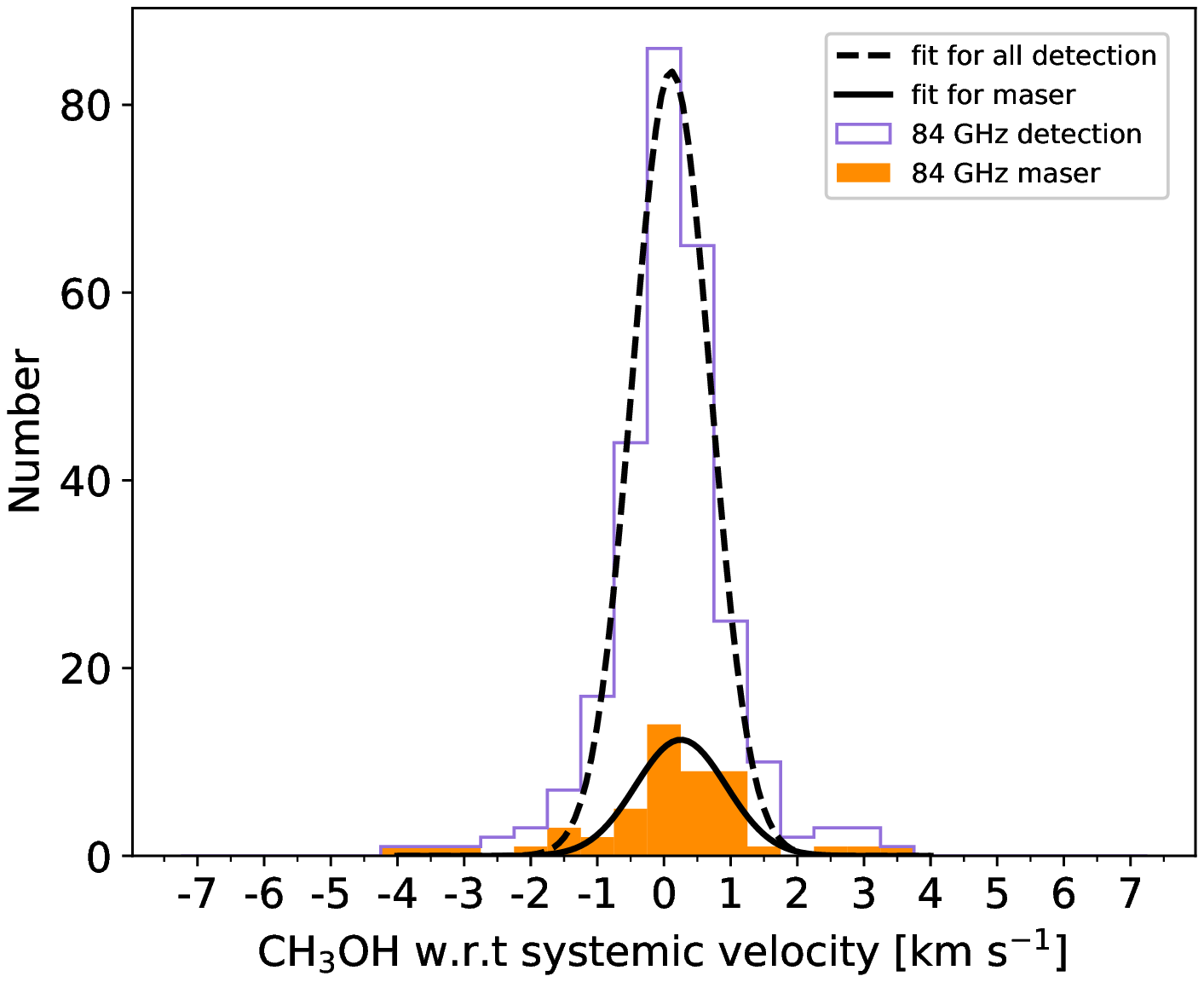}
\includegraphics[width=0.4\textwidth]{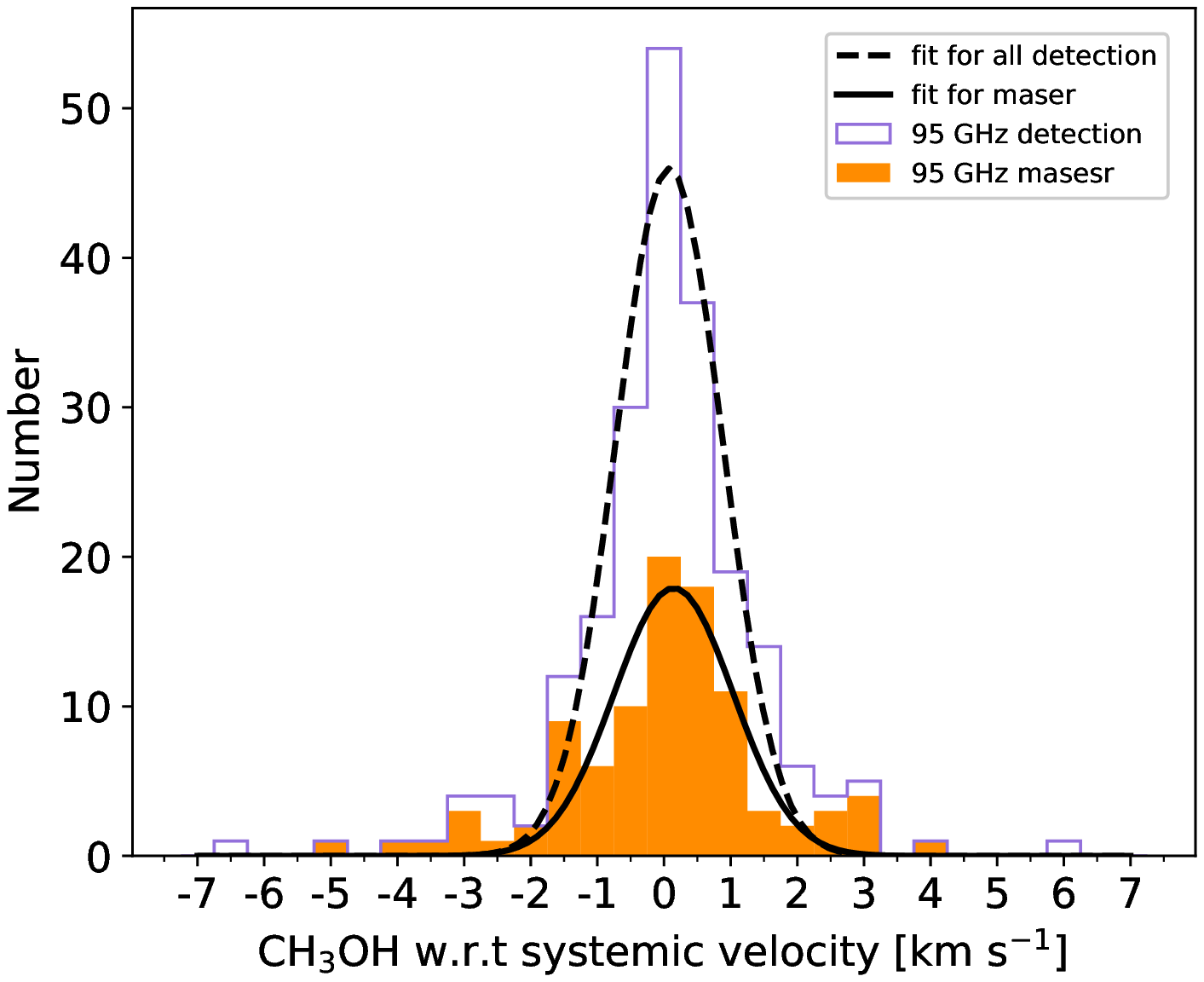}
\includegraphics[width=0.4\textwidth]{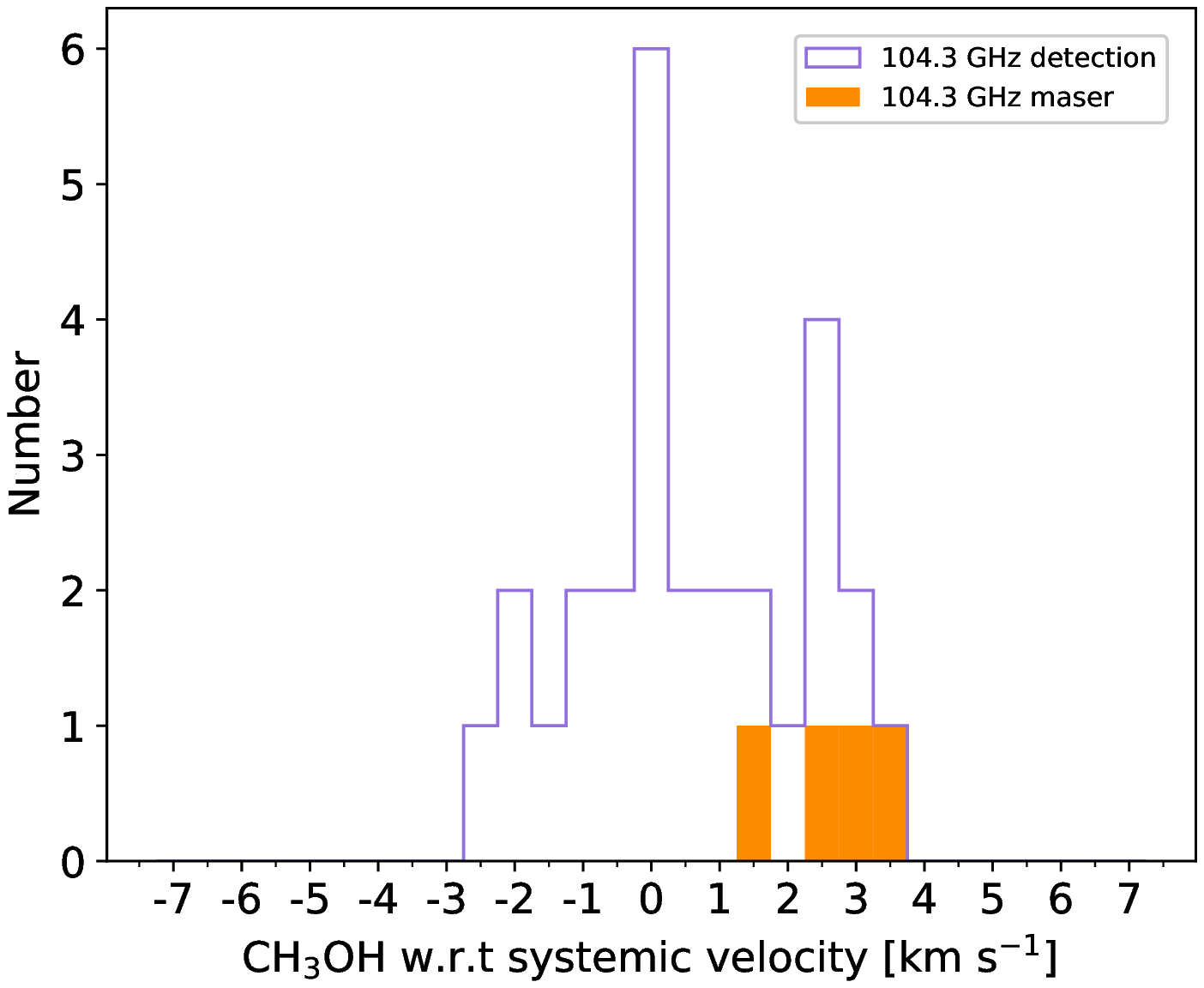}
\caption{Histograms of peak velocity of methanol emission with respect to the systemic velocity \cite[taken from][]{2022MNRAS.510.3389U} for the 84 GHz (top), 95 GHz (middle) and 104.3 GHz (bottom) lines. The orange filling and purple hollow histograms represent methanol maser and all methanol detections (including maser and maser candidate), respectively. The black solid and dashed lines depict the Gaussian fitting results for maser and all methanol detection, respectively.
\label{fig:ch3oh-velo}}
\end{figure}

Several studies have revealed that class I methanol masers are a good tracer of systemic velocities \citep[e.g.][]{2017MNRAS.471.3915J}. In this work, 
we adopt $V_{\rm LSR}$ from \cite{2022MNRAS.510.3389U} as the ATLASGAL clump systemic velocities to study the relationship between the velocity of class I methanol masers and systemic velocity.

Table~\ref{Tab:componets} gives the numbers of maser sources and all methanol detections showing single and multiple components in our three methanol transitions, as well as the fractions of masers (i.e. the number of masers divided by the number of all detection).
It is clear that the majority of the  methanol sources detected at 84, 95 and 104.3\,GHz shows a single component, while the detection rate of masers in multi-component sources is higher than that in the single component sources. This is expected, because multiple components can be present distributed in different spots within a telescope beam.

\begin{table}[!htbp]
\caption{Single or multiple components of three methanol transitions.}\label{Tab:componets}
\small
\centering
\setlength{\tabcolsep}{2.3pt}
\begin{tabular}{lcccccc}
\hline \hline
 & \multicolumn{2}{c}{84 GHz}  & \multicolumn{2}{c}{95 GHz}  & \multicolumn{2}{c}{104.3 GHz} \\
\cmidrule(lr){2-3} \cmidrule(lr){4-5} \cmidrule(lr){6-7}
 & single & multiple  & single & multiple  & single & multiple \\
\hline 
Masers & 16 & 38 & 35 & 65 & 2 & 2\\
All detections & 204 & 78 & 135 & 89 & 22 & 7\\
Fraction of masers & 8\% & 49\% & 26\% & 73\% & 9\% & 29\% \\
\hline \hline
\end{tabular}
\normalsize
\end{table}


Figure~\ref{fig:ch3oh-velo} shows the histograms of peak velocity of methanol emission with respect to the systemic velocity ($V_{\rm CH_3OH}-V_{\rm sys}$) at 84 GHz, 95 GHz and 104.3 GHz.
Apart from some outliers (see later description in this Section), the distributions of the velocity difference show Gaussian profiles for both the 84 and the 95~GHz transitions.
The $V_{\rm CH_3OH}-V_{\rm sys}$ ranges from $-$3.83~\kms\, to 3.31~\kms\, for the 84 GHz masers, and ranges from $-$5.09~\kms\, to 4.23~\kms\, for the 95 GHz masers.
The mean and median velocity difference of the 84~GHz (95~GHz) maser sample is 0.17~\kms\, (0.12~\kms), and 0.03~\kms\,(0.01~\kms), with a standard error of 0.19~\kms\, (0.16~\kms).
For all methanol detections (i.e. including maser and maser candidate) at 84 and 95~GHz, the mean and median values are similar to those of our maser samples. 
The standard deviations (1$\sigma$) of the velocity difference are 1.33 and 0.92~\kms\,for methanol masers and all detections at 84\,GHz, and 1.54 and 1.40~\kms\,for the masers and all detections at 95\,GHz.
All of the 84\,GHz masers and 99\% of the 95\,GHz masers (with the only exception of G23.44$-$0.18 showing the greatest velocity difference) are consistent with the systemic velocity within 3$\sigma$.
That, overall, the velocities of 84 and 95 GHz methanol masers are consistent with the systemic velocity, confirms that the velocity of the 84 and 95 GHz methanol maser emission well traces the systemic velocity \citep[e.g.][]{2017MNRAS.471.3915J}.


There are some outliers (84 and 95\,GHz methanol emission in G43.17+0.01, and 95\,GHz maser emission in G31.02+0.26), showing large ($>$~10~\kms) velocity differences relative to the systemic velocity listed by \cite{2022MNRAS.510.3389U}. 
However, after examining our own 3-mm spectral line survey data, we found that the velocity of methanol emission is consistent with that of C$^{18}$O (1--0) emission.
G43.17+0.01 is located in the massive and luminous W49 region, so that  complex velocity profiles are not surprising.
Both the 84 and 95\,GHz lines in this source show two blended broad ($\Delta V>$~4~\kms) components. In the case of the 84\,GHz line, the stronger one peaks at 13~\kms, and the weaker and broader feature peaks at 7.3~\kms\,(4.6~\kms\,in case of the 95\,GHz transition), which is similar to what is found for the double peak profile of C$^{18}$O (1--0) emission, whose two components peak at 4.4 and 11.9~\kms. 
In G31.02+0.26, only a 95\,GHz maser single component  is detected. 
Its velocity coincides  with that of the second strongest C$^{18}$O (1--0) emission component (at $\sim$78~\kms) but not with that of the strongest one (at $\sim$96~\kms). Therefore, the consistence between the maser velocities and systemic velocities still holds. Previous observations of 84 and 95~GHz masers towards DR21(OH) suggest that they are formed in the interface between outflows and ambient dense gas \citep{1988ApJ...329L.117B,1990ApJ...364..555P}. The masers are thus expected at the cloud systemic velocities. Our statistical results indicate that this scenario should apply to most of class I methanol masers in the Milky Way. 

Similar to class I methanol masers, water masers are also believed to be caused by collisional pumping in star formation regions \citep[e.g.][]{1992ARA&A..30...75E}, but water masers show a different behavior in that they can exhibit multiple maser features within a wide velocity range, in extreme cases such as W49N, the velocity of water masers deviates from the systemic velocity more than 250~\kms\, \citep[e.g.][]{1976ApJ...210..100M,2018IAUS..336..279K}, and they are thought to trace motions of gas bullets powered by stellar winds or jets emitted from massive YSOs \citep[e.g.][]{2010A&A...517A..71S}.

Due to the limited number of methanol masers detected at 104.3~GHz, the velocity differences shown in Fig.~\ref{fig:ch3oh-velo} do not follow a Gaussian profile.
It is worth noting that the four detected 104.3\,GHz methanol masers are all clearly redshifted with respect to the systemic velocity (see also Fig.~\ref{fig:ch3oh-104}). 
G27.37$-$0.17 shows the maximum offset from the systemic velocity of 3.3~\kms, and G28.20$-$0.05 shows the minimum offset of 1.7~\kms.
A faint and broad feature at 104.3\,GHz is also detected in G10.34$-$0.14 and G28.20$-$0.05, that could be of thermal origin and is aligned with the systemic velocity (see Figs.~\ref{fig:ch3oh-104}a and \ref{fig:ch3oh-104}d).
All four 104.3\,GHz maser sources have nearby 22.2\,GHz H$_2$O maser associations \citep[e.g.][]{2011MNRAS.416..178B,2014MNRAS.442.2240W}. 

\section{Discussion}\label{Sec:discuss}

\subsection{The relationship between class I methanol masers and SiO emission}

SiO is one of the best tracers of shocks in star formation regions.
The abundance of SiO is enhanced in shocked regions \citep[e.g.][]{2008A&A...482..549J} and the emission profile varies with shock velocity.
In low-velocity shocks, the SiO emission shows a narrow profile \citep[with a FWHM line width of $\le$ 1--2~\kms,][]{2009ApJ...695..149J}, while the SiO profiles are broader in high-velocity shocks.

Cross-matching with the SiO data of \citet{2016A&A...586A.149C}, we find that 9 sources showing narrow and strong methanol maser characteristics are not associated with SiO emission. Among these SiO non-detections, four sources (G10.75+0.02, G13.87+0.28, G14.18$-$0.53 and G17.64+0.15) harbor 84 GHz masers and seven sources (G08.05$-$0.24, G10.75+0.02, G13.87+0.28, G27.55$-$0.94, G31.02+0.26, G31.10+0.26 and G37.48$-$0.10) harbor 95 GHz masers.
The abundance of gas phase methanol may be enhanced by low-velocity shocks in which the high-velocity shock tracer SiO is harder to detect.
G10.75+0.02, G14.18$-$0.53 and G31.02+0.26 are classified as sources in a quiescent stage of evolution \citep{2022MNRAS.510.3389U}, suggesting that class I methanol masers could be a unique signpost of protostellar activity in extremely embedded objects at the earliest evolutionary stage.


\begin{figure*}[htbp]
\centering
\includegraphics[width=0.33\textwidth]{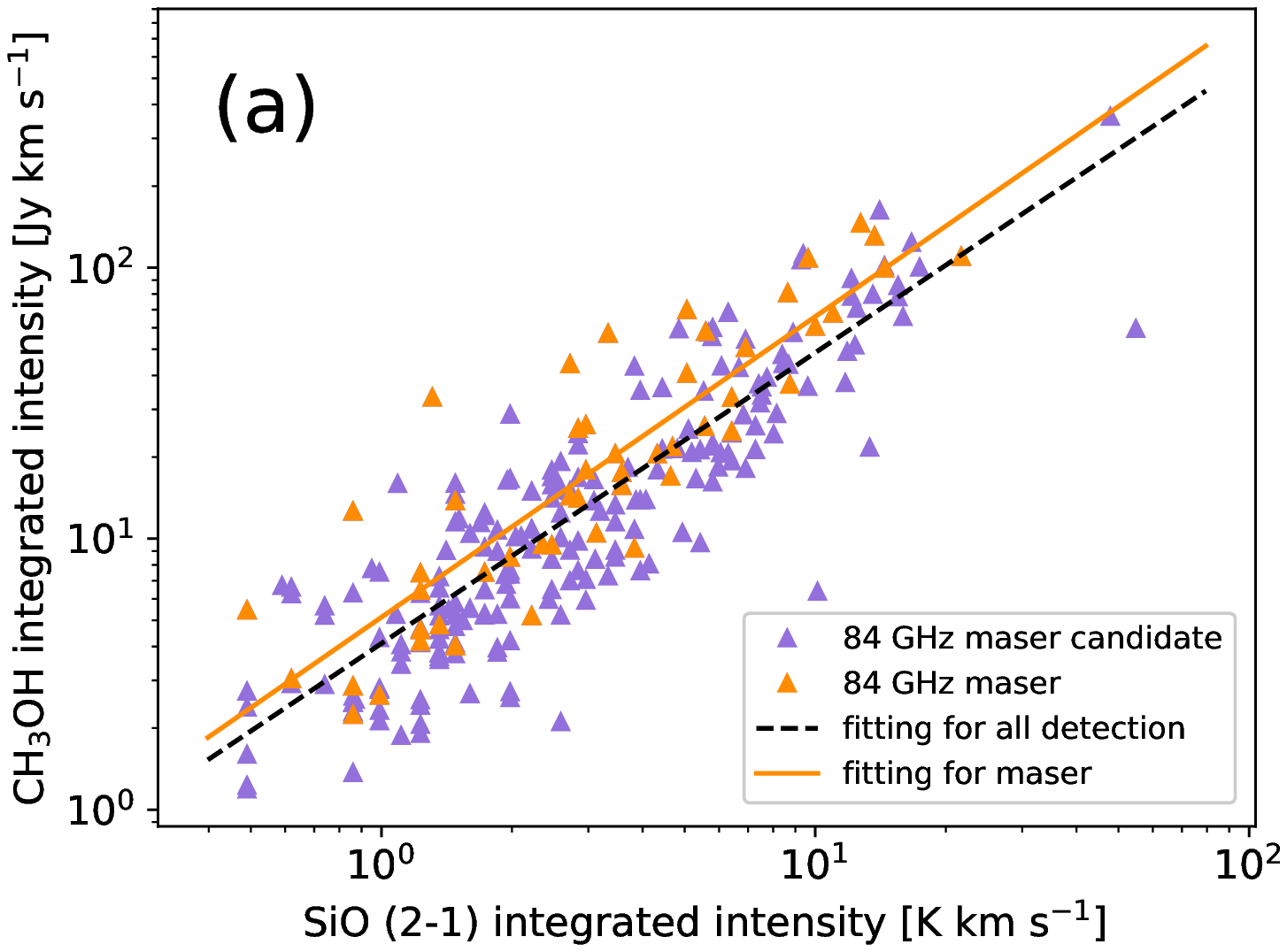}
\includegraphics[width=0.33\textwidth]{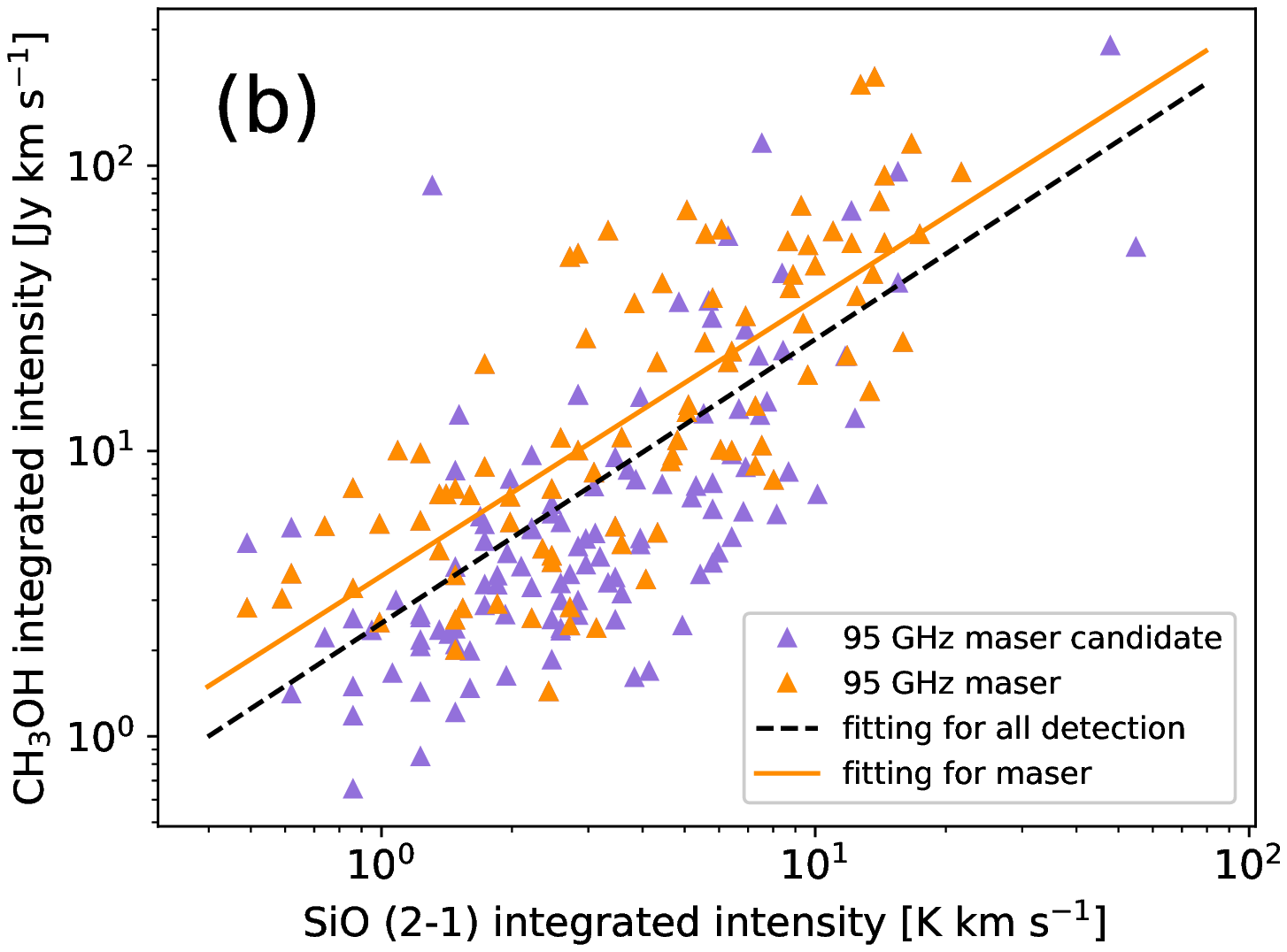}
\includegraphics[width=0.33\textwidth]{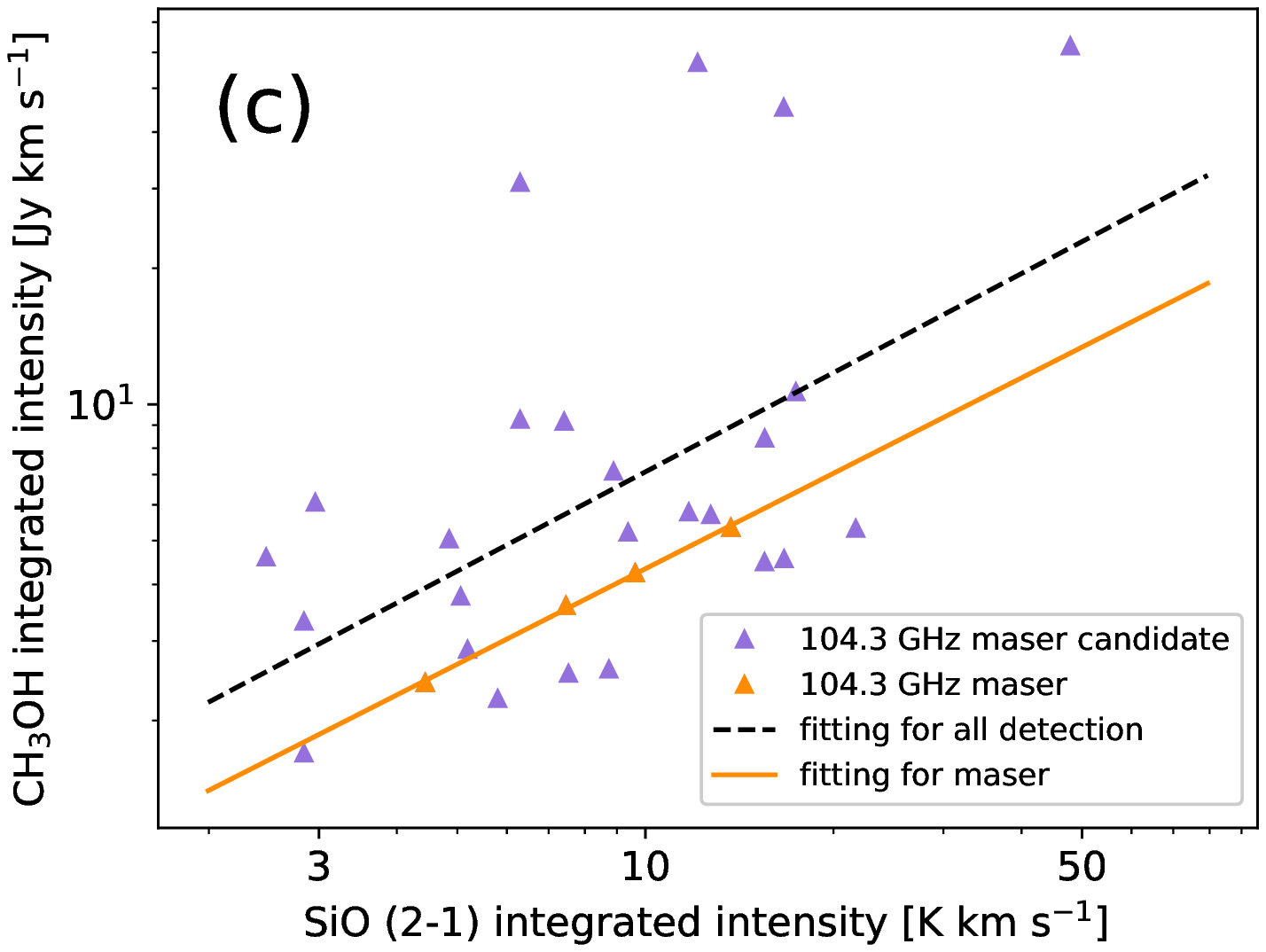}
\includegraphics[width=0.33\textwidth]{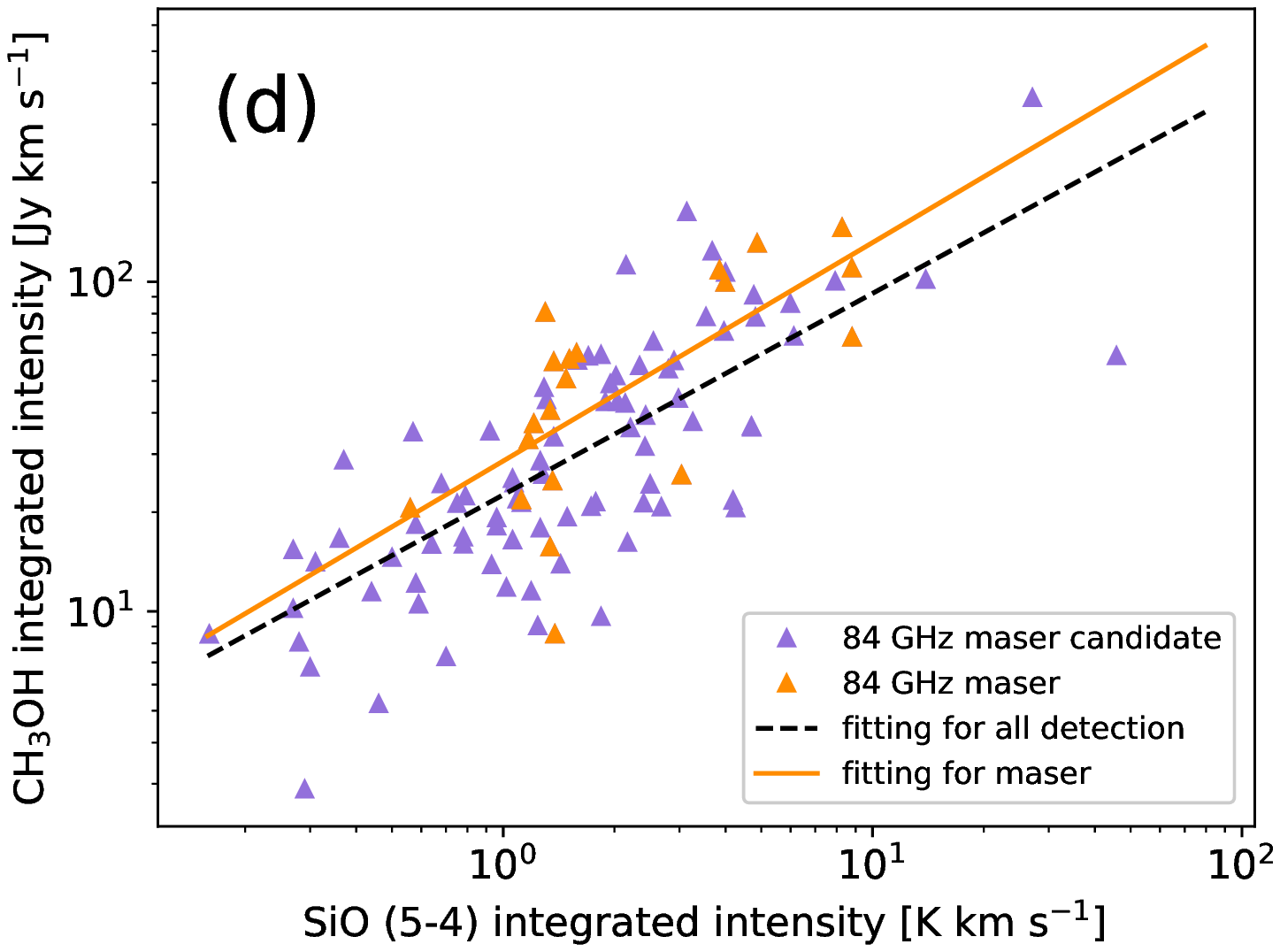}
\includegraphics[width=0.33\textwidth]{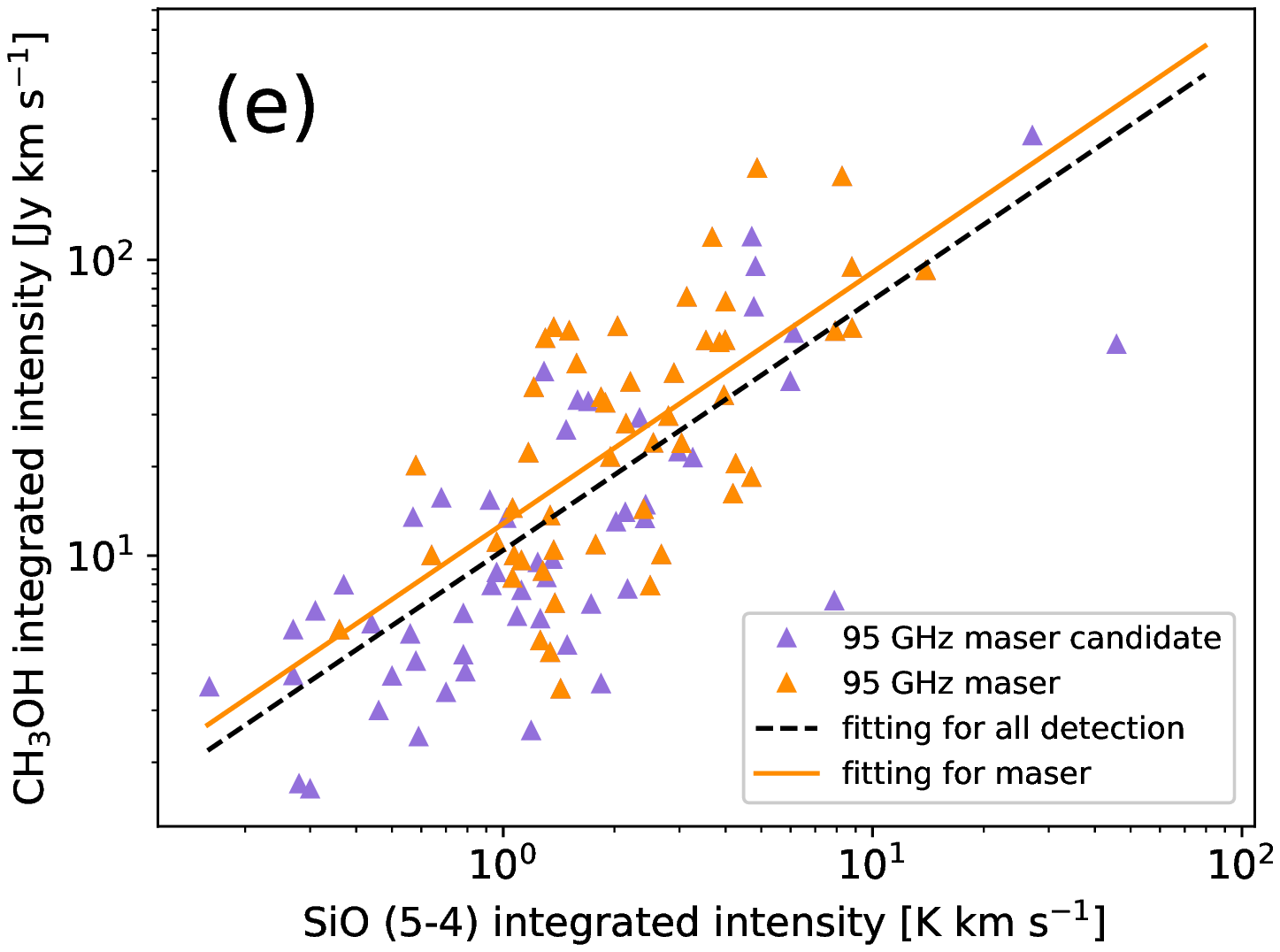}
\includegraphics[width=0.33\textwidth]{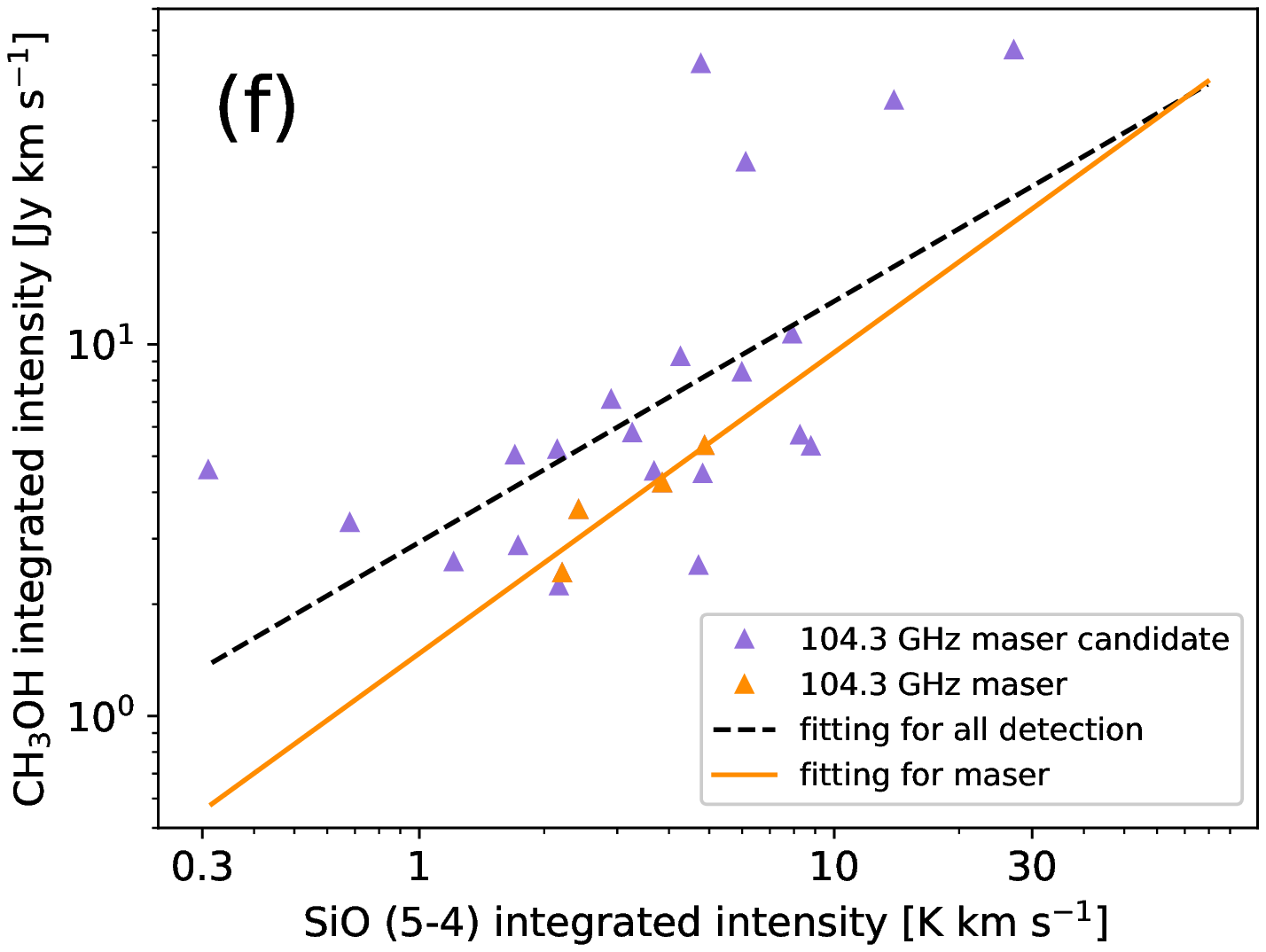}
\caption{Methanol integrated intensity as a function of SiO integrated intensity. Top panels (a,b,c) show the integrated 84, 95 and 104.3 GHz methanol emission versus SiO (2--1) integrated intensity, respectively. Bottom panels (d,e,f) show the integrated 84, 95 and 104.3 GHz methanol emission versus SiO (5--4) integrated intensity, respectively. 
The orange and purple triangles represent the clumps hosting masers and maser candidates, respectively. The orange solid lines and black dashed lines depict the least-square fitting results for methanol maser and all methanol detections, respectively.
\label{fig:ch3oh-sio-inte}}
\end{figure*}

\begin{figure}[htbp]
\centering
\includegraphics[width=0.45\textwidth]{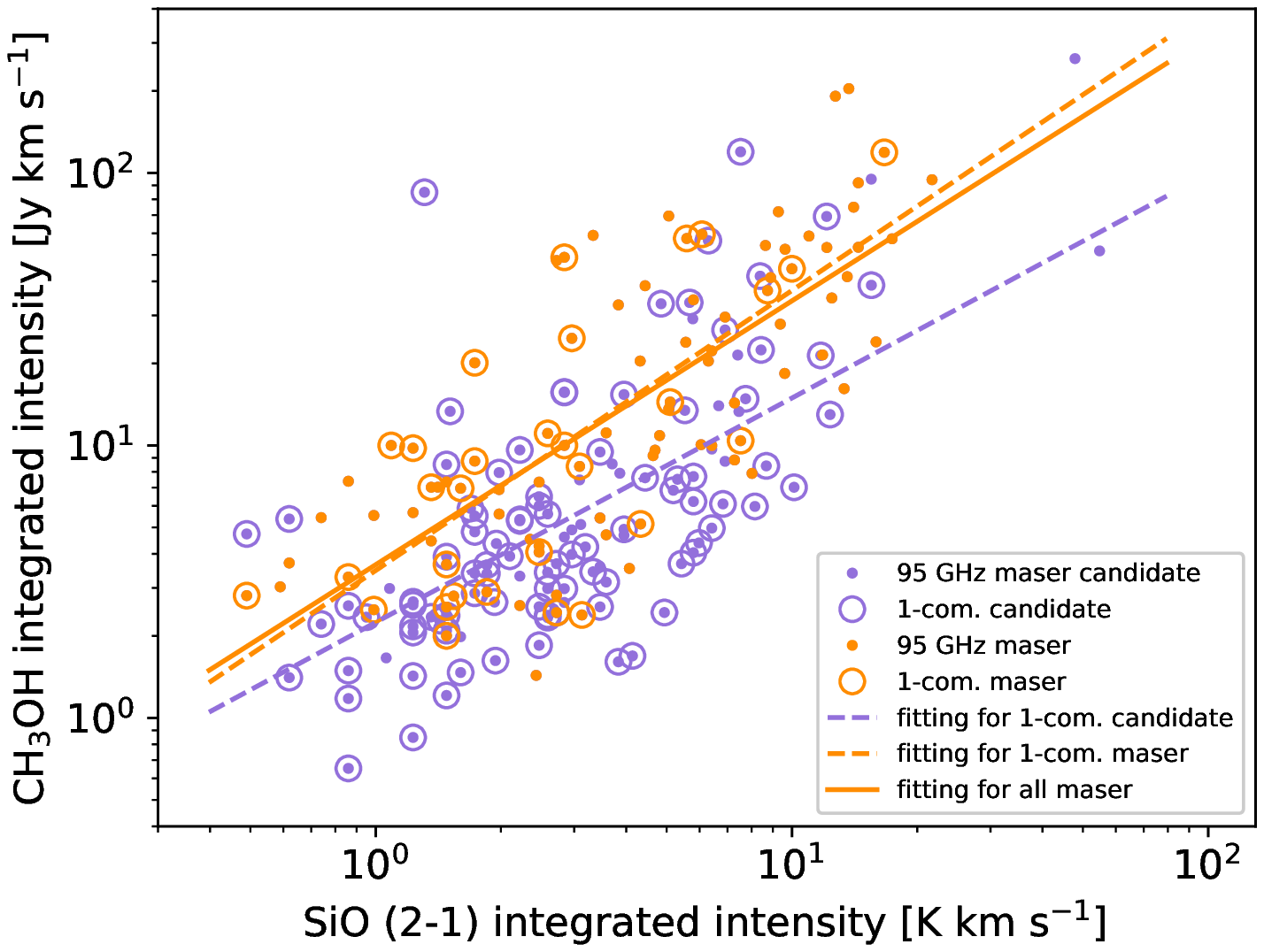}
\caption{95\,GHz methanol integrated intensity versus SiO (2--1) integrated intensity.
The orange and purple dots represent the clumps hosting masers and maser candidates, respectively. The orange and purple circle mark the source with a single component.
The orange and purple dashed lines depict the least-square fitting results for methanol maser and maser candidates with single component, respectively.
The orange solid line represents the least-square fitting results for methanol masers as in Fig.~\ref{fig:ch3oh-sio-inte}b.
\label{fig:95ch3oh-sio-inte-single}}
\end{figure}



Figure~\ref{fig:ch3oh-sio-inte} shows 84, 95 and 104.3\,GHz integrated intensities of methanol masers and maser candidates plotted against integrated SiO (2--1) and SiO (5--4) intensities.
Their linear fitting results in log-log form for three methanol transitions are summarised in Table~\ref{Tab:sio-fit}. Given that the correlation coefficients and slopes of the fitting lines are almost identical within the errors, we infer that there is no significant difference between masers and maser candidates in the 84 and 95\,GHz lines.
On the other hand, the positive slopes and correlation coefficients for the two lines indicate that stronger class I methanol maser emission is typically associated with higher SiO integrated intensity. Despite the small number of sources showing the 104.3~GHz emission, the integrated intensities of the 104.3~GHz emission appear to be also linearly correlated with the integrated SiO (2--1) and SiO (5--4) intensities. 
This is consistent with previous studies reporting similarly positive correlations between class I methanol masers at 36, 44 and 84\,GHz and SiO emission \citep{2017MNRAS.471.3915J,2019MNRAS.484.5072B}. In order to minimise the contribution of thermal emission of a maser source, we pick out the 95\,GHz maser sources and maser candidates with a single component, and do least-square fitting separately. 
Figure~\ref{fig:95ch3oh-sio-inte-single} shows no significant difference between least-square fitting results for all masers and those for masers with a single component, suggesting that the potential contribution of thermal emission is negligible. Hence, the scaling relationships support the close link between the three class I methanol masers and SiO emission.

\begin{table*}[hbt]
\caption{Summary of linear fitting results in log-log form of methanol integrated intensity versus the SiO intensity and the FWZP of SiO (2--1) emission for the 84, 95 and 104.3\,GHz transitions.
}\label{Tab:sio-fit} 
\small
\centering
\setlength{\tabcolsep}{4pt}
\renewcommand\arraystretch{1.15}
\begin{tabular}{lcccccc}
\hline \hline
\multirow{2}{*}{Parameters}  &  \multicolumn{3}{c}{Maser} &  \multicolumn{3}{c}{Detection (i.e. maser+candidate)}\\
\cmidrule(lr){2-4} 
\cmidrule(lr){5-7} 
 & slope & $r$  & $p$-value  & slope & $r$  & $p$-value \\
\hline
\multicolumn{7}{c}{84\,GHz} \\
\hline
$\int S_{84}{\rm d}V$ vs. $\int S_{\rm SiO (2-1)}{\rm d}V$  & 1.11$\pm$0.09 & 0.88 & 7.7$\times$10$^{-17}$  & 1.07$\pm$0.04 & 0.87 & 3.6$\times$10$^{-77}$  \\
$\int S_{84}{\rm d}V$ vs. $\int S_{\rm SiO (5-4)}{\rm d}V$  & 0.66$\pm$0.17 & 0.68 & 9.4$\times$10$^{-4}$ & 0.61$\pm$0.06 & 0.70 & 5.9$\times$10$^{-17}$ \\
$\int S_{84}{\rm d}V$ vs. FWZP of SiO (2--1)  & 1.72$\pm$0.33 & 0.61 & 3.2$\times$10$^{-6}$ & 1.41$\pm$0.13 & 0.56 & 3.2$\times$10$^{-22}$ \\
\hline
\multicolumn{7}{c}{95\,GHz}\\
\hline
$\int S_{95}{\rm d}V$ vs. $\int S_{\rm SiO (2-1)}{\rm d}V$  & 0.97$\pm$0.09 & 0.75 & 1.2$\times$10$^{-17}$ & 0.99$\pm$0.07 & 0.72 & 3.8$\times$10$^{-34}$  \\
$\int S_{95}{\rm d}V$ vs. $\int S_{\rm SiO (5-4)}{\rm d}V$  & 0.85$\pm$0.14 & 0.65 & 2.6$\times$10$^{-7}$ & 0.85$\pm$0.08 & 0.74 & 4.8$\times$10$^{-19}$ \\
$\int S_{95}{\rm d}V$ vs. FWZP of SiO (2--1)  & 1.14$\pm$0.23 & 0.46 & 4.2$\times$10$^{-6}$ & 1.13$\pm$0.18 & 0.41 & 8.7$\times$10$^{-10}$ \\
\hline
\multicolumn{7}{c}{104.3\,GHz}\\
\hline
$\int S_{104.3}{\rm d}V$ vs. $\int S_{\rm SiO (2-1)}{\rm d}V$  & 0.70$\pm$0.02 & 0.999 & 9.4$\times$10$^{-4}$ & 0.73$\pm$0.32 & 0.53 & 2.8$\times$10$^{-3}$ \\
$\int S_{104.3}{\rm d}V$ vs. $\int S_{\rm SiO (5-4)}{\rm d}V$  & 0.81$\pm$0.25 & 0.92 & 8.3$\times$10$^{-2}$ & 0.65$\pm$0.17 & 0.63 & 8.3$\times$10$^{-4}$ \\
$\int S_{104.3}{\rm d}V$ vs. FWZP of SiO (2--1)  & 1.21$\pm$0.52 & 0.85 & 0.15 & 0.36$\pm$0.51 & 0.14 & 0.49 \\
\hline \hline
\end{tabular}
\normalsize
\note{Columns 2--7 give the slopes of the linear fitting results, Pearson correlation coefficients $r$, and $p$-value for masers and all methanol detections.}
\end{table*}



\begin{figure}[htbp]
\centering
\includegraphics[width=0.4\textwidth]{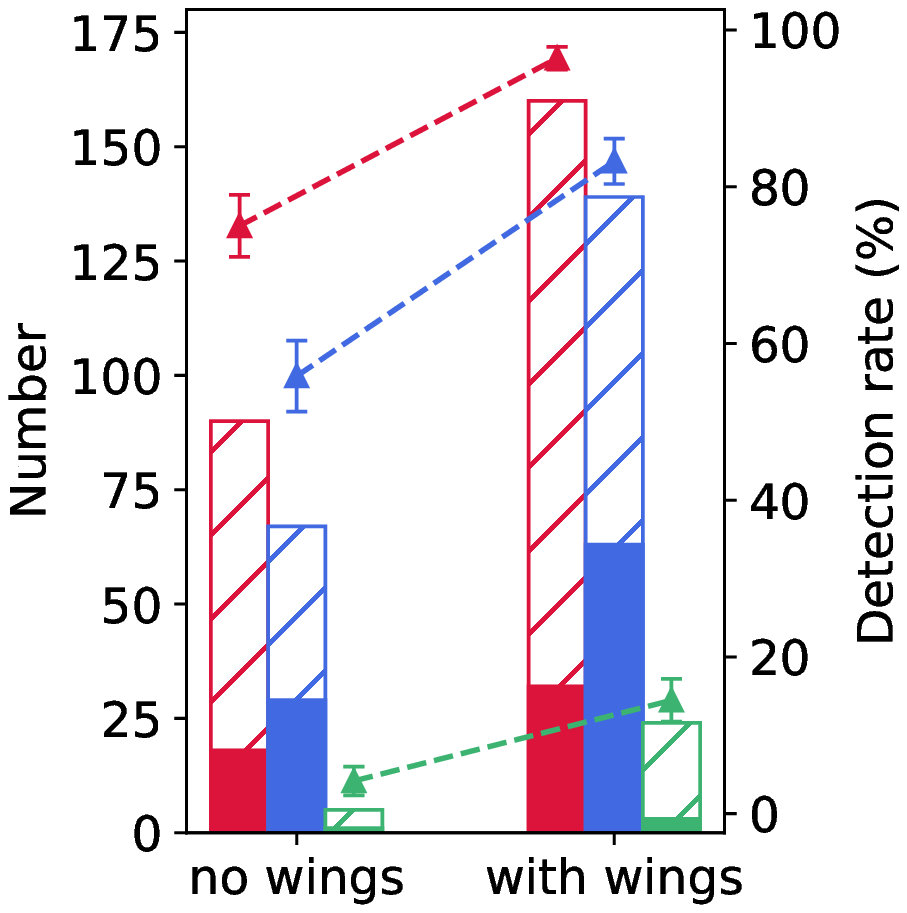}
\caption{Distribution of detection number and detection rate of three methanol transitions for SiO (2--1) emission with or without wings.
The labels for the detection numbers are on the left side, while those for the detection rates are on the right side.
The red, blue and green bins present the detection numbers of methanol emission at 84, 95 and 104.3\,GHz, respectively, where the filled and hatched parts represent masers and maser candidates. 
The red, blue and green triangles indicate the detection rates of methanol emission at 84, 95 and 104.3\,GHz. 
\label{fig:hist-wing}}
\end{figure}
\begin{figure}[htbp]
\centering
\includegraphics[width=0.242\textwidth]{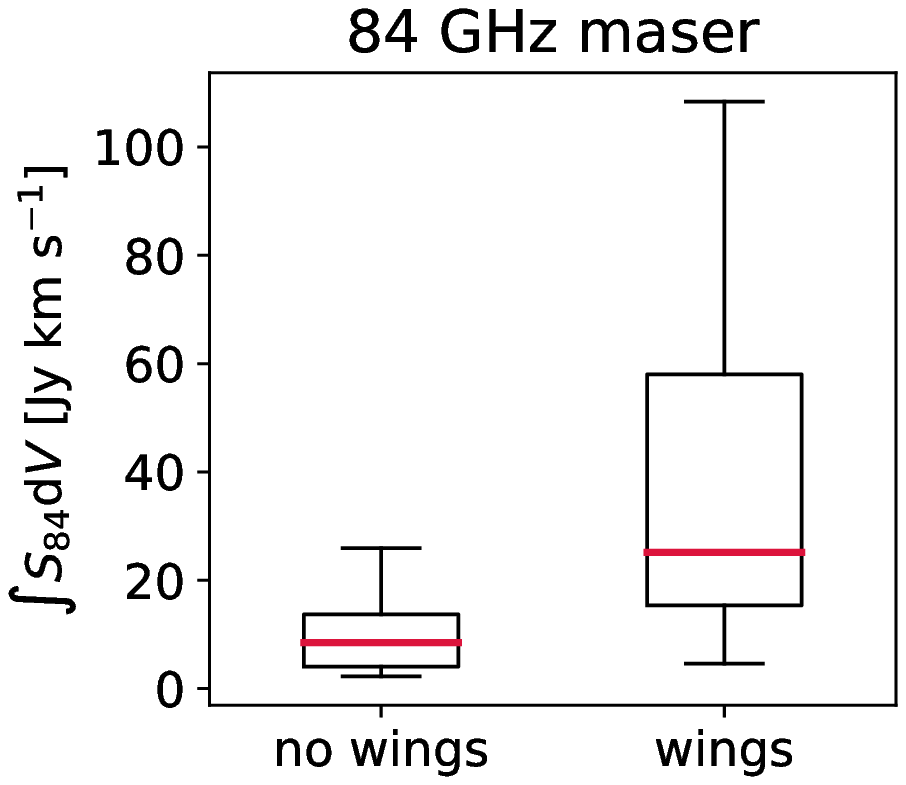}
\includegraphics[width=0.242\textwidth]{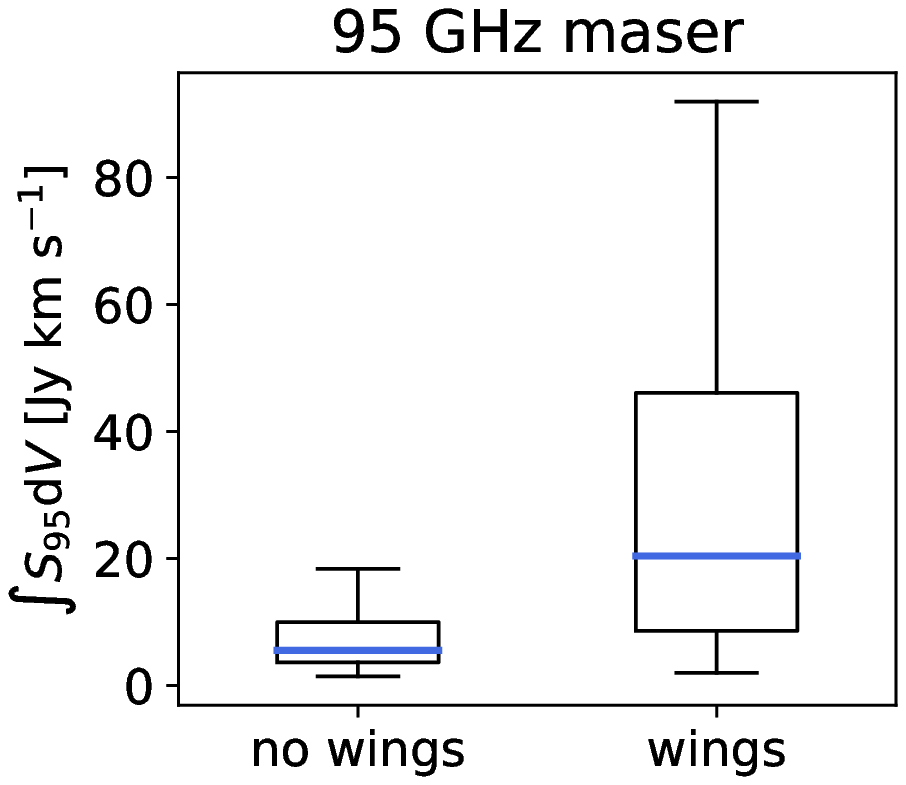}
\caption{Box plots of the integrated intensity for the 84 (left) and 95 (right) GHz masers that are divided into two groups with and without SiO (2-1) wings.
In each box plot, the horizontal colored line represents the median value, the box represents the interquartile range between the 25th and 75th percentiles, and the vertical lines (the `whiskers') show the ranges from the minimum value to the 25th percentile and from the 75th percentile to the maximum value.
\label{fig:wing-box}}
\end{figure}

SiO emission with high-velocity wings indicates shocked gas due to fast material ejection.  \cite{2016A&A...586A.149C} detected high-velocity SiO (2--1) wings towards 167 sources in our sample, while 120 sources do not show such wings. 
Figure~\ref{fig:hist-wing} reveals that there are more detections and higher detection rates of the three class I methanol transitions (as well as for methanol maser emission) towards sources showing wings than sources without wings.
The uncertainties of the detection rates in this work are calculated under the assumption that the detection rates satisfy the binomial distribution, that is $\sqrt{p \times (1-p)/n}$, where $p$ is the detection rate and $n$ is the sample size.
Figure~\ref{fig:wing-box} shows that methanol masers have higher integrated intensities in the sources showing SiO (2--1) wings than in the sources without SiO (2--1) wings. This indicates that brighter class I methanol masers are produced by higher shock velocities. In order to further investigate the relationship, we adopt the full width at zero power (FWZP) for this analysis in the following. 


\begin{figure*}[!htbp]
\centering
\includegraphics[width=0.33\textwidth]{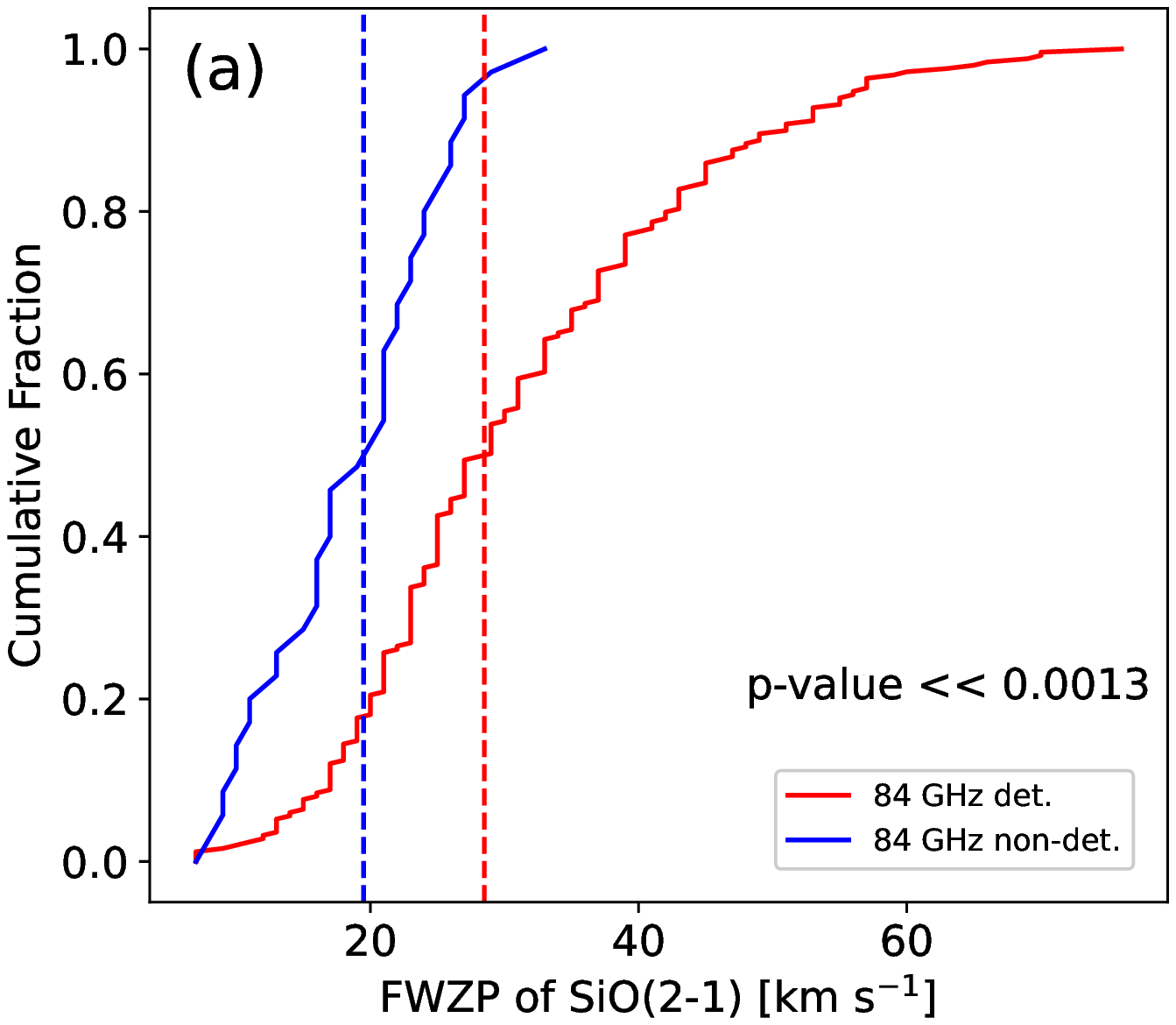}
\includegraphics[width=0.33\textwidth]{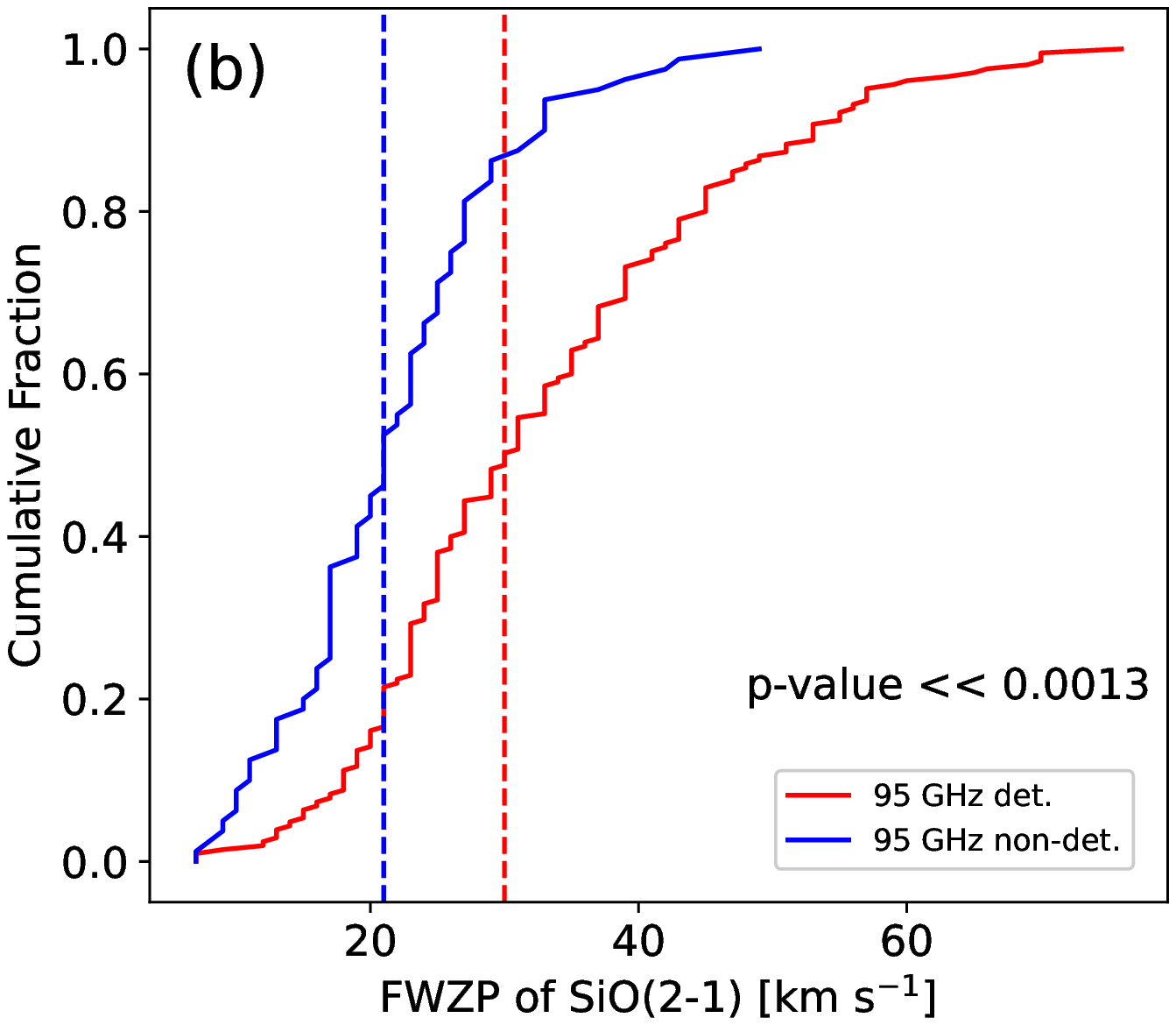}
\includegraphics[width=0.33\textwidth]{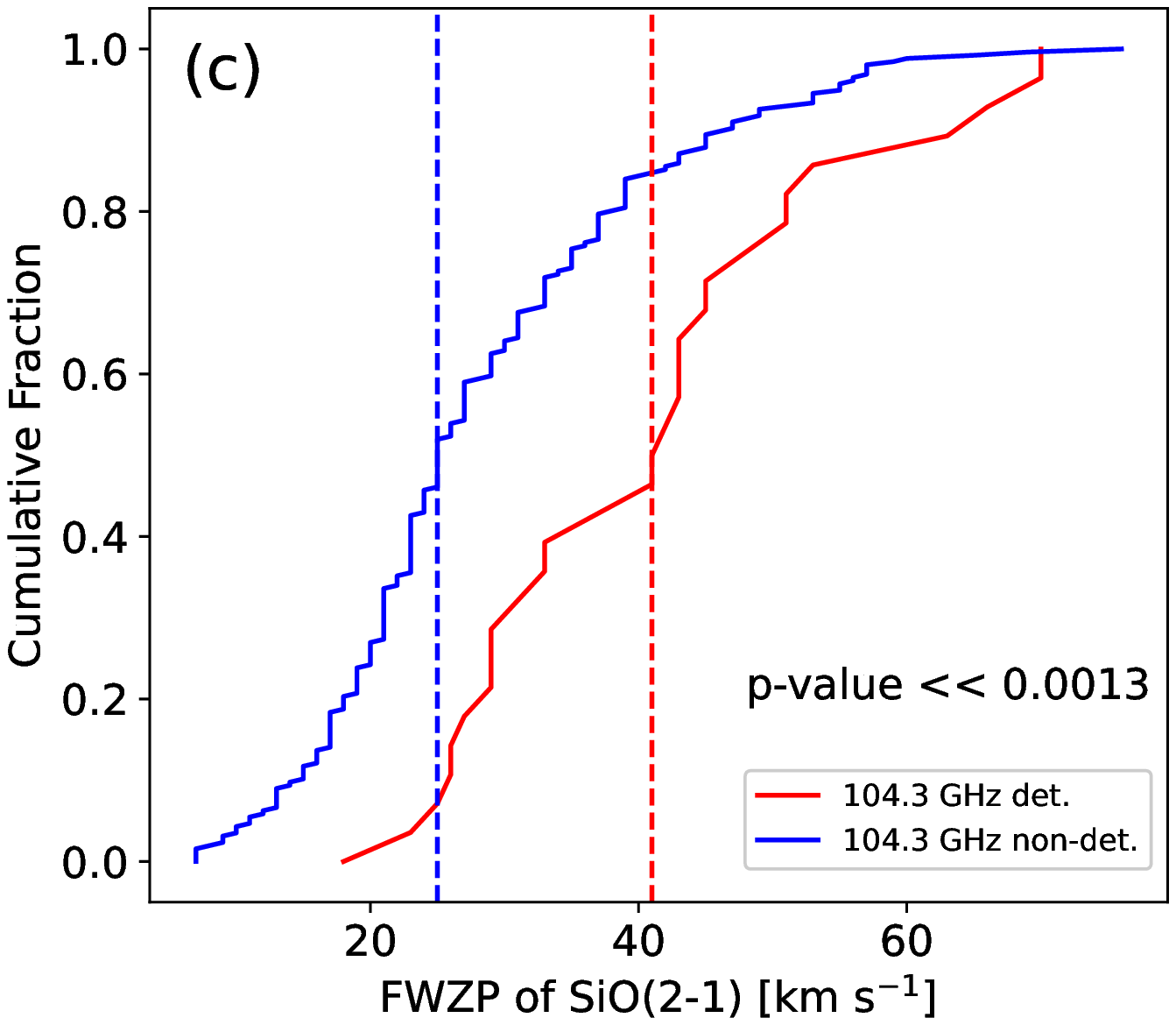}
\caption{Cumulative distribution functions of the FWZP of SiO (2--1) emission for sources with and without methanol detections. 
From panel a to c, the distributions are shown for 84, 95 and 104.3\,GHz methanol detections (red lines) and non-detections (blue lines), respectively.
The vertical dashed lines in the corresponding colors depict the median values of the two samples. 
The $p$-values from the K-S test are presented in the panel.
\label{fig:ks-fwzp}}
\end{figure*}

\begin{figure*}[htbp]
\includegraphics[width=0.33\textwidth]{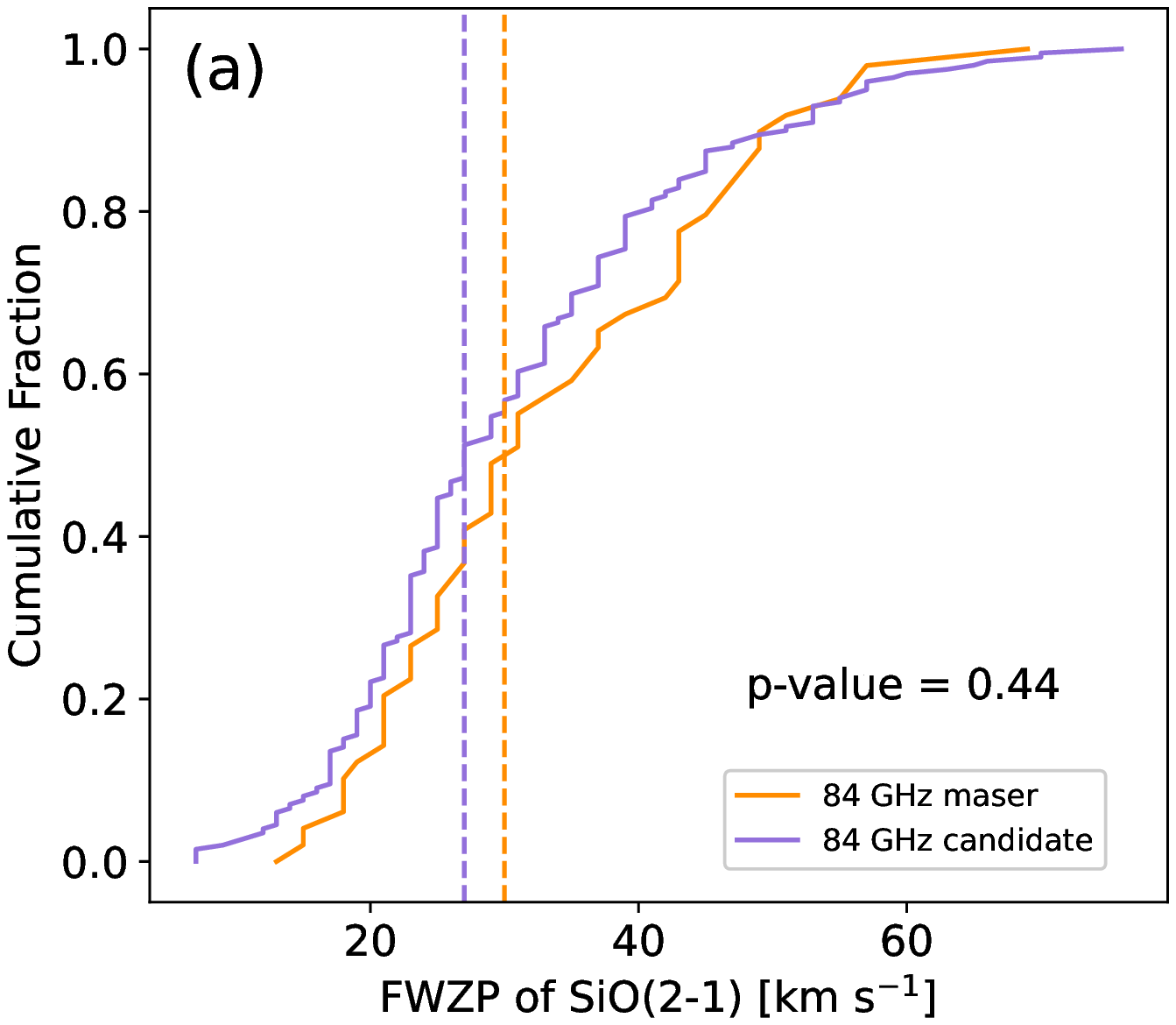}
\includegraphics[width=0.33\textwidth]{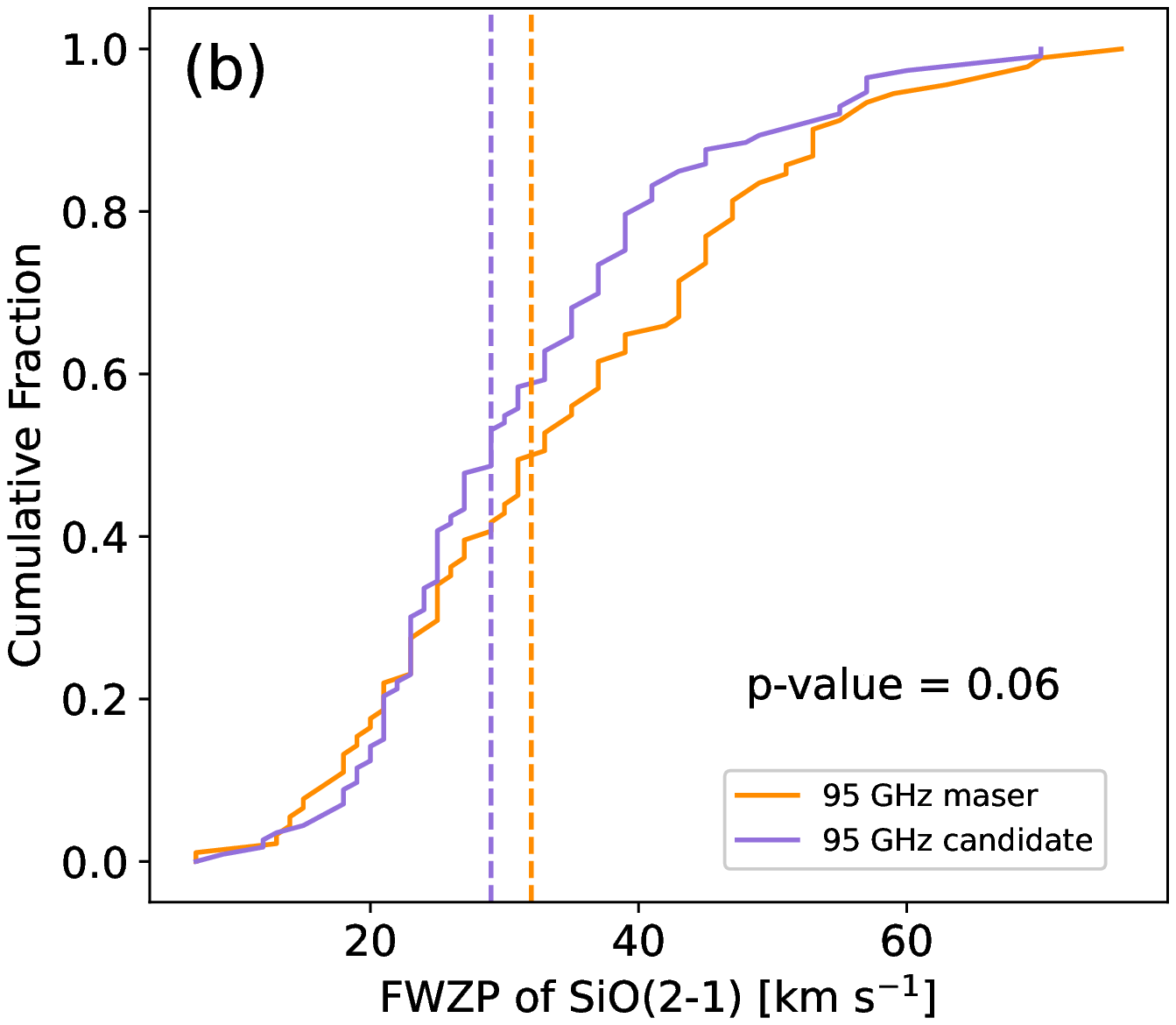}
\caption{Cumulative distribution functions of the FWZP of SiO (2--1) emission for masers and maser candidates. 
The panels of (a) and (b) show the distributions for 84, 95\,GHz methanol maser (orange lines) and maser candidates (purple lines), respectively.
The vertical dashed lines in the corresponding colors depict the median values of the two samples. 
The $p$-values from the K-S test are presented in the panel.
\label{fig:ks-fwzp-maser}}
\end{figure*}

\cite{2007ApJ...668..348B} and \cite{2019ApJ...878...29L} divided their observed SiO spectra by the FWZP into several regimes, and considered that sources with FWZP $>$20~\kms\,indicate gas at relatively high velocities with respect to the ambient gas.
We use the FWZP of SiO (2--1) which is taken from \cite{2016A&A...586A.149C} to roughly characterise shock speed, and consider that the broader FWZP traces the faster shock speed.
We perform a Kolmogorov-Smirnov (K-S) test to study the FWZP distribution for the sources with and without methanol detection.
Figure~\ref{fig:ks-fwzp} shows that the FWZP of SiO emission for sources with and without methanol detection are significantly different, indicated by a very small $p$-value ($\ll$ 0.0013) from the K-S test.
This difference is shown in all of our three methanol transitions. Moreover, the sources with methanol detection show broader FWZPs of SiO emission than sources without methanol detection. 
The median value of the FWZP of SiO emission for sources showing 84 or 95\,GHz emission is about 30~\kms, while the SiO emission generally exhibits a larger FWZP with a median of about 40~\kms\,in sources detected in the 104.3\,GHz transition.
This suggests that the excitation of 104.3\,GHz methanol masers requires faster shock velocities than the 84 and 95\,GHz masers.
Figure~\ref{fig:ks-fwzp-maser} shows that the differences of SiO (2--1) FWZP between maser and maser candidates are not statistically significant at both 84 and 95\,GHz. 
This insignificant difference indicates that the shock speeds, at which maser and maser candidate emission are created, are similar on clump scales. 
It may also be due to our conservative classification that a significant portion of maser candidates host masers.

\begin{figure}[htbp]
\centering
\includegraphics[width=0.49\textwidth]{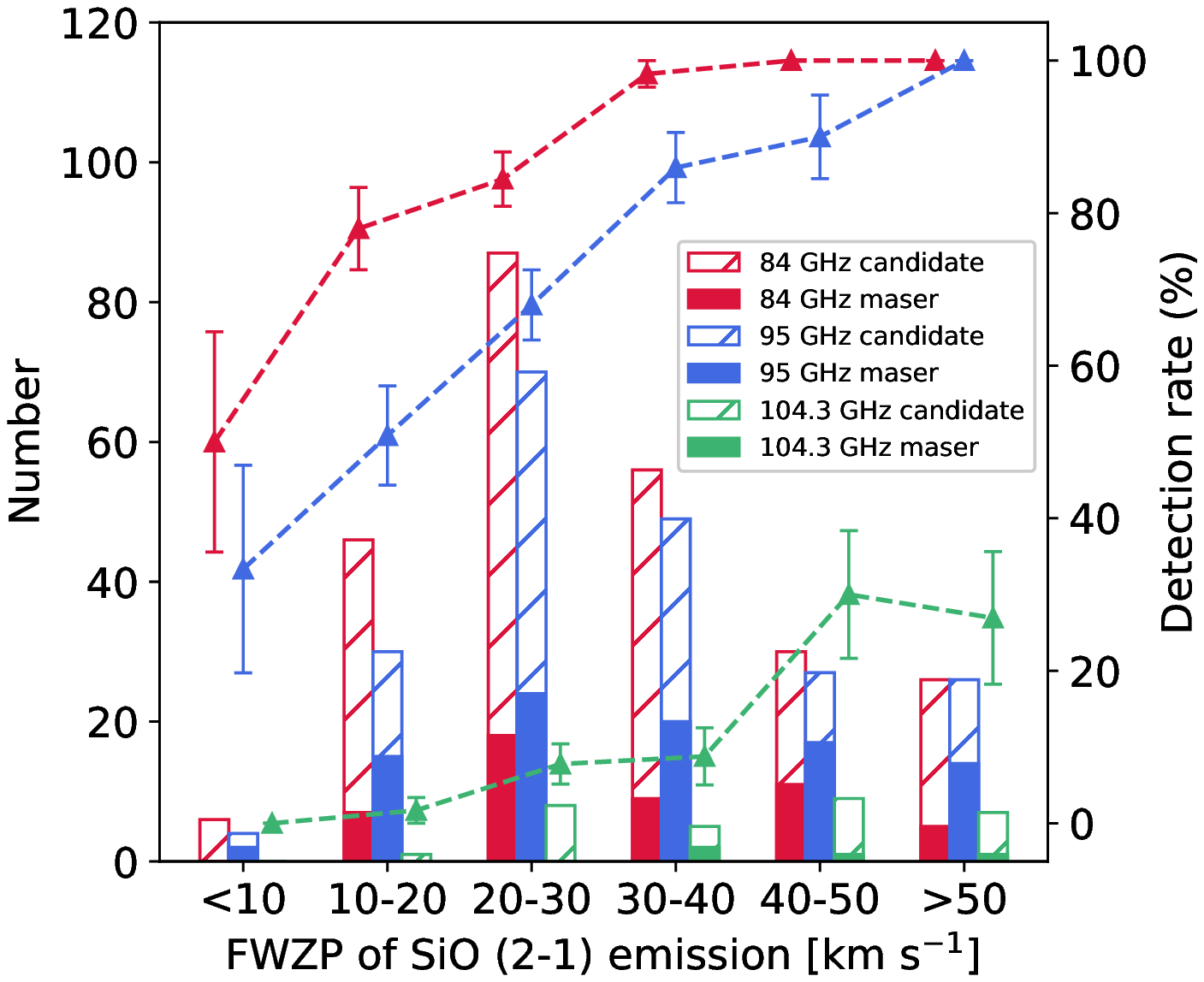}
\caption{Distribution of detection number and detection rate of three methanol transitions under different FWZP of SiO (2--1) emission.
The labels for the detection numbers are on the left side, while those for the detection rates are on the right side.
The red, blue and green bins present the detection numbers of methanol emission at 84, 95 and 104.3\,GHz, respectively, where the solid and hatched parts represent masers and maser candidates. 
The red, blue and green triangles indicate the detection rates of methanol emission at 84, 95 and 104.3\,GHz. 
\label{fig:hist-fwzp}}
\end{figure}

Figure~\ref{fig:hist-fwzp} shows the detection numbers and detection rates of the three methanol transitions in different FWZP ranges of SiO (2--1) emission.
For the 84 and 95\,GHz transitions, the detection numbers first increase with increasing FWZP of SiO emission and peak at a FWZP range of 20--30~\kms, then decrease as the FWZP increases.
When the FWZP $<$10~\kms, few sources show methanol emission, and only two out of four 95\,GHz detections show maser features.
With FWZP $>$30~\kms, the numbers of detected 84 and 95\,GHz emission gradually decrease, the detection rates for both lines keep rising (which due to a smaller number of sources), and reaching a detection rate of 100\% for FWZP $>$50~\kms. For the 104.3\,GHz transition, the detection rate also shows a tendency to increase with increasing FWZP, when FWZP $<$ 50~\kms. With FWZP $>$ 50~\kms, the detection rate drops slightly. 
This indicates that the faster shock speeds lead to the higher detection rates of methanol emission when shocks have not yet completely destroyed SiO and methanol molecules. This is consistent with the astrochemical modeling results that the methanol abundance is correlated with shock velocities \citep[e.g., see Fig. 3 in ][]{2017AJ....154...38H}. 

\begin{figure*}[htbp]
\centering
\begin{minipage}[b]{6cm}
\includegraphics[width=1.05\textwidth]{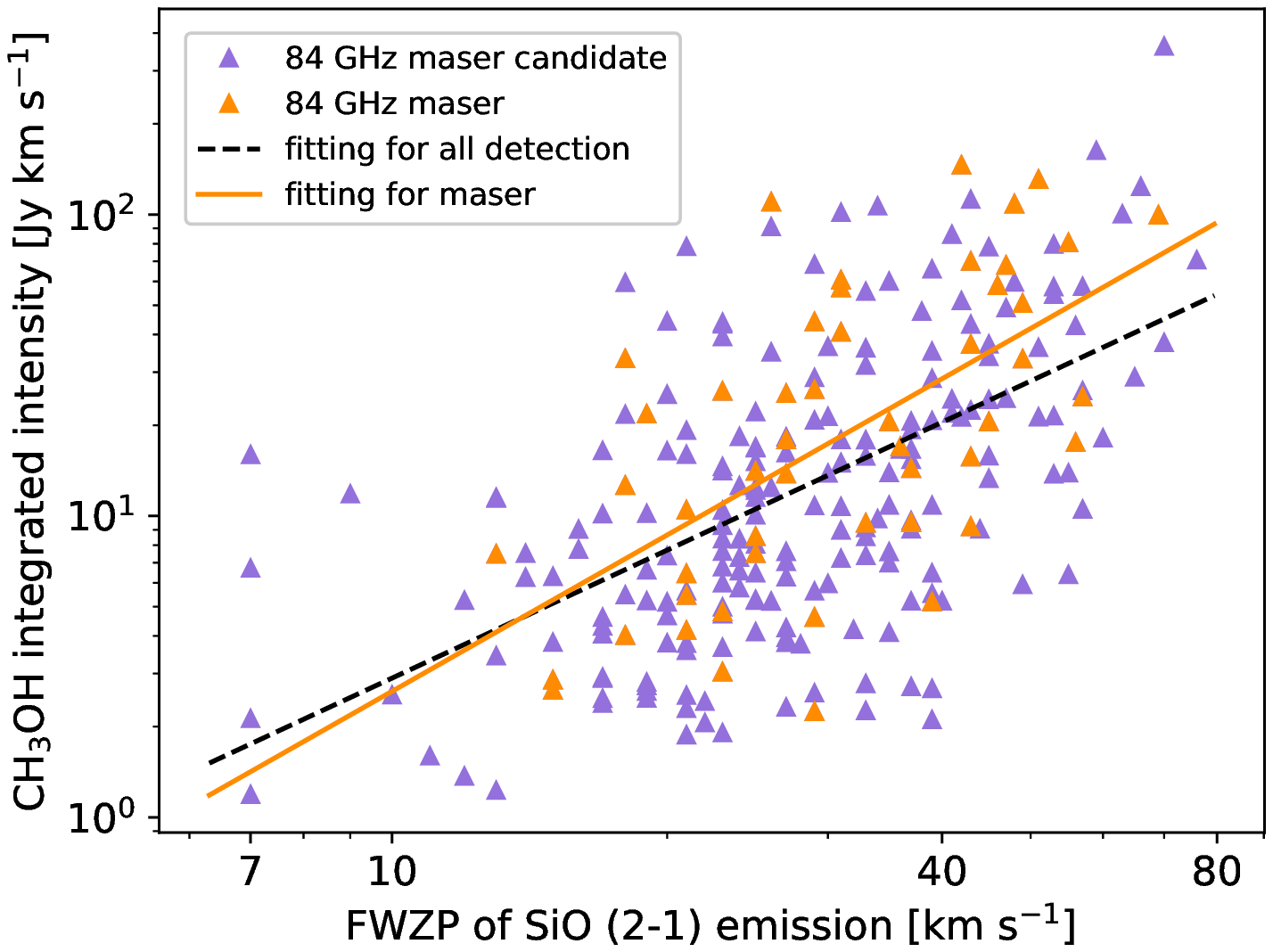}
\centerline{(a)}
\end{minipage}
\begin{minipage}[b]{6cm}
\includegraphics[width=1.05\textwidth]{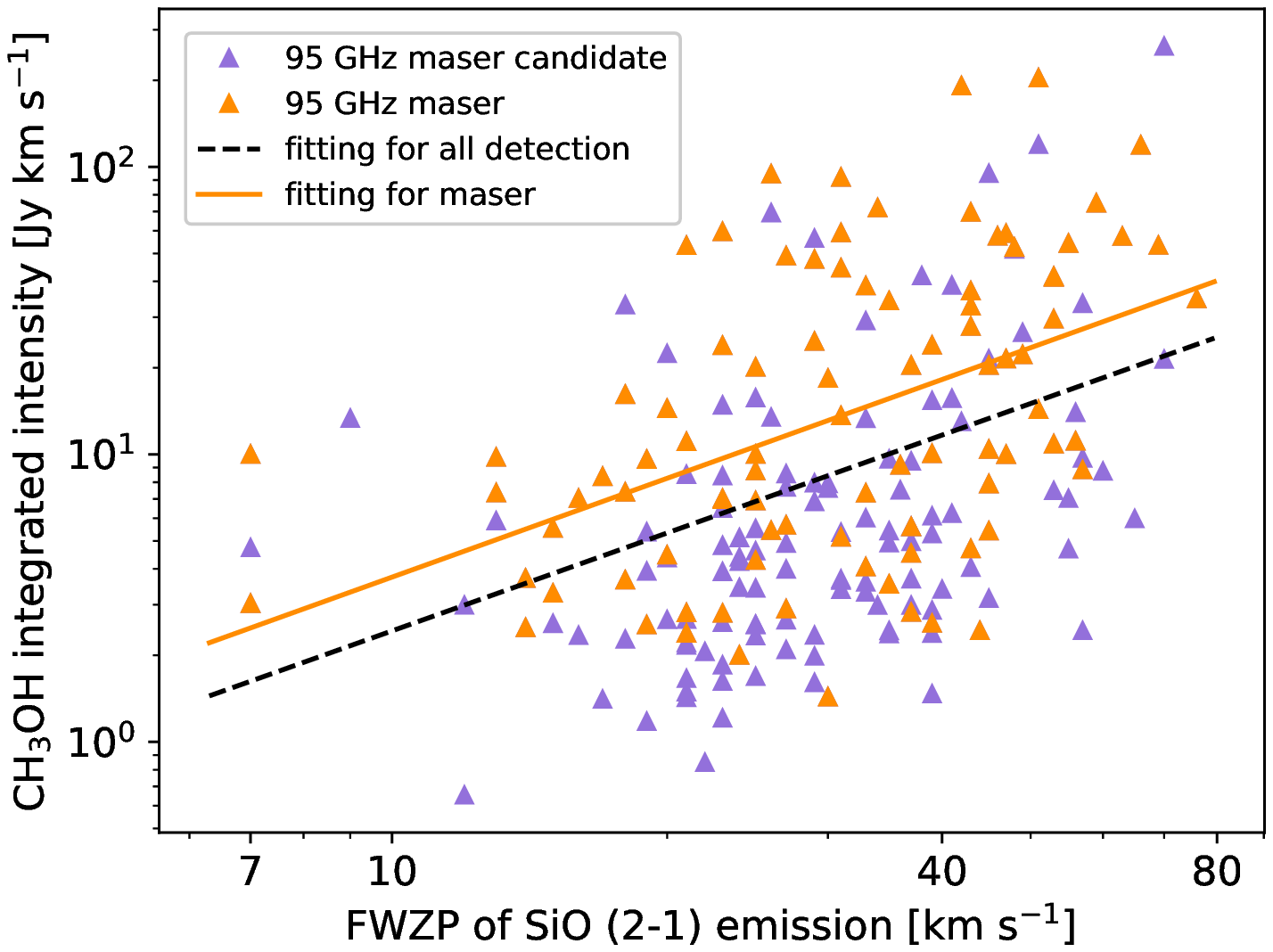}
\centerline{(b)}
\end{minipage}
\begin{minipage}[b]{6cm}
\includegraphics[width=1.05\textwidth]{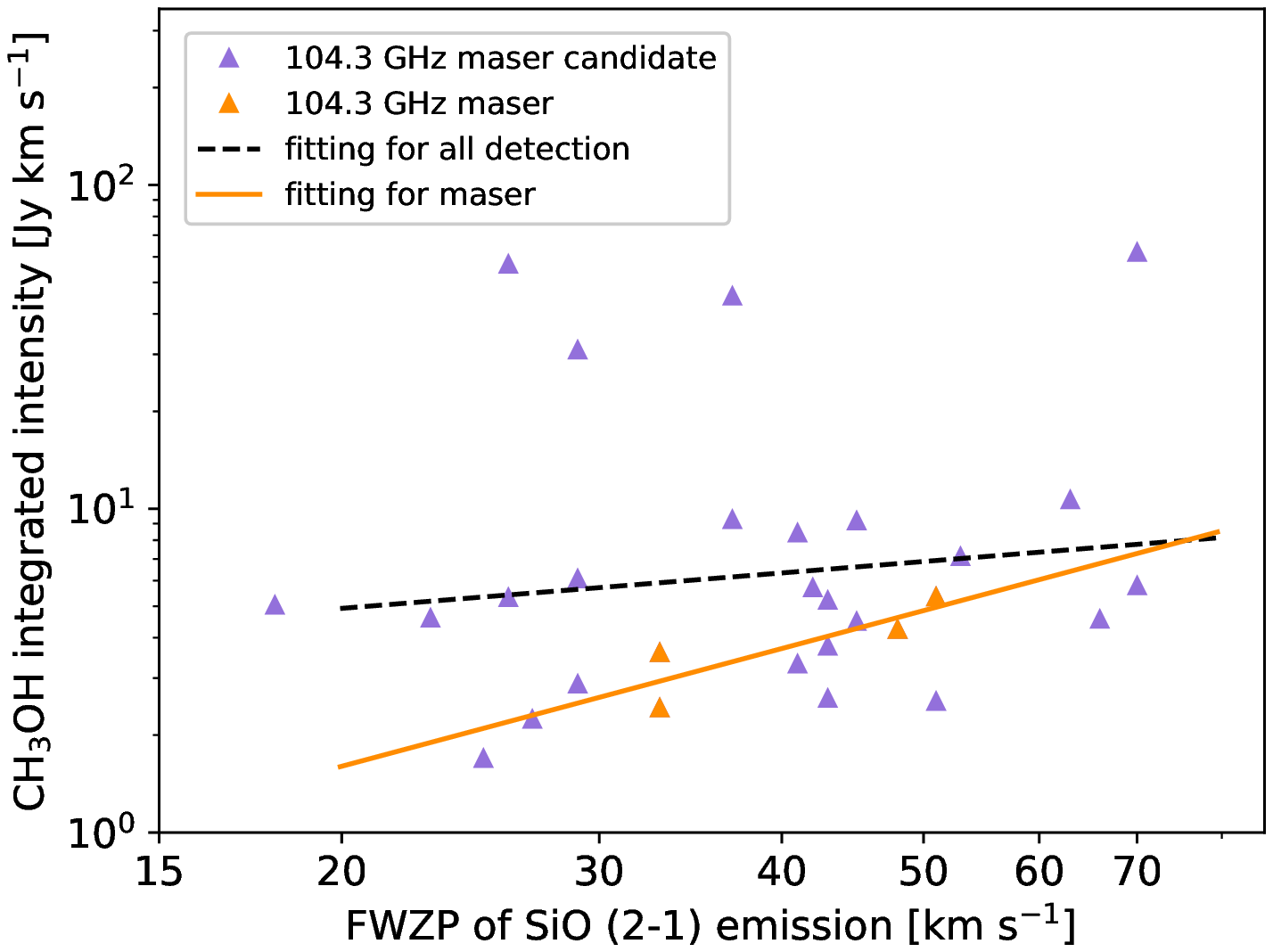}
\centerline{(c)}
\end{minipage}
\caption{Methanol integrated intensity as a function of the FWZP of SiO (2--1) for the 84, 95 and 104.3 GHz detection, respectively. The orange and purple triangles represent the clumps hosting masers and maser candidates, respectively. The orange solid lines and black dashed lines depict the least-square fitting results for methanol maser and all methanol detections, respectively.
\label{fig:fwzp-inte}}
\end{figure*}

We further quantify the relationship between methanol emission and the SiO (2--1) FWZP in Figure~\ref{fig:fwzp-inte}, which shows upward trends of 84, 95 and 104.3\,GHz methanol integrated intensity with increasing FWZP values of the SiO (2--1) emission. This further supports that the integrated intensities of class I methanol emission increase with increasing shock speeds. 
The linear fitting results in log-log form are summarised in Table~\ref{Tab:sio-fit}.



\subsection{Physical properties on a clump scale} \label{sec:atlasgal}

\begin{figure*}[!htbp]
\centering
\mbox{
\begin{minipage}[b]{6cm}
\includegraphics[width=1.05\textwidth]{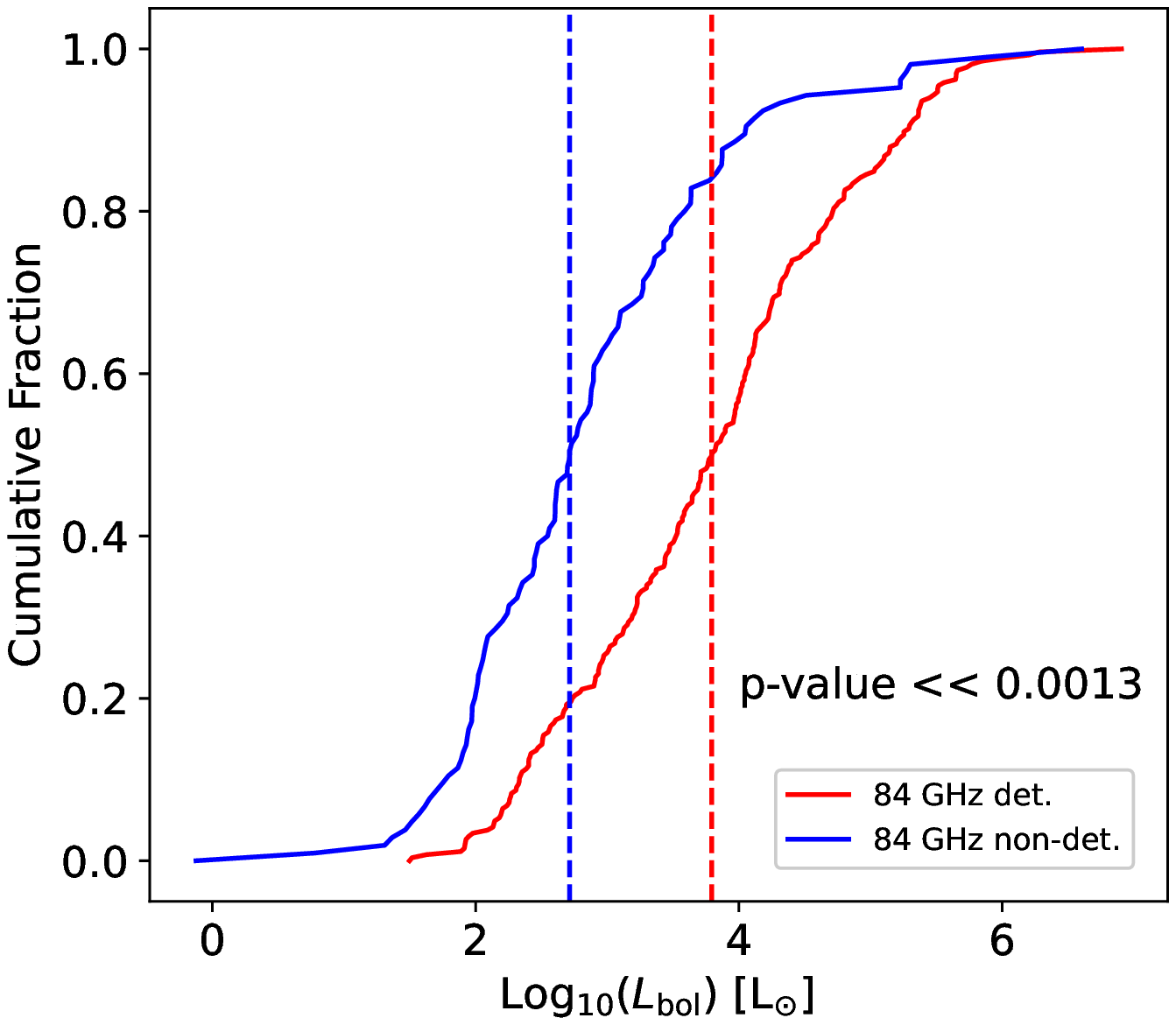}
\centerline{(a)}
\vspace{-3.5mm}
\end{minipage}
\begin{minipage}[b]{6cm}
\includegraphics[width=1.05\textwidth]{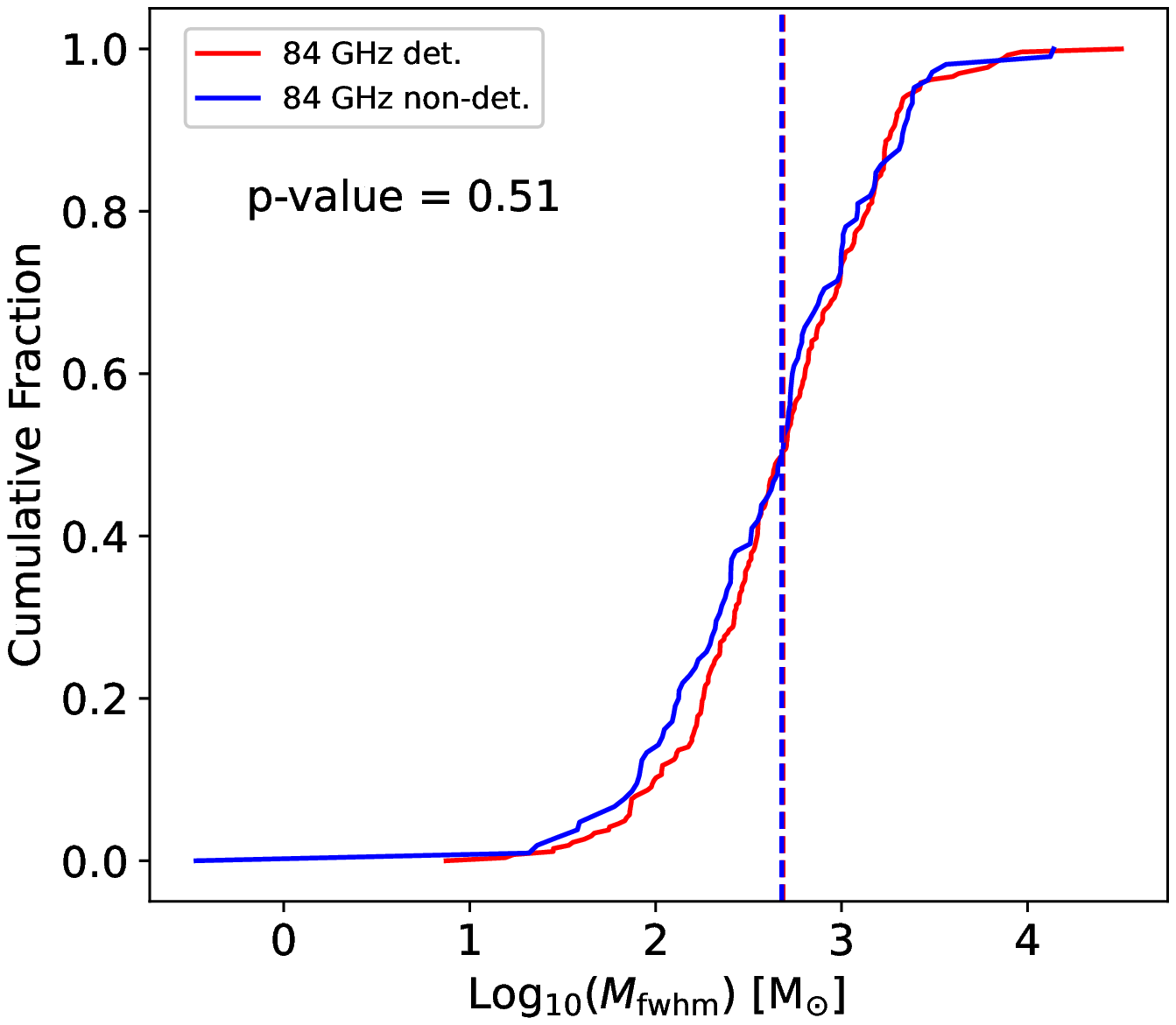}
\centerline{(b)}
\vspace{-3.5mm}
\end{minipage}
\begin{minipage}[b]{6cm}
\includegraphics[width=1.05\textwidth]{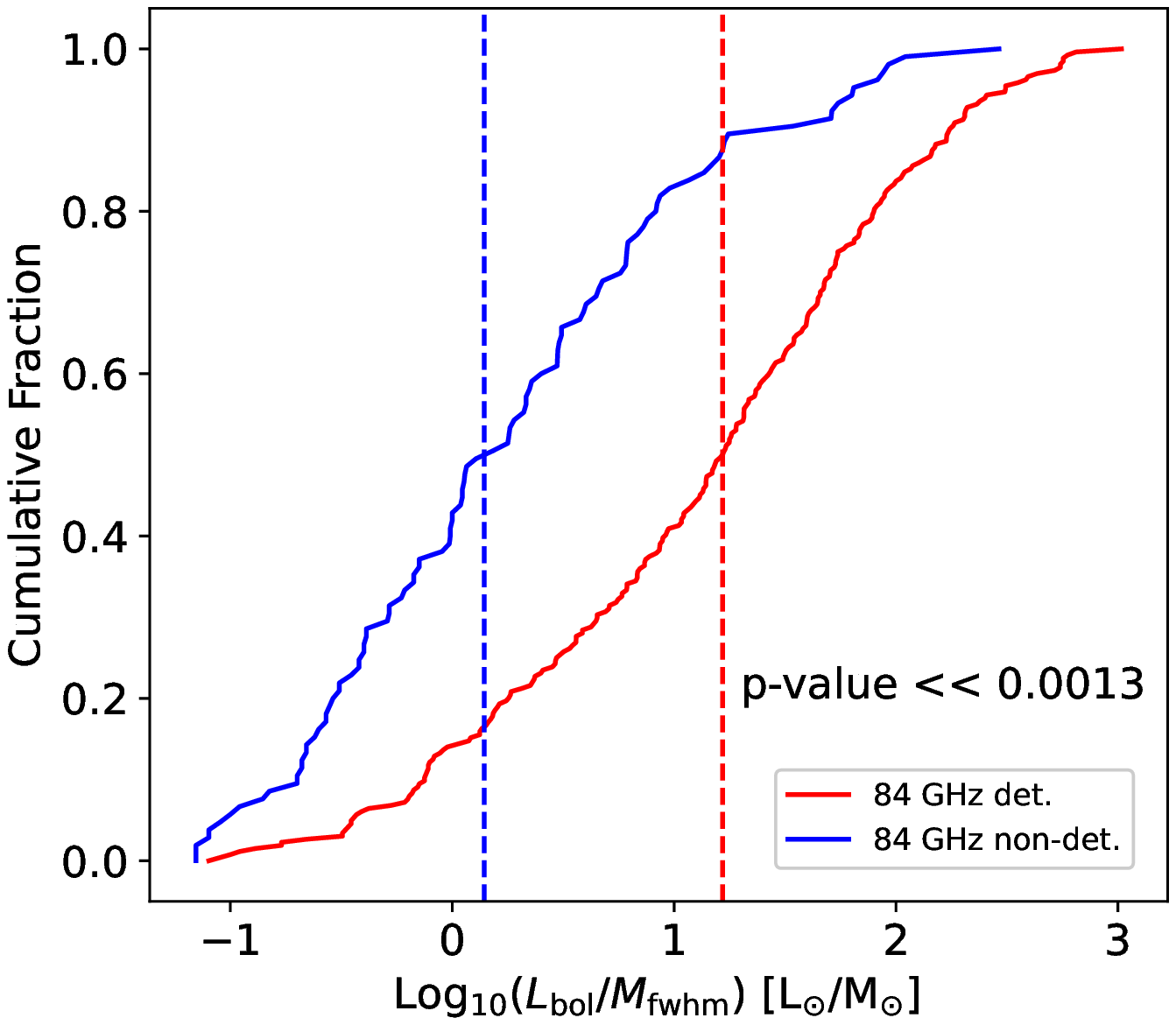}
\centerline{(c)}
\vspace{-3.5mm}
\end{minipage}
}
\mbox{
\begin{minipage}[b]{6cm}
\includegraphics[width=1.05\textwidth]{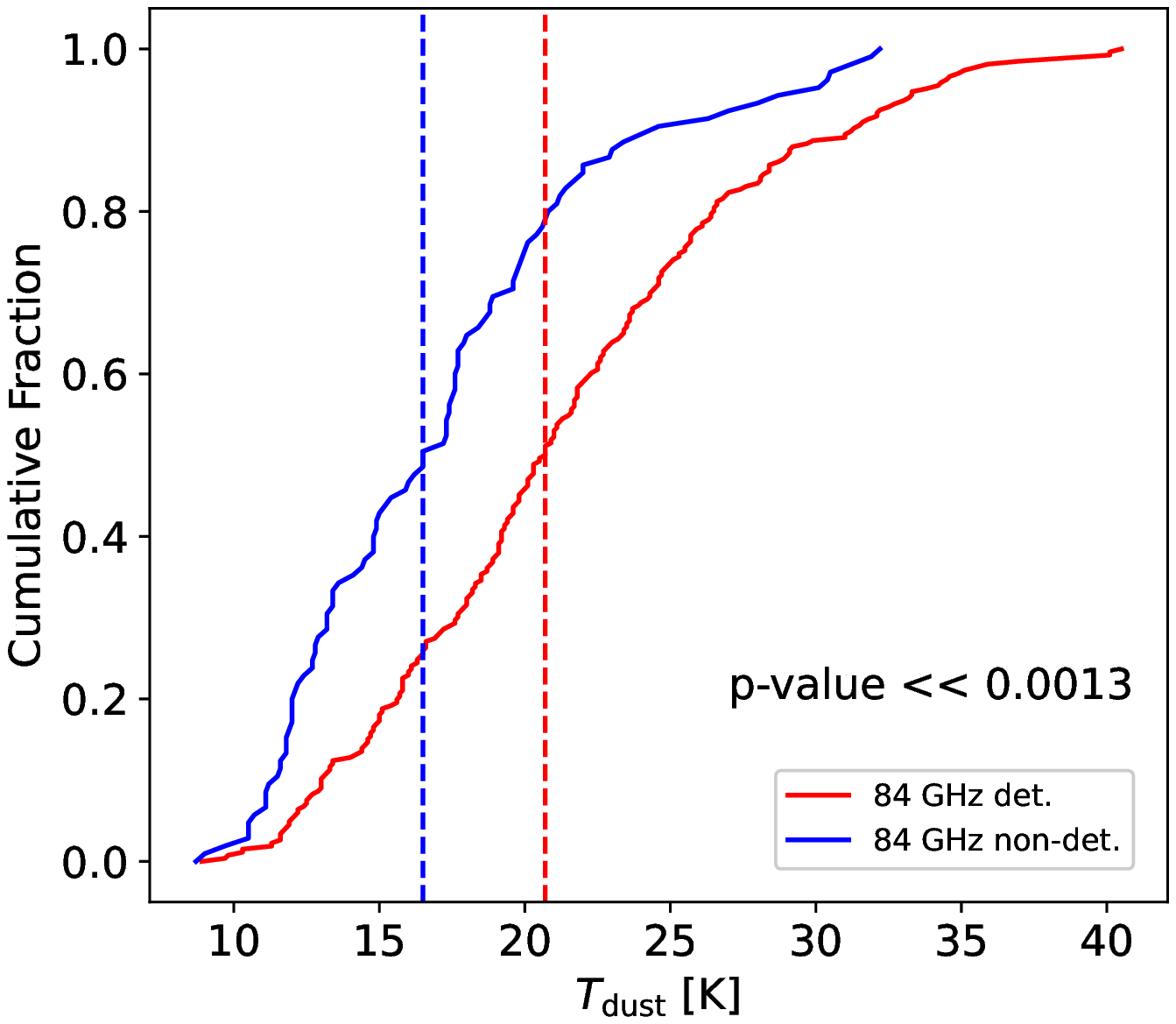}
\centerline{(d)}
\vspace{-3.5mm}
\end{minipage}
\begin{minipage}[b]{6cm}
\includegraphics[width=1.05\textwidth]{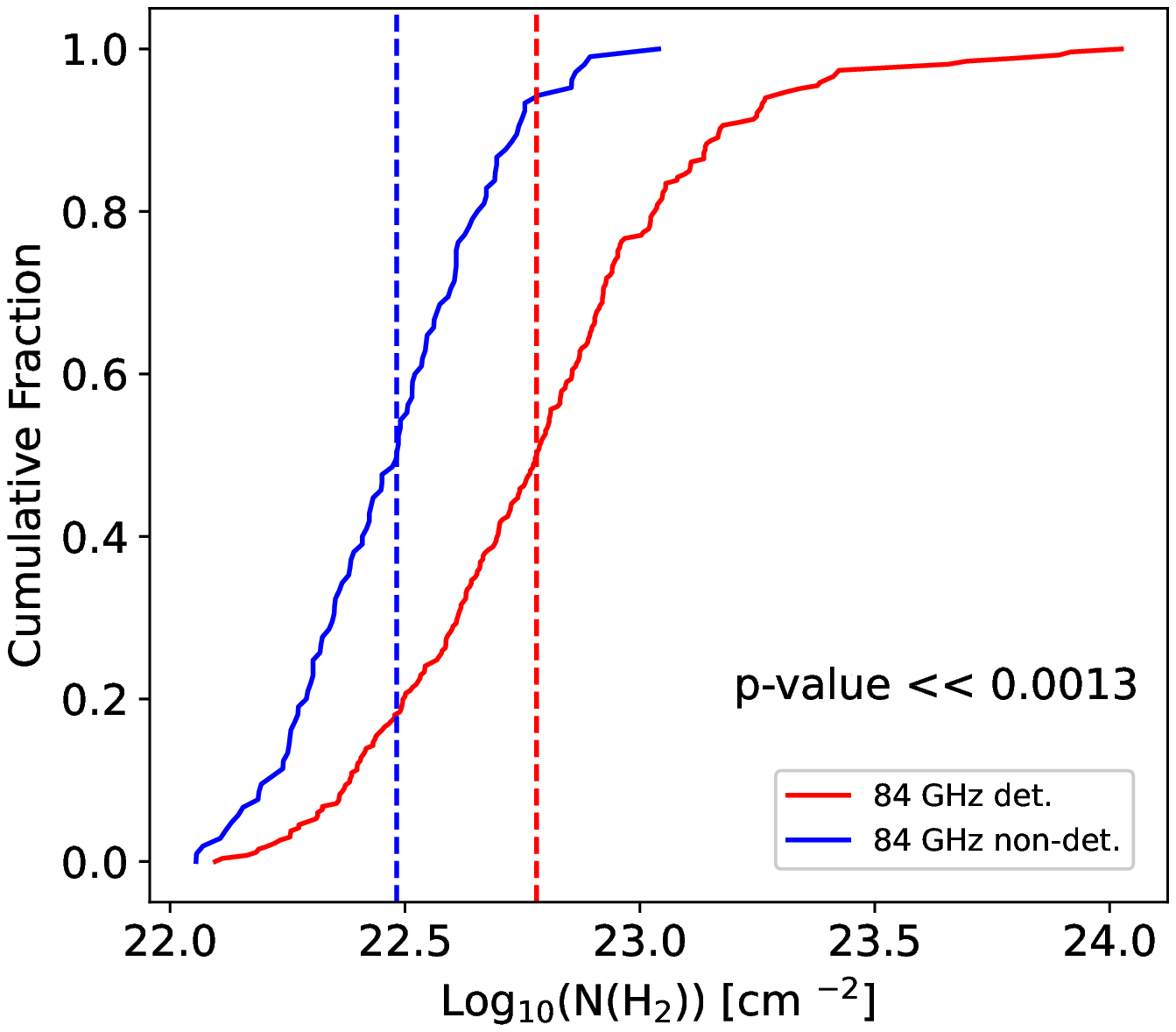}
\centerline{(e)}
\vspace{-3.5mm}
\end{minipage}
\begin{minipage}[b]{6cm}
\includegraphics[width=1.05\textwidth]{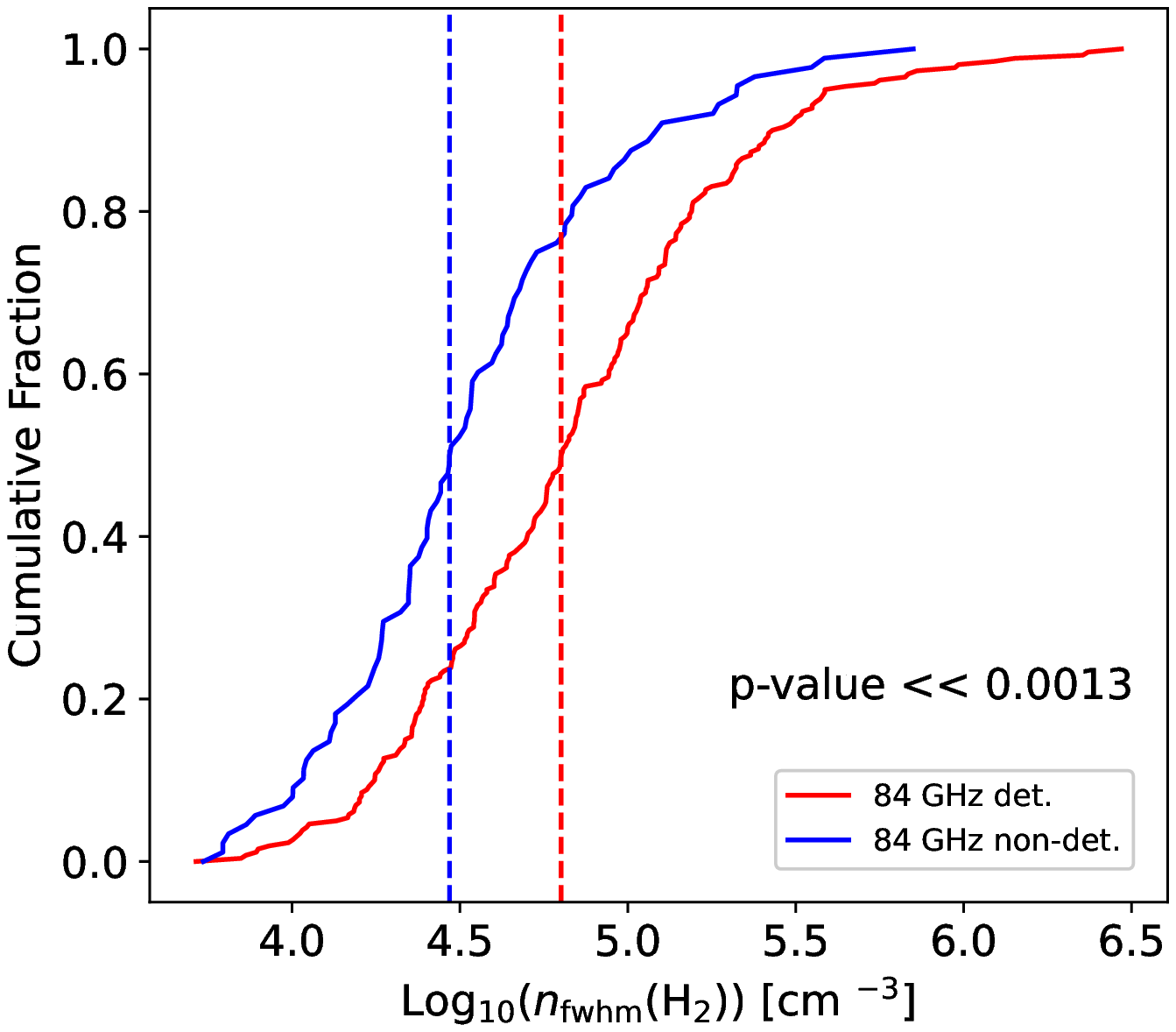}
\centerline{(f)}
\vspace{-3.5mm}
\end{minipage}
}
\caption{Cumulative distribution functions of the physical properties for the clumps with and without 84 GHz methanol detections. From (a) to (f), cumulative distributions of the bolometric luminosity, the FWHM clump mass, the luminosity-to-mass ratio, the dust temperature, the peak H$_2$ column density, the mean H$_2$ FWHM volume density for 84 GHz methanol detections (red lines) and non-detections (blue lines), respectively. The vertical dashed lines in the corresponding colors depict the median values of the two samples. The p-values from the K-S tests are presented in each panel.
\label{fig:84-ks-atlasgal}}
\end{figure*}

\begin{figure*}[!htbp]
\centering
\mbox{
\begin{minipage}[b]{6cm}
\includegraphics[width=1.05\textwidth]{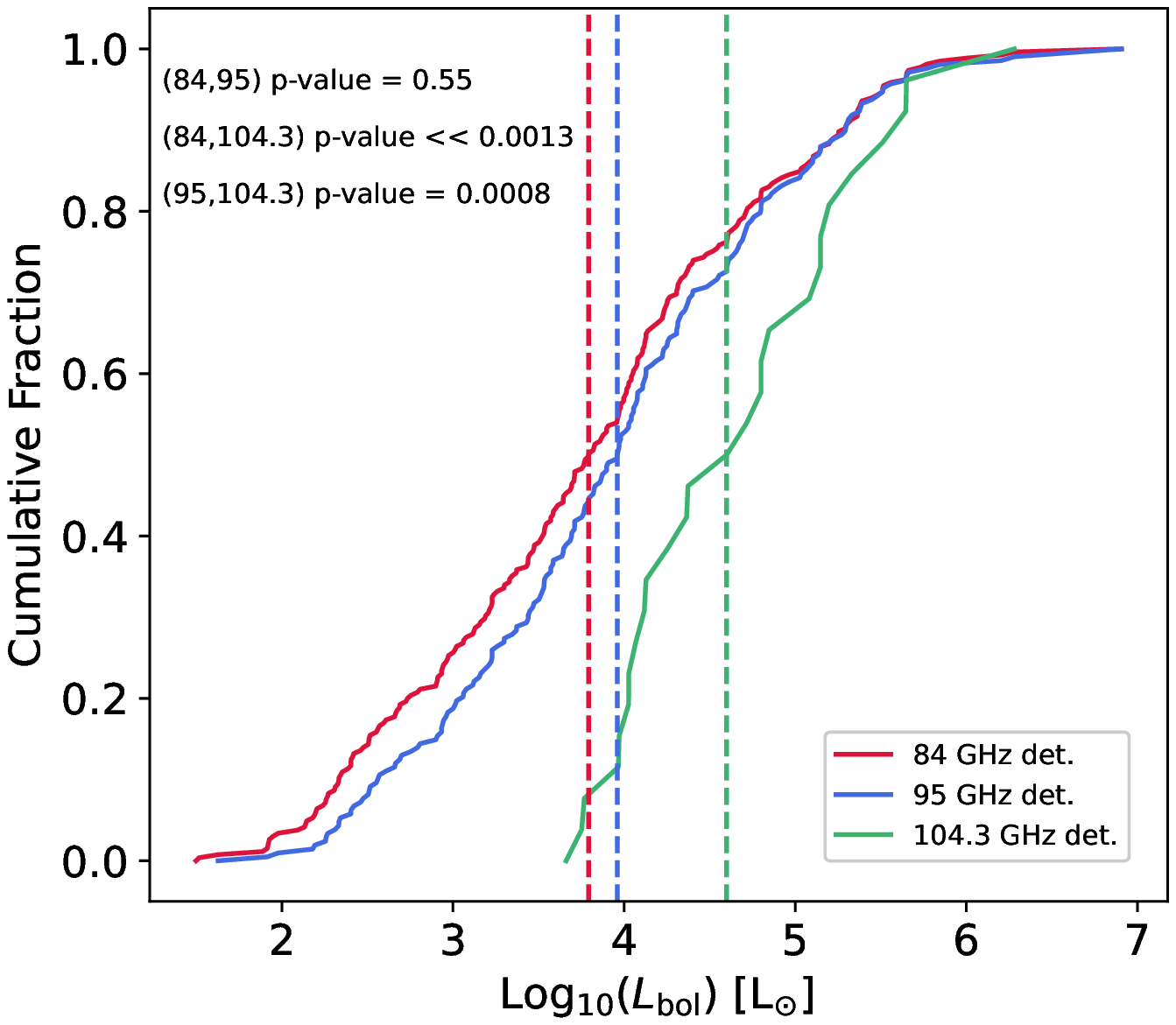}
\centerline{(a)}
\vspace{-3.5mm}
\end{minipage}
\begin{minipage}[b]{6cm}
\includegraphics[width=1.05\textwidth]{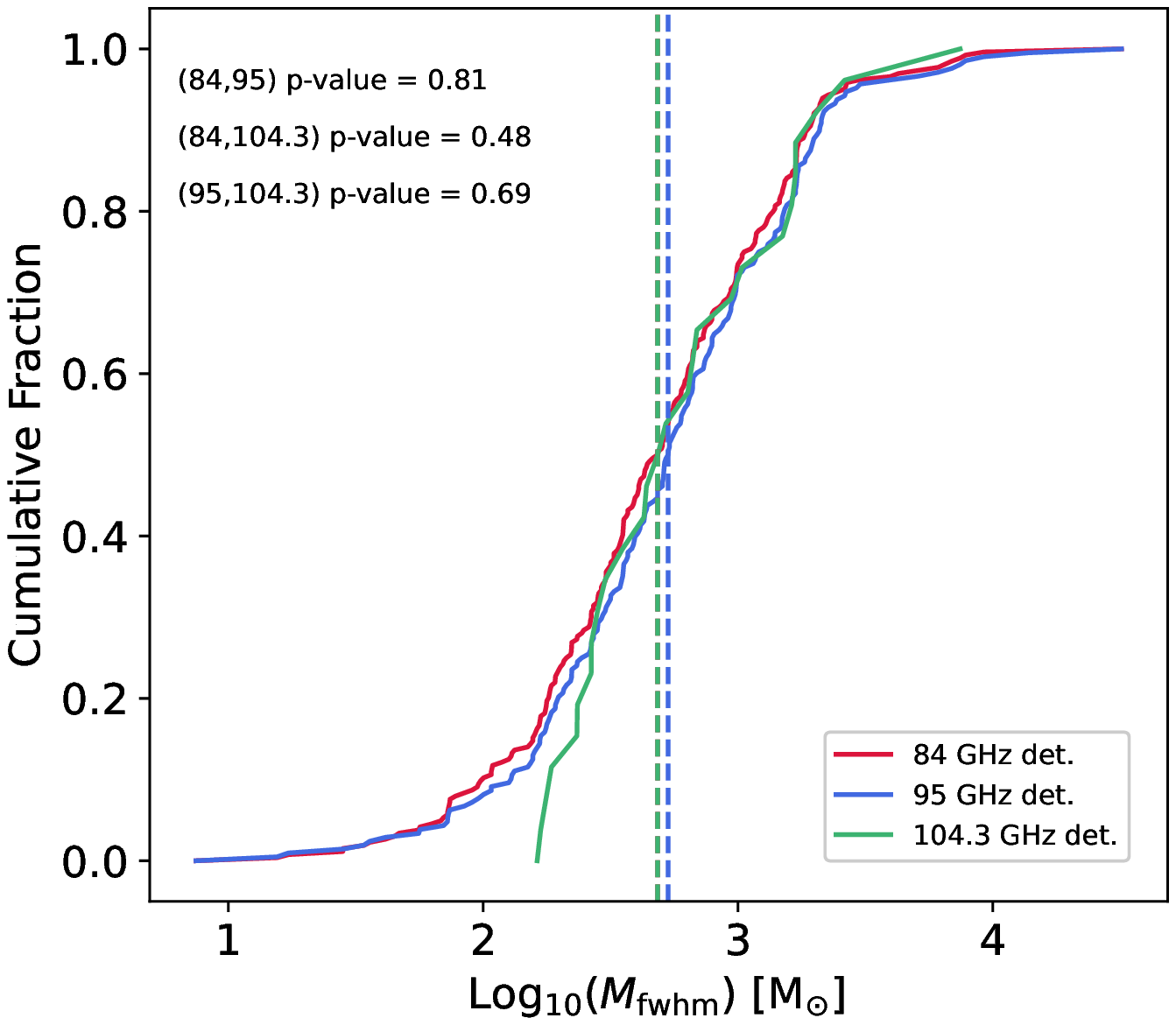}
\centerline{(b)}
\vspace{-3.5mm}
\end{minipage}
\begin{minipage}[b]{6cm}
\includegraphics[width=1.05\textwidth]{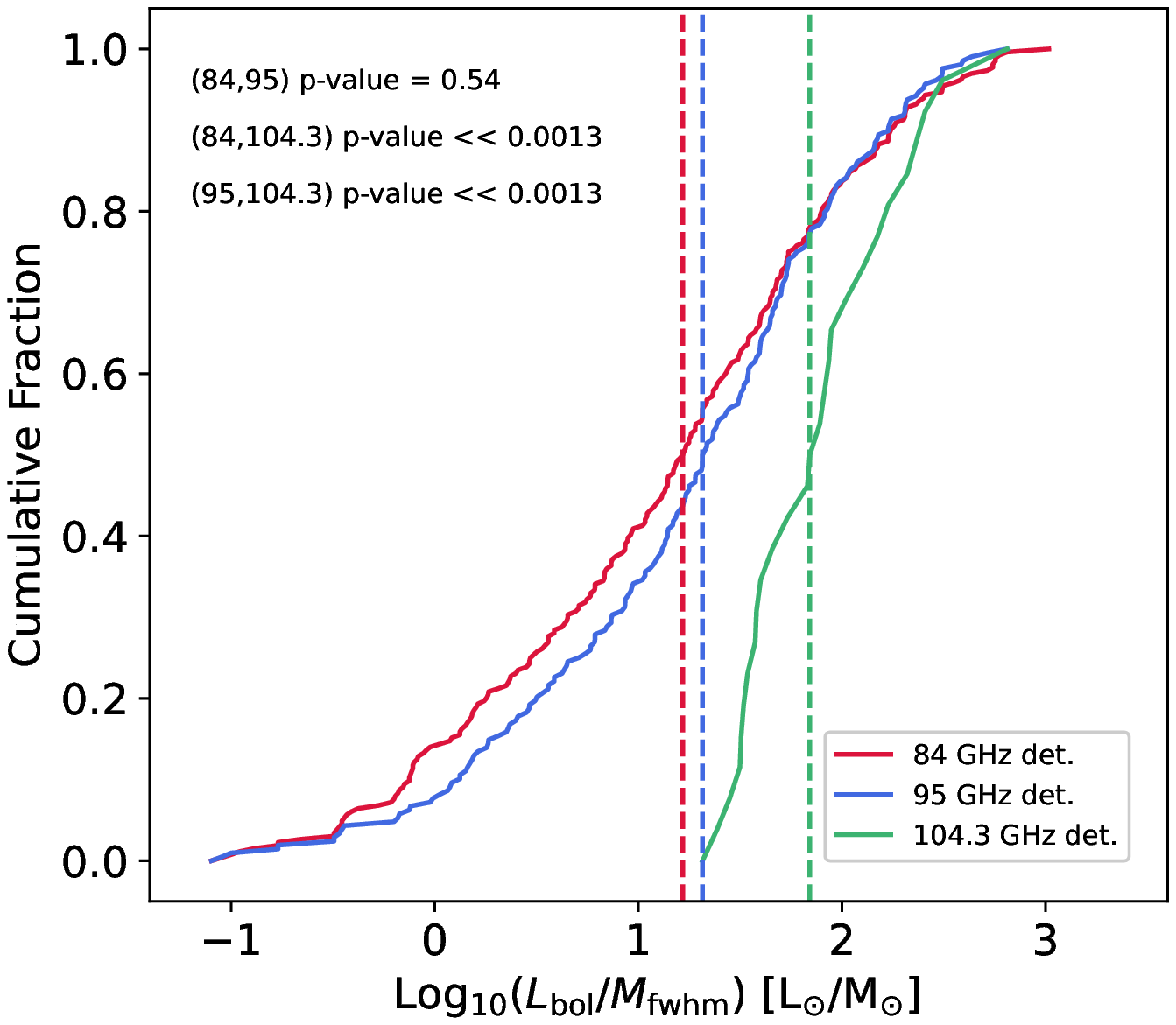}
\centerline{(c)}
\vspace{-3.5mm}
\end{minipage}
}
\mbox{
\begin{minipage}[b]{6cm}
\includegraphics[width=1.05\textwidth]{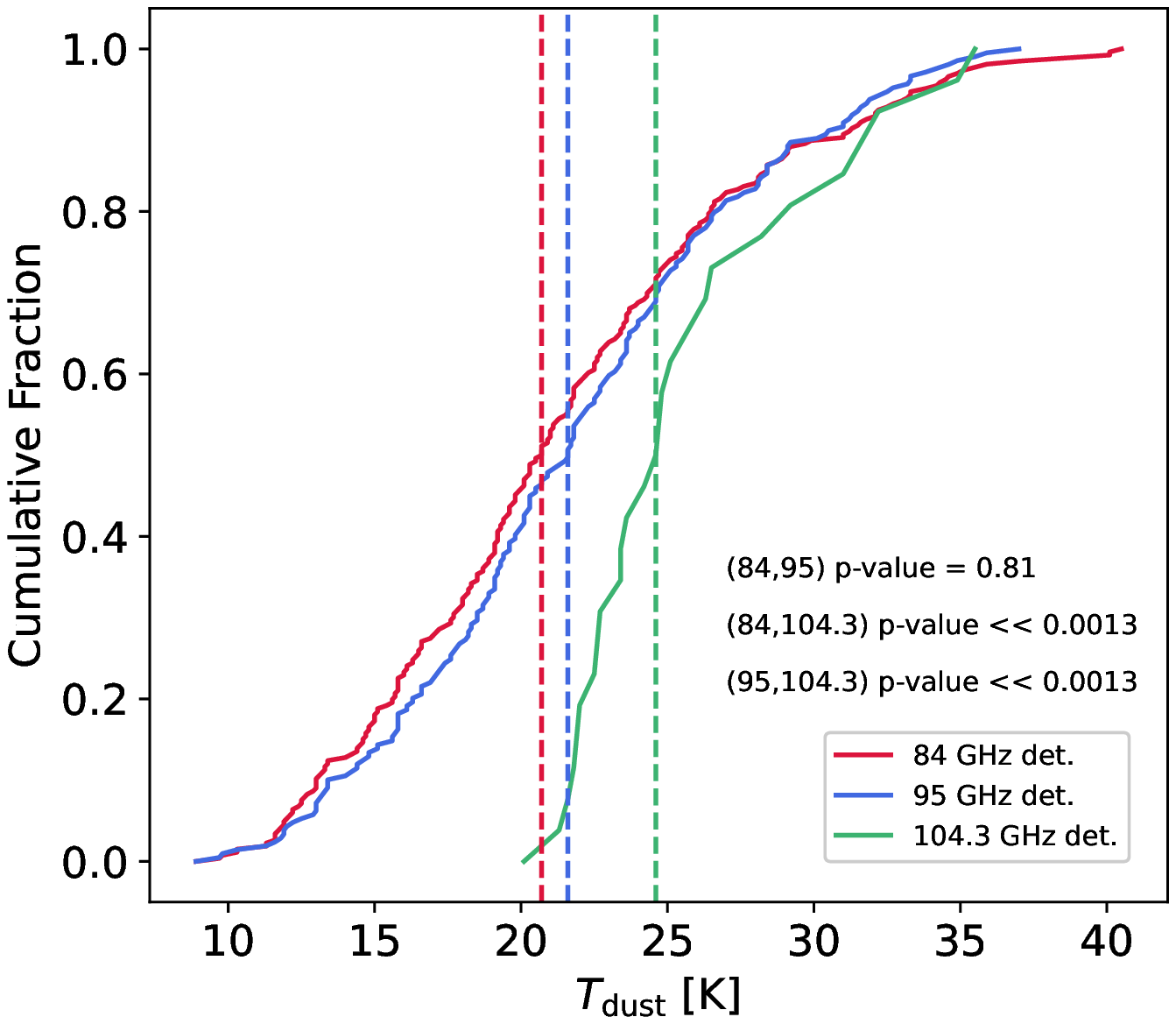}
\centerline{(d)}
\vspace{-3.5mm}
\end{minipage}
\begin{minipage}[b]{6cm}
\includegraphics[width=1.05\textwidth]{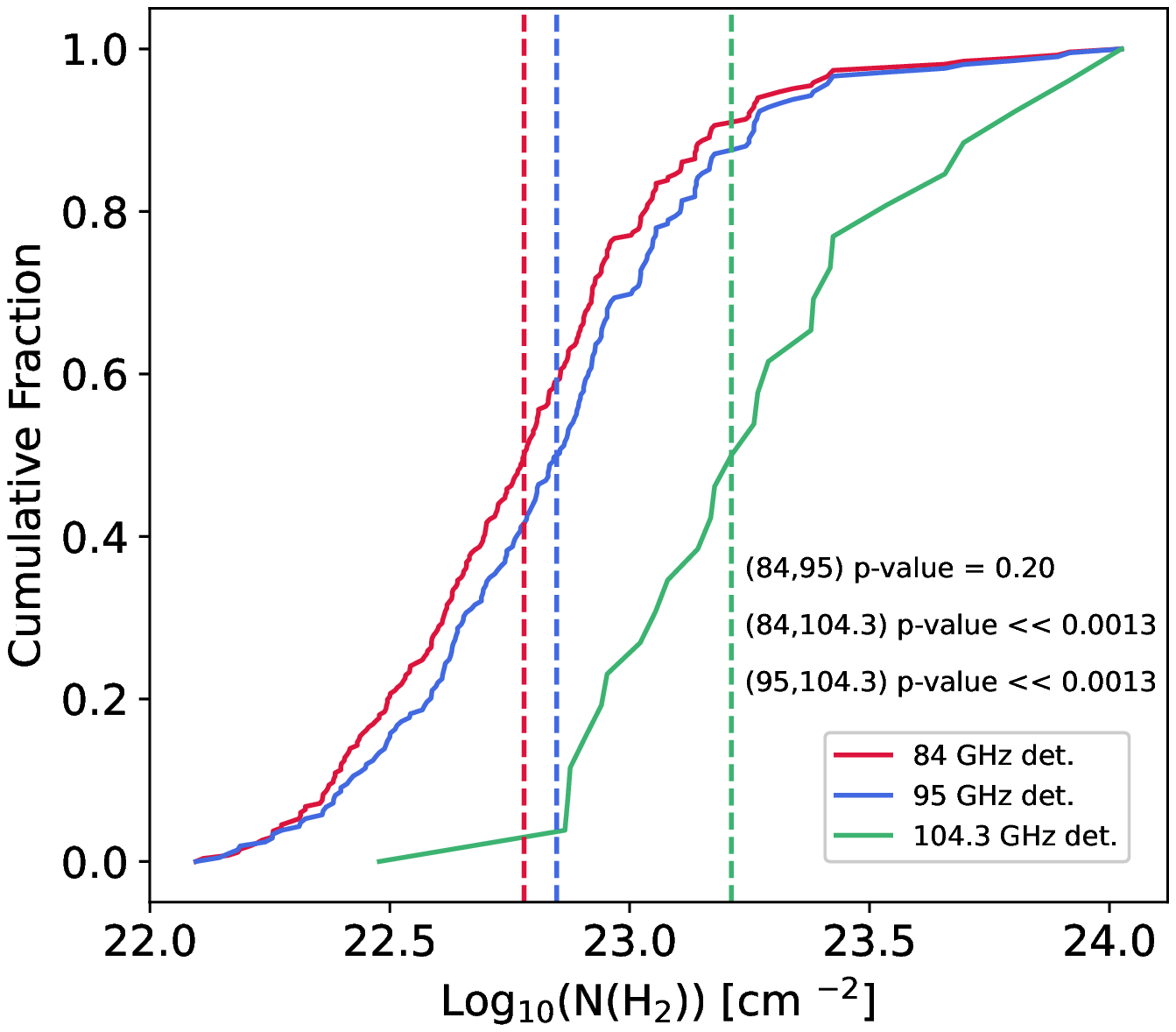}
\centerline{(e)}
\vspace{-3.5mm}
\end{minipage}
\begin{minipage}[b]{6cm}
\includegraphics[width=1.05\textwidth]{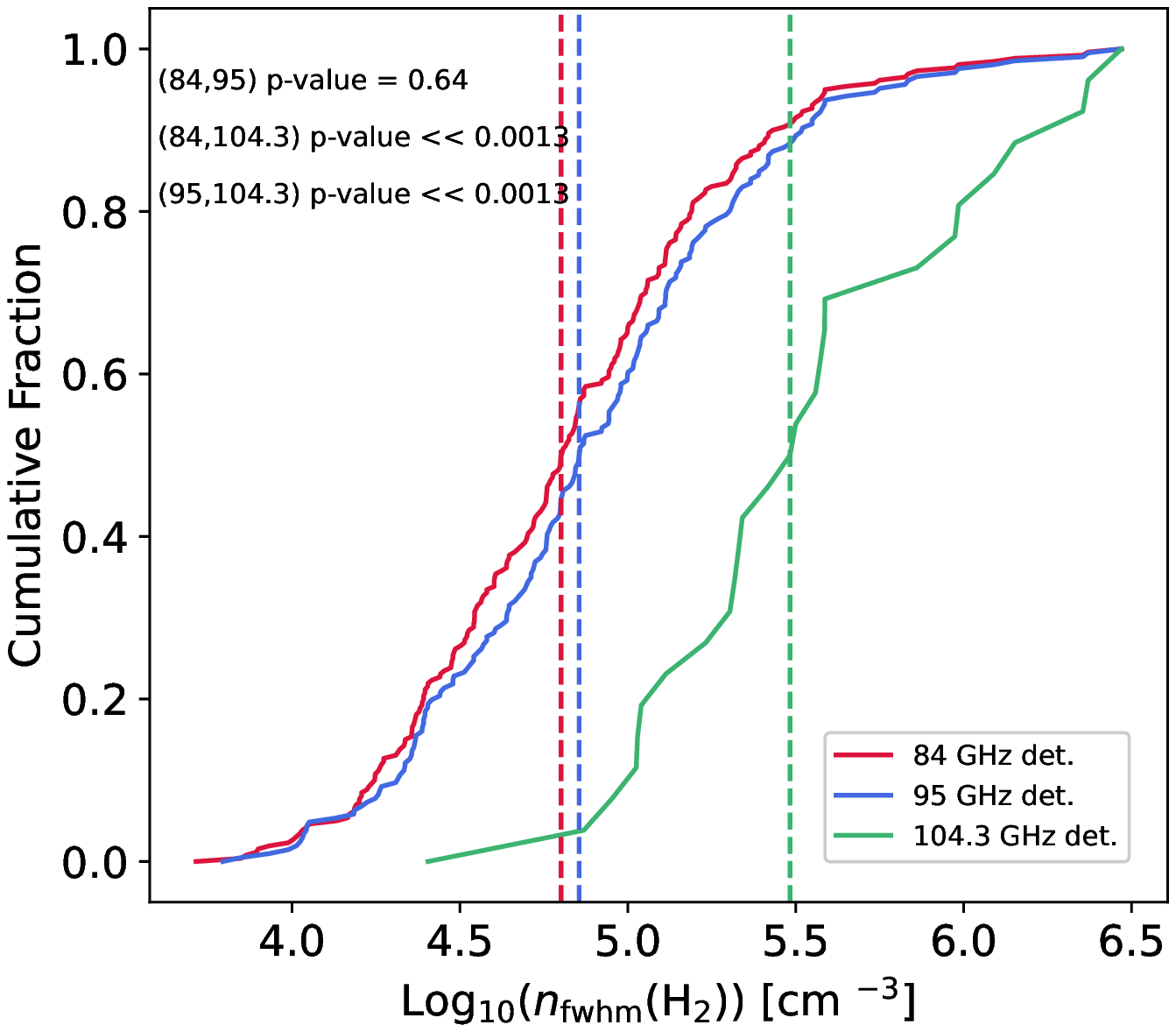}
\centerline{(f)}
\vspace{-3.5mm}
\end{minipage}
}
\caption{Cumulative distribution functions of the physical properties for the clumps with 84, 95 and 104.3 GHz methanol detections. Panels (a) to (f) are similar to the corresponding one in Fig.~\ref{fig:84-ks-atlasgal}. The vertical dashed lines in the corresponding colors depict the median values of each sample. 
The p-values from the K-S tests between each two methanol emission are presented in each panel.
\label{fig:all-ks-atlasgal}}
\end{figure*}

We make use of the latest catalog of ATLASGAL physical clump properties \citep{2022MNRAS.510.3389U} to study the relationship between the physical parameters of clumps and associated class I methanol (maser) emission detections.
The clump parameters used for analysis include the bolometric luminosity, $L_{\rm bol}$, the FWHM clump mass, $M_{\rm fwhm}$, the luminosity-to-mass ratio, $L_{\rm bol}/M_{\rm fwhm}$, the dust temperature ,$T_{\rm dust}$,  the mean H$_2$ FWHM volume density, $n_{\rm fwhm}(\rm H_2)$, as well as the peak H$_2$ column density, $N(\rm H_2)$ 
\citep{2018MNRAS.473.1059U}.
The FWHM clump mass is calculated using the integrated 870-$\mu$m flux density within the 50 per cent contour of the peak of the ATLASGAL continuum emission, and the mean H$_2$ FWHM volume density is calculated using the $M_{\rm fwhm}$ and the FWHM clump size, which is determined by the number of pixels within the 50 per cent contour
(see details in Equations 3 and 5 in \citealt{2022MNRAS.510.3389U}).
These FWHM parameters eliminate observational bias due to source evolution, as demonstrated by \cite{2019MNRAS.490.2779B}. 

Figure~\ref{fig:84-ks-atlasgal} presents the cumulative distributions of the clump physical properties for sources with and without 84\,GHz methanol detection. 
The K-S tests reveal that except for the FWHM clump mass, the properties of clumps with and without 84\,GHz methanol detection are significantly different.
The situation is similar in our sample of 95 and 104.3\,GHz sources; their cumulative distributions are shown in Figs.~\ref{fig:95-ks-atlasgal} and \ref{fig:104-ks-atlasgal}. Previous research towards 6.7\,GHz class II methanol masers \citep{2019MNRAS.490.2779B} and 95\,GHz class I methanol masers \citep{2020AJ....160..213L} also shows a very similar trend in that only the FWHM clump mass shows no difference between maser detection and non-maser detection or for the entire sample.

Table~\ref{Tab:atlasgal-stat} provides a statistical summary of the physical properties for the maser subsample, the detection subsample (i.e. maser and maser candidate) and the non-detections for the 84, 95 and 104.3\,GHz transitions.
It is clear to see that the $L_{\rm bol}$, $L_{\rm bol}/M_{\rm fwhm}$, $T_{\rm dust}$, $n_{\rm fwhm}(\rm H_2)$, and $N(\rm H_2)$ of clumps with methanol detection typically show higher values (i.e. the minimum, maximum, median and mean) than those without a detection for all three methanol transitions, as also suggested in Figs.~\ref{fig:84-ks-atlasgal}, \ref{fig:95-ks-atlasgal} and \ref{fig:104-ks-atlasgal}. 
While for the clumps with methanol maser and with methanol detections, the values of these properties are comparable.
Figures~\ref{fig:84maser-ks-atlasgal} and \ref{fig:95maser-ks-atlasgal} show the results of K-S tests of the cumulative distributions of the clump physical properties for sources with maser and maser candidates for 84 and 95\,GHz transitions. The $p$-values of the K-S tests for each property are not smaller than 0.0013, indicating that the two subsamples belong to the same distribution. 
We conclude that the warmer, brighter and denser clumps are more easily detectable in methanol and more likely to host the class I methanol masers. 

Figure~\ref{fig:all-ks-atlasgal} shows the cumulative distributions of the clump physical properties for sources with 84, 95 and 104.3 GHz methanol detection, which are similar to Fig.~\ref{fig:84-ks-atlasgal}. 
The K-S tests reveal that except for the FWHM clump mass, the properties of clumps with 104.3~GHz methanol detection are significantly different from the clumps with 84 or 95~GHz ($p$-values$\ll$0.0013). In contrast, the properties of clumps with 84 and 95~GHz methanol detection are similar ($p$-values$\gg$0.0013), due to the fact that a large portion of sources have both maser species. This supports a significant overlap in the physical conditions that excite 84 and 95\,GHz masers.

As shown in Fig.~\ref{fig:all-ks-atlasgal}, the clumps with 104.3\,GHz masers generally show brighter luminosities, warmer dust temperatures, larger luminosity-to-mass ratios, and denser environments than the clumps with 84 and 95\,GHz masers (see also Table~\ref{Tab:atlasgal-stat}). 
The narrower coverage (i.e. maximum minus minimum in Table~\ref{Tab:atlasgal-stat}) of these parameters for the clumps hosting 104.3\,GHz masers also indicate that this maser species occurs only under more demanding physical conditions.
The minimum values of these parameters of the clumps hosting 104.3\,GHz masers suggest that this maser species would not arise in environments (on clump scales) where $L_{\rm bol}$ $\lesssim$ 10$^4$~\Lsun, $M_{\rm fwhm}$ $\lesssim$ 200~\Msun, $T_{\rm dust}$ $\lesssim$ 22~K, $N(\rm H_2)$ $\lesssim$ 10$^{23}$~cm$^{-2}$ and $n_{\rm fwhm}(\rm H_2)$ $\lesssim$ 10$^5$~cm$^{-3}$.

\begin{figure*}[!htbp]
\centering
\mbox{
\begin{minipage}[b]{6cm}
\includegraphics[width=1.04\textwidth]{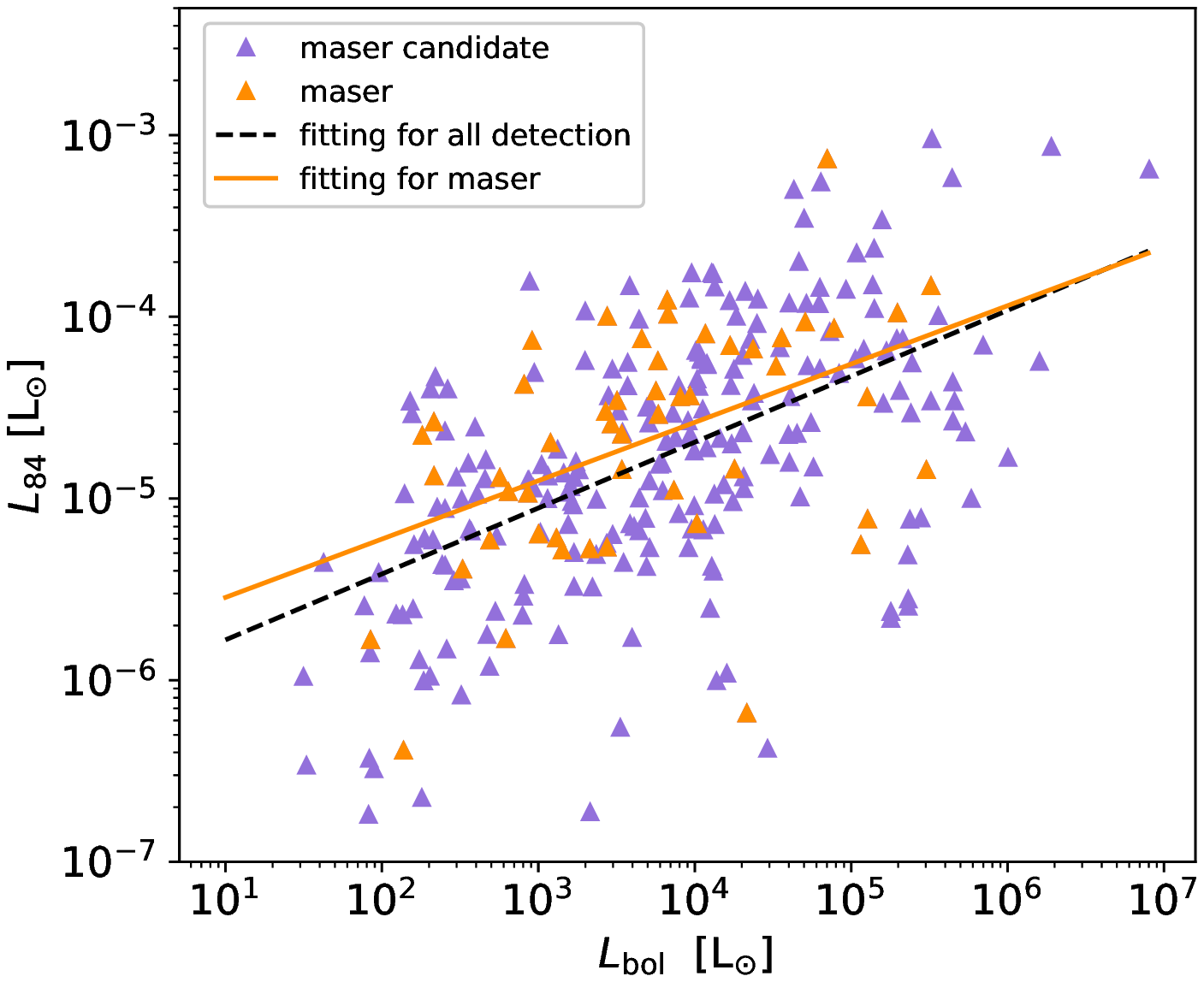}
\centerline{(a)}
\vspace{-3mm}
\end{minipage}
\begin{minipage}[b]{6cm}
\includegraphics[width=1.04\textwidth]{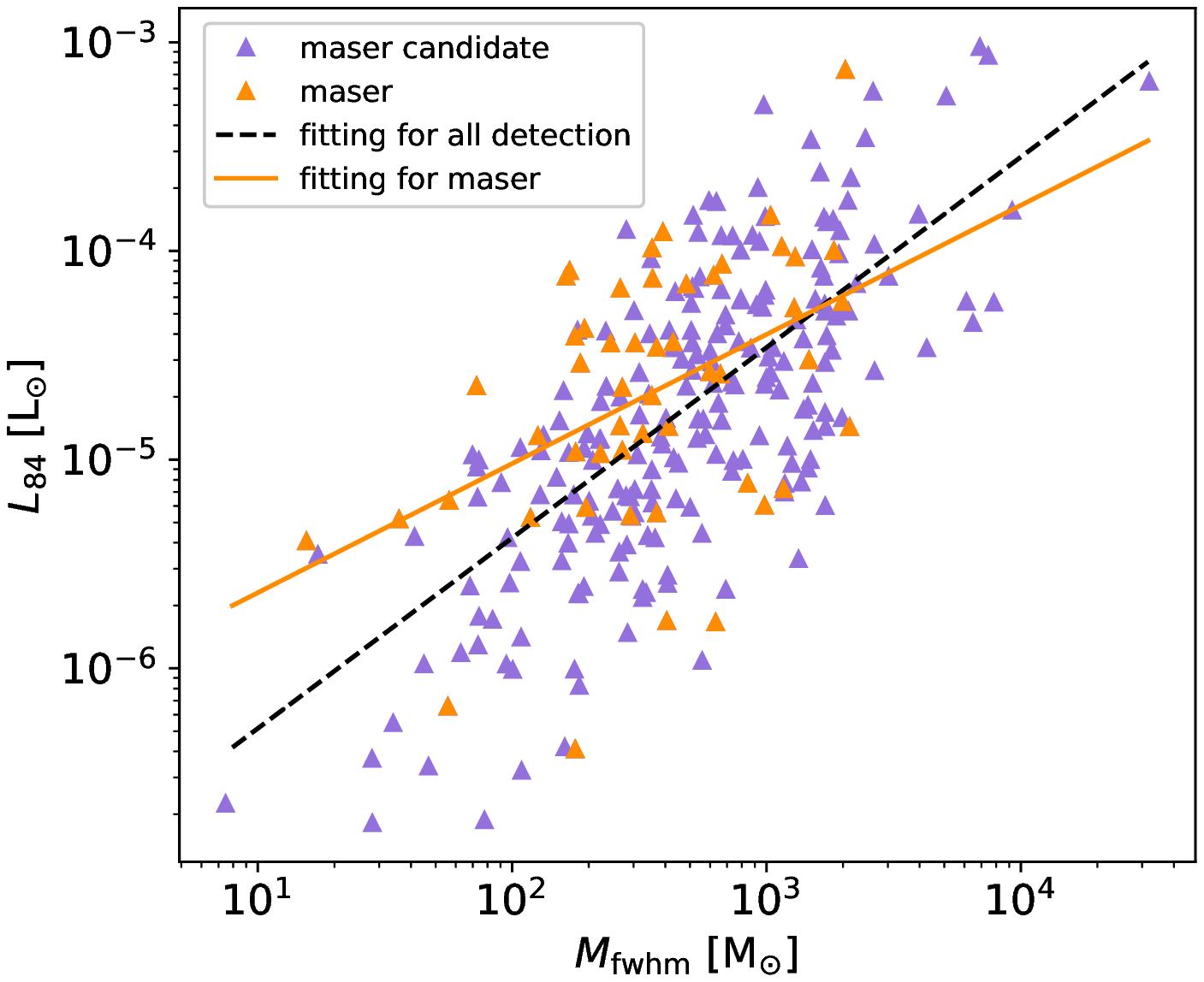}
\centerline{(b)}
\vspace{-3mm}
\end{minipage}
\begin{minipage}[b]{6cm}
\includegraphics[width=1.04\textwidth]{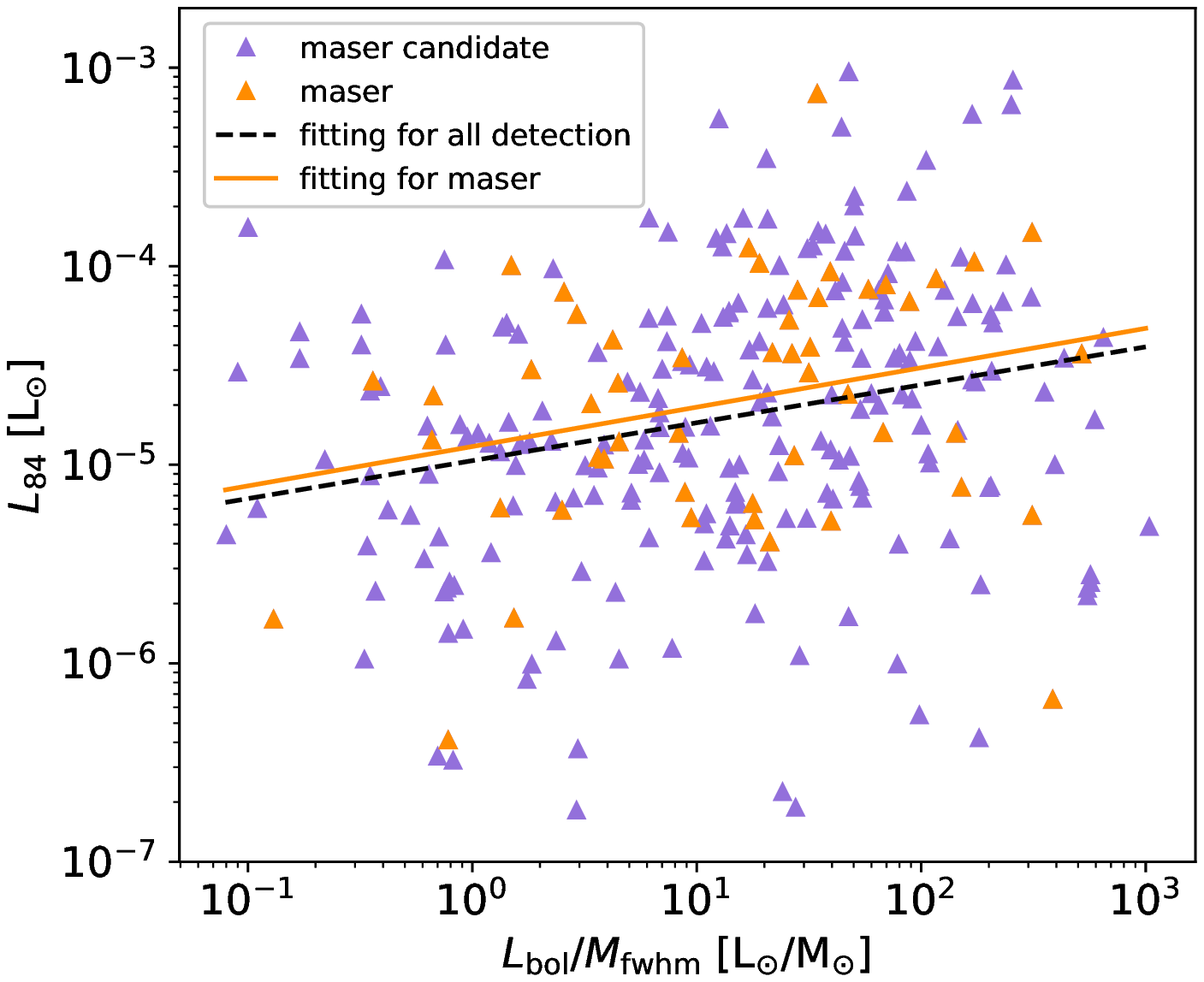}
\centerline{(c)}
\vspace{-3mm}
\end{minipage}
}
\mbox{
\begin{minipage}[b]{6cm}
\includegraphics[width=1.04\textwidth]{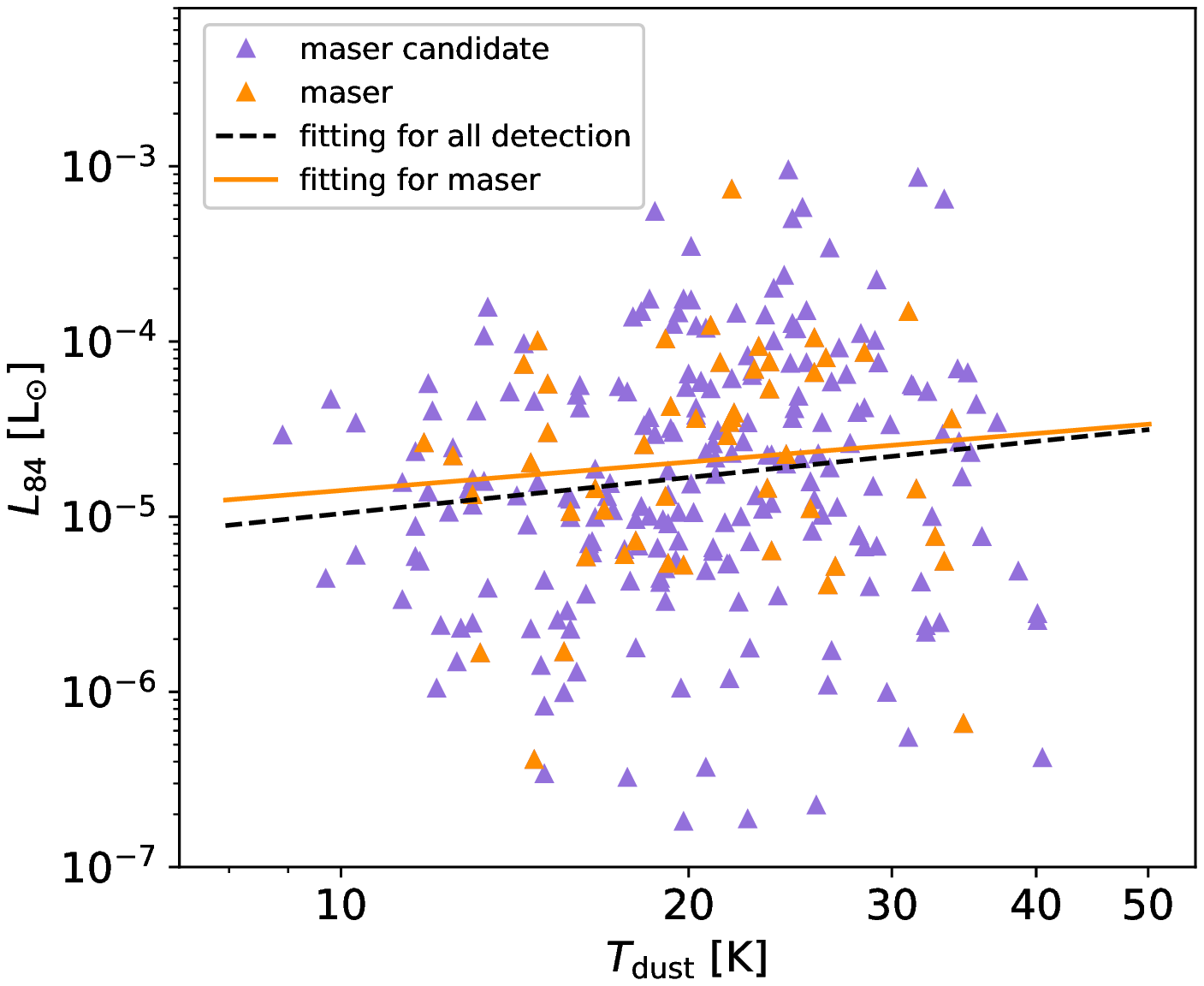}
\centerline{(d)}
\vspace{-3mm}
\end{minipage}
\begin{minipage}[b]{6cm}
\includegraphics[width=1.04\textwidth]{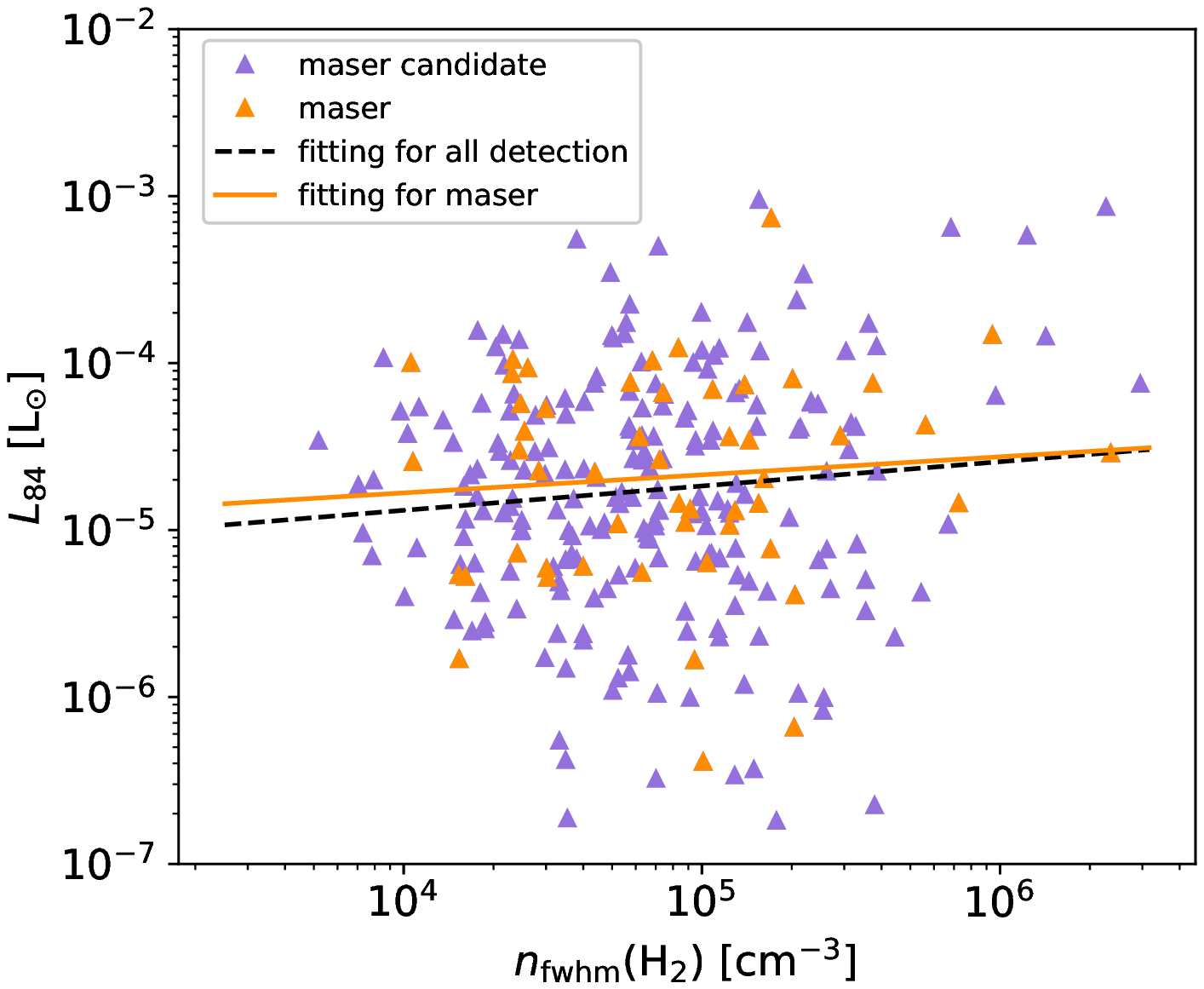}
\centerline{(e)}
\vspace{-3mm}
\end{minipage}
\begin{minipage}[b]{6cm}
\includegraphics[width=1.04\textwidth]{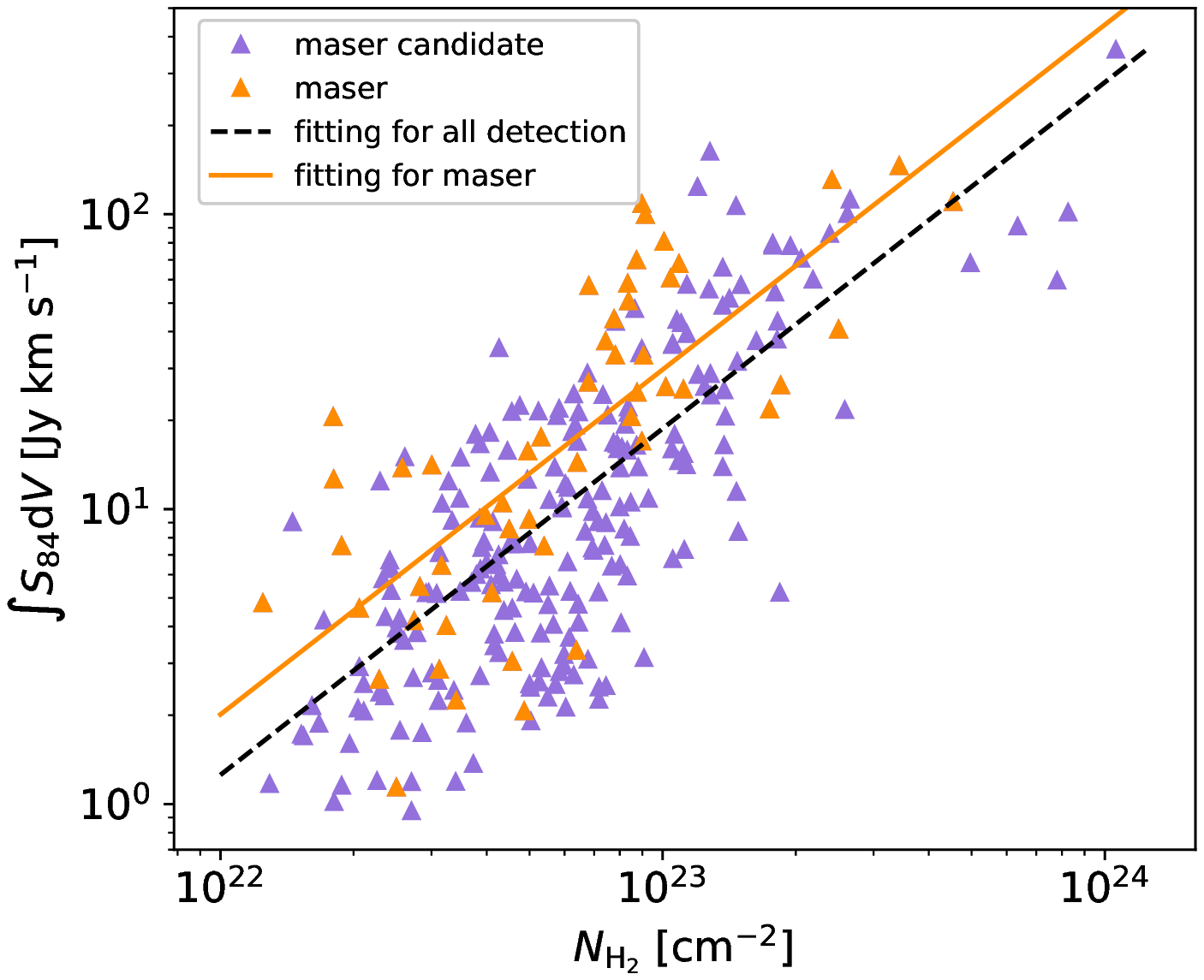}
\centerline{(f)}
\vspace{-3mm}
\end{minipage}
}
\caption{Distribution of 84\,GHz methanol luminosity or integrated intensity against properties of ATLASGAL clumps.
From panel (a) to (e), methanol isotropic luminosities are plotted against the bolometric luminosity, the FWHM clump mass, the luminosity-to-mass ratio, the dust temperature and the mean H$_2$ FWHM volume density. In panel (f), total integrated intensities of the 84\,GHz methanol transition are plotted against the peak H$_2$ column density.
The orange and purple triangles represent the ATLASGAL clumps host maser and maser candidates, respectively.
The orange solid lines and black dashed lines depict the least-square fitting results for methanol maser and all methanol detections, respectively.
\label{fig:84-atlasgal}}
\end{figure*}

The isotropic luminosity of CH$_3$OH masers can be calculated by the following equations:
\begin{equation}
L_{\rm 84} 
= 8.81 \times 10^{-8} L_{\odot}\,\left(\frac{d}{\rm 1~kpc} \right)^2 \left(\frac{\int S_{84}{\rm d}V}{\rm 1~Jy~km~s^{-1}} \right), 
\end{equation}

\begin{equation}
L_{\rm 95} = 9.92 \times 10^{-8} L_{\odot}\,\left(\frac{d}{\rm 1~kpc} \right)^2 \left(\frac{\int S_{95}{\rm d}V}{\rm 1~Jy~km~s^{-1}} \right),
\end{equation}

\begin{equation}
L_{\rm 104.3} = 1.09 \times 10^{-7} L_{\odot}\,\left(\frac{d}{\rm 1~kpc} \right)^2 \left(\frac{\int S_{104.3}{\rm d}V}{\rm 1~Jy~km~s^{-1}} \right). 
\end{equation}
where $d$ is the distance in units of kpc, and $\int S_{\nu}{\rm d}V$ is the total integrated flux density in units of Jy~\kms\, at the corresponding frequency $\nu$.
Figure~\ref{fig:84-atlasgal} shows the correlations between 84\,GHz methanol (maser) luminosity or integrated intensity and ATLASGAL properties. Similarly, 
the correlations for the 95 and 104.3 GHz lines are shown in Figs.~\ref{fig:95-atlasgal} and \ref{fig:104-atlasgal}. Their linear fitting results in log-log form for three methanol transitions are summarised in Table~\ref{Tab:atlasgal-fit}.

From these figures and Table~\ref{Tab:atlasgal-fit}, it can be seen that (1) for all three methanol transitions, the class I maser luminosity shows significant positive correlations with the bolometric luminosity of the embedded protostellar objects and the FWHM clump mass, which is consistent with previous studies of the 95 GHz masers towards BGPS  sources \citep{2011ApJS..196....9C,2012ApJS..200....5C,2020ApJS..248...18Y}; (2) for 84 and 95\,GHz methanol transitions, the total integrated intensity of class I methanol masers has a significant positive correlation with peak H$_2$ column density; (3) for all three methanol transitions, the methanol luminosity of class I methanol masers has a very weak, or no statistically significant correlation with luminosity-to-mass ratio, dust temperature or the mean H$_2$ FWHM volume density. The methanol luminosity independence on the mean H$_2$ FWHM volume density agrees with the results in \citet{2012ApJS..200....5C}.


\begin{table*}[!hbt]
\caption{Summary of physical parameters of ATLASGAL clumps with masers, methanol detections and non-detections for 84, 95 and 104.3 GHz transitions. }\label{Tab:atlasgal-stat} 
\small
\centering
\setlength{\tabcolsep}{2.5pt}
\renewcommand\arraystretch{1.15}
\begin{tabular}{lcccccccccccccccc}
\hline \hline
\multirow{2}{*}{ATLASGAL Parameters}  &  \multicolumn{5}{c}{Maser} &  \multicolumn{5}{c}{Detection (i.e. maser+candidate)} &  \multicolumn{5}{c}{Non-detection}\\
\cmidrule(lr){2-6} 
\cmidrule(lr){7-11} 
\cmidrule(lr){12-16} 
 & min & max  & median  & mean  & $\sigma$ & min & max  & median  & mean  & $\sigma$ & min & max  & median  & mean  & $\sigma$ \\
\hline
\multicolumn{16}{c}{84\,GHz} \\
\hline
log$_{10}$($L_{\rm bol}$) [\Lsun] & 1.93  & 5.51  & 3.66 & 3.69  & 0.92 & 1.50  & 6.91  & 3.79  & 3.78  & 1.06   
& $-$0.12  & 6.60  & 2.71  & 2.84  & 1.06   \\
log$_{10}$($M_{\rm fwhm}$) [\Msun]  & 1.19  & 3.33  & 2.55  & 2.55  & 0.46  & 0.87  & 4.51  & 2.69  & 2.67  & 0.53   
& $-$0.47  & 4.14  & 2.68  & 2.60  & 0.63   \\
$L_{\rm bol}$/$M_{\rm fwhm}$ [\Lsun/\Msun]  & 0.13 & 518.68  & 19.01  & 57.86  & 106.16  & 0.08  & 1037.23  & 16.53  & 63.96  & 127.43  
& 0.07  & 291.03  & 1.40  & 11.97  & 34.47  \\
$T_{\rm dust}$ [K] & 11.8  & 34.6  & 21.1  & 21.2  & 6.0 
 & 8.9  & 40.5  & 20.7  & 21.4  & 6.6 
 & 8.7  & 32.2  & 16.5  & 17.2  & 5.5   \\
log$_{10}$($N({\rm H_2})$) [cm $^{-2}$]  & 22.10  & 23.66  & 22.81  & 22.78  & 0.34
 & 22.10  & 24.03  & 22.78  & 22.79  & 0.33  
 & 22.06  & 23.04  & 22.48  & 22.47  & 0.29   \\
log$_{10}$($n_{\rm fwhm}(\rm H_2)$) [cm $^{-3}$]  & 4.02  & 6.37  & 4.92  & 4.89  & 0.51
 & 3.71  & 6.47  & 4.80  & 4.83  & 0.49
 & 3.74  & 5.85  & 4.47  & 4.53  & 0.43   \\
\hline
\multicolumn{16}{c}{95\,GHz}\\
\hline
log$_{10}$($L_{\rm bol}$) [\Lsun]  & 1.63  & 5.65  & 3.77  & 3.76  & 0.91
 & 1.63   & 6.91  & 3.96 & 3.94  & 0.99  
 & $-$0.12  & 6.00  & 2.79  & 3.00  & 1.11   \\
log$_{10}$($M_{\rm fwhm}$) [\Msun]  & 0.87  & 3.97  & 2.61  & 2.62  & 0.52 & 0.87  & 4.51  & 2.73  & 2.72  & 0.54  
 & $-$0.47  & 4.12  & 2.60  & 2.56  & 0.57  \\
$L_{\rm bol}$/$M_{\rm fwhm}$ [\Lsun/\Msun]  & 0.08  & 647.86  & 17.73  & 47.54  & 95.62
 & 0.08  & 647.86 & 20.66  & 57.78  & 94.09 
 & 0.07 & 1037.23  & 2.10  & 40.40  & 132.47   \\
$T_{\rm dust}$ [K]  & 8.9  & 35.9  & 20.9  & 21.3  & 5.6 
 & 8.9  & 37.0  & 21.6  & 21.8  & 6.1  
 & 8.7  & 40.5  & 17.3  & 18.3  & 6.8   \\
log$_{10}$($N({\rm H_2})$) [cm $^{-2}$]  & 22.10  & 23.92  & 22.89  & 22.87  & 0.36 
 & 22.10  & 24.03  & 22.85  & 22.85  & 0.34  
 & 22.06  & 23.04  & 22.52  & 22.52  & 0.21   \\
log$_{10}$($n_{\rm fwhm}(\rm H_2)$) [cm $^{-3}$]  & 3.93  & 6.47  & 5.02  & 5.01  & 0.49
 & 3.79  & 6.47  & 4.86  & 4.89  & 0.50  
 & 3.71  & 5.85  & 4.52  & 4.55 & 0.40  \\
\hline
\multicolumn{16}{c}{104.3\,GHz}\\
\hline
log$_{10}$($L_{\rm bol}$) [\Lsun]  & 4.03  & 5.15  & 4.46  & 4.52  & 0.56 
 & 3.66  & 6.28  & 4.60  & 4.64  & 0.69  
 & $-$0.12  & 6.91  & 3.36  & 3.42  & 1.12   \\
log$_{10}$($M_{\rm fwhm}$) [\Msun]  & 2.23  & 3.31  & 2.67  & 2.72  & 0.51
 & 2.21  & 3.87  & 2.69  & 2.76  & 0.43  
 & $-$0.47  & 4.51  & 2.68  & 2.64  & 0.57   \\
$L_{\rm bol}$/$M_{\rm fwhm}$ [\Lsun/\Msun]  & 34.40  & 149.56  & 57.55  & 74.77  & 51.98 
 & 20.66  & 647.86 & 69.56  & 114.73  & 132.35  
 & 0.07  & 1037.23  & 6.46  & 43.95  & 108.47   \\
$T_{\rm dust}$ [K]  & 21.8  & 28.2  & 25.5  & 25.3  & 2.7 
 & 20.1 & 35.5  & 24.6  & 25.6  & 4.2  
 & 8.7  & 40.5  & 18.9  & 19.8  & 6.5   \\
log$_{10}$($N({\rm H_2})$) [cm $^{-2}$]  & 22.95  & 23.38  & 23.10  & 23.13  & 0.19 
 & 22.48  & 24.03  & 23.21  & 23.25  & 0.36  
 & 22.06  & 23.89  & 22.64  & 22.65  & 0.29   \\
log$_{10}$($n_{\rm fwhm}(\rm H_2)$) [cm $^{-3}$]  & 5.04  & 5.33  & 5.27 & 5.23  & 0.13 
 & 4.41  & 6.47  & 5.48  & 5.52  & 0.51 
 & 3.71  & 5.85  & 4.67  & 4.69  & 0.43   \\
\hline \hline
\end{tabular}
\normalsize
\note{Columns 2--16 give the minimum, maximum, median and standard deviation values of the ATLASGAL parameters given in column 1 for subsamples hosting 84, 95 and 104.3\,GHz masers.}
\end{table*}

\begin{table*}[hbt]
\caption{Summary of linear fitting results in log-log form of  methanol luminosity or integrated intensity versus the properties of ATLASGAL clumps for 84, 95 and 104.3\,GHz transitions.
}\label{Tab:atlasgal-fit} 
\small
\centering
\setlength{\tabcolsep}{4pt}
\renewcommand\arraystretch{1.15}
\begin{tabular}{lcccccc}
\hline \hline
\multirow{2}{*}{Parameters}  &  \multicolumn{3}{c}{Maser} &  \multicolumn{3}{c}{Detection}\\
\cmidrule(lr){2-4} 
\cmidrule(lr){5-7} 
 & slope & $r$  & $p$-value  & slope & $r$  & $p$-value \\
\hline
\multicolumn{7}{c}{84\,GHz} \\
\hline
$L_{84}$ vs. $L_{\rm bol}$ & 0.32$\pm$0.09 & 0.47 & 5.3$\times$10$^{-4}$ & 0.36$\pm$0.03 & 0.55 & 4.2$\times$10$^{-22}$ \\
$L_{84}$ vs. $M_{\rm fwhm}$ & 0.62$\pm$0.17 & 0.45 & 8.4$\times$10$^{-4}$ & 0.91$\pm$0.16 & 0.68 & 1.1$\times$10$^{-37}$ \\
$L_{84}$ vs. $L_{\rm bol}$/$M_{\rm fwhm}$ & 0.20$\pm$0.10 & 0.26 & 0.06 & 0.12$\pm$0.03 & 0.24 & 6.2$\times$10$^{-5}$ \\
$L_{84}$ vs. $T_{\rm dust}$ & 0.54$\pm$0.74 & 0.10 & 0.47 & 0.69$\pm$0.32 & 0.13 & 0.03 \\
$L_{84}$ vs. $n_{\rm fwhm}(\rm H_2)$ &  0.11$\pm$0.18 & 0.09 & 0.54 & 0.15$\pm$0.09 & 0.10 & 0.11 \\
$\int S_{84}{\rm d}V$ vs. $N({\rm H_2})$ & 1.17$\pm$0.15 & 0.74 & 4.1$\times$10$^{-10}$ & 1.17$\pm$0.07 & 0.74 & 1.1$\times$10$^{-47}$ \\
\hline
\multicolumn{7}{c}{95\,GHz}\\
\hline
$L_{95}$ vs. $L_{\rm bol}$ & 0.40$\pm$0.06 & 0.57 & 8.4$\times$10$^{-10}$ & 0.36$\pm$0.04 & 0.57 & 1.5$\times$10$^{-19}$ \\
$L_{95}$ vs. $M_{\rm fwhm}$ & 0.63$\pm$0.11 & 0.52 & 5.6$\times$10$^{-8}$ & 0.68$\pm$0.07 & 0.57 & 1.3$\times$10$^{-19}$ \\
$L_{95}$ vs. vs. $L_{\rm bol}$/$M_{\rm fwhm}$ & 0.25$\pm$0.08 & 0.32 & 1.3$\times$10$^{-3}$ & 0.19$\pm$0.04 & 0.32 & 2.3$\times$10$^{-6}$ \\
$L_{95}$ vs. $T_{\rm dust}$  & 1.25$\pm$0.52 & 0.24 & 0.02 & 1.23$\pm$0.34 & 0.24 & 3.7$\times$10$^{-4}$ \\
$L_{95}$ vs. $n_{\rm fwhm}(\rm H_2)$ & 0.11$\pm$0.13 & 0.09 & 0.39 & 0.20$\pm$0.09 & 0.16 & 0.02 \\
$\int S_{95}{\rm d}V$ vs. $N({\rm H_2})$  & 1.04$\pm$0.11 & 0.69 & 3.4$\times$10$^{-15}$ & 1.06$\pm$0.08 & 0.66 & 4.1$\times$10$^{-28}$ \\
\hline
\multicolumn{7}{c}{104.3\,GHz}\\
\hline
$L_{104.3}$ vs. $L_{\rm bol}$ & 0.76$\pm$0.32 & 0.86 & 0.14 & 0.66$\pm$0.11 & 0.78 & 1.6$\times$10$^{-6}$ \\
$L_{104.3}$ vs. $M_{\rm fwhm}$ & 0.96$\pm$0.11 & 0.99 & 0.01 & 1.19$\pm$0.13 & 0.88 & 2.1$\times$10$^{-9}$ \\
$L_{104.3}$ vs. $L_{\rm bol}$/$M_{\rm fwhm}$  & $-$0.13$\pm$1.26 & $-$0.07 & 0.93 & 0.64$\pm$0.28 & 0.42 & 0.03 \\
$L_{104.3}$ vs. $T_{\rm dust}$ & $-$4.48$\pm$6.71 & $-$0.43 & 0.57 & 2.51$\pm$1.65 & 0.29 & 0.14 \\
$L_{104.3}$ vs. $n_{\rm fwhm}(\rm H_2)$  & $-$2.18$\pm$2.20 &  $-$0.57 & 0.43 & 0.41$\pm$0.21 & 0.36 & 0.07 \\
$\int S_{104.3}{\rm d}V$ vs. $N({\rm H_2})$ & 0.45$\pm$0.43 & 0.60 & 0.40 & 1.01$\pm$0.11 & 0.87 & 3.1$\times$10$^{-9}$ \\
\hline \hline
\end{tabular}
\normalsize
\note{Columns 2--7 give the slopes of the linear fitting results and Pearson correlation coefficients $r$ and $p$-values for the parameters given in Column 1 of maser and methanol detections.}
\end{table*}

\subsection{Class I methanol masers in different evolutionary stages} \label{sec:evo}

\begin{figure}[!htbp]
\centering
\includegraphics[width=0.5\textwidth]{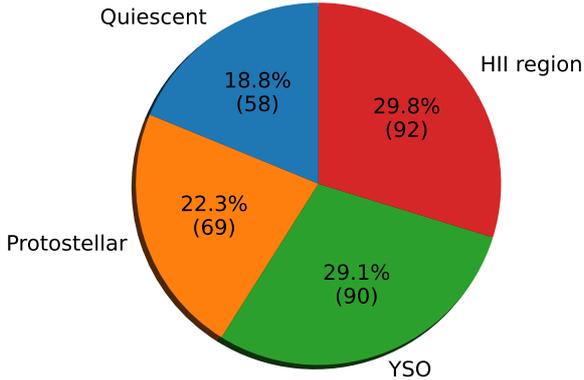}
\caption{Proportions of 309 ATLASGAL clumps classified into the four evolutionary stages. 
\label{fig:pie-evo}}
\end{figure}

\begin{figure}[!htbp]
\centering
\includegraphics[width=0.5\textwidth]{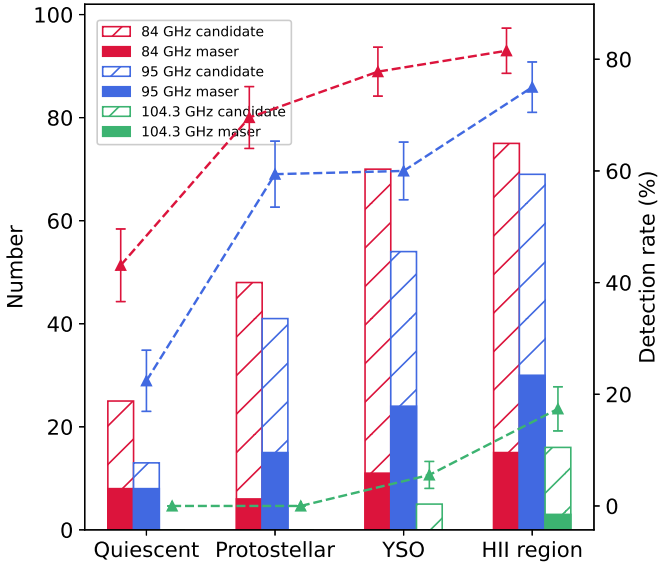}
\caption{Distribution of methanol detection number and detection rate of three methanol transitions under four evolutionary stages. The labels for the detection numbers are on the left side, while those for the detection rates are on the right side. The red, blue and green bins present the detection numbers of methanol emission at 84, 95 and 104.3\,GHz, respectively, where the solid and hatched parts represent masers and maser candidates. The red, blue and green triangles indicate the detection rates of methanol emission at 84, 95 and 104.3\,GHz. 
\label{fig:evo}}
\end{figure}

\begin{figure*}[!htbp]
\centering
\includegraphics[width=0.95\textwidth]{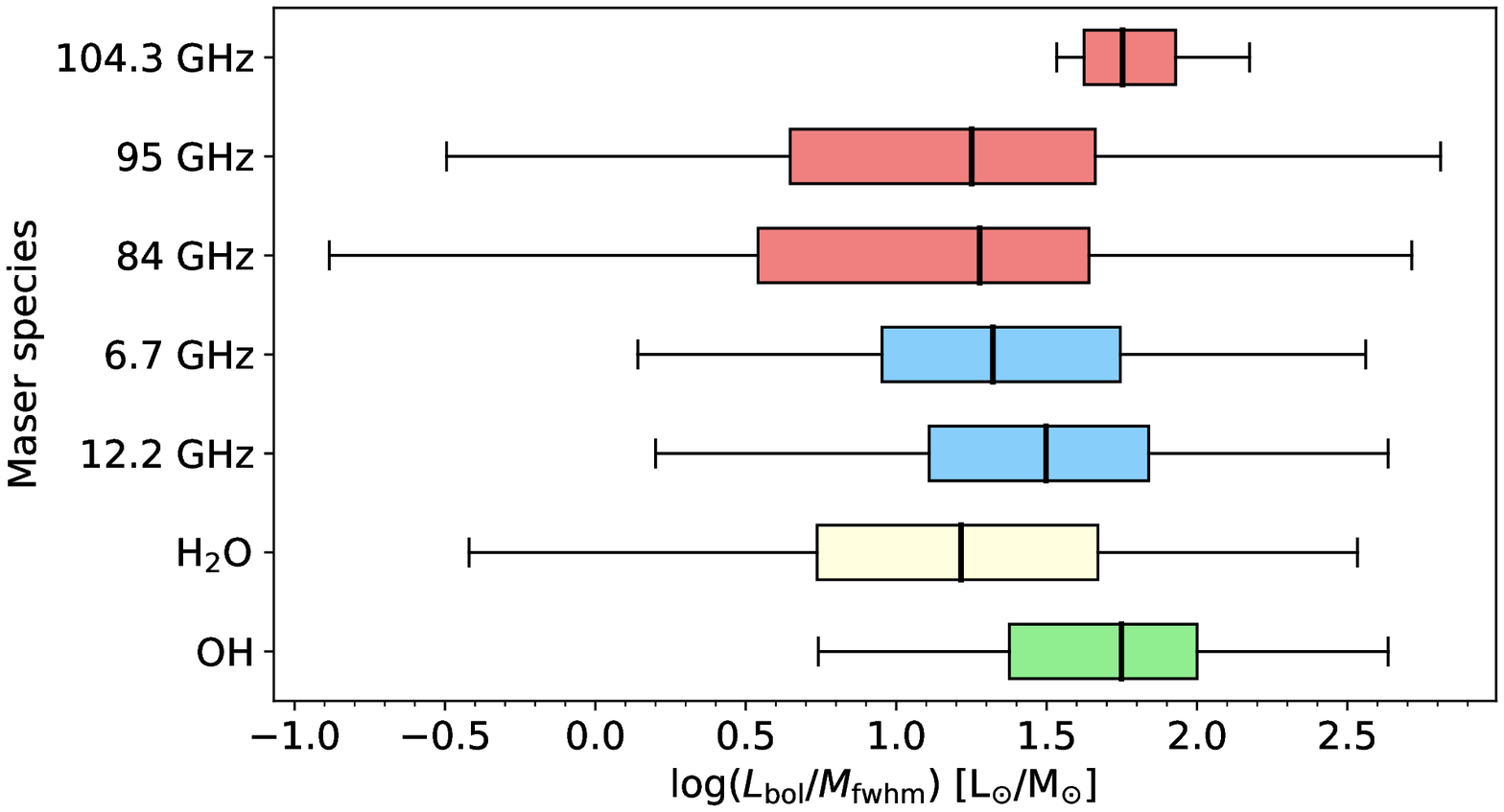}
\caption{Box plot shows the distributions of bolometric luminosity-to-mass ratios for clumps associated with different masers.
Except for class I maser data, the plotting data for H$_2$O, OH, and class II methanol masers at 6.7 and 12.2 GHz are taken from Fig. 7(A) in \cite{2022ApJS..261...14L}.
The boxes filled in red, blue, light yellow and green  indicate the class I and II methanol masers, H$_2$O, and OH masers, respectively.
Each box represents the interquartile range between 25th and 75th percentile with a thick black line denoting the median value. The whiskers show the full range of the data. 
\label{fig:lm}}
\end{figure*}

As massive proto-stars evolve, their surrounding material can be dramatically changed due to powerful stellar radiation and jet/outflow activities, and the changing physical and chemical properties may successively meet the excitation conditions for different maser species.
\cite{2007MNRAS.377..571E} suggested to use the presence and absence of interstellar masers to infer an evolutionary timeline for high-mass star formation regions, and proposed a possible evolutionary sequence for common maser species, including class I and II methanol, water, and OH masers.
\cite{2010MNRAS.401.2219B} then refined and quantified the maser-based evolutionary timeline (see their Fig.~6).
Due to the fact that infrared colors are redder in sources solely associated with class I methanol masers than in sources associated with both class I and II masers, class I methanol masers are generally believed to trace an earlier stage than class II masers \citep{2006ApJ...638..241E}.
However, some studies have revealed that in some cases class I methanol masers can also be associated with sources in more evolved stages \citep[e.g.][]{2010MNRAS.405.2471V,2014MNRAS.439.2584V,2016ApJS..222...18G}, for example with expanding 
H{\sc ii} regions \citep{2010MNRAS.405.2471V,2014MNRAS.439.2584V}. 
Here, we use the properties of the ATLASGAL clumps to study the evolutionary stages in which class I methanol masers may arise.

ATLASGAL clumps have been classified into four evolutionary stages, namely quiescent, protostellar, young stellar objects (YSOs), and H{\sc ii} regions \citep{2017A&A...599A.139K,2017A&A...603A..33G,2018MNRAS.473.1059U,2022MNRAS.510.3389U}.
In this work, we adopt the latest evolutionary stage classification given by \cite{2022MNRAS.510.3389U} for each source.
Table~\ref{Tab:source} lists the evolutionary stage of each target. 
Out of 408 ATLASGAL sources, a total of 309 ($\sim$76\%) are assigned to one of the four evolutionary stages.
Fifty-eight sources are starless, that is, the quiescent stage, 69 sources belong to a protostellar stage, 90 sources are classified as YSOs, and 92 sources host H{\sc ii} regions.
The other 99 sources are classified as ``ambiguous", ``complicated" and ``PDR+embedded source", or remain undefined in \cite{2022MNRAS.510.3389U}, due to the complexity of the submillimeter/infrared emission from their environments, which precludes a reliable determination of the current level of star formation in them. Consequently, these 99 sources are excluded from the following statistical analysis.
Figure~\ref{fig:pie-evo} shows the proportions of the clumps in the four evolutionary stages.

Figure~\ref{fig:evo} shows that the detection numbers and detection rates increase with evolution for all three of our studied class I methanol detections (including maser sources and candidates). Especially, we find that class I methanol masers are favored to be detected at the H{\sc ii} stage on a clump scale. Previous VLA observations result in a detection rate (54\%) towards UCH{\sc ii} regions \citep{2016ApJS..222...18G}, supporting our findings on an even smaller scale. 
Different from the 84 GHz and 95 GHz masers that are seen in four evolutionary stages, methanol emission at 104.3\,GHz never occurs before the YSO stage, and the rare 104.3\,GHz maser emission only arises at the H{\sc ii} stage. This is expected because the 104.3\,GHz line has an upper energy level of 158.6~K and is difficult to be excited in cold environments. 


In addition, the bolometric luminosity-to-mass ($L_{\rm bol}$/$M_{\rm fwhm}$) ratio can be used as an evolutionary indicator in the star formation process \citep[e.g.][]{2016ApJ...826L...8M,2021MNRAS.504.2742E,2022MNRAS.510.3389U}. High $L_{\rm bol}$/$M_{\rm fwhm}$ ratios are related to evolved sources where the luminosity increases and the envelope mass is consumed, while low ratios are related to young sources.
Our observing targets cover a wide range of $L_{\rm bol}/M_{\rm fwhm}$ values from 0.07 to 1037~\Lsun/\Msun\,spanning more than four orders of magnitude.

\cite{2020MNRAS.499.2744B} presented a box plot of the central 95 percent of $L_{\rm bol}$/$M_{\rm fwhm}$ ratios for 6.7, 12.2 GHz class II methanol masers, OH maser and H$_2$O masers (see their Fig. 14). \cite{2022ApJS..261...14L} updated this analysis with larger samples for these masers.
Figure~\ref{fig:lm} shows a similar box plot of the $L_{\rm bol}$/$M_{\rm fwhm}$ ratio distributions for ATLASGAL clumps associated with class I masers, comparing to the four maser species studies by \cite{2022ApJS..261...14L}. From the figure, we can see that class I masers at 84 and 95~GHz masers have a great degree of overlap in the $L_{\rm bol}$/$M_{\rm fwhm}$ ratios with H$_2$O, OH, and class II methanol masers. 
These two maser species can trace a similar evolutionary stage as H$_2$O maser, and appear prior to 6.7 and 12.2 GHz methanol and OH masers. 
Despite the small number, the 104.3~GHz class I masers appear to trace a short and more evolved stage compared to the other class I maser species.




\subsection{Physical conditions of class I masers}

\begin{figure*}[htbp]
\centering
\mbox{
\begin{minipage}[b]{8.5cm}
\includegraphics[width=1.1\textwidth]{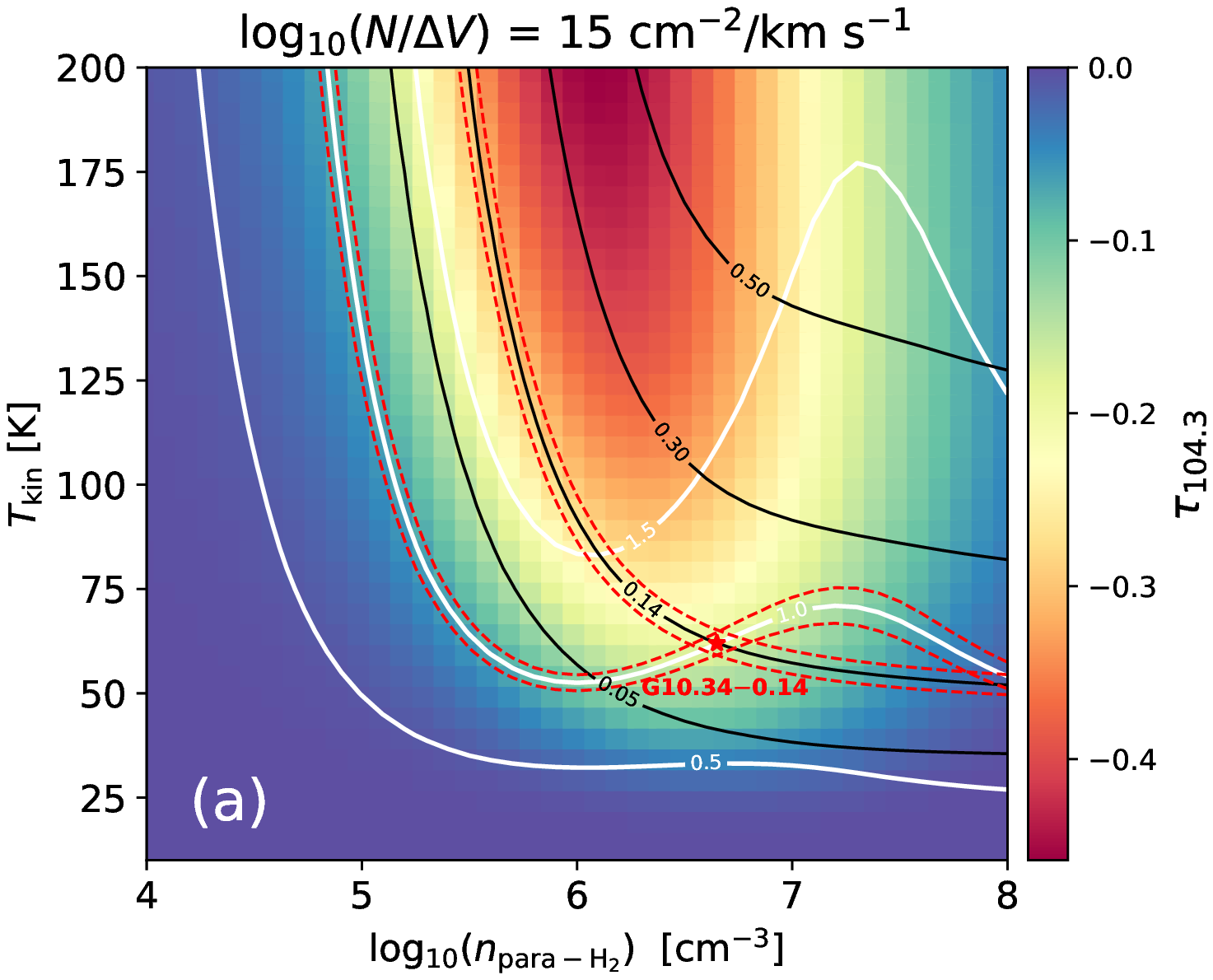}
\end{minipage}
\hspace{3mm}
\begin{minipage}[b]{8.5cm}
\includegraphics[width=1.1\textwidth]{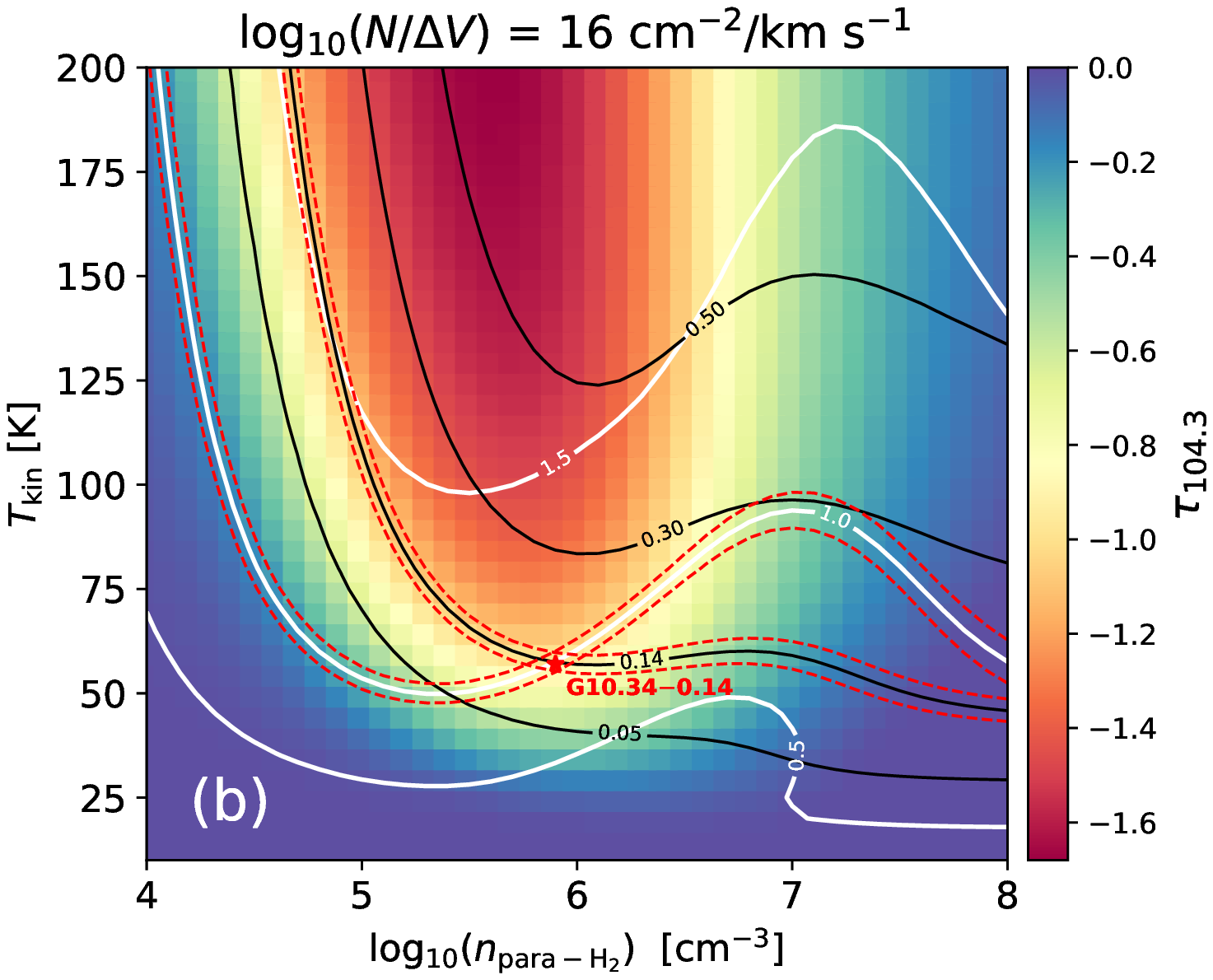}
\end{minipage}
}
\caption{Statistical equilibrium calculations of optical depths and line ratios.
The left and right panels show that the calculations are performed under the specific column densities of 10$^{15}$ and 10$^{16}$~cm$^{-2}$~km$^{-1}$~s, respectively.
The colored background shows the regions with negative optical depths of the 104.3\,GHz methanol transition where masers can be generated.
The black and white contours represent the peak intensity ratios of $T_{\rm r,104.3}/T_{\rm r,84}$ and $T_{\rm r,95}/T_{\rm r,84}$, respectively. 
The red star is the intersection of the observed two pairs of ratios in G10.34$-$0.14, and the red dashed lines represent the 1$\sigma$ uncertainties of the observed ratios.
\label{fig:radex}}
\end{figure*}

As indicated in \citet{2016A&A...592A..31L}, the intensity ratio of two maser transitions usually gives much stronger constraints on physical conditions than one single transition.
If there are three coincident methanol maser transitions detected in one source, theoretical calculations of the maser intensity ratios would put stronger constraints on the physical conditions (such as kinetic temperature and H$_2$ volume density) where all masers are excited. 

We use RADEX \citep{2007A&A...468..627V} and myRadex\footnote{\url{https://github.com/fjdu/myRadex}} \citep{2022ascl.soft05011D} to create model grids for the methanol lines. MyRadex solves the same problem as RADEX, except that a different approach is used to solve the statistical equilibrium problem (i.e. using an ordinary differential equation solver to evolve the system towards equilibrium under an initial distribution). 
In our work, we show the modeling results from the myRadex code instead of RADEX, because RADEX calculations do not reach convergence when the specific methanol column densities ($N/\Delta V$) are higher than 10$^{14}$~cm$^{-2}$~km$^{-1}$~s, and such high specific methanol column densities are required to generate bright class I methanol masers  \citep{2016A&A...592A..31L}.

We make use of the molecular data of CH$_{3}$OH from the Leiden Atomic and Molecular Database \citep[LAMDA\footnote{\url{https://home.strw.leidenuniv.nl/~moldata/}};][]{2005A&A...432..369S}, where the energy levels, transition frequencies and Einstein A coefficients are adopted from the CDMS \citep[CDMS;][]{2005JMoSt.742..215M,ENDRES201695} and the latest collisional rates are obtained from \citet{2010MNRAS.406...95R}. 
The collisional rates for the torsional ground states of $A$- and $E$-type methanol are only known for collisions with para H$_2$ in a temperature range from 10 to 200~K and including the rotational states up to $J$=15 for collisions \citep{2010MNRAS.406...95R}. 
We note that myRadex is able to account for a variable ortho-to-para ratio, which can be calculated under a local thermodynamic equilibrium assumption and only depends on local temperature \citep[e.g. see Equ. 1 in ][]{2001ApJ...561..254T}.
However, significant conversion from para- to ortho-H only starts at 700~K in C-shocks and at even higher temperatures in J-shocks \citep{2000A&A...356.1010W}. That means that the abundance of ortho-H$_2$ is very low at a kinetic temperature of $<$200~K.
Thus, it is reasonable to consider that collisions with ortho-H$_2$ are negligible at kinetic temperatures of $<$200~K.

Methanol molecules can exist in one of two different symmetry species,  $A$- and $E$- type CH$_3$OH. 
Since the very long timescale of proton exchange reactions to interconvert the two types of methanol, we can consider the $A$- and $E$-type methanol as two different molecules.
The $A/E$ methanol abundance ratio is related to the temperature where methanol was formed. The value of the ratio is about unity at a spin temperature (characterizing the relative population of non-interacting spin types of a molecule) of 30--40~K and increasing at lower temperatures \citep[see][]{2011A&A...533A..24W}. 
Based on previous studies \citep{2005IAUS..227..174S,2006MNRAS.373..411V,2018IAUS..336...17L}, class I methanol masers are usually excited at kinetic temperatures higher than 40~K. Therefore, it is reasonable to assume a methanol $A/E$ type ratio of unity in our calculations.


Figure~\ref{fig:radex} shows myRadex calculations for 84, 95 and 104.3\,GHz methanol lines over an equally-spaced grid with 39 temperatures from 10 to 200~K and 41 para-H$_2$ densities from 10$^4$ to 10$^8$~cm$^{-3}$. 
The specific column densities of methanol are fixed at 10$^{15}$~cm$^{-2}$~km$^{-1}$~s (Fig.~\ref{fig:radex}a) and 10$^{16}$~cm$^{-2}$~km$^{-1}$~s (Fig.~\ref{fig:radex}b). 
A plane-parallel slab geometry is adopted, which is appropriate for shocks, and the corresponding escape probability $\beta$ = (1-e$^{-3 \tau}$)/3$\tau$, where $\tau$ is the optical depth. 
No external radiation field is taken into account except for the cosmic microwave background, because class I methanol masers are usually offset from infrared sources and UCH{\sc ii} regions, and interferometric observations reveal that they reside in the interface regions between outflows and ambient dense material \citep{1990ApJ...364..555P,2006MNRAS.373..411V}. 
The black and white contours represent the peak intensity ratios of $T_{\rm r,104.3}/T_{\rm r,84}$ and $T_{\rm r,95}/T_{\rm r,84}$, respectively. 
We assume that the maser components in these three transitions with the same velocity are co-spatial, and have the same small masing area (i.e. emission size $\ll$ beam size). 
In the calculations of peak intensity ratios, different beam dilution factors of each transition have been taken into account. For example, $T_{\rm r,104.3}/T_{\rm r,84}$ equals to $(T_{\rm mb,104.3}/T_{\rm mb,84}) \cdot (\theta_{\rm beam, 104.3}^{2}/\theta_{\rm beam, 84}^{2})$, where $T_{\rm mb}$ and $\theta_{\rm beam}$ are the observed brightness temperature and the beam size at the corresponding frequency.

From our four detected 104.3\,GHz masers, only G10.34$-$0.14 shows distinguishable and velocity-aligned (at $\sim$14.5~\kms) maser features in the three transitions. 
We adopt this feature as an example to find out under what physical conditions the three masers can be excited simultaneously.
The red star in Fig.~\ref{fig:radex} is the intersection of the observed two pairs of ratios, and the red dashed lines represent the uncertainties of the observed ratios. 
Under an assumed specific column density of methanol of 10$^{15}$~cm$^{-2}$~km$^{-1}$~s, a kinetic temperature of 62$\pm$3~K and a para-H$_2$ number density of 4.4($\pm$1.5)$\times$10$^6$~cm$^{-3}$ are needed to excited the three class I methonal masers. When the assumed specific column density of methanol increases to 10$^{16}$~cm$^{-2}$~km$^{-1}$~s, the masering region in G10.34$-$0.14 should have a kinetic temperature of 57$\pm$3~K and a para-H$_2$ number density of 7.9($\pm$2.5)$\times$10$^5$~cm$^{-3}$.
Comparing with the two sub-figures in Fig.~\ref{fig:radex}, it can be seen that (1) the 104.3\,GHz methanol masers are usually produced in a dense ($\gtrsim$10$^5$~cm$^{-3}$) environment, (2) with increasing specific column densities of methanol, the absolute values of optical depth become higher, which means stronger maser emission, and the strongest maser occurs in a less dense environment. 

\section{Summary}\label{Sec:sum}

We performed a survey of three class I methanol masers in the 84, 95 and 104.3\,GHz transitions using the IRAM-30m telescope towards 408 ATLASGAL clumps. The main results are summarized as follows.

\begin{itemize}

\item[1.] We detect in 282 (70\%), 224 (55\%) and 29 (7\%) sources with methanol emission at 84, 95 and 104.3 GHz, respectively. The 104.3 GHz emission is only found in sources with both 84 and 95\,GHz detections. 
A total of 54, 100 and 4 sources show maser-like features at 84, 95 and 104.3 GHz, respectively. 
Among them, fifty 84 GHz masers, twenty nine 95 GHz masers, and four 104.3\,GHz masers are new discoveries. Our work increases the number of known 104.3\,GHz masers from five to nine.

\item[2.] In our sample, the 95\,GHz class I methanol maser is generally stronger than its 84\,GHz maser counterpart. For both the 84 and 95\,GHz class I masers, the relative velocities between maser velocity and systemic velocity are within $\sim$5~\kms, confirming that the class I maser velocity can well trace the systemic velocity. 

\item[3.] All detected 104.3\,GHz methanol masers are redshifted with respect to the systemic velocity. Based on the properties of their associated ATLASGAL clumps, this maser species would not arise in environments where $L_{\rm bol}$ $\lesssim$ 10$^4$~\Lsun, $M_{\rm fwhm}$ $\lesssim$ 200~\Msun, $T_{\rm dust}$ $\lesssim$ 22~K, $N(\rm H_2)$ $\lesssim$ 10$^{23}$~cm$^{-2}$ and $n_{\rm fwhm}(\rm H_2)$ $\lesssim$ 10$^5$~cm$^{-3}$.

\item[4.] We find 9 sources that show class I methanol masers but no SiO emission, indicating that class I methanol masers might be the only signpost of protostellar activity in extremely embedded objects at the earliest evolutionary stage. More and stronger class I masers were detected towards sources showing SiO line wings than towards sources without SiO wings.
The total integrated intensity of class I masers is positively correlated with SiO integrated intensity and FWZP of SiO (2--1) emission. These facts strongly suggest that the properties of class I masers are regulated by shock properties also traced by SiO.

\item[5.] The properties of class I methanol masers show positive correlations with the following properties of associated ATLASGAL clumps: bolometric luminosity, clump mass and peak H$_2$ column density. 
There is no statistically significant correlation between the luminosity of class I methanol masers and the luminosity-to-mass ratio, dust temperature, or mean H$_2$ volume density.

\item[6.] Our results show that the 84 GHz and 95 GHz methanol masers exist in the quiescent, protostellar, YSO, and H{\sc ii} region stages. In contrast, the 104.3 GHz methanol masers are only detected in the H{\sc ii} region stage. Based on the distribution of bolometric luminosity-to-mass ratios associated with different masers, we suggest that class I methanol masers at 84 and 95~GHz can trace a similar evolutionary stage as H$_2$O maser, and appear prior to 6.7 and 12.2 GHz methanol and OH masers. 
The 104.3~GHz class I masers appear to trace a short and more evolved stage compared to the other class I maser species.

\item[7.]
Our calculations show that physical conditions can be better constrained in region in which multiple class I methanol masers arise. 
With an assumed specific column density of methanol of 10$^{15}$~cm$^{-2}$~km$^{-1}$~s (10$^{16}$~cm$^{-2}$~km$^{-1}$~s), a kinetic temperature of 62$\pm$3~K (57$\pm$3~K), and a para-H$_2$ number density of 4.4($\pm$1.5)$\times$10$^6$~cm$^{-3}$ (7.9($\pm$2.5)$\times$10$^5$~cm$^{-3}$) are needed to simultaneously excite the three class I methanol masers in G10.34$-$0.14.

\end{itemize}  

We present a systemic study of class I methanol masers in the inner Galaxy and their relation with SiO emission on clump scales. 
In future work, more in-depth statistical analysis such as mathematical classification models can be performed, to understand which properties of the ATLASGAL clump are the best predictors of an associated class I methanol maser. This analysis can then be used to predict which ATLASGAL clumps have not been observed should be targeted for class I maser searches.
In addition, higher angular follow-up observations will be able to pinpoint the location of class I methanol masers on smaller scales, which will provide more constraints on the relation between the masers and their driving stars.

\section*{ACKNOWLEDGMENTS}\label{sec.ack}
T. Cs. has received financial support from the French State in the framework of the IdEx Universit\'e de Bordeaux Investments for the future Program.
This work made use of Python libraries including Astropy\footnote{\url{https://www.astropy.org/}} \citep{2013A&A...558A..33A}, NumPy\footnote{\url{https://www.numpy.org/}} \citep{5725236}, SciPy\footnote{\url{https://www.scipy.org/}} \citep{jones2001scipy}, Matplotlib\footnote{\url{https://matplotlib.org/}} \citep{Hunter:2007}.
This research has made use of the VizieR catalogue, operated at CDS, Strasbourg, France.

\bibliographystyle{aa}
\bibliography{references}

\begin{appendix}
\section{Spectra of nine methanol maser transitions for all observed 408 ATLASGAL sources}\label{sec:appendix-a}

\begin{figure*}[!htbp]
\centering
\mbox{
\begin{minipage}[b]{9.5cm}
\includegraphics[width=0.9\textwidth]{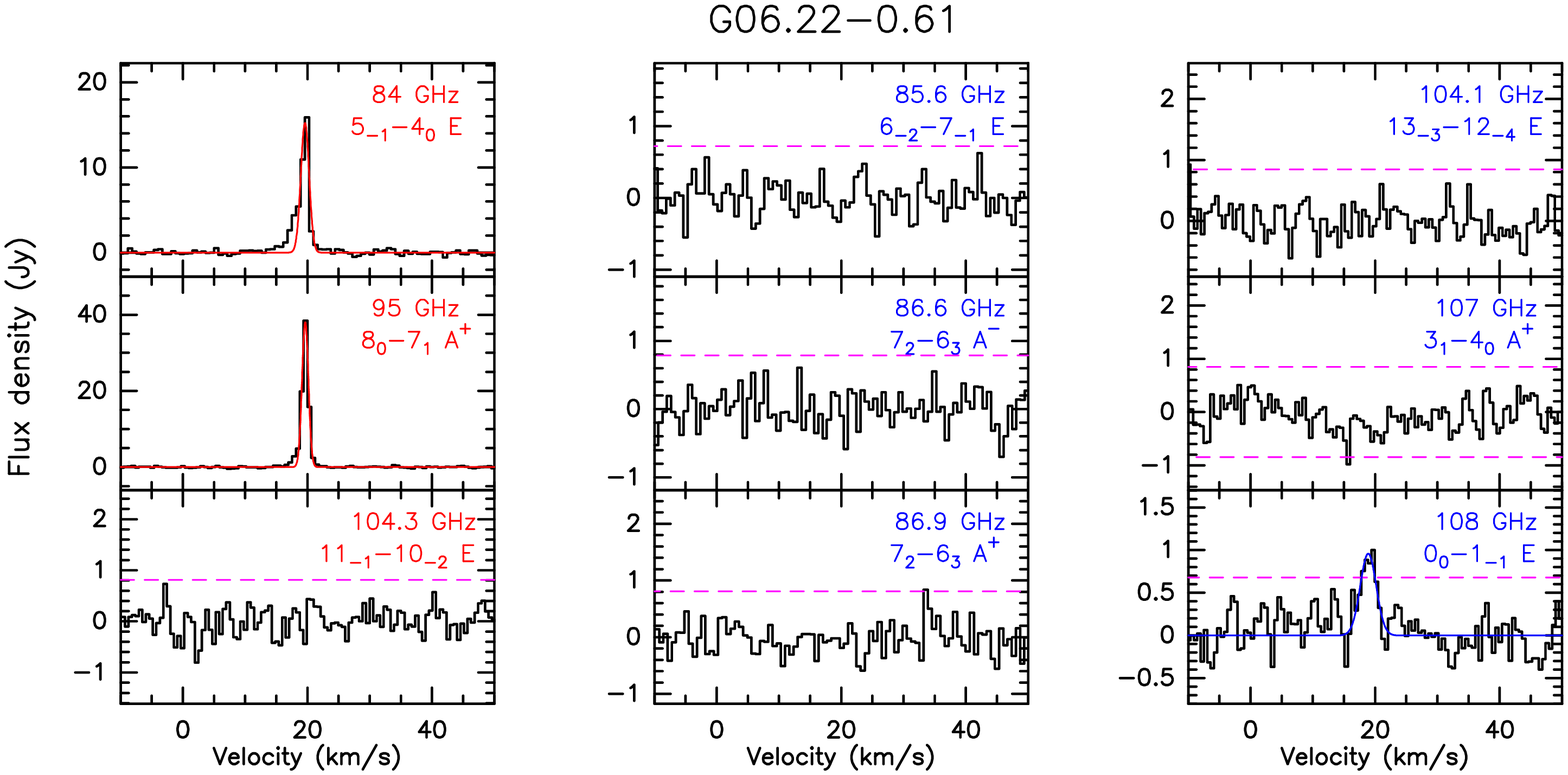}
\vspace{3.5mm}
\end{minipage}
\begin{minipage}[b]{9.5cm}
\hspace{-4mm}
\includegraphics[width=0.9\textwidth]{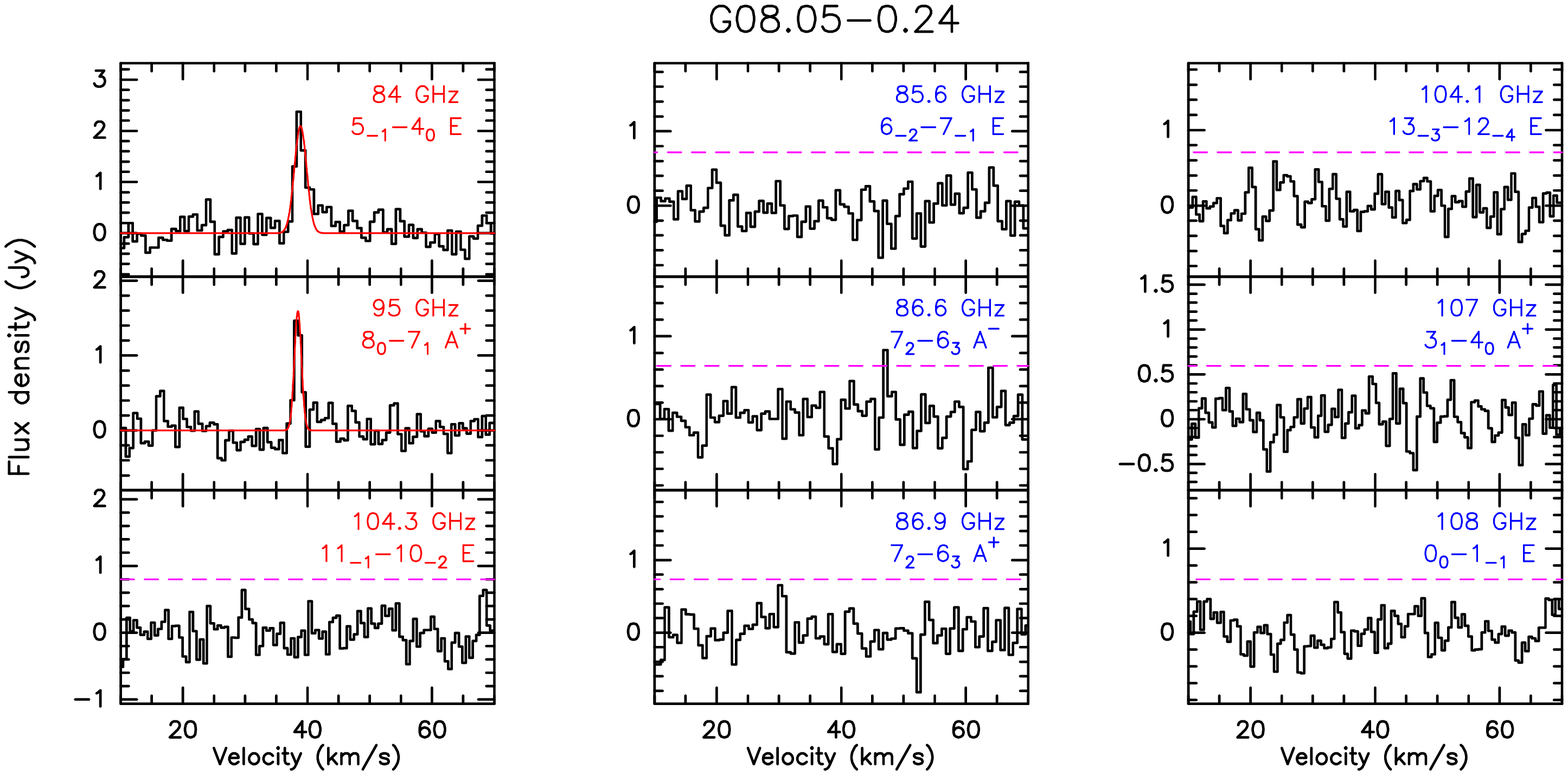}
\vspace{3.5mm}
\end{minipage}
}
\mbox{
\begin{minipage}[b]{9.5cm}
\includegraphics[width=0.9\textwidth]{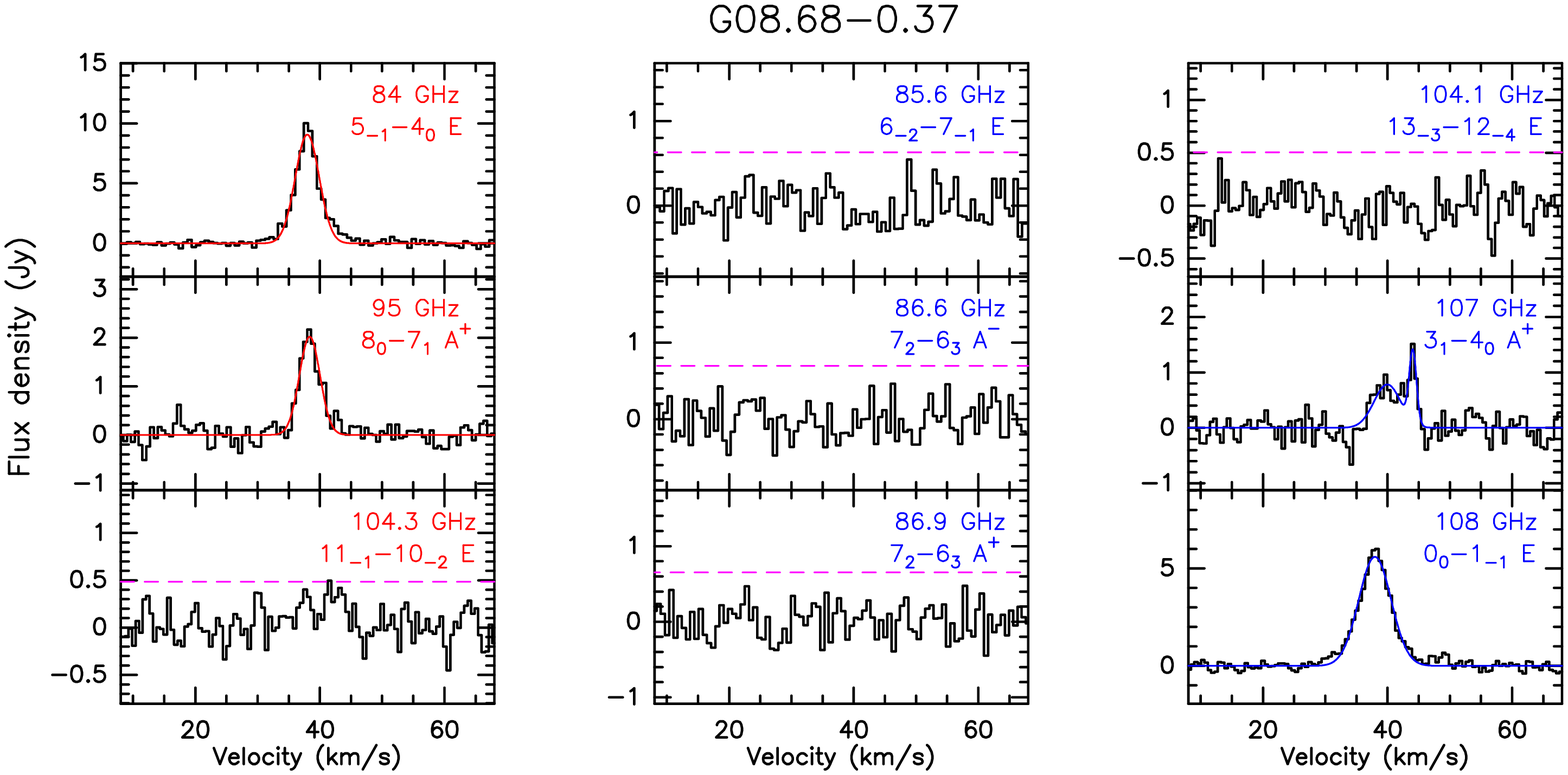}
\vspace{3.5mm}
\end{minipage}
\begin{minipage}[b]{9.5cm}
\hspace{-4mm}
\includegraphics[width=0.9\textwidth]{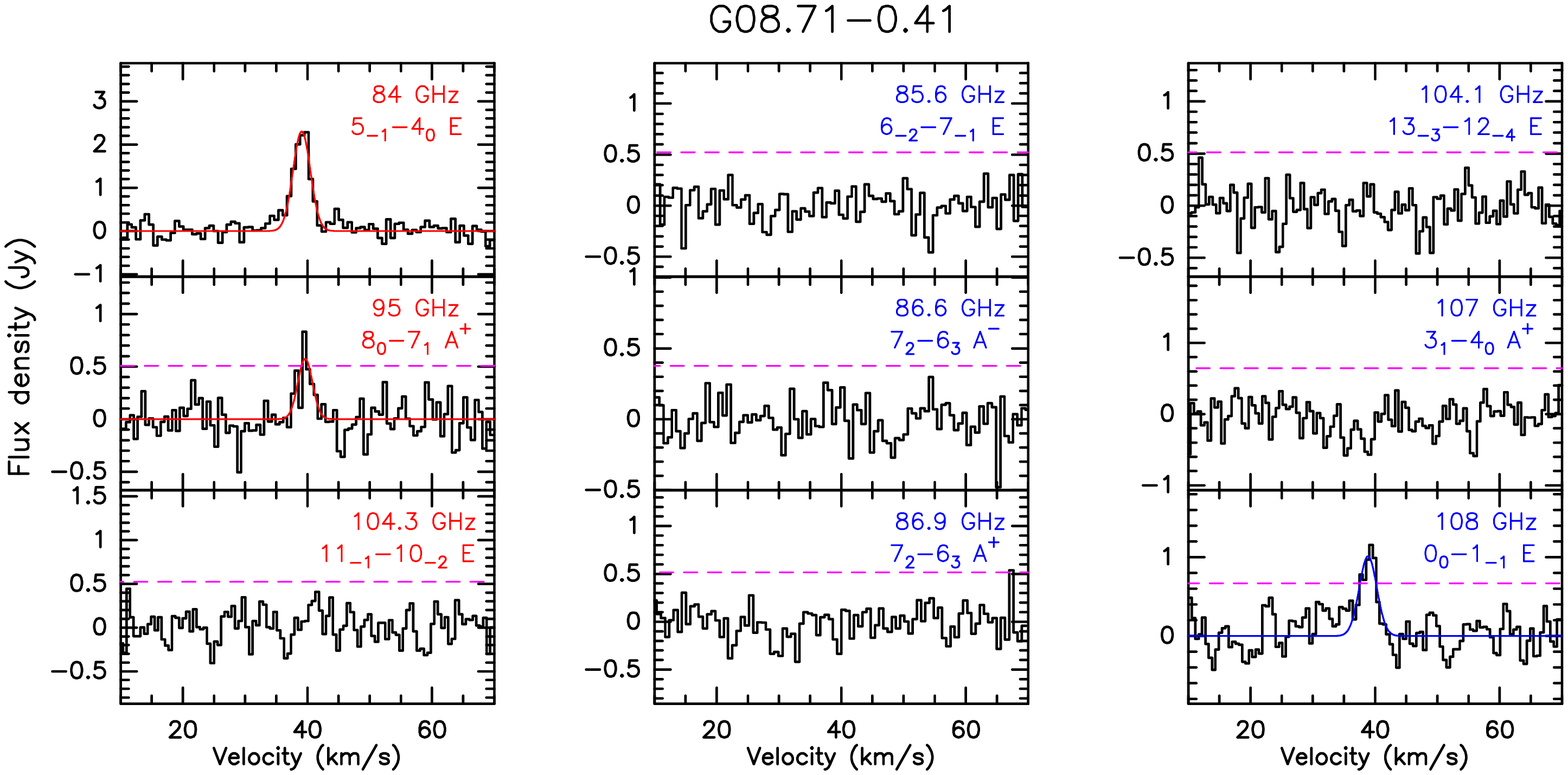}
\vspace{3.5mm}
\end{minipage}
}
\mbox{
\begin{minipage}[b]{9.5cm}
\includegraphics[width=0.9\textwidth]{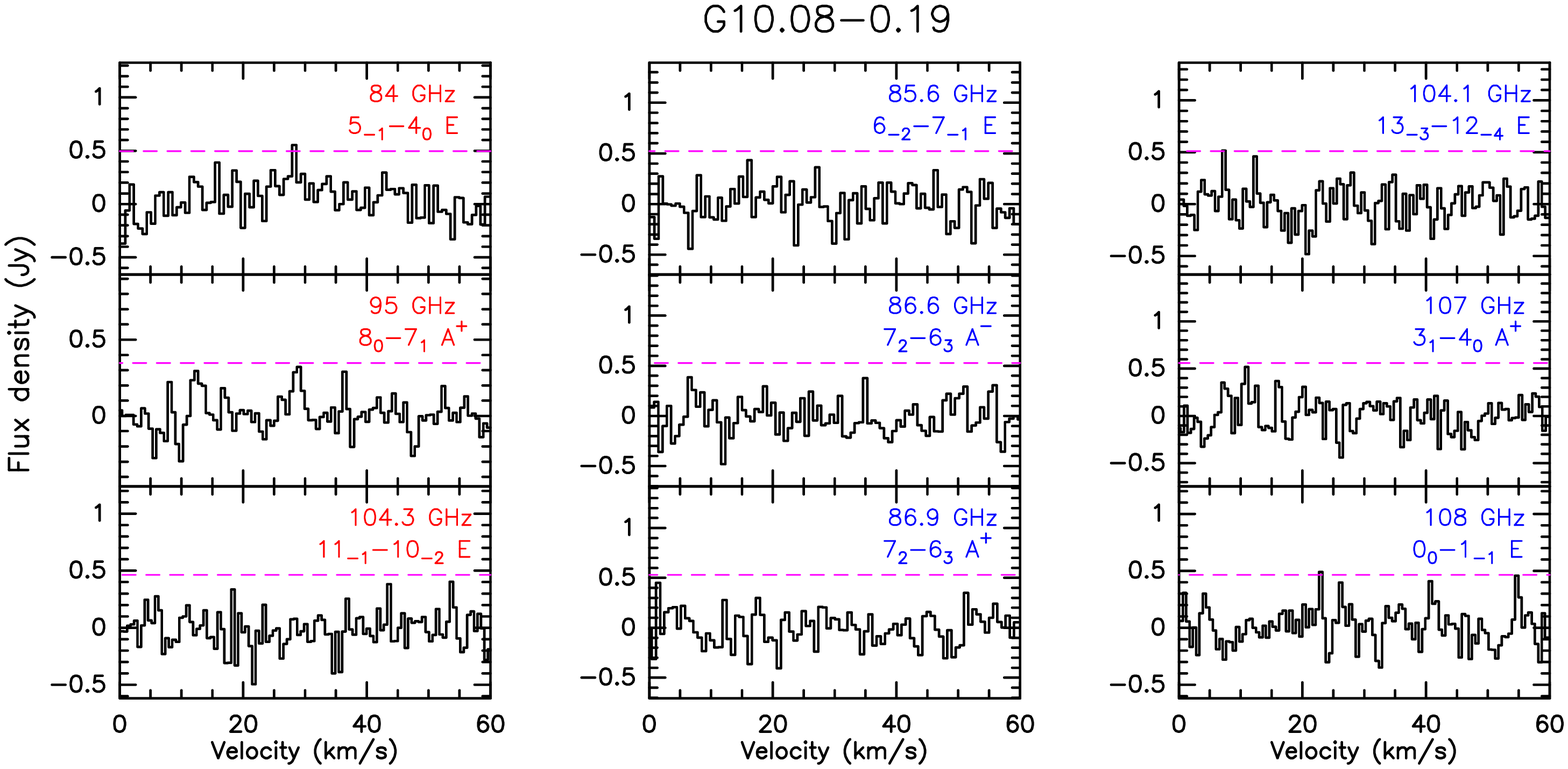}
\vspace{3.5mm}
\end{minipage}
\begin{minipage}[b]{9.5cm}
\hspace{-4mm}
\includegraphics[width=0.9\textwidth]{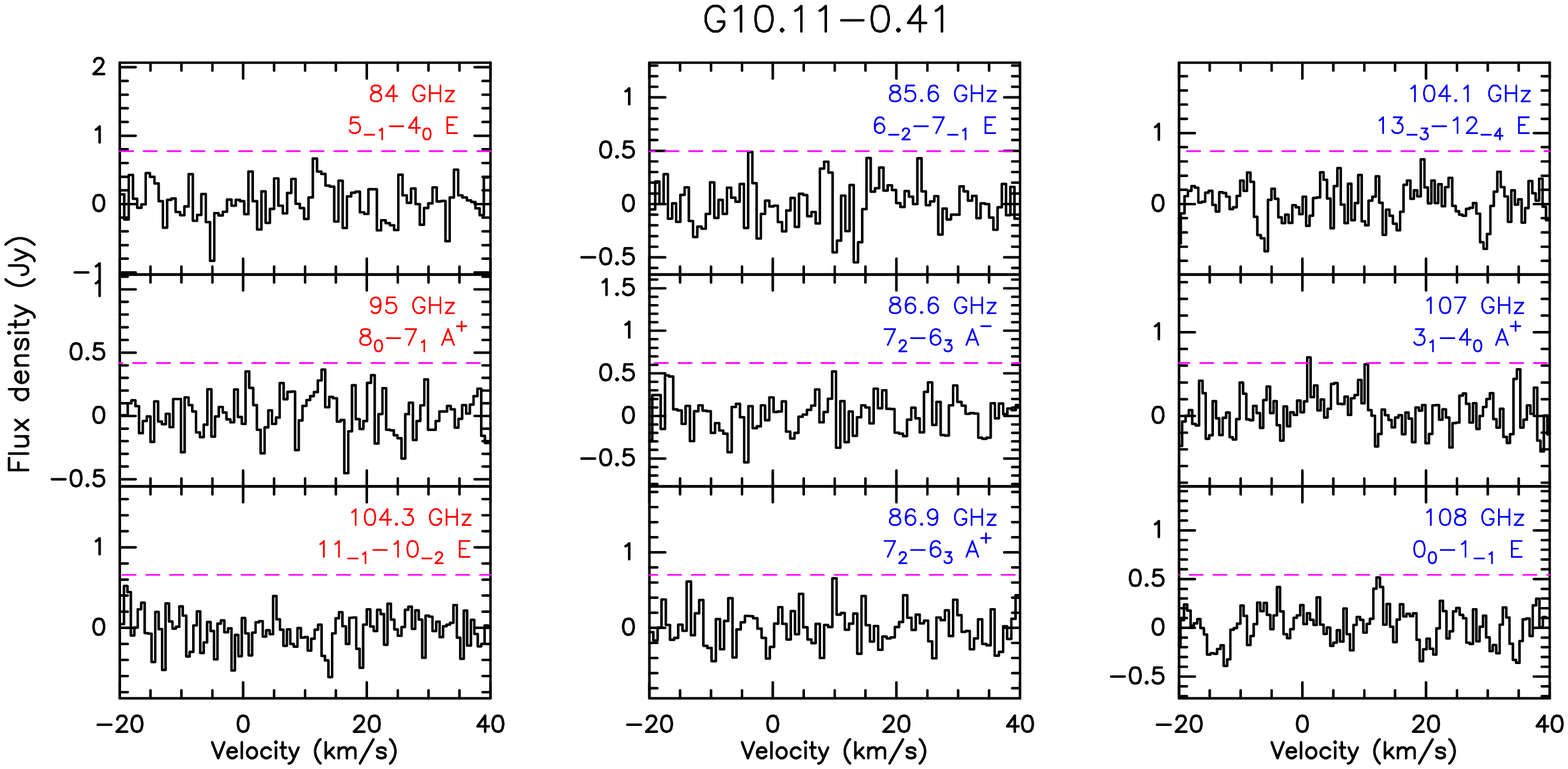}
\vspace{3.5mm}
\end{minipage}
}
\caption{Observed spectra of nine methanol maser transitions at 3~mm towards the first six ATLASGAL sources in our sample.
The frequency and the quantum numbers of class I and II CH$_3$OH maser transitions are labelled in red and blue, respectively, and their Gaussian fitting results are plotted in the corresponding colors.
The horizontal magenta dashed lines represent the 3$\sigma$ noise level for the non-detection transitions. For marginal detections with S/N ratios of $\sim$3, both Gaussian fitting results and 3$\sigma$ levels are presented.
The spectra of the whole sample are available in electronic form at the Zenodo via: \url{https://doi.org/10.5281/zenodo.7442831}. 
\label{fig:408spec}}
\end{figure*}

\section{K-S test results of ATLASGAL properties}\label{sec:appendix-b}
\begin{figure*}[!htbp]
\centering
\mbox{
\begin{minipage}[b]{5.5cm}
\includegraphics[width=1\textwidth]{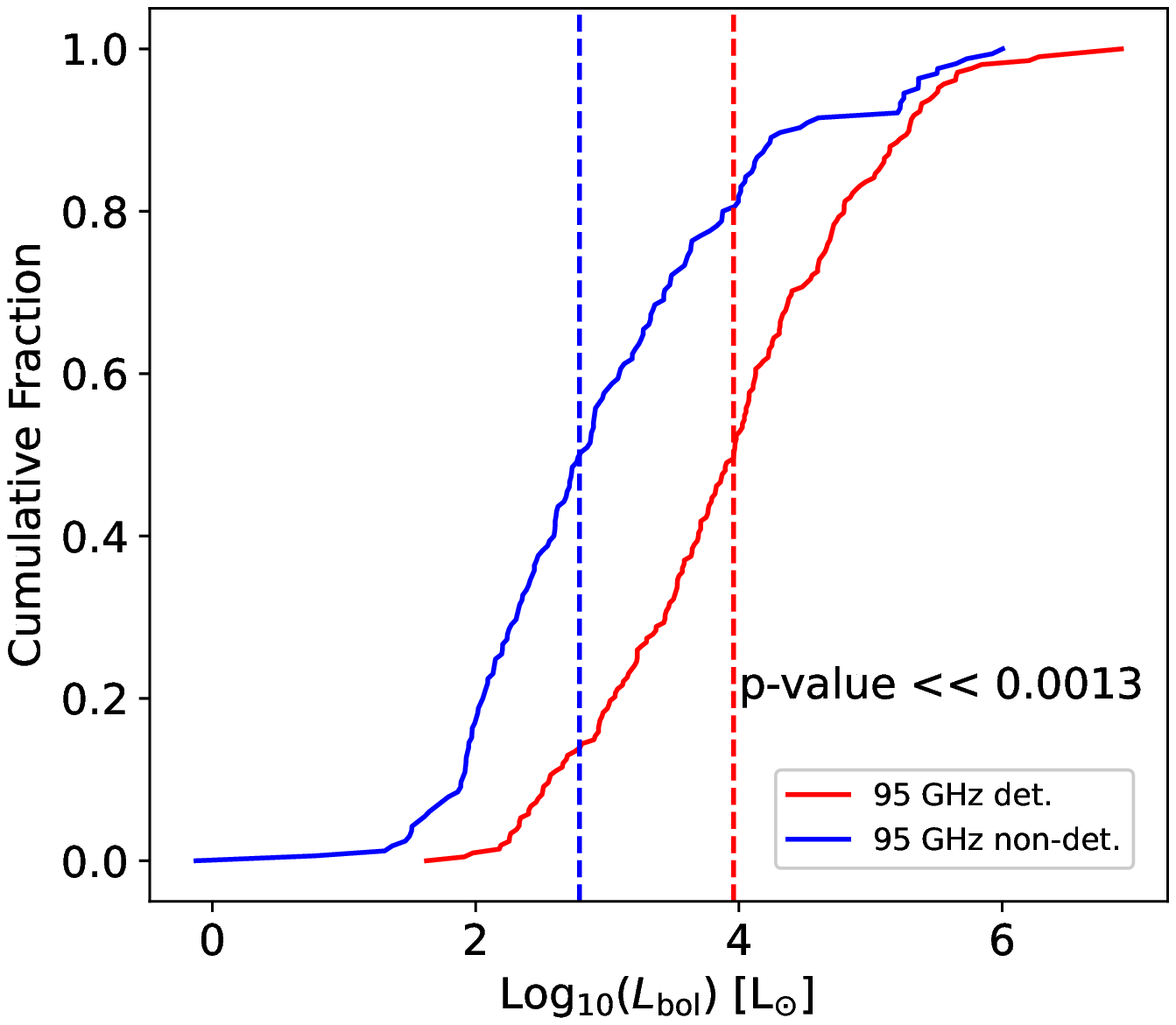}
\centerline{(a)}
\vspace{-3.5mm}
\end{minipage}
\begin{minipage}[b]{5.5cm}
\includegraphics[width=1\textwidth]{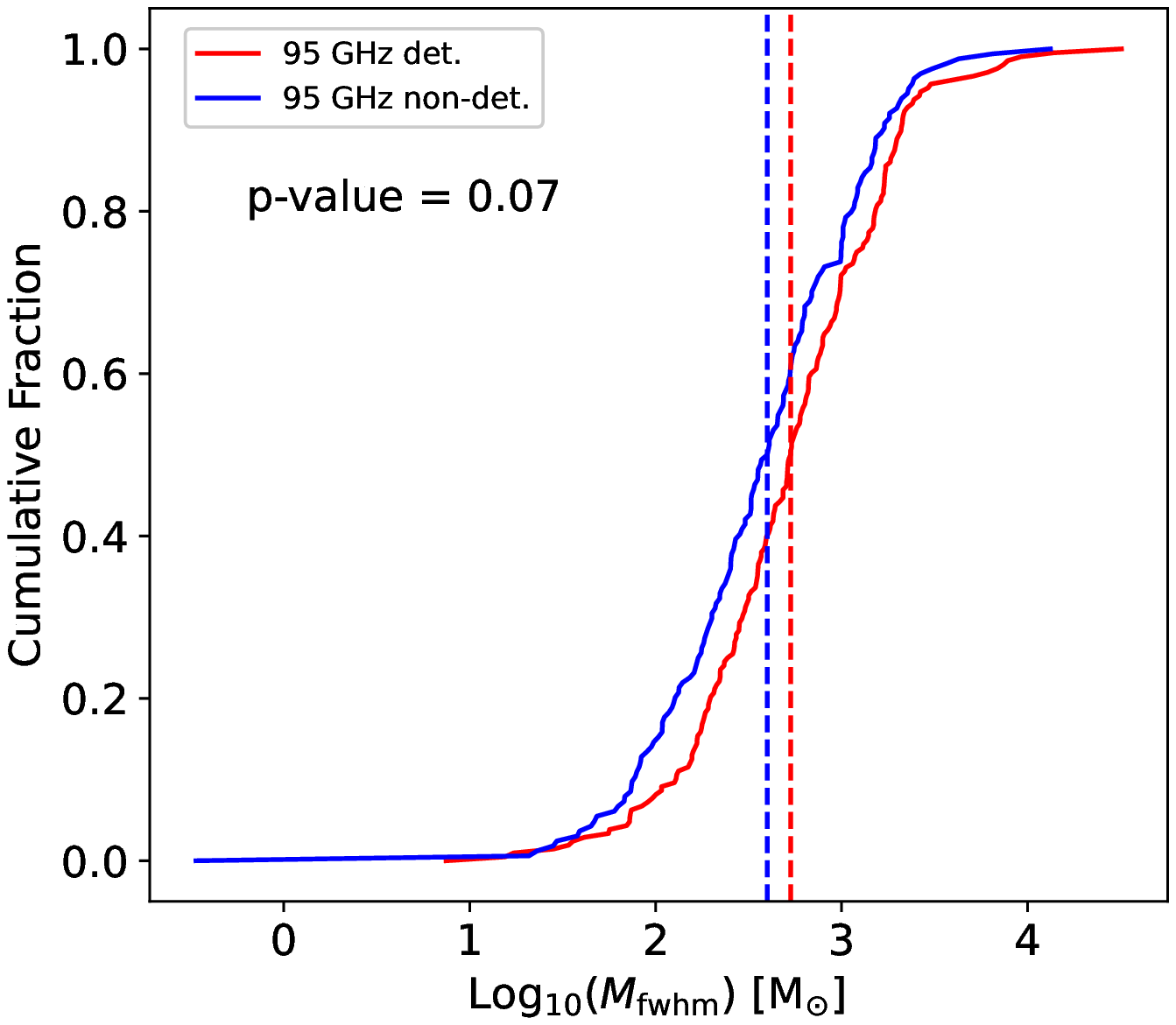}
\centerline{(b)}
\vspace{-3.5mm}
\end{minipage}
\begin{minipage}[b]{5.5cm}
\includegraphics[width=1\textwidth]{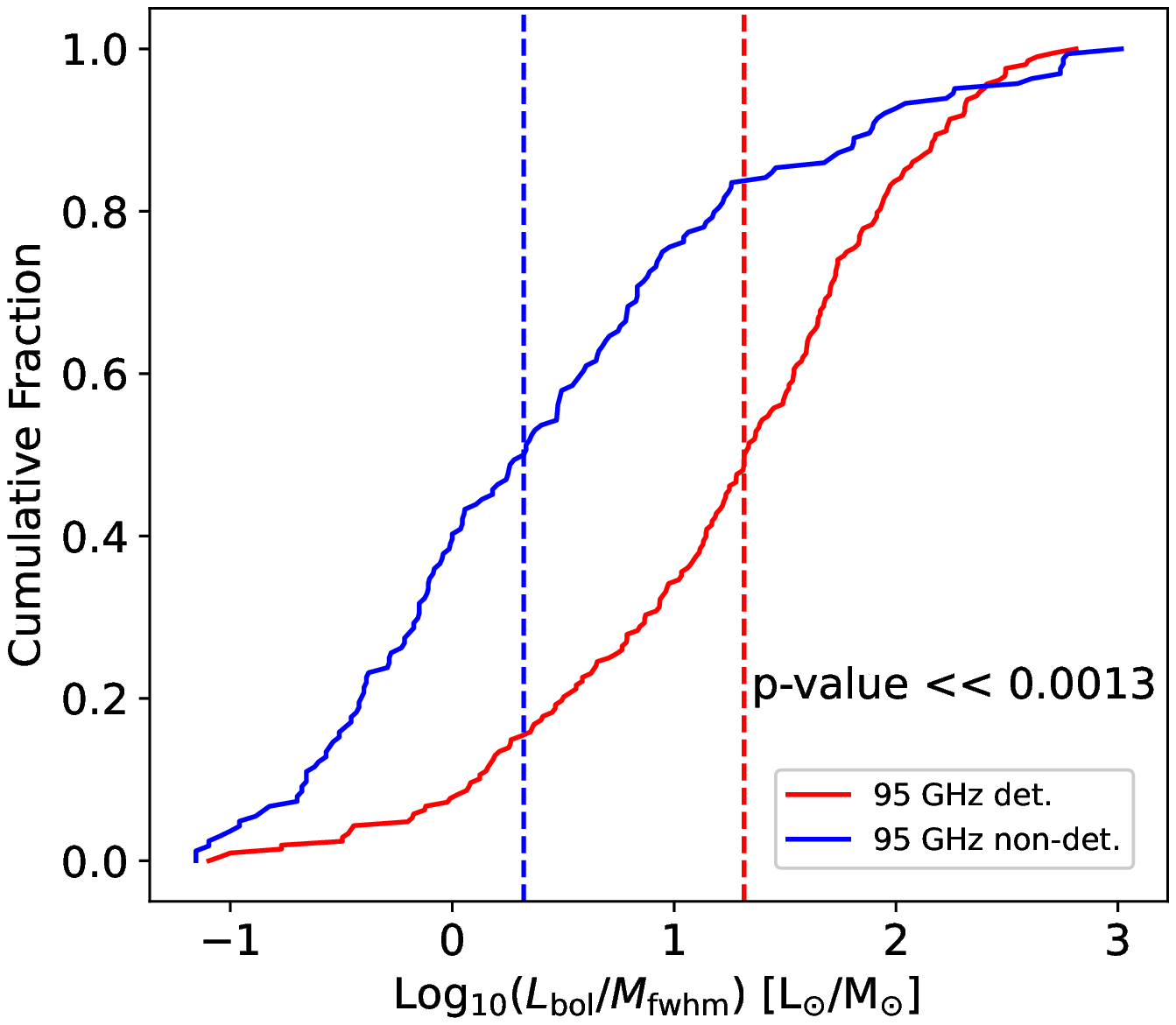}
\centerline{(c)}
\vspace{-3.5mm}
\end{minipage}
}
\mbox{
\begin{minipage}[b]{5.5cm}
\includegraphics[width=1\textwidth]{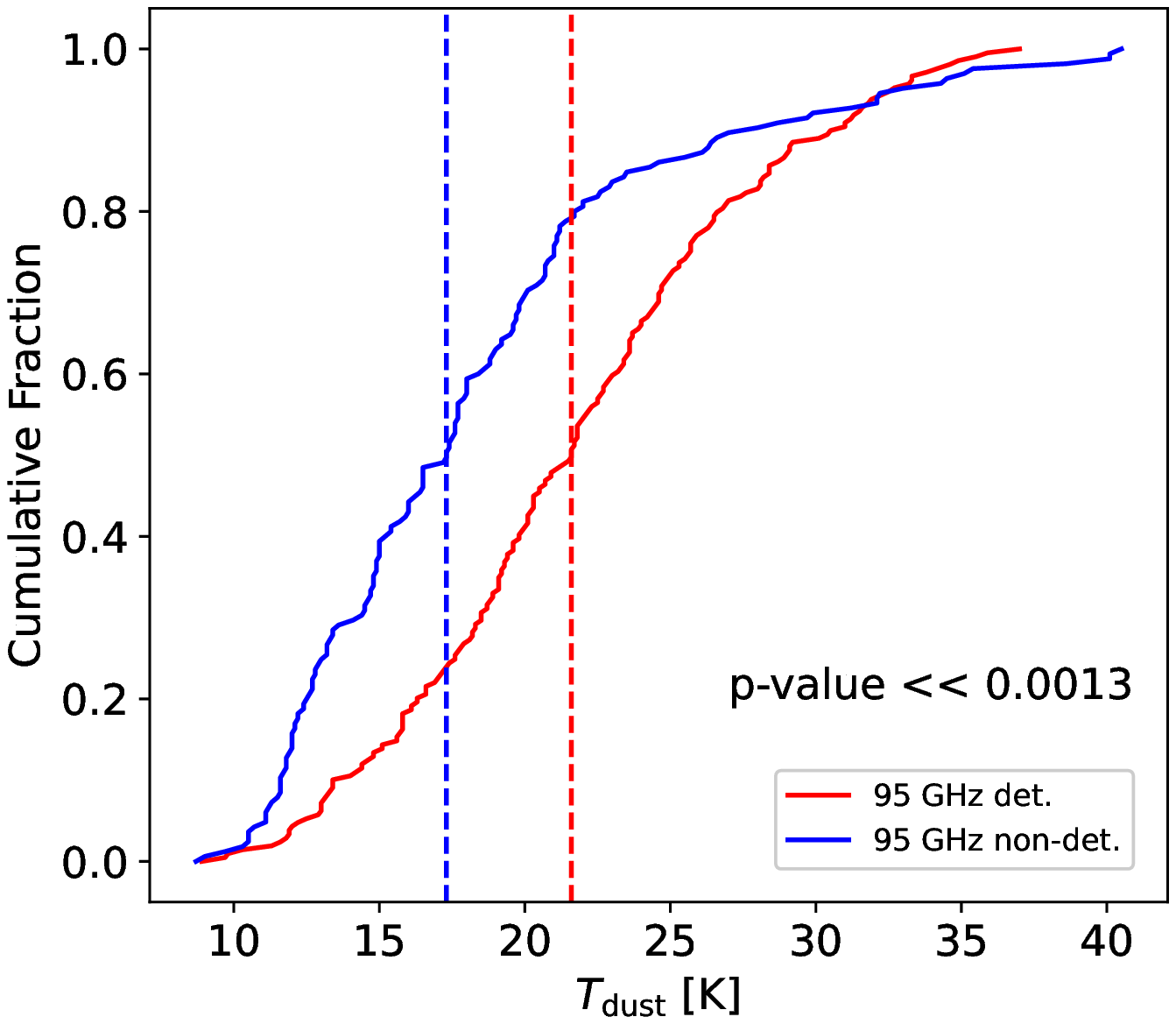}
\centerline{(d)}
\vspace{-3.5mm}
\end{minipage}
\begin{minipage}[b]{5.5cm}
\includegraphics[width=1\textwidth]{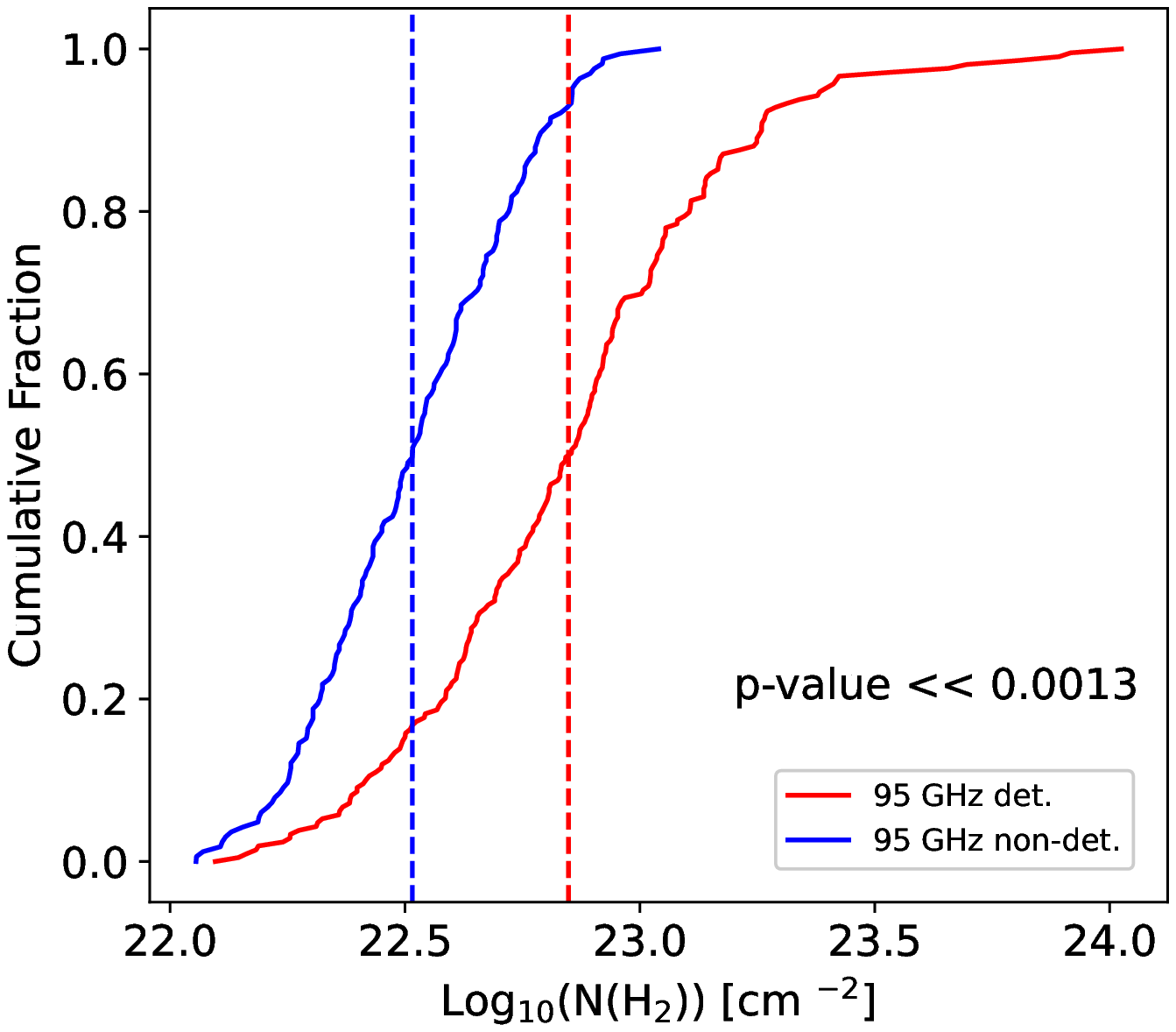}
\centerline{(e)}
\vspace{-3.5mm}
\end{minipage}
\begin{minipage}[b]{5.5cm}
\includegraphics[width=1\textwidth]{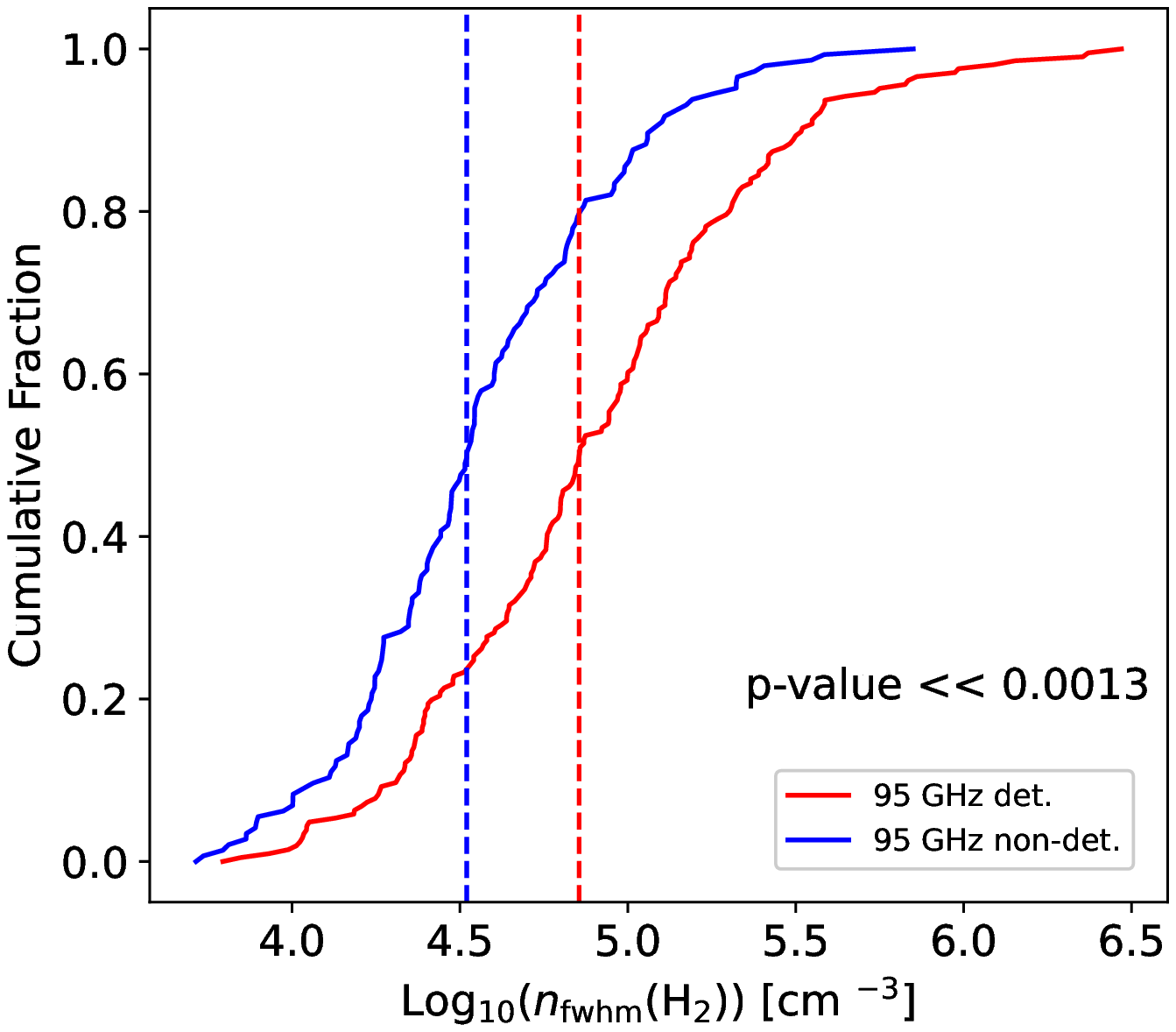}
\centerline{(f)}
\vspace{-3.5mm}
\end{minipage}
}
\caption{Cumulative distribution functions of the physical properties for the clumps with and without 95 GHz methanol detections. From (a) to (f), cumulative distributions of the bolometric luminosity, the FWHM clump mass, the luminosity-to-mass ratio, the dust temperature, the peak H$_2$ column density, the mean H$_2$ FWHM volume density for 95 GHz methanol detections (red lines) and non-detections (blue lines), respectively. The vertical dashed lines in the corresponding colors depict the median values of the two samples. The p-values from the K-S tests are presented in each panel.
\label{fig:95-ks-atlasgal}}
\end{figure*}

\begin{figure*}[!htbp]
\centering
\mbox{
\begin{minipage}[b]{5.5cm}
\includegraphics[width=1\textwidth]{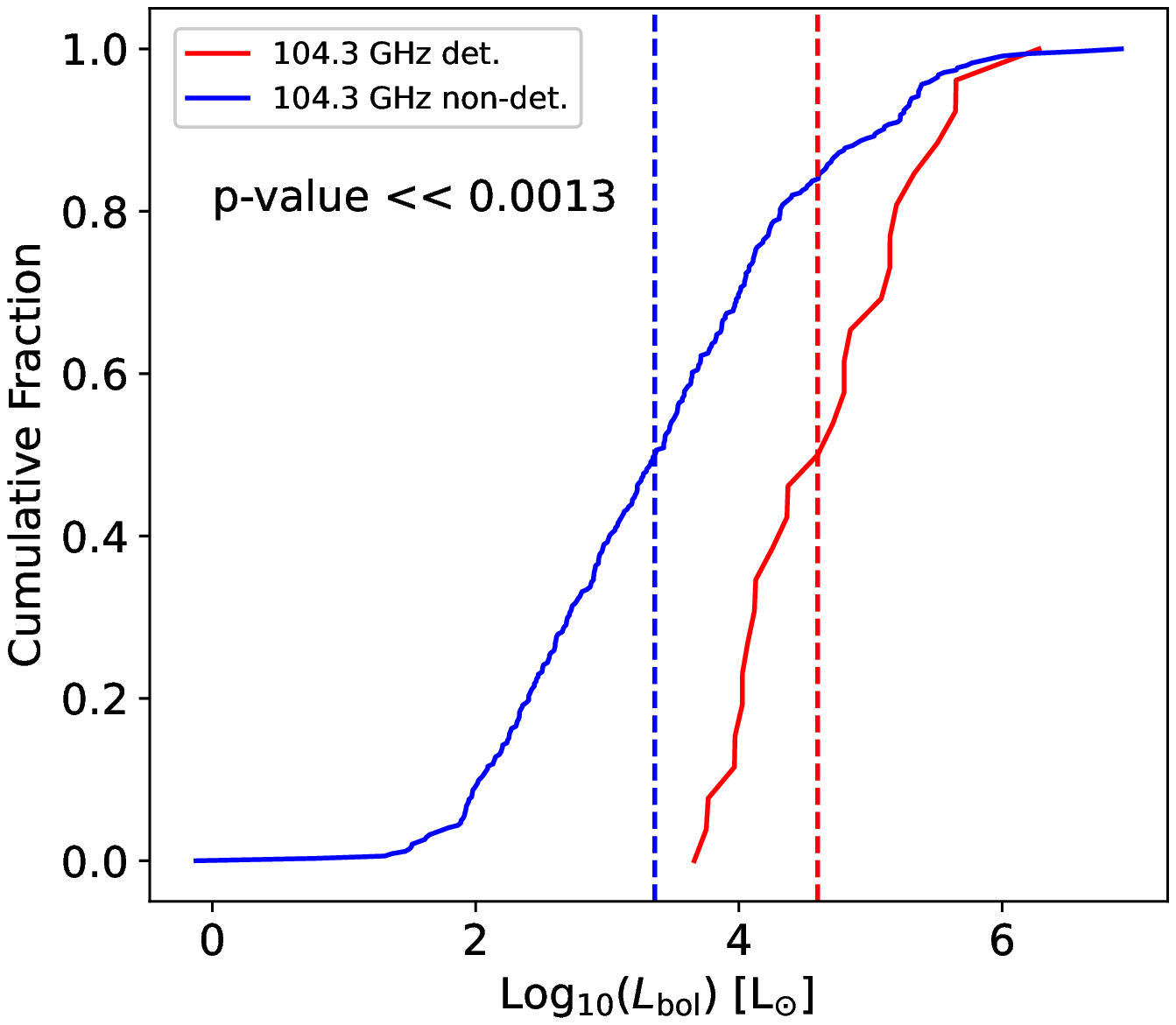}
\centerline{(a)}
\vspace{-3.5mm}
\end{minipage}
\begin{minipage}[b]{5.5cm}
\includegraphics[width=1\textwidth]{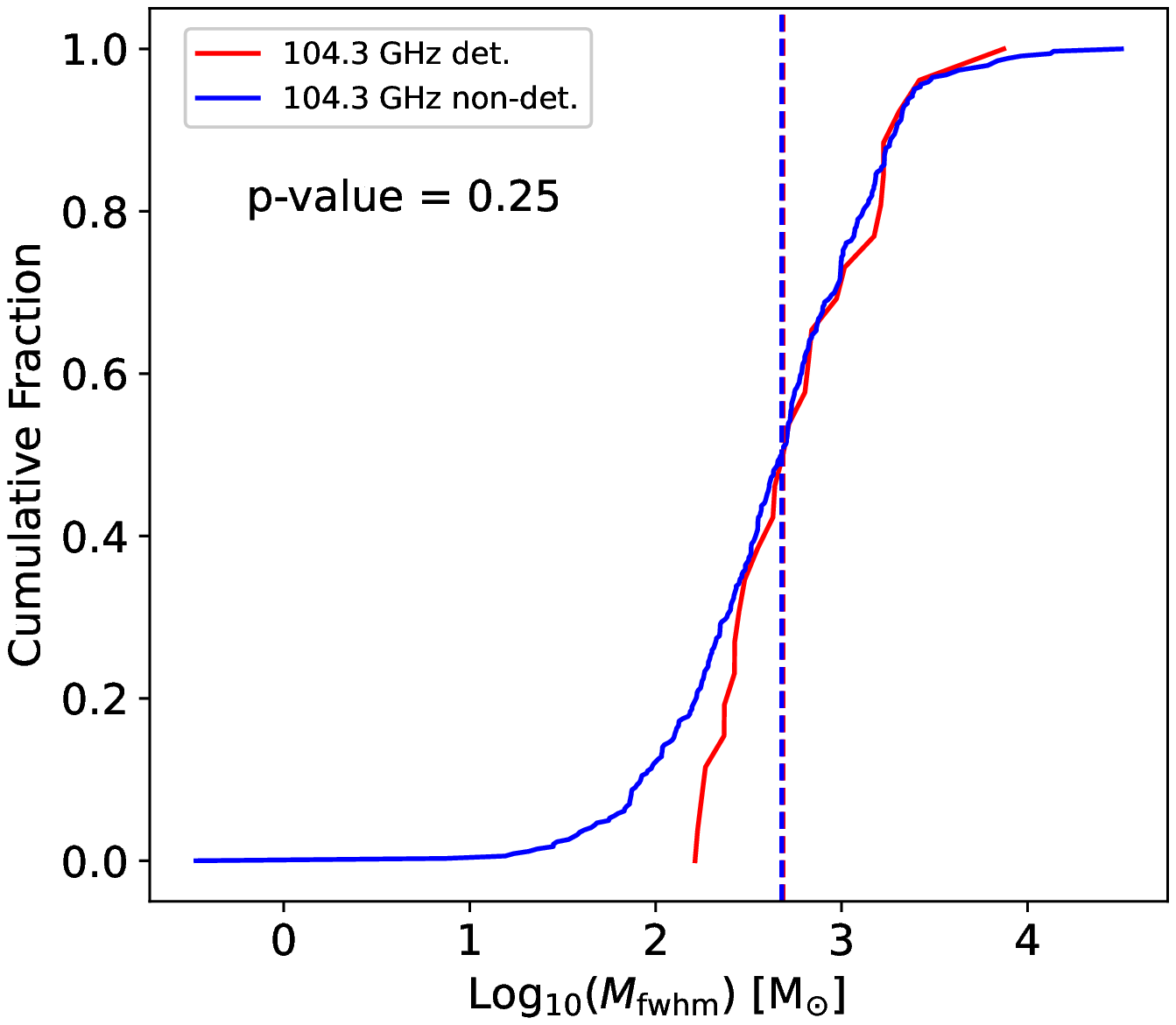}
\centerline{(b)}
\vspace{-3.5mm}
\end{minipage}
\begin{minipage}[b]{5.5cm}
\includegraphics[width=1\textwidth]{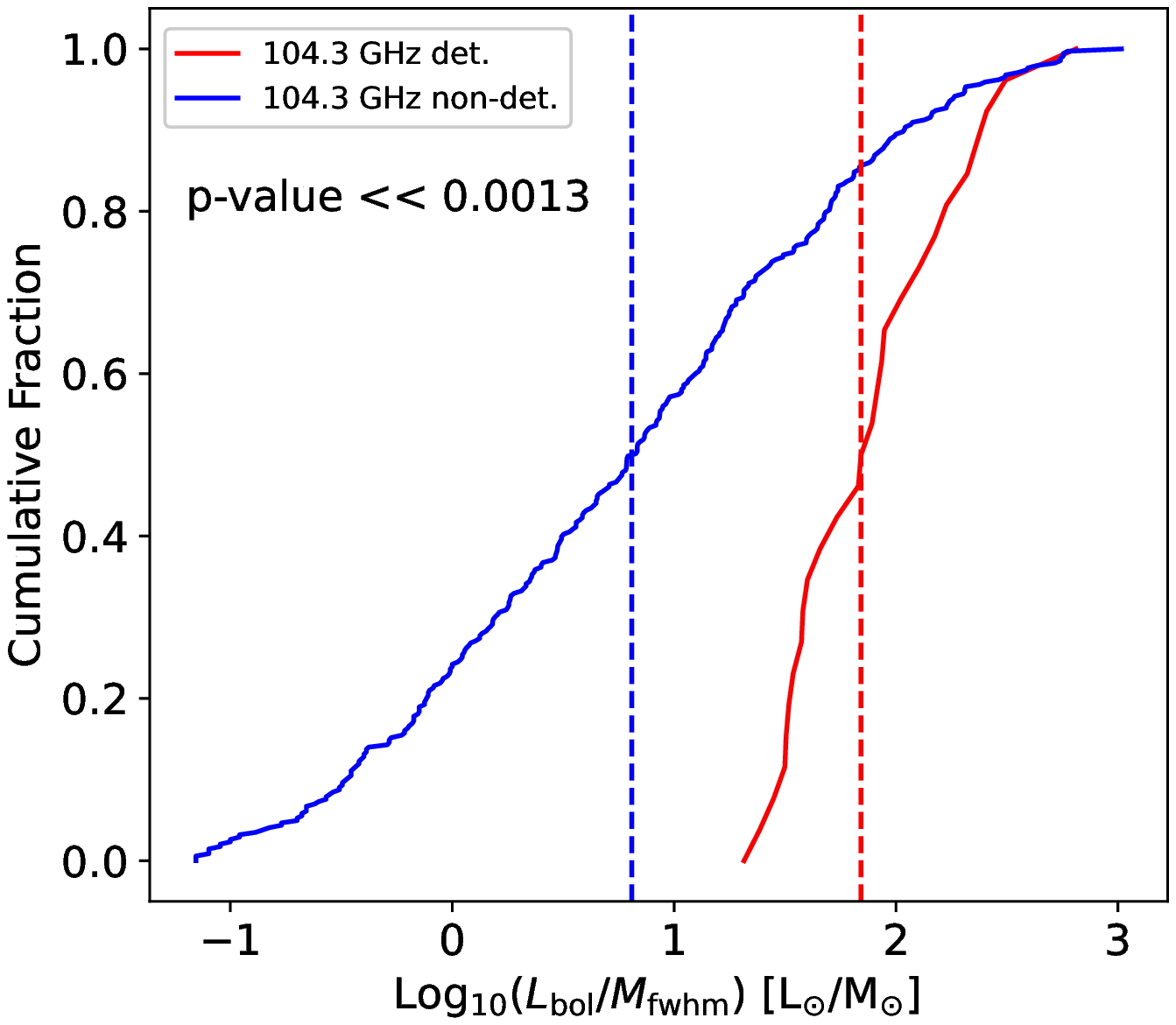}
\centerline{(c)}
\vspace{-3.5mm}
\end{minipage}
}
\mbox{
\begin{minipage}[b]{5.5cm}
\includegraphics[width=1\textwidth]{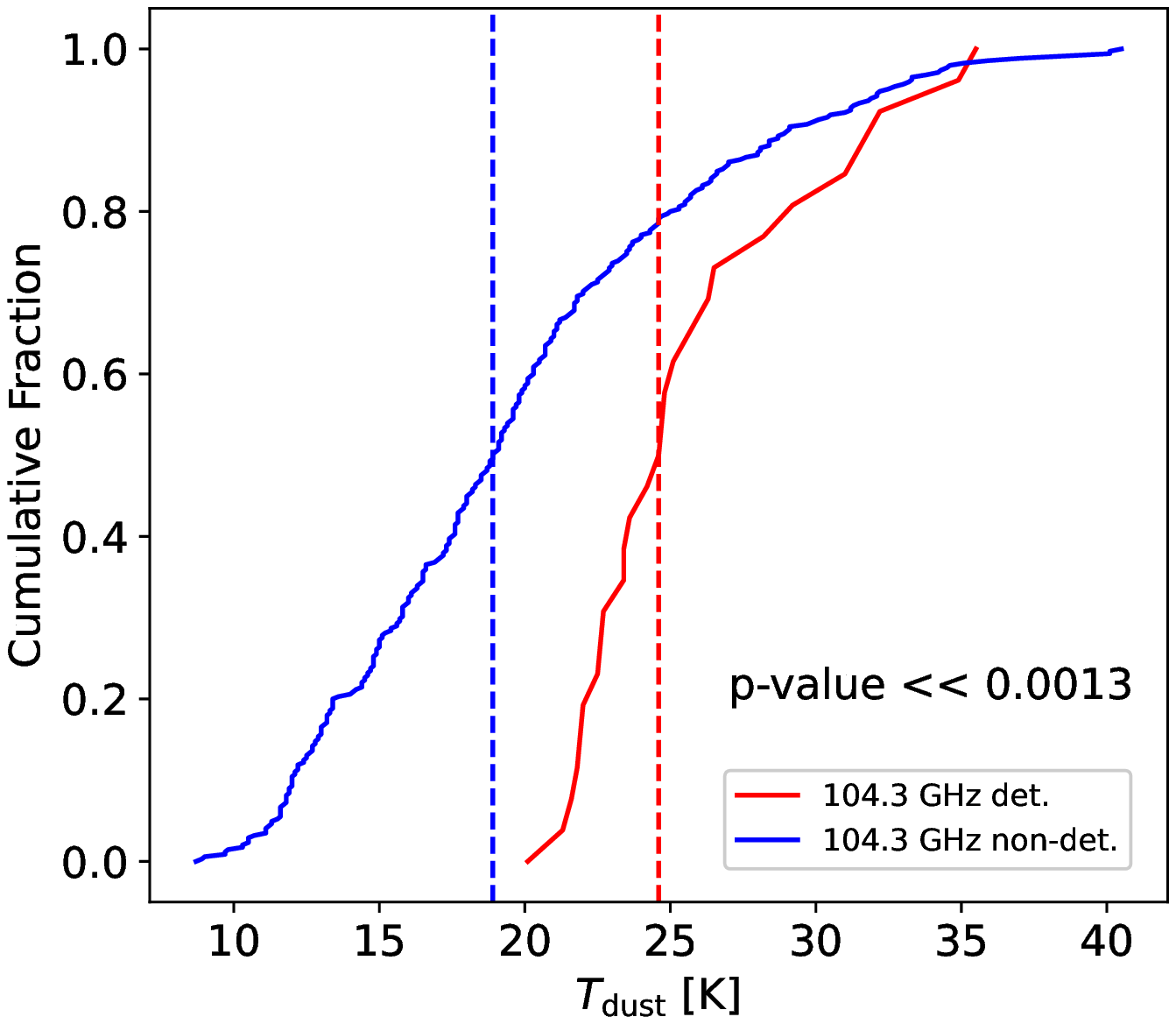}
\centerline{(d)}
\vspace{-3.5mm}
\end{minipage}
\begin{minipage}[b]{5.5cm}
\includegraphics[width=1\textwidth]{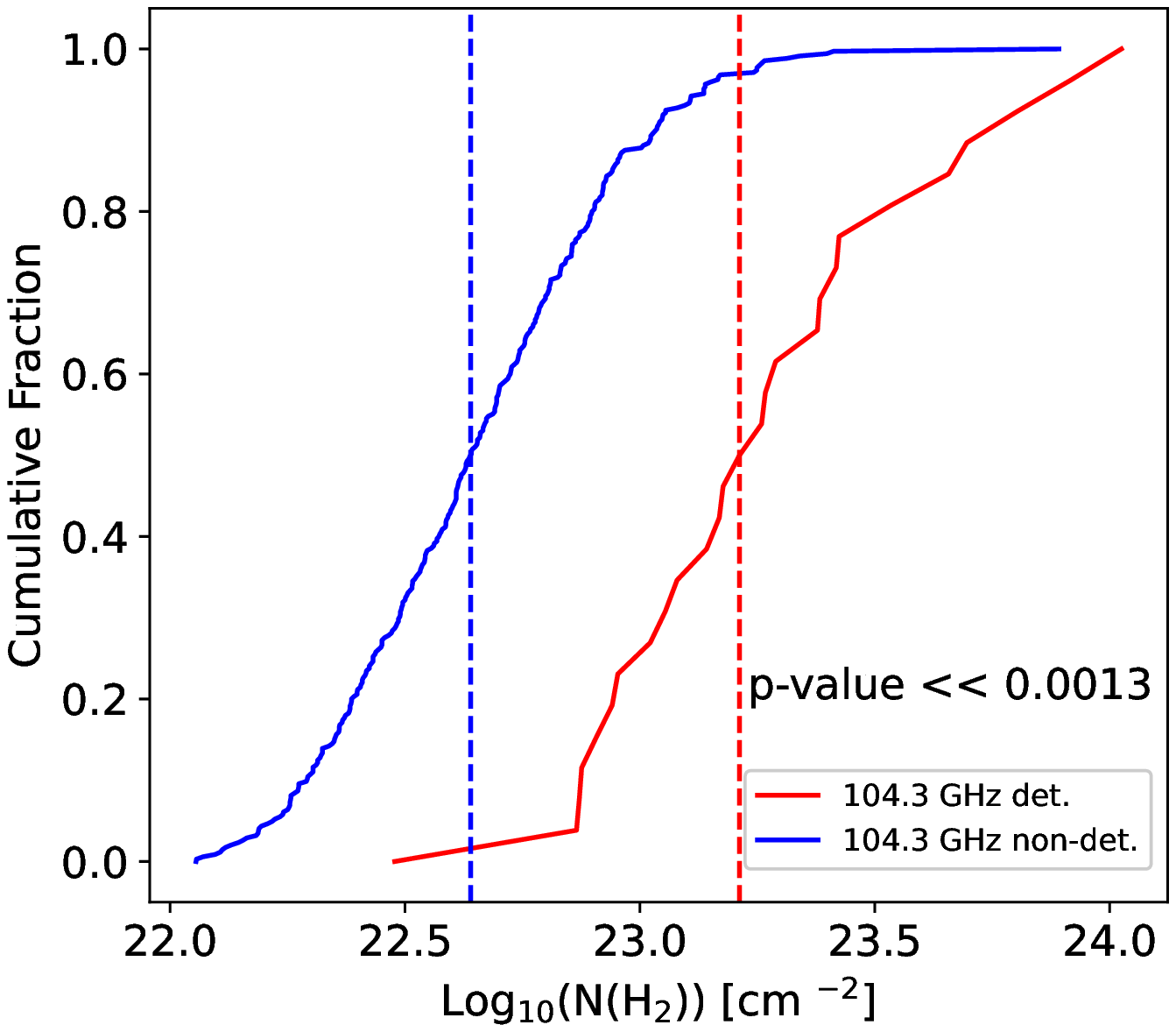}
\centerline{(e)}
\vspace{-3.5mm}
\end{minipage}
\begin{minipage}[b]{5.5cm}
\includegraphics[width=1\textwidth]{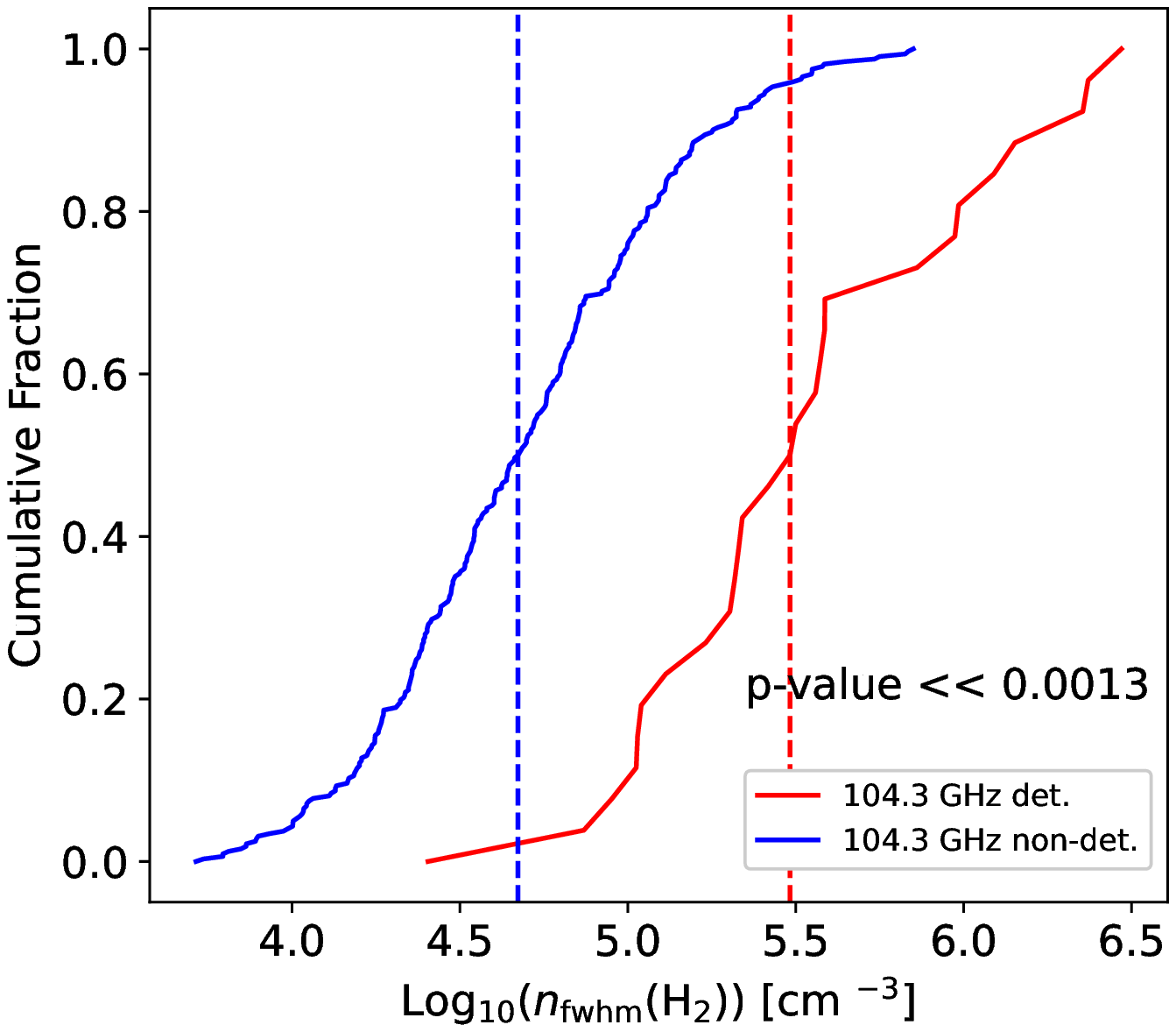}
\centerline{(f)}
\vspace{-3.5mm}
\end{minipage}
}
\caption{Cumulative distribution functions of the physical properties for the clumps with and without 104.3 GHz methanol detections. Panels (a) to (f) are similar to the corresponding one in Fig.~\ref{fig:95-ks-atlasgal}.
\label{fig:104-ks-atlasgal}}
\end{figure*}

\begin{figure*}[!htbp]
\centering
\mbox{
\begin{minipage}[b]{5.5cm}
\includegraphics[width=1\textwidth]{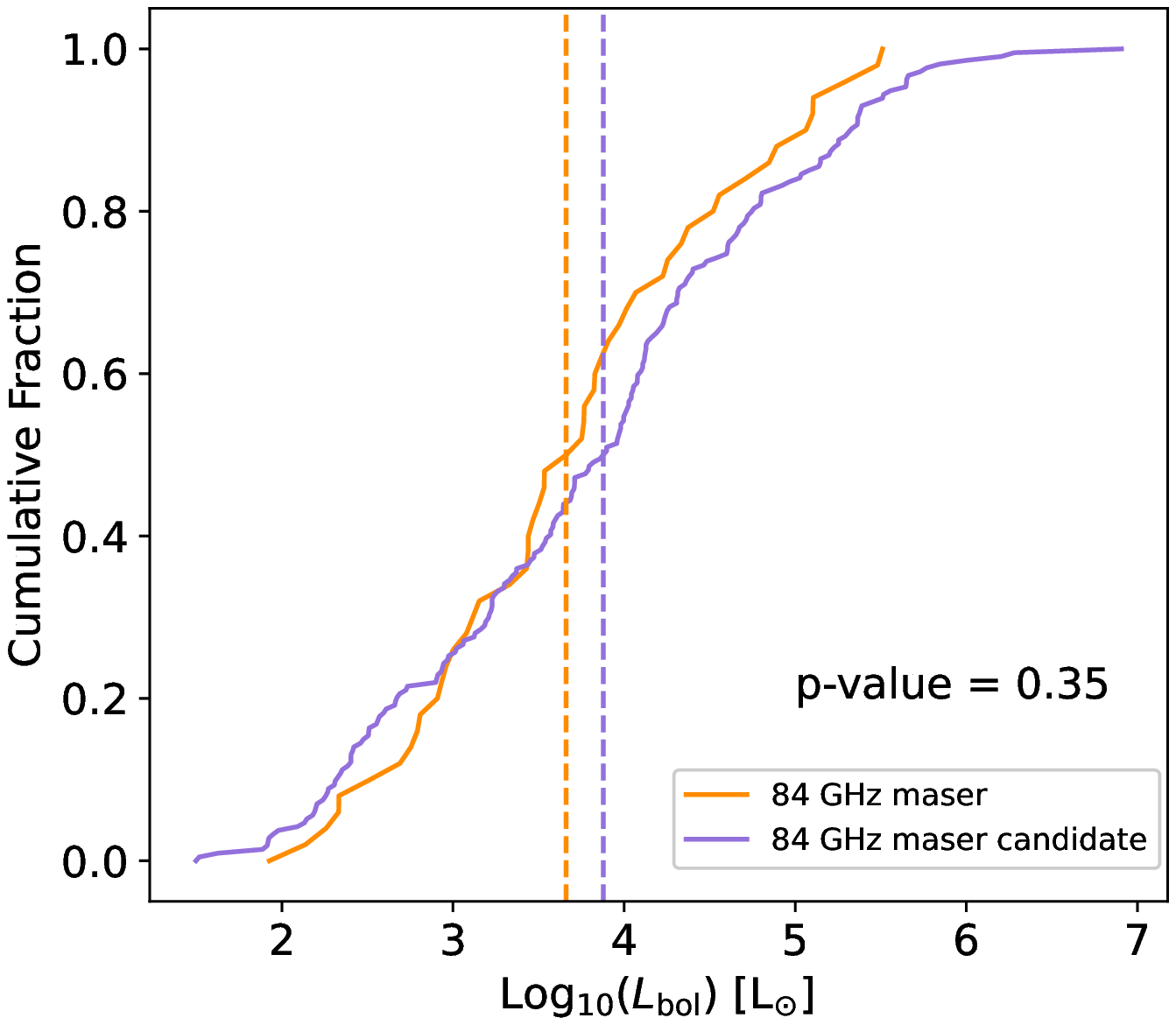}
\centerline{(a)}
\vspace{-3.5mm}
\end{minipage}
\begin{minipage}[b]{5.5cm}
\includegraphics[width=1\textwidth]{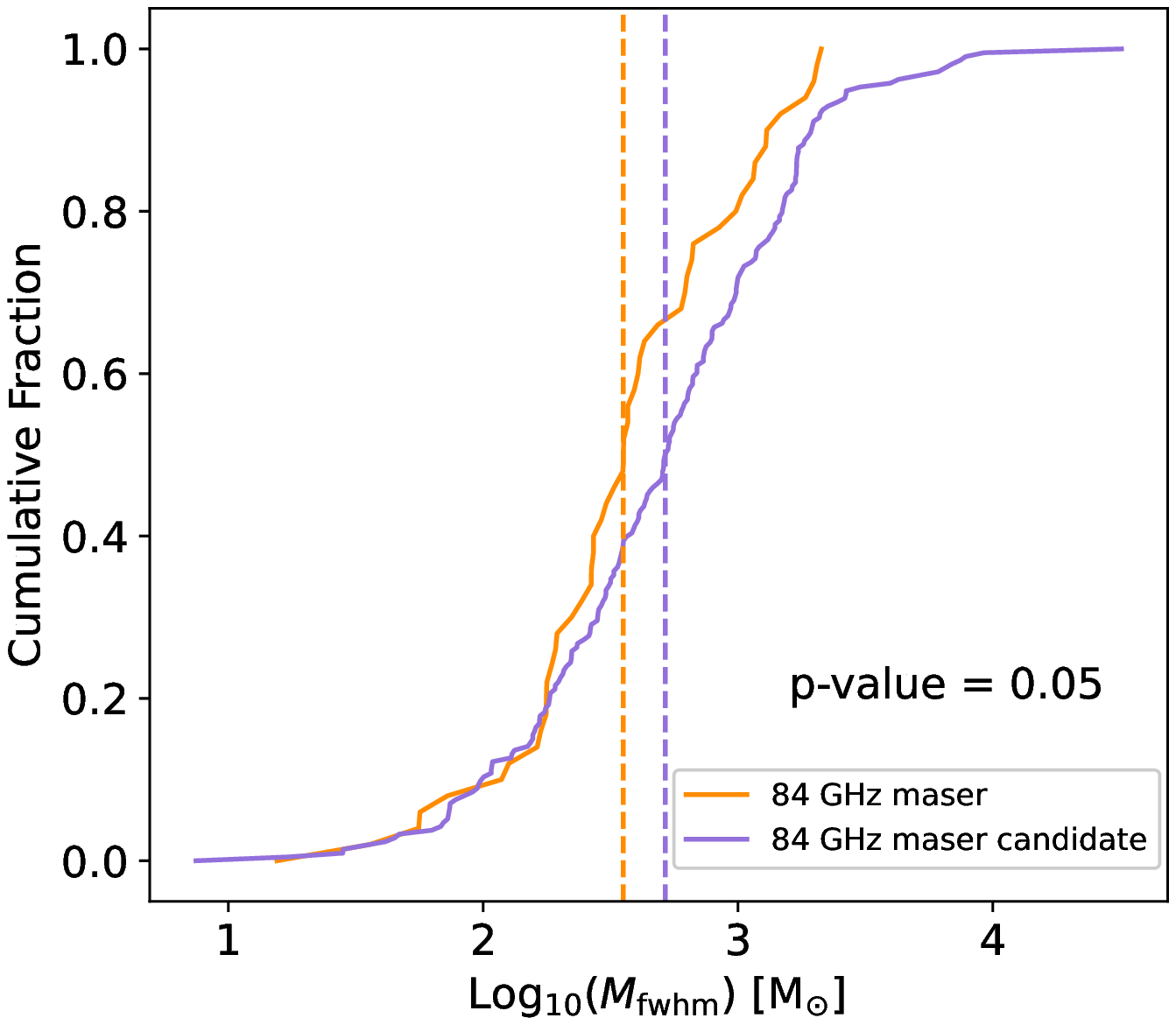}
\centerline{(b)}
\vspace{-3.5mm}
\end{minipage}
\begin{minipage}[b]{5.5cm}
\includegraphics[width=1\textwidth]{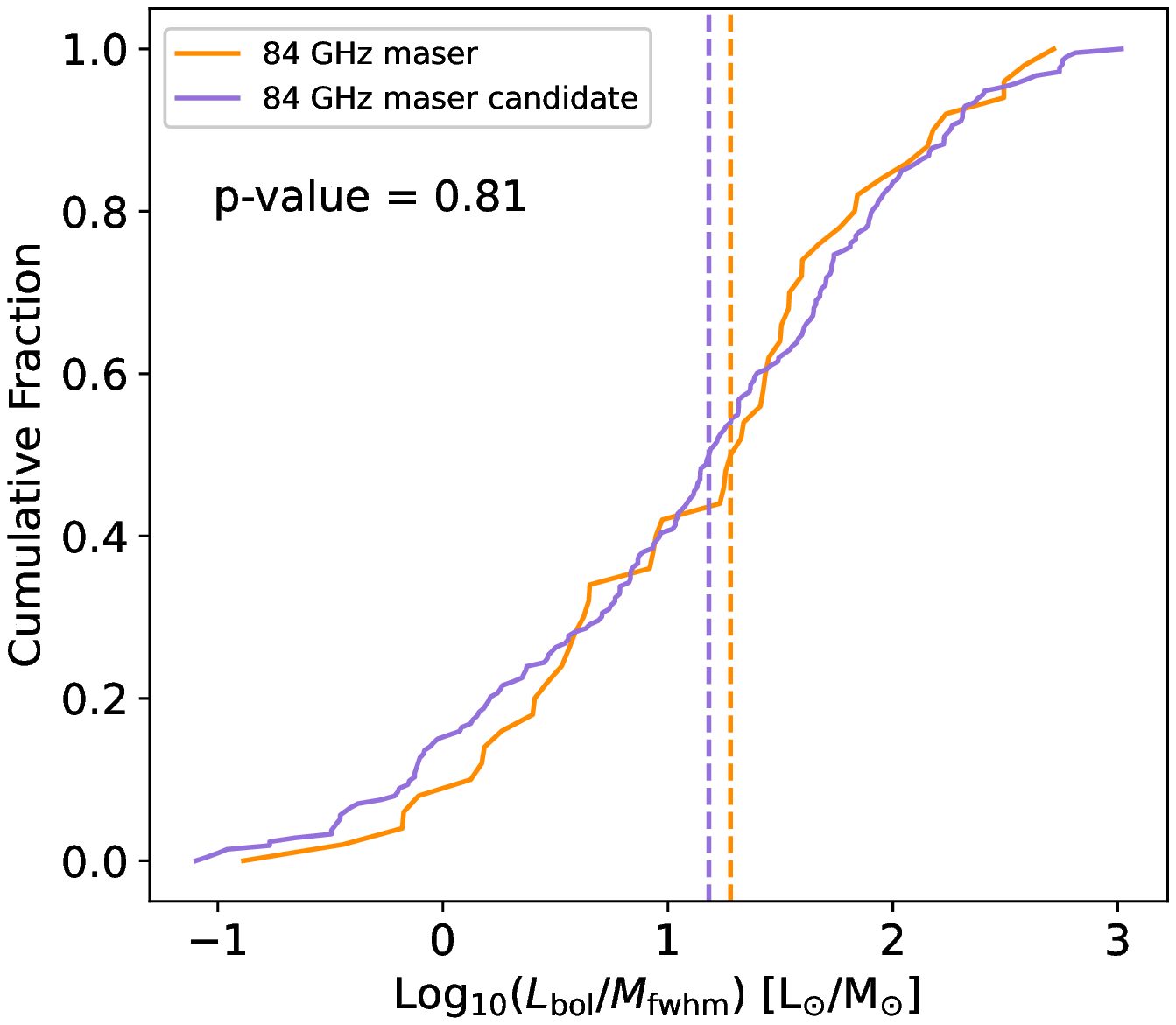}
\centerline{(c)}
\vspace{-3.5mm}
\end{minipage}
}
\mbox{
\begin{minipage}[b]{5.5cm}
\includegraphics[width=1\textwidth]{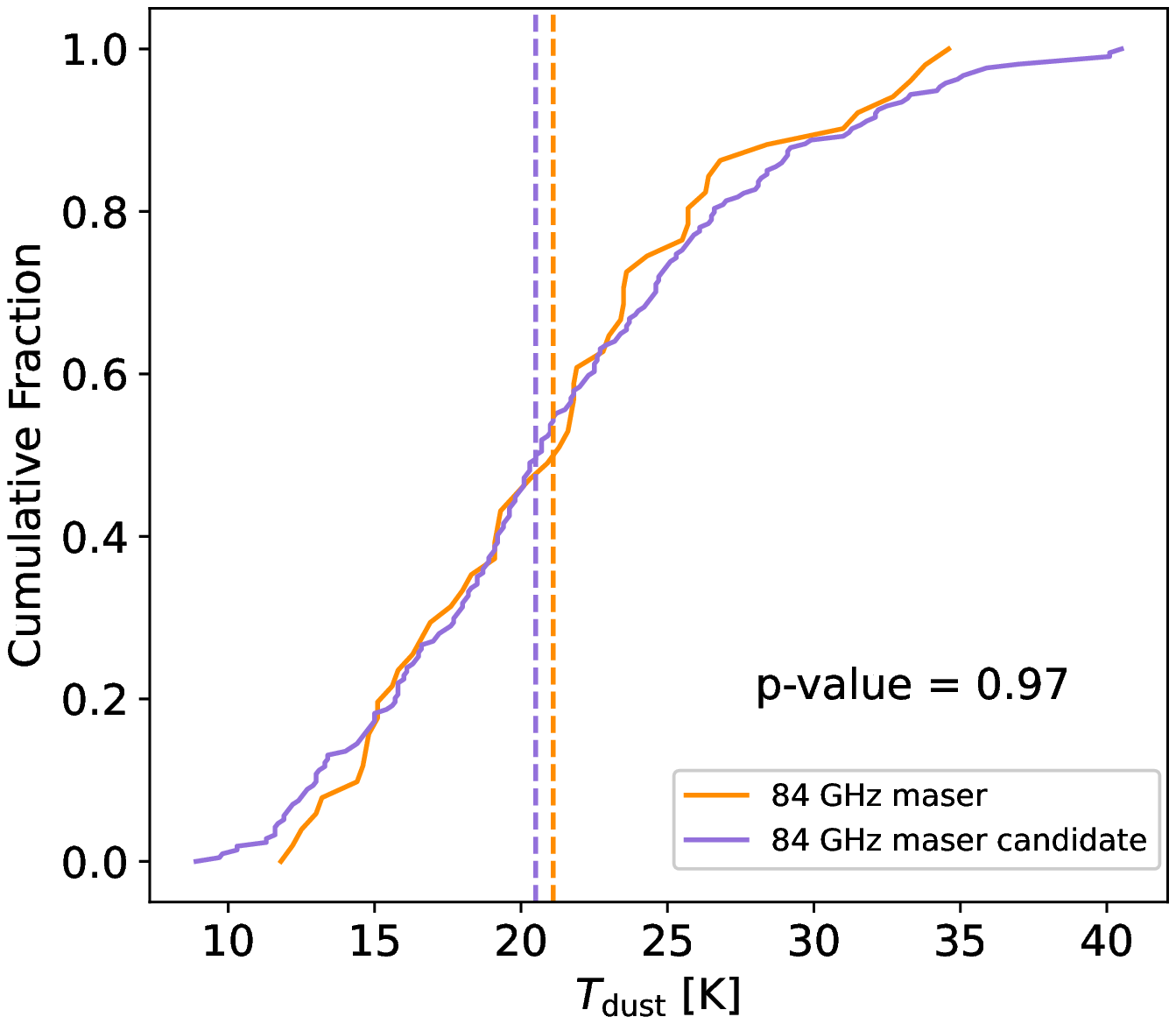}
\centerline{(d)}
\vspace{-2.5mm}
\end{minipage}
\begin{minipage}[b]{5.5cm}
\includegraphics[width=1\textwidth]{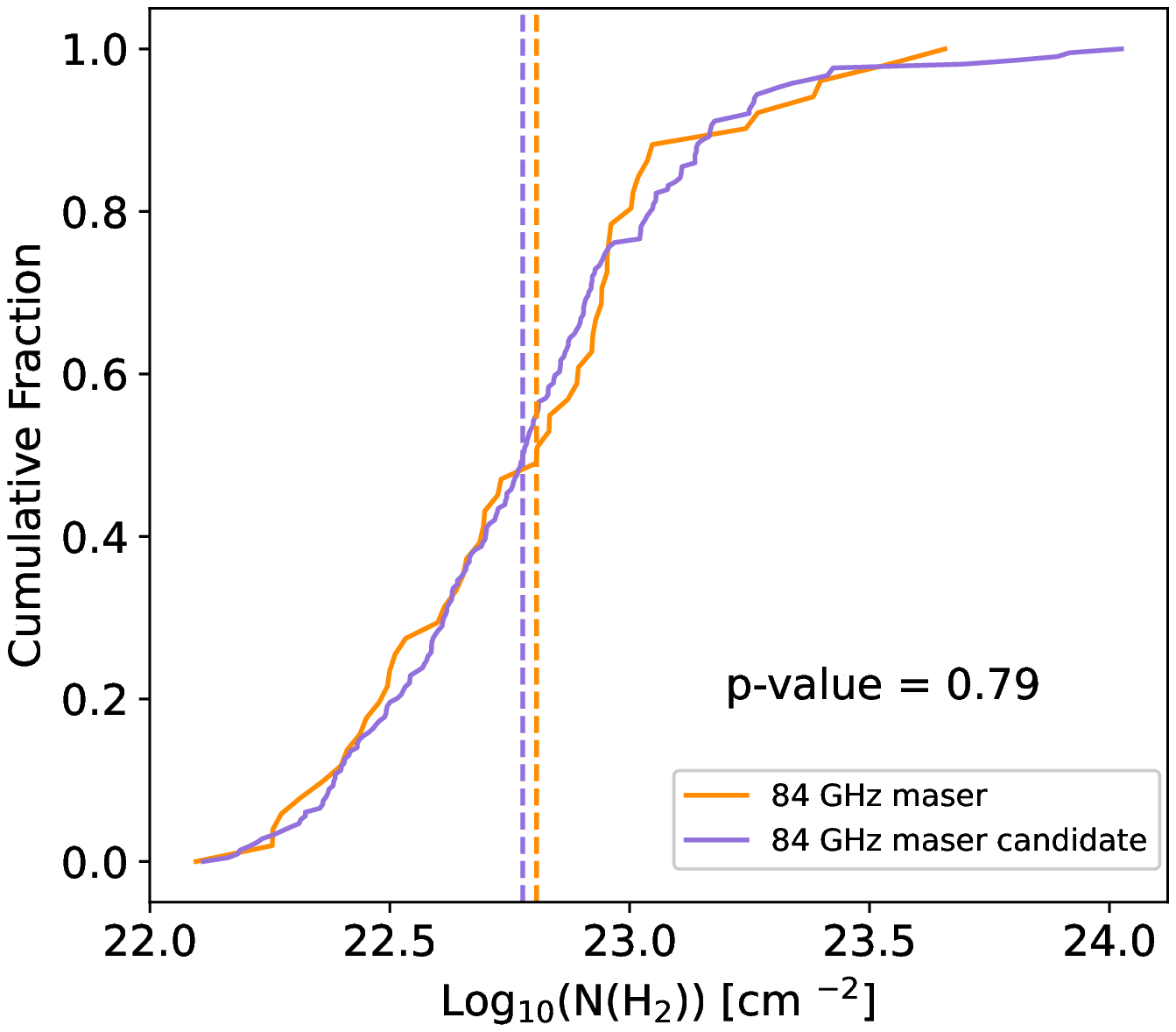}
\centerline{(e)}
\vspace{-2.5mm}
\end{minipage}
\begin{minipage}[b]{5.5cm}
\includegraphics[width=1\textwidth]{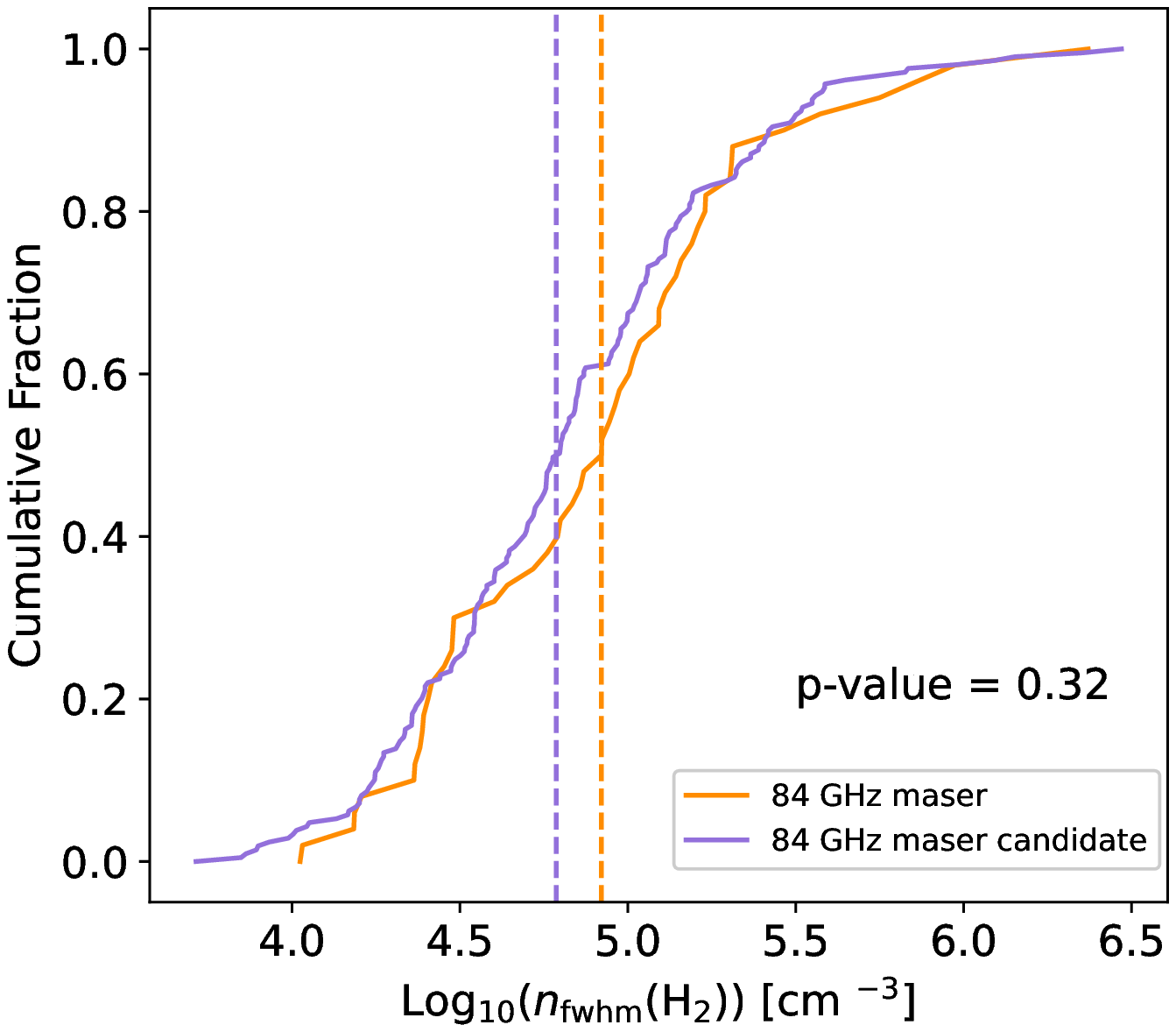}
\centerline{(f)}
\vspace{-3.5mm}
\end{minipage}
}
\caption{Cumulative distribution functions of the physical properties for the clumps with 84\,GHz methanol masers and maser candidates. 
From (a) to (f), cumulative distributions of the bolometric luminosity, the FWHM clump mass, the luminosity-to-mass ratio, the dust temperature, the peak H$_2$ column density, the mean H$_2$ FWHM volume density for 84 GHz methanol masers (orange lines) and maser candidates (purple lines), respectively. The vertical dashed lines in the corresponding colors depict the median values of the two samples. The p-values from the K-S tests are presented in each panel.
\label{fig:84maser-ks-atlasgal}}
\end{figure*}

\begin{figure*}[!htbp]
\centering
\mbox{
\begin{minipage}[b]{5.5cm}
\includegraphics[width=1\textwidth]{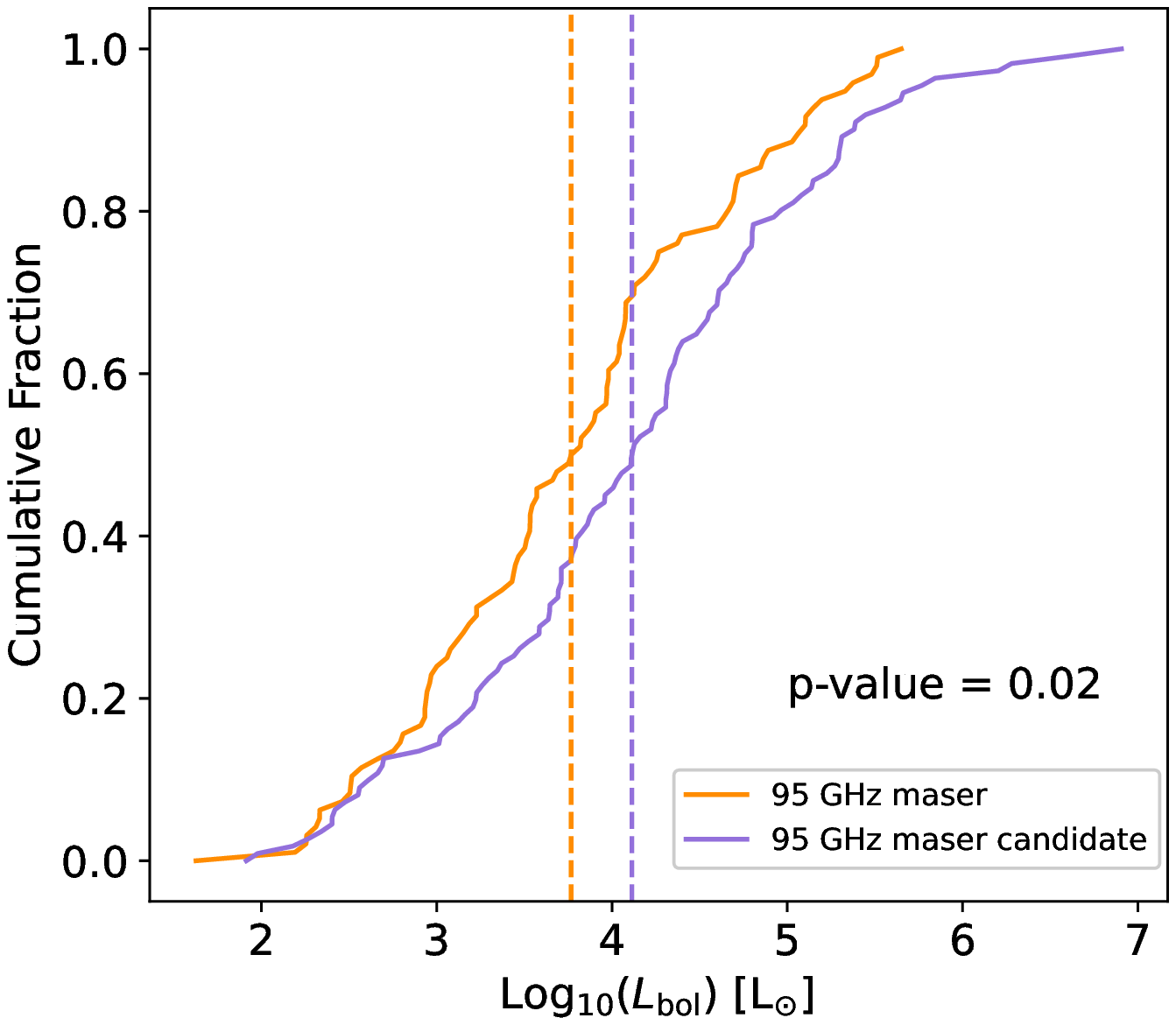}
\centerline{(a)}
\vspace{-3.5mm}
\end{minipage}
\begin{minipage}[b]{5.5cm}
\includegraphics[width=1\textwidth]{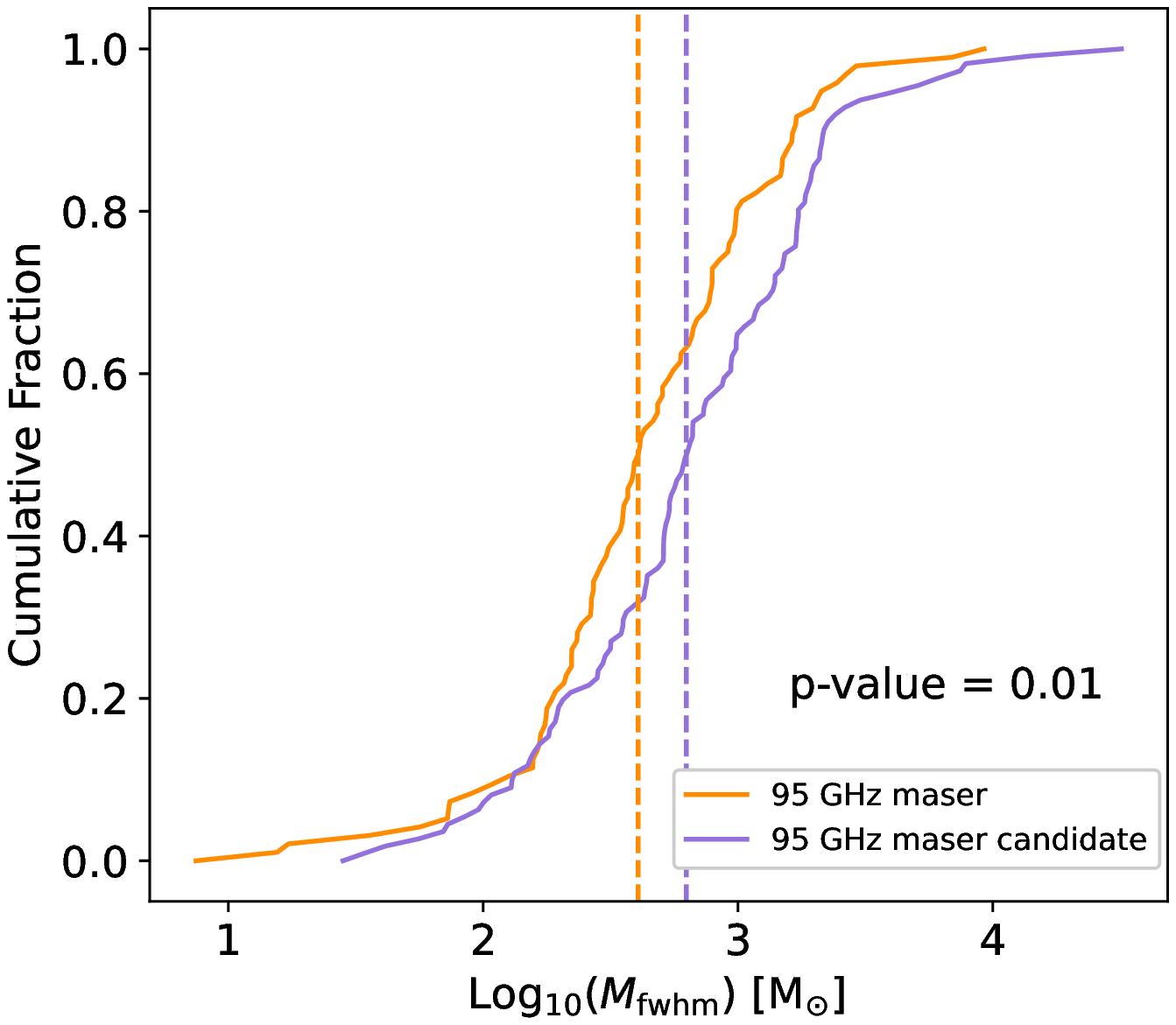}
\centerline{(b)}
\vspace{-3.5mm}
\end{minipage}
\begin{minipage}[b]{5.5cm}
\includegraphics[width=1\textwidth]{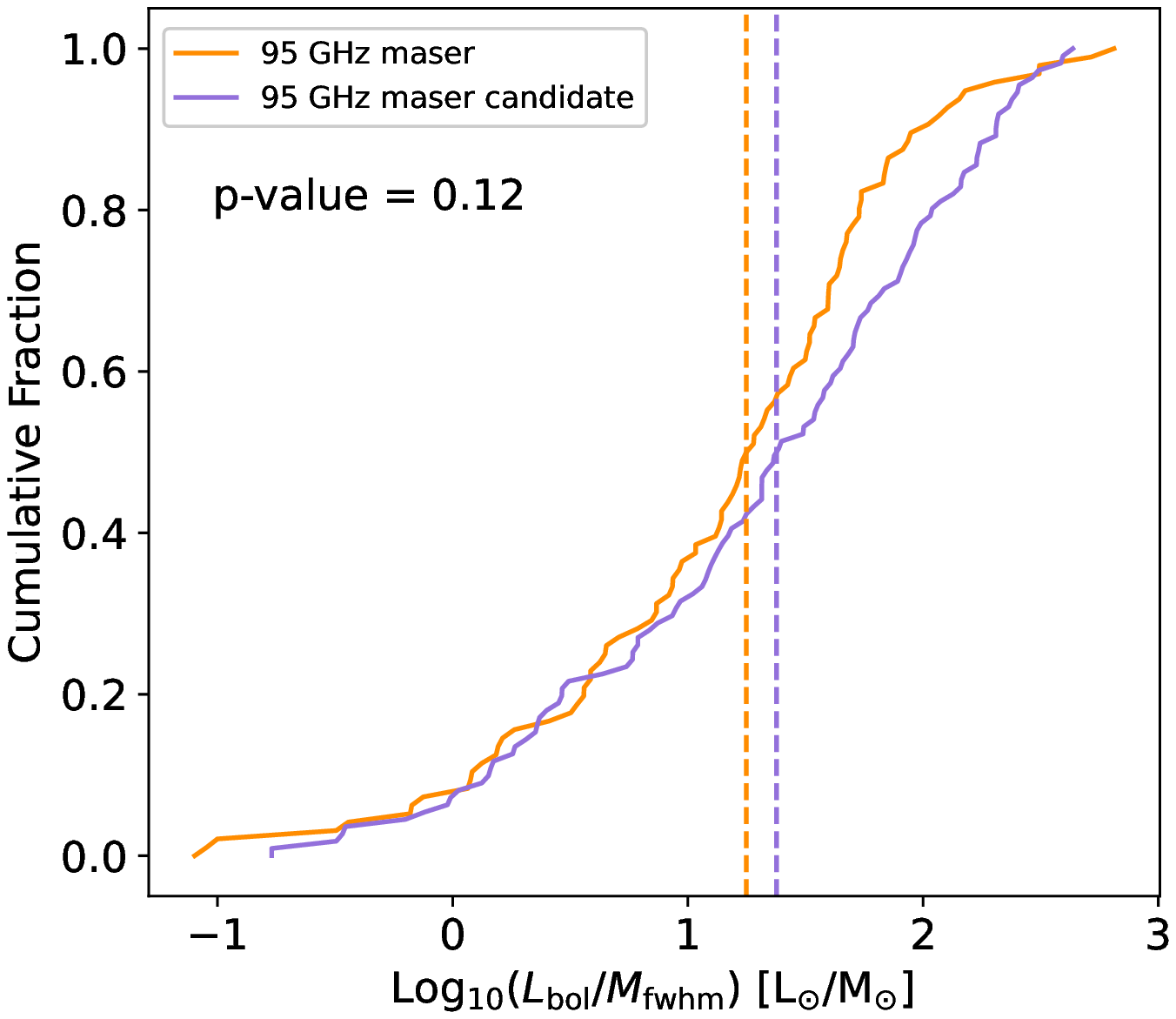}
\centerline{(c)}
\vspace{-3.5mm}
\end{minipage}
}
\mbox{
\begin{minipage}[b]{5.5cm}
\includegraphics[width=1\textwidth]{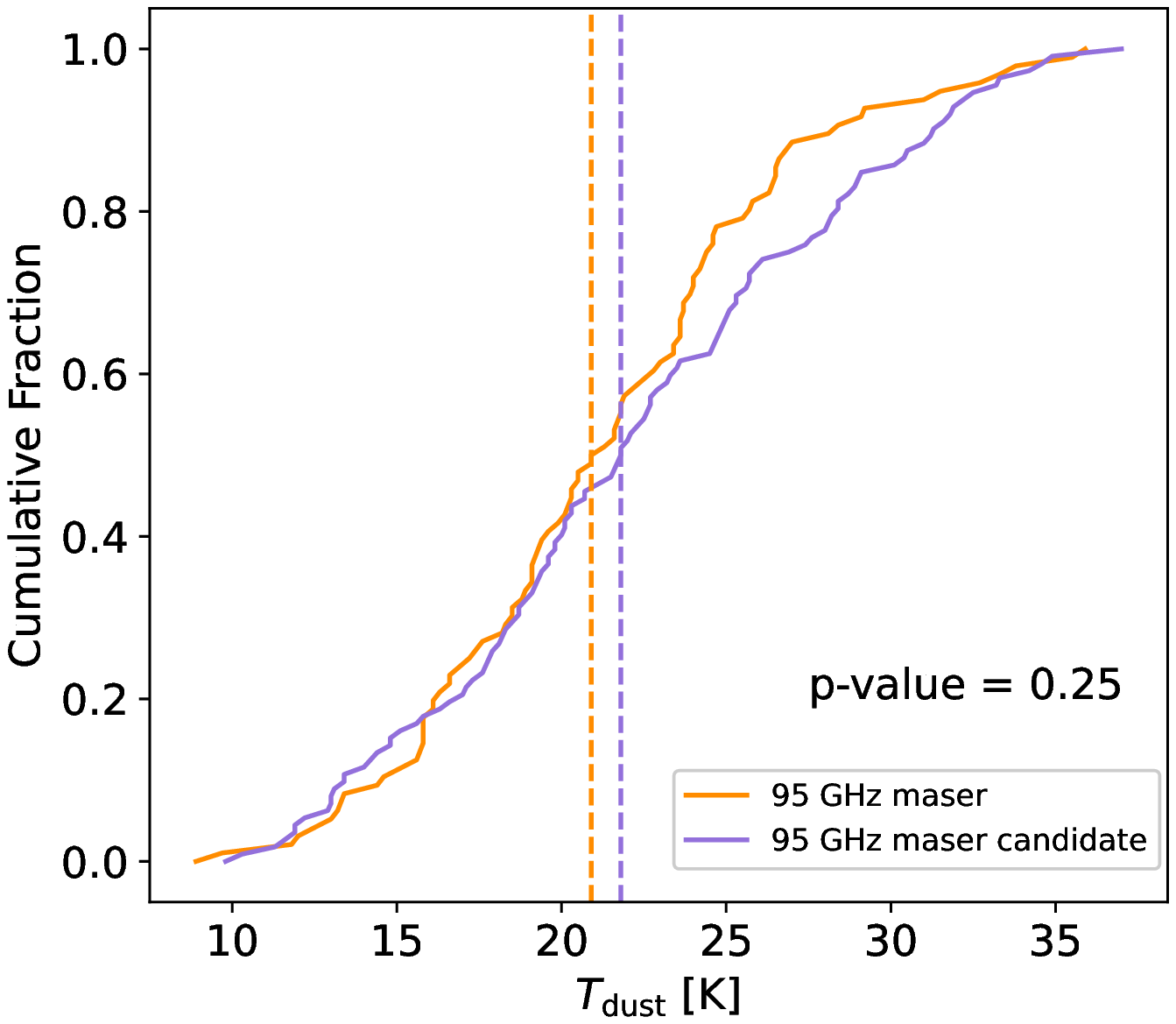}
\centerline{(d)}
\vspace{-3.5mm}
\end{minipage}
\begin{minipage}[b]{5.5cm}
\includegraphics[width=1\textwidth]{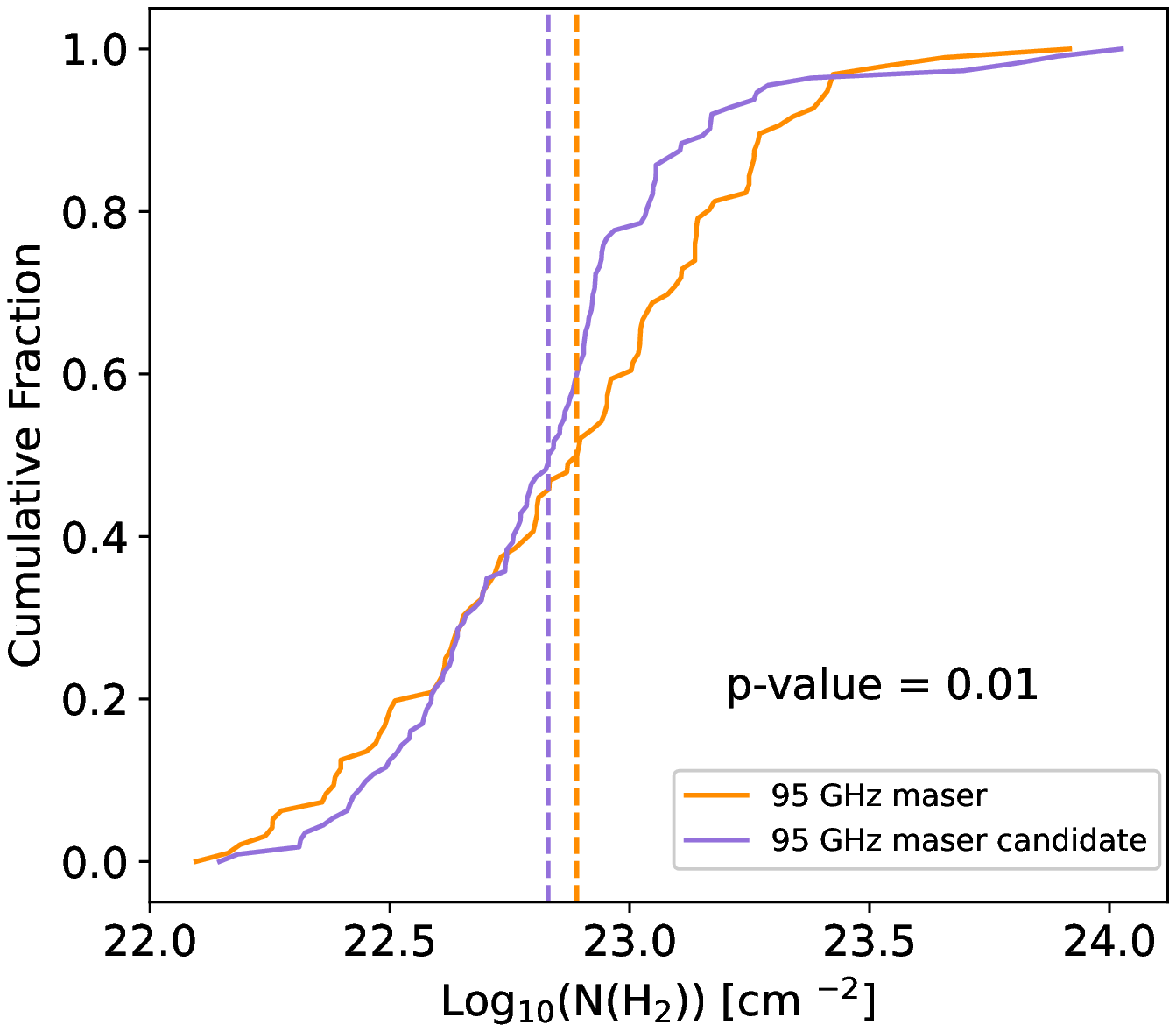}
\centerline{(e)}
\vspace{-3.5mm}
\end{minipage}
\begin{minipage}[b]{5.5cm}
\includegraphics[width=1\textwidth]{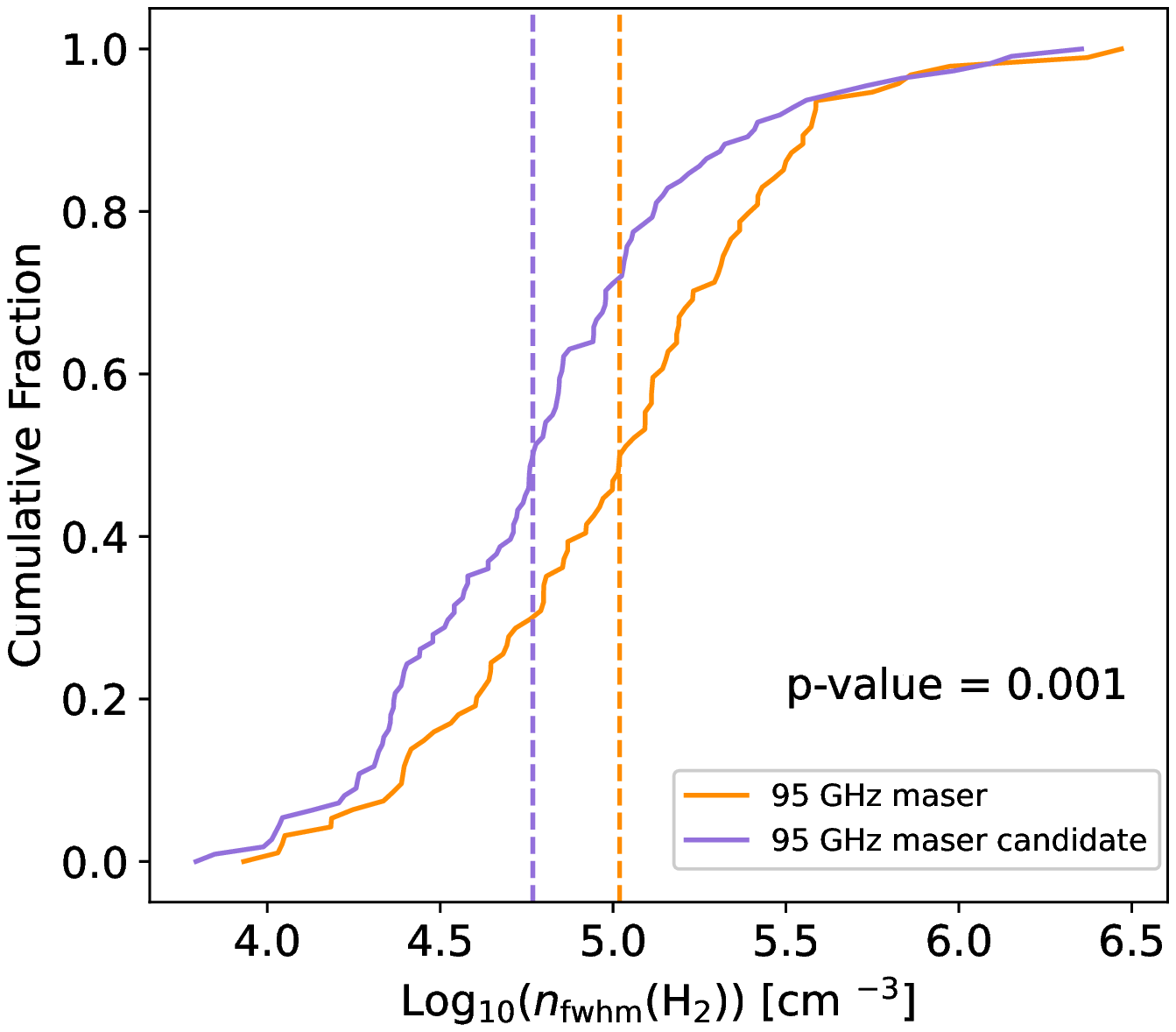}
\centerline{(f)}
\vspace{-3.5mm}
\end{minipage}
}
\caption{Cumulative distribution functions of the physical properties for the clumps with 95\,GHz methanol masers and maser candidates. 
Panels (a) to (f) are similar to the corresponding one in Fig.~\ref{fig:84maser-ks-atlasgal}.
\label{fig:95maser-ks-atlasgal}}
\end{figure*}

\section{Distribution of 95 and 104.3\,GHz methanol luminosity or integrated intensity against properties of ATLASGAL clumps.}\label{sec:appendix-c}

\begin{figure*}[!htbp]
\centering
\mbox{
\begin{minipage}[b]{5.5cm}
\includegraphics[width=1.03\textwidth]{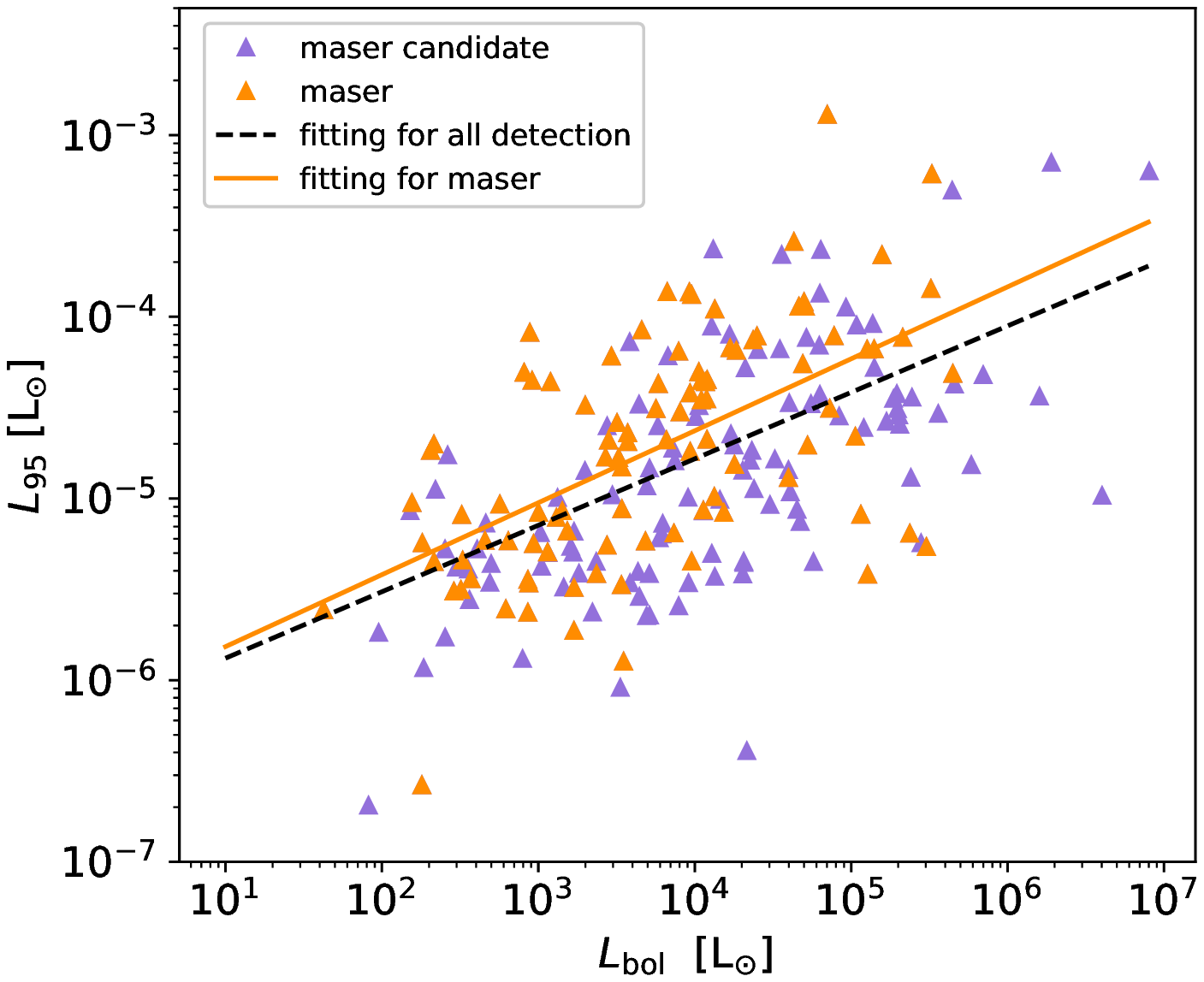}
\centerline{(a)}
\vspace{-3mm}
\end{minipage}
\begin{minipage}[b]{5.5cm}
\includegraphics[width=1.03\textwidth]{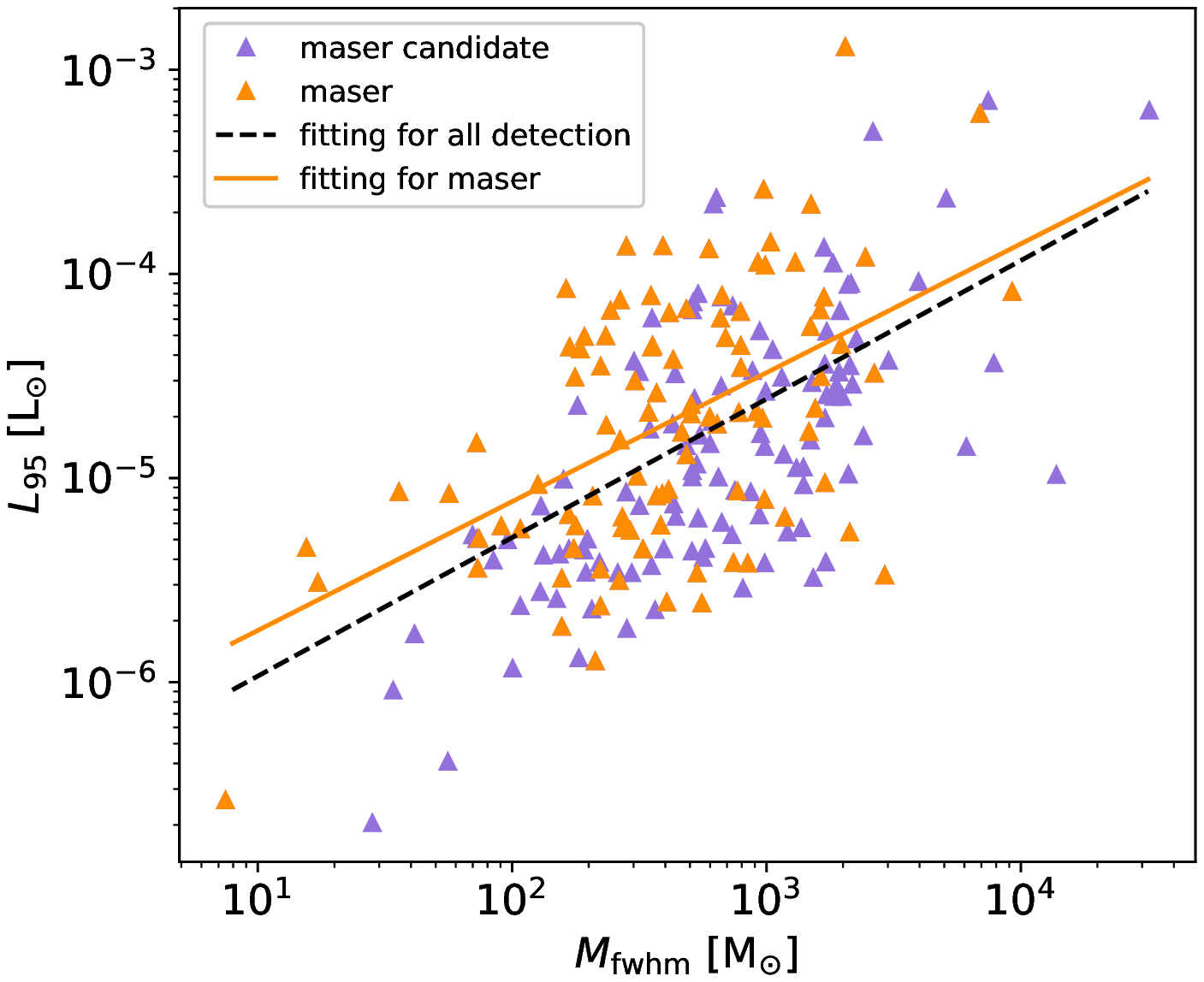}
\centerline{(b)}
\vspace{-3mm}
\end{minipage}
\begin{minipage}[b]{5.5cm}
\includegraphics[width=1.03\textwidth]{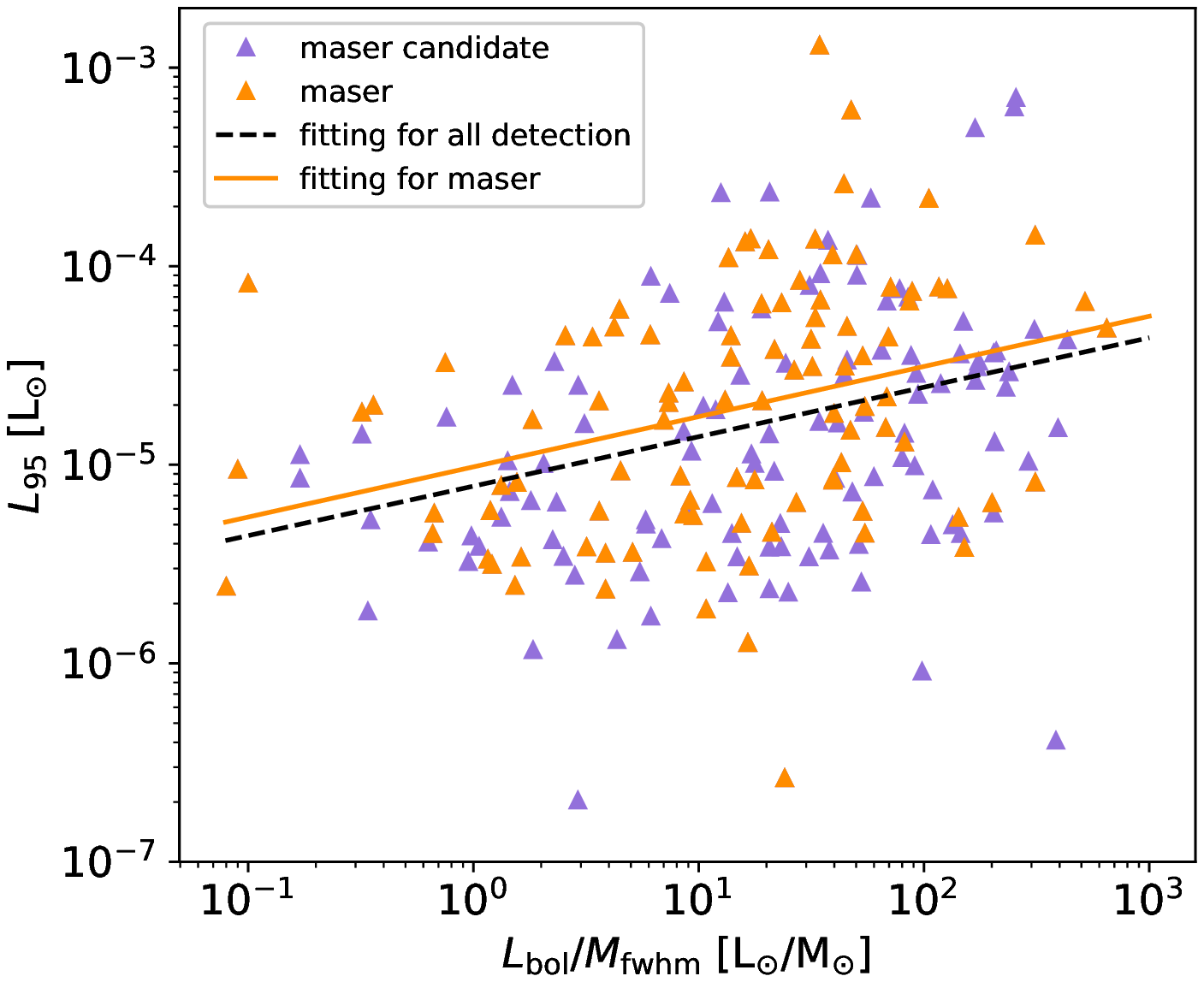}
\centerline{(c)}
\vspace{-3mm}
\end{minipage}
}
\mbox{
\begin{minipage}[b]{5.5cm}
\includegraphics[width=1.03\textwidth]{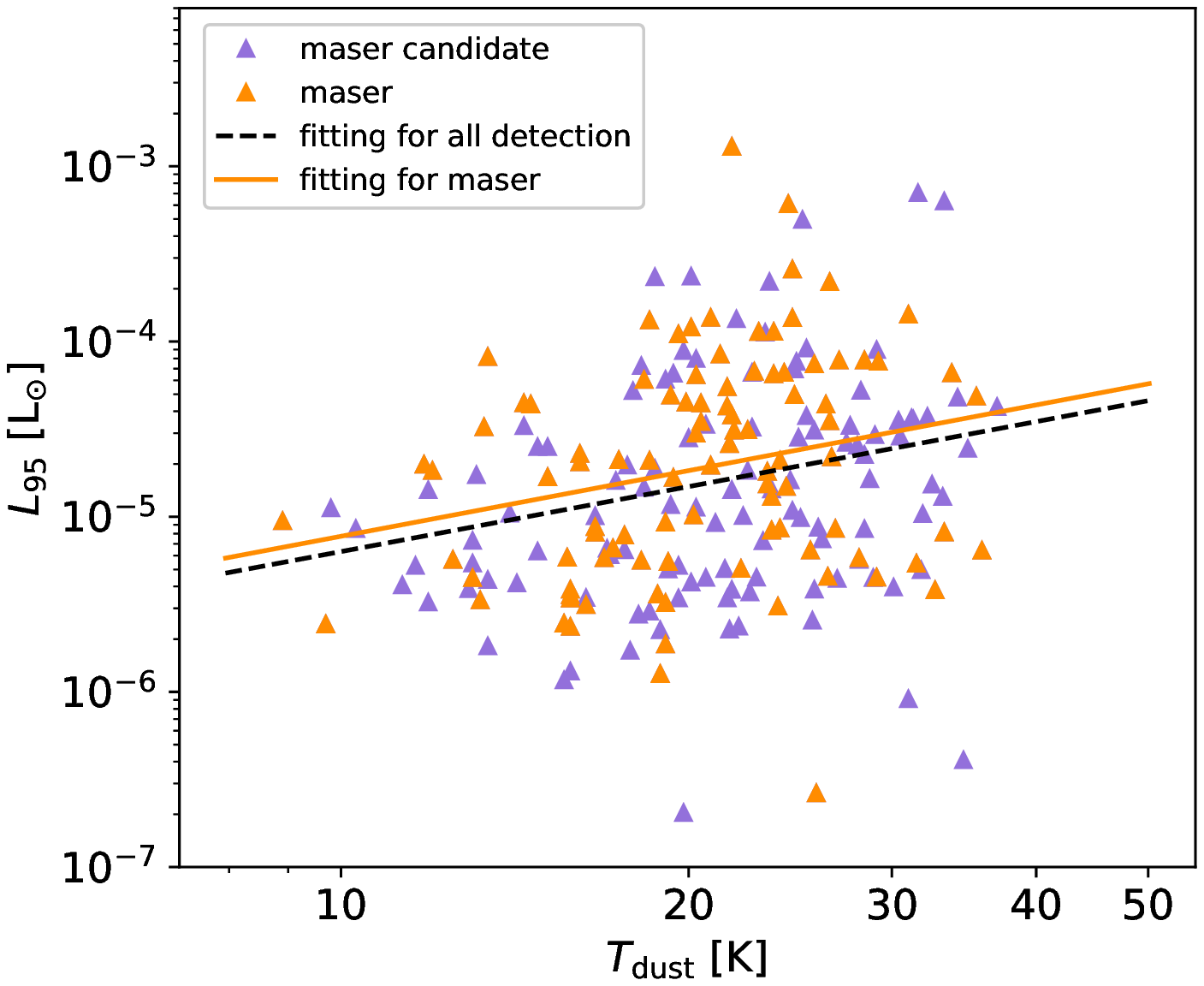}
\centerline{(d)}
\vspace{-3mm}
\end{minipage}
\begin{minipage}[b]{5.5cm}
\includegraphics[width=1.03\textwidth]{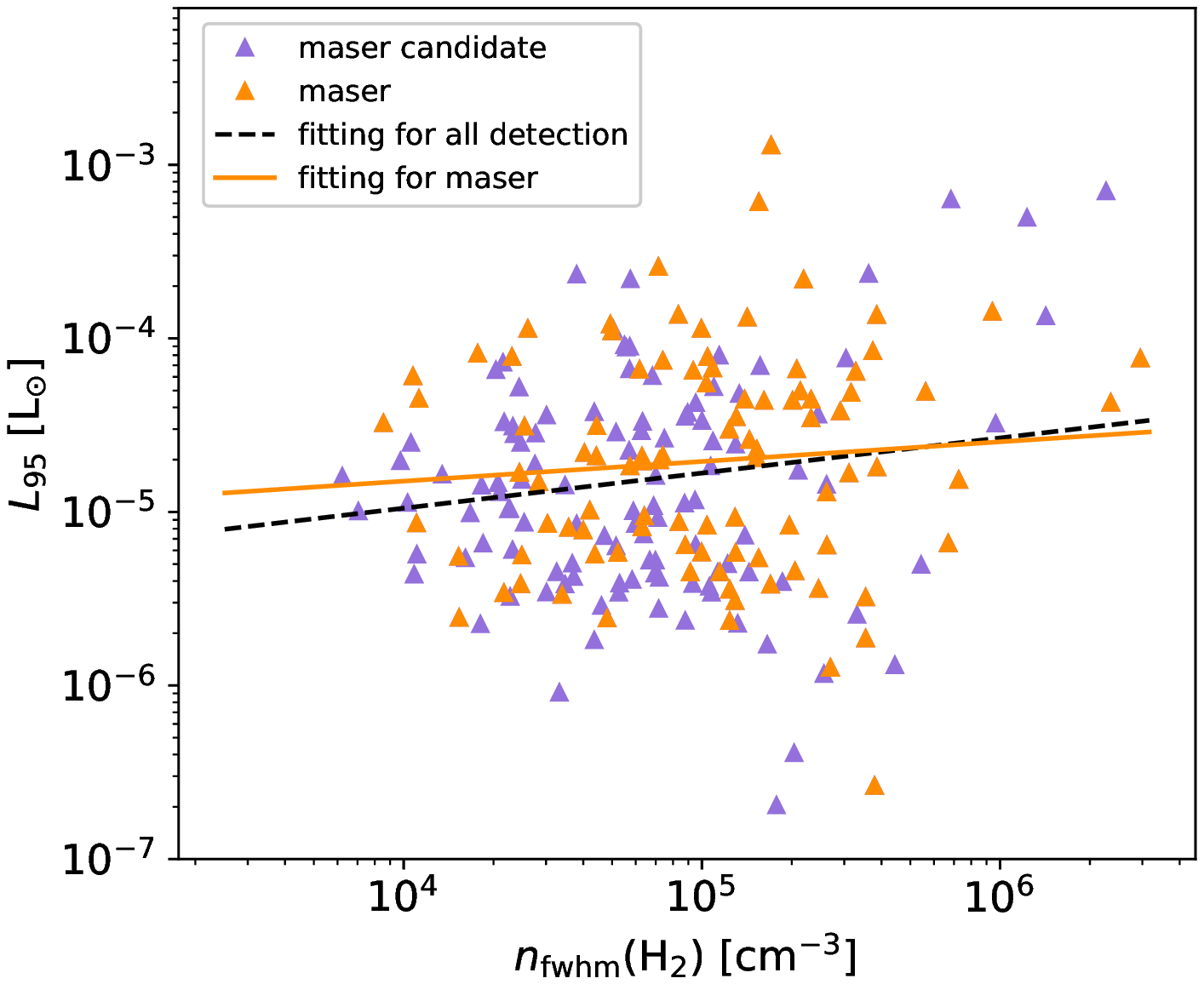}
\centerline{(e)}
\vspace{-3mm}
\end{minipage}
\begin{minipage}[b]{5.5cm}
\includegraphics[width=1.03\textwidth]{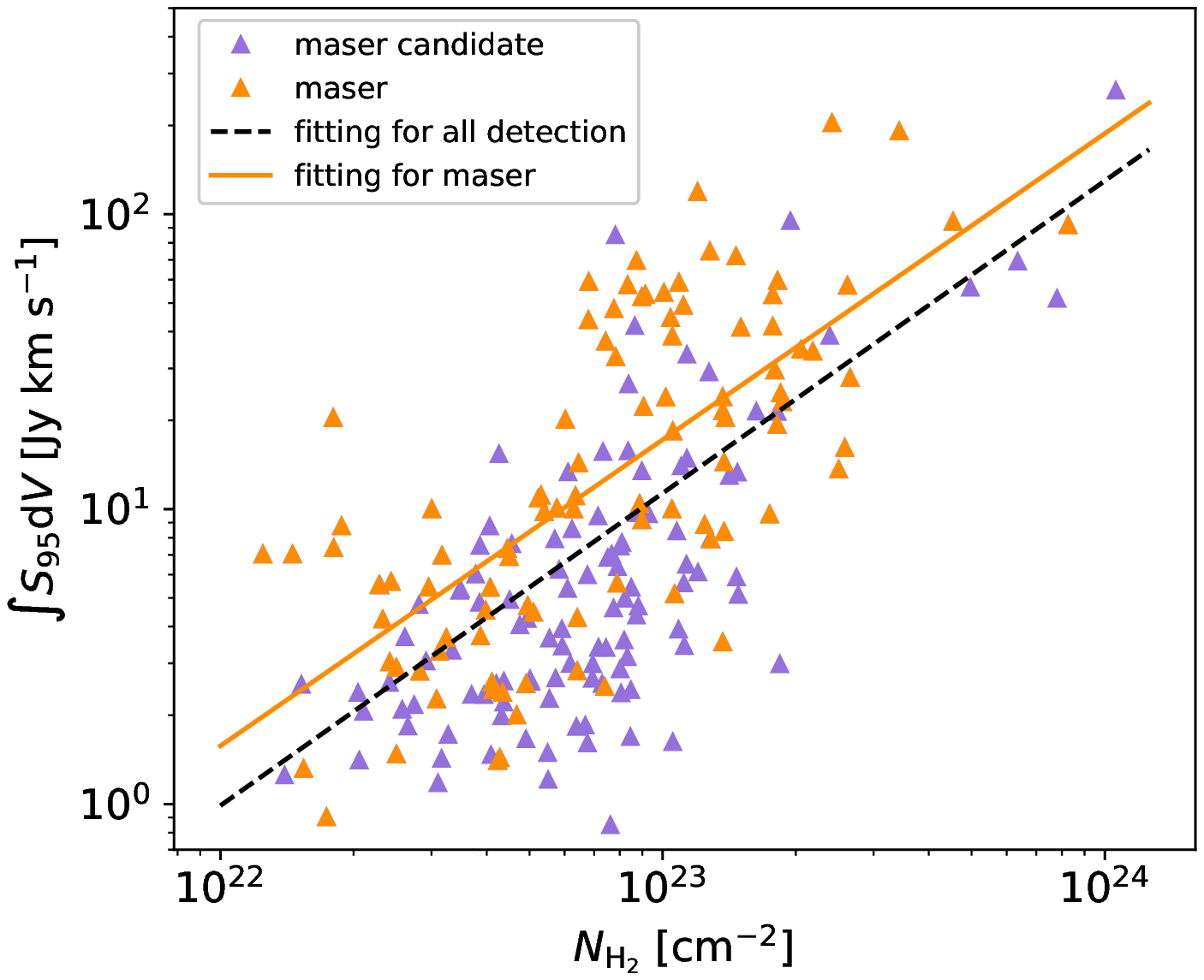}
\centerline{(f)}
\vspace{-3mm}
\end{minipage}
}
\caption{Distribution of 95\,GHz methanol luminosity or integrated intensity against properties of ATLASGAL clumps.
From panel (a) to (e), total methanol isotropic luminosities are plotted against the bolometric luminosity, the FWHM clump mass, the luminosity-to-mass ratio, the dust temperature and the mean H$_2$ FWHM volume density. In panel (f), total integrated intensities of the 95\,GHz methanol transition are plotted against the peak H$_2$ column density.
The orange and purple triangles represent the ATLASGAL clumps host maser and maser candidates, respectively.
The orange solid lines and black dashed lines depict the least-square fitting results for methanol maser and all methanol detections, respectively.
\label{fig:95-atlasgal}}
\end{figure*}

\begin{figure*}[!htbp]
\centering
\mbox{
\begin{minipage}[b]{5.5cm}
\includegraphics[width=1.03\textwidth]{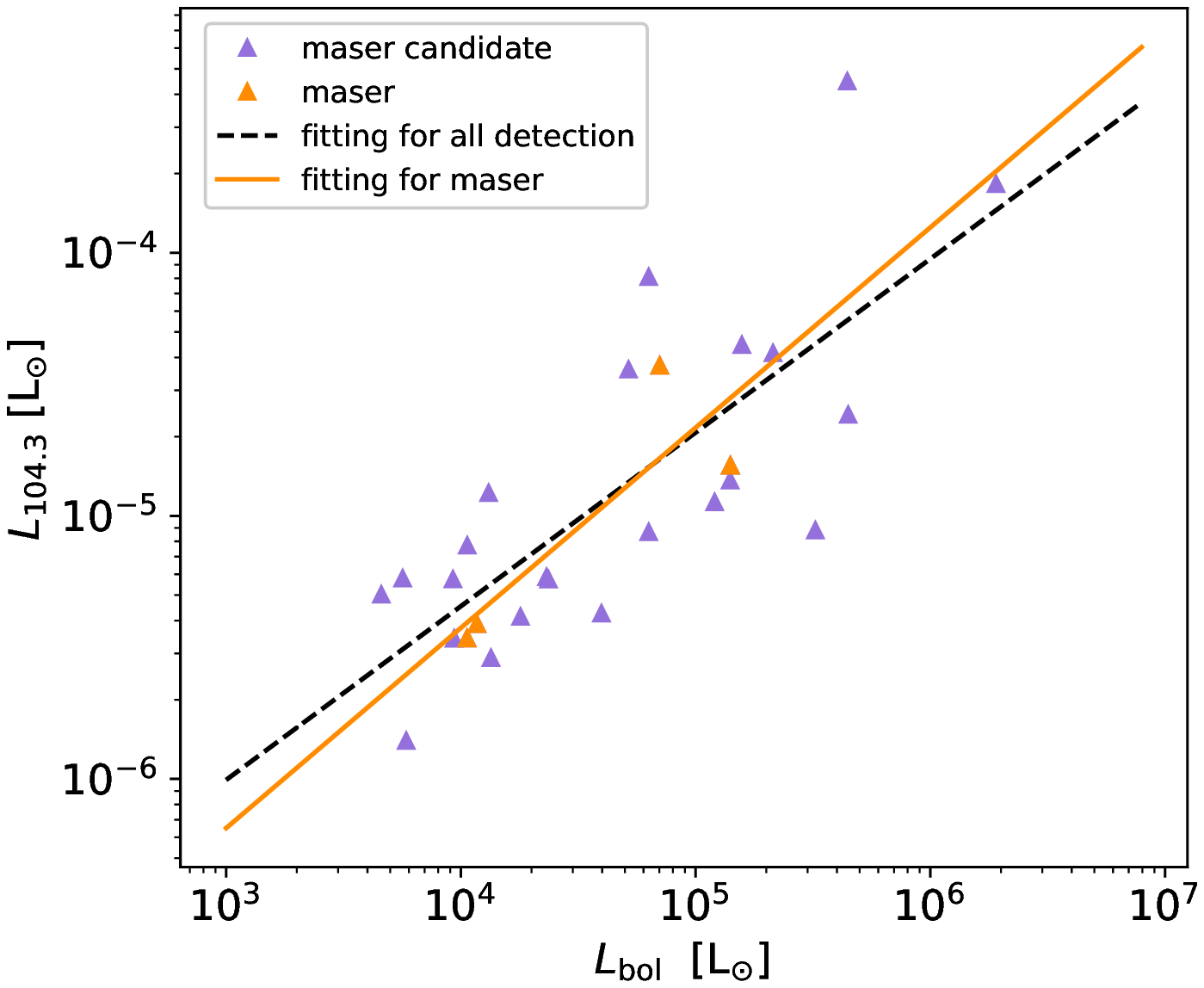}
\centerline{(a)}
\vspace{-3mm}
\end{minipage}
\begin{minipage}[b]{5.5cm}
\includegraphics[width=1.03\textwidth]{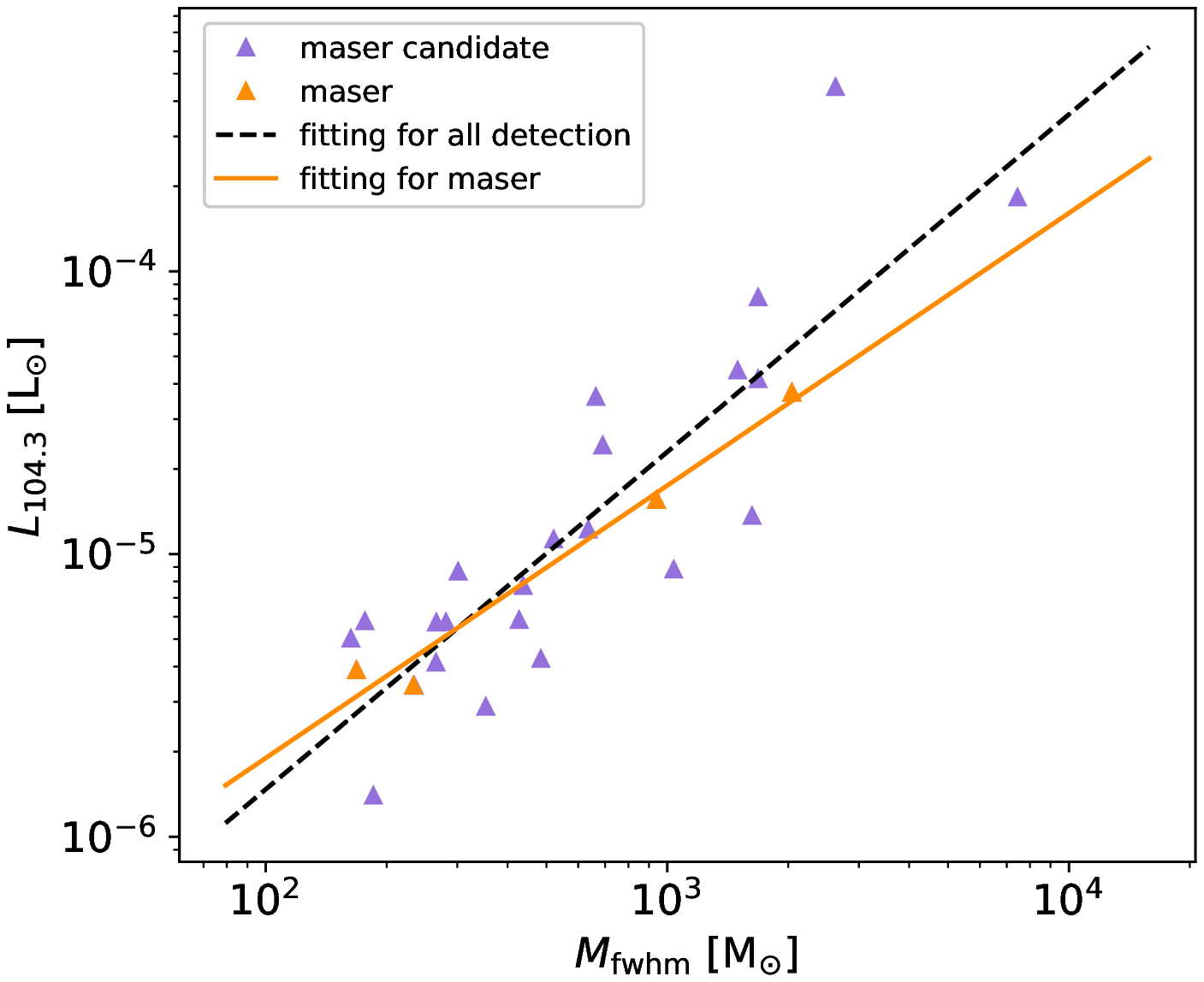}
\centerline{(b)}
\vspace{-3mm}
\end{minipage}
\begin{minipage}[b]{5.5cm}
\includegraphics[width=1.03\textwidth]{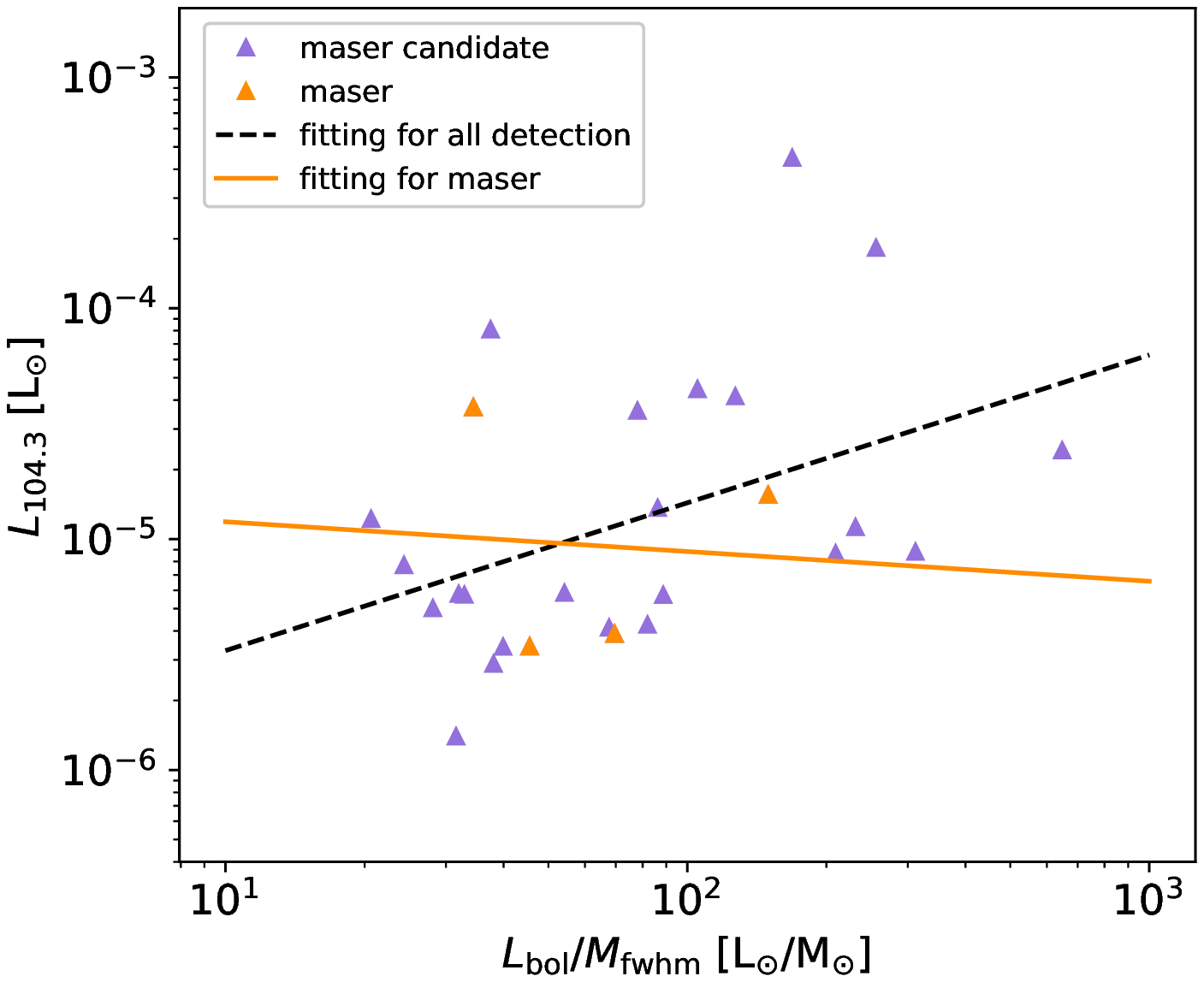}
\centerline{(c)}
\vspace{-3mm}
\end{minipage}
}
\mbox{
\begin{minipage}[b]{5.5cm}
\includegraphics[width=1.03\textwidth]{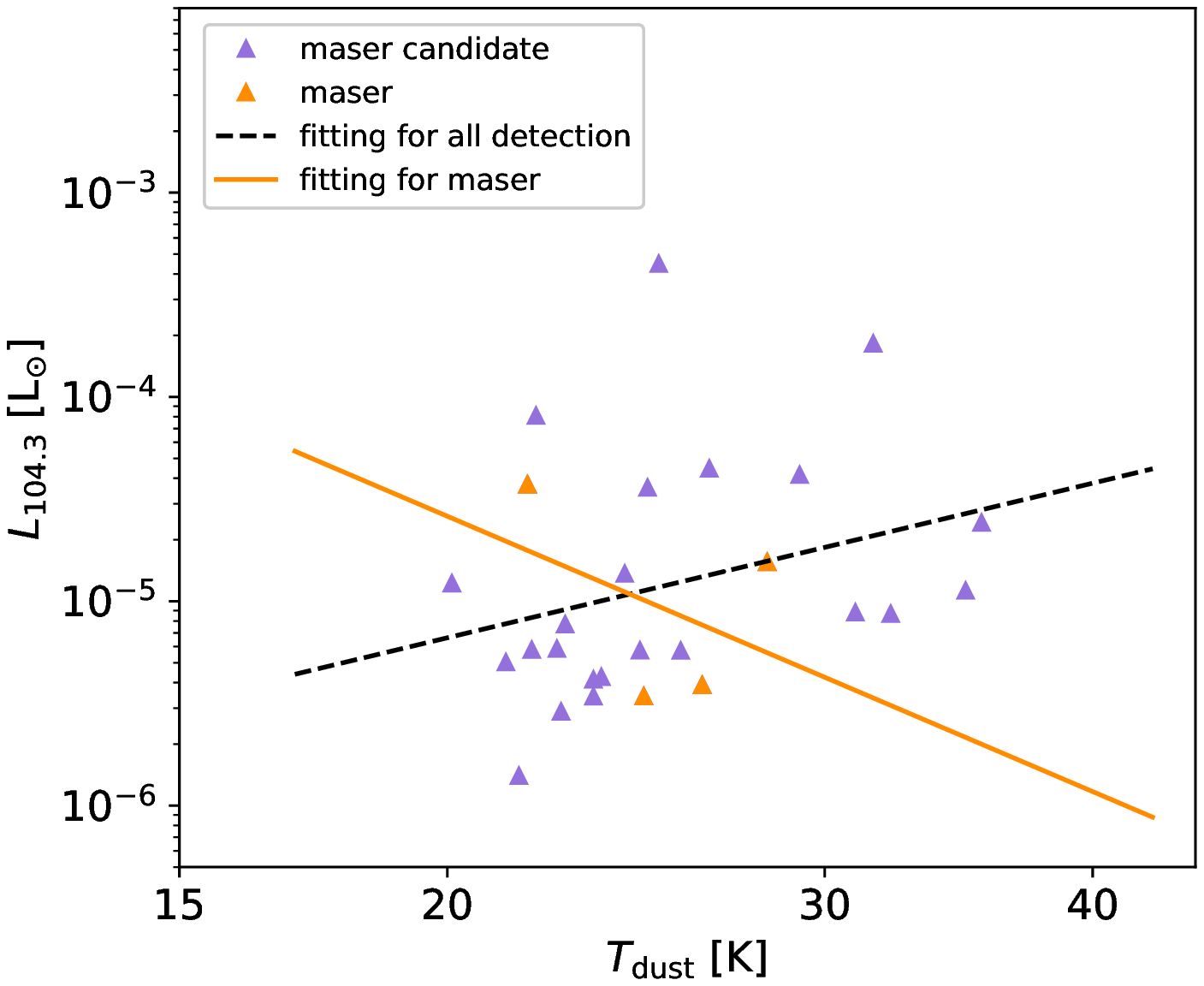}
\centerline{(d)}
\vspace{-3mm}
\end{minipage}
\begin{minipage}[b]{5.5cm}
\includegraphics[width=1.03\textwidth]{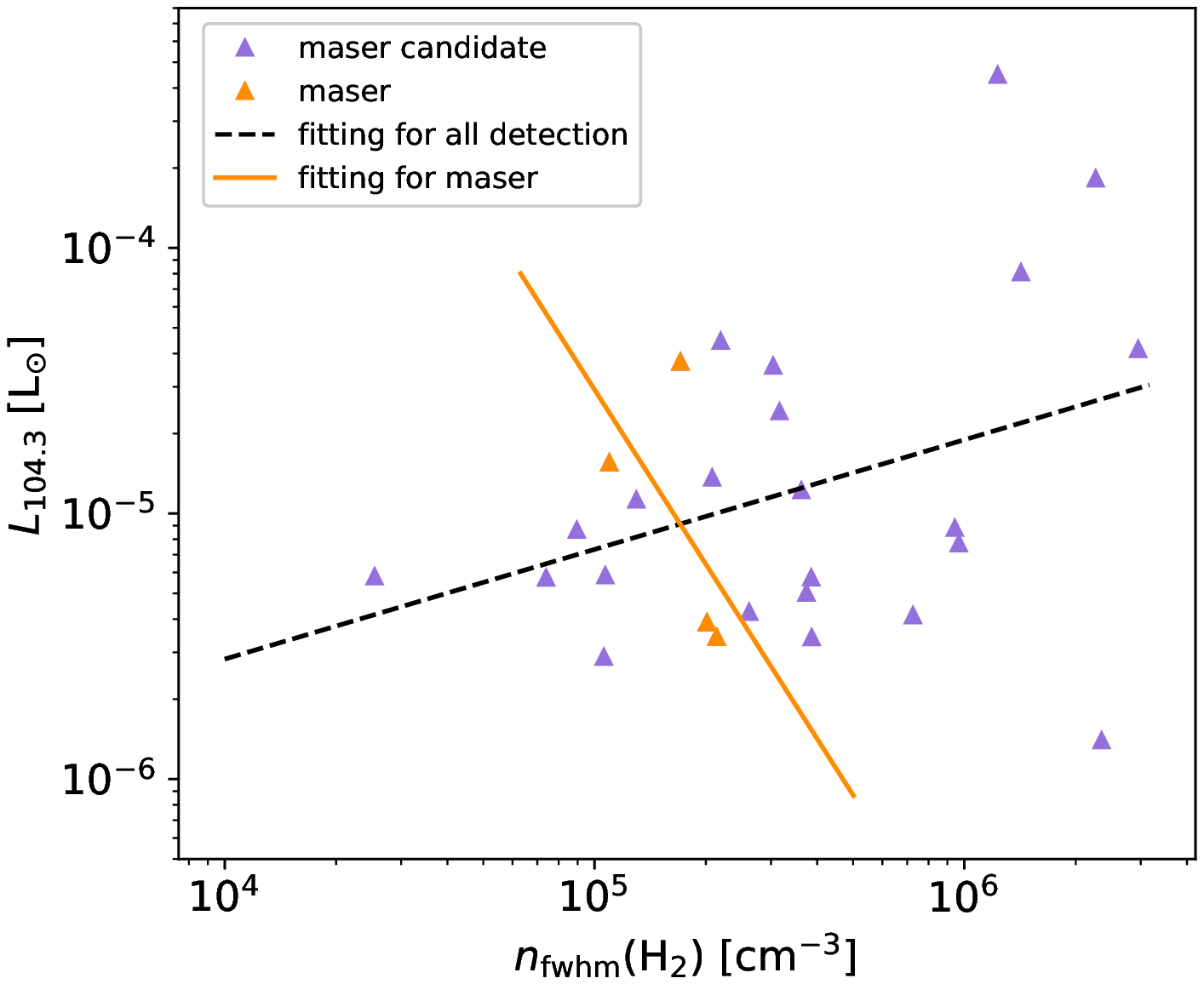}
\centerline{(e)}
\vspace{-3mm}
\end{minipage}
\begin{minipage}[b]{5.5cm}
\includegraphics[width=1.03\textwidth]{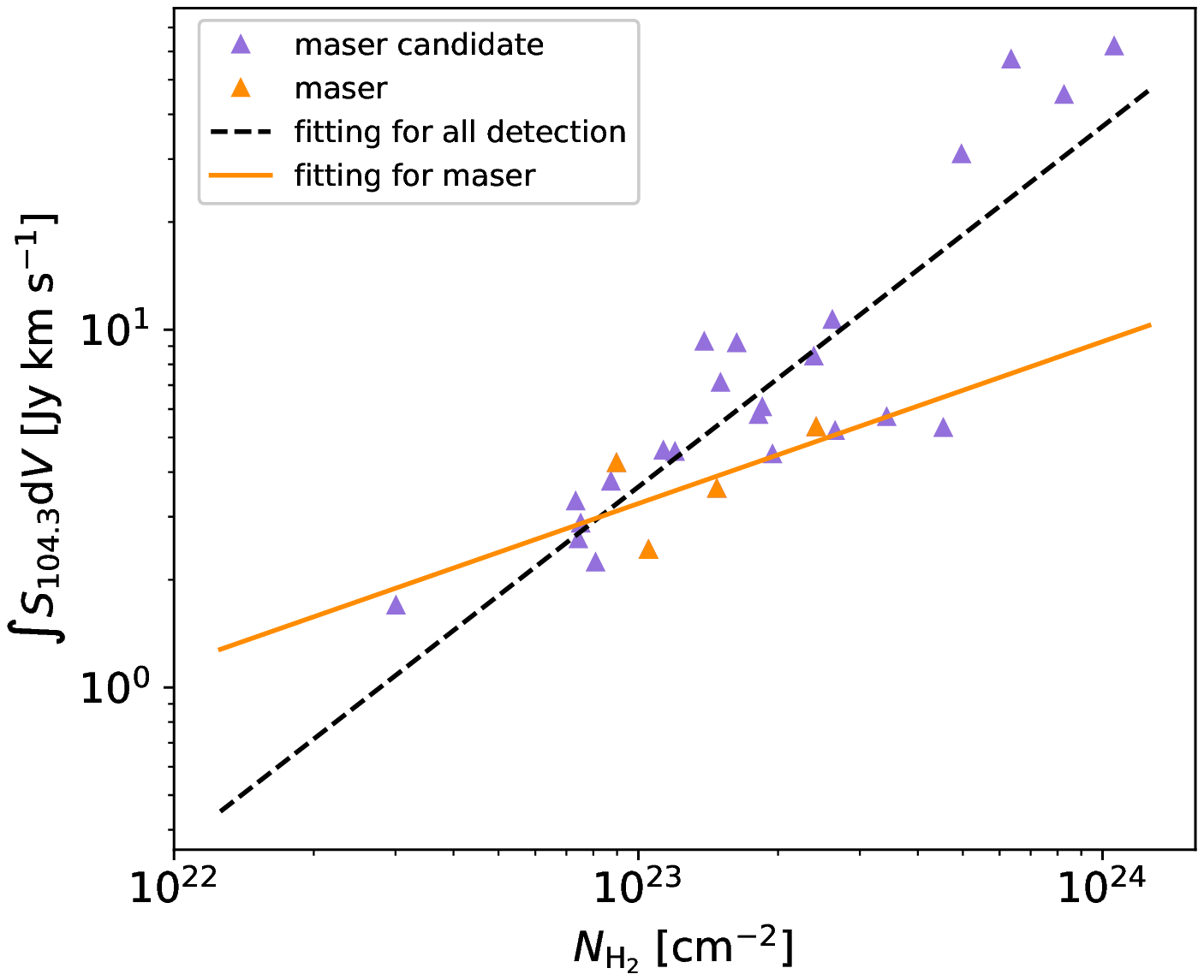}
\centerline{(f)}
\vspace{-3mm}
\end{minipage}
}
\caption{Distribution of 104.3\,GHz methanol luminosity or integrated intensity against properties of ATLASGAL clumps.
Panels (a) to (f) are similar to the corresponding one in Fig.~\ref{fig:95-atlasgal}.
\label{fig:104-atlasgal}}
\end{figure*}

\end{appendix}

\end{document}